\documentclass[conference]{IEEEtran}
\IEEEoverridecommandlockouts

\usepackage{cite}
\usepackage{amsmath,amssymb,amsfonts}
\usepackage{algorithmic}
\usepackage{graphicx}
\usepackage{textcomp}
\usepackage{xcolor}
\def\BibTeX{{\rm B\kern-.05em{\sc i\kern-.025em b}\kern-.08em
    T\kern-.1667em\lower.7ex\hbox{E}\kern-.125emX}}

\usepackage{booktabs} 
\usepackage{pifont}
\usepackage{color} 
\usepackage{balance} 
\usepackage{url} 
\usepackage[all]{nowidow}
\usepackage{tabularx}
\usepackage{siunitx} 
\usepackage{tabularx,subfigure,multirow, graphicx}
\usepackage{epsfig}
\usepackage{epstopdf} 
\usepackage{enumitem}
\usepackage[a-2b]{pdfx}

\usepackage{diagbox}
 
\usepackage{amsthm,amsmath}  

\usepackage{relsize}

\newcolumntype{L}[1]{>{\raggedright\arraybackslash}p{#1}}
\newcolumntype{C}[1]{>{\centering\arraybackslash}p{#1}}
\newcolumntype{R}[1]{>{\raggedleft\arraybackslash}p{#1}}

\usepackage[linesnumbered,ruled,vlined]{algorithm2e}
\usepackage{soul}

\soulregister\eqref7
\soulregister\ref7

\SetKwRepeat{Do}{do}{while}
\SetCommentSty{mycommfont}

\long\def\comment#1{}

\DeclareMathOperator*{\argmin}{arg\,min}

\newcommand{\nop}[1]{}

\newtheorem{theorem}{\bf Theorem}[section]
\newtheorem{lemma}{\bf Lemma}[section]

\newtheorem{example}{\bf Example}

\theoremstyle{remark}

\theoremstyle{definition}
\newtheorem{definition}{\bf Definition}

\newcommand{\mysplit}[1]{\\ \textcolor{red}{----------------------------------------------split----------------------------------------------}\\}

\newcommand{\CheckmarkBold}{\ding{51}}     

\newcommand{\entity}[1]{\mathcal{#1}}

\newcommand{\constvar}[1]{\mathbb{#1}}

\newcommand{\vectorfont}[1]{\boldsymbol{#1}}

\newcommand{\eqbold}[1]{\boldsymbol{\mathrm{#1}}}

\newcommand{\shrink}[1]{\hbox{#1}}
\newcommand{\shrinkem}[1]{\hbox{\em #1}}
\newcommand{\rbindexem}[1]{\hbox{\scriptsize\em #1}}
\newcommand{\rbindex}[1]{\hbox{\scriptsize #1}}

\newcommand{\rbindexsmallem}[1]{\hbox{\tiny\em #1}}
\newcommand{\rbindexsmall}[1]{\hbox{\tiny #1}}

\newcommand{\rbindexmuchsmallem}[1]{\hbox{\tiny\smaller\em #1}}
\newcommand{\rbindexmuchsmall}[1]{\hbox{\tiny\smaller #1}}

\newcommand{\scriptmath}[1]{\ensuremath{{\scriptstyle #1}}}

\usepackage{xparse}
\NewDocumentCommand{\simpleexpectation}{g}{%
	\mathbb{E}%
	\IfValueT{#1}{\left[#1\right]}%
}
\NewDocumentCommand{\simpleexpectationinner}{g}{%
	\mathbb{E}%
	\IfValueT{#1}{[#1]}%
}

\NewDocumentCommand{\var}{g}{%
	\operatorname{\shrink{Var}}%
	\IfValueT{#1}{\left[#1\right]}%
}
\NewDocumentCommand{\varinner}{g}{%
	\operatorname{\shrink{Var}}%
	\IfValueT{#1}{[#1]}%
}

\NewDocumentCommand{\varem}{g}{%
	\operatorname{\shrink{\em Var}}%
	\IfValueT{#1}{\left[#1\right]}%
}
\NewDocumentCommand{\varstar}{g}{%
	\operatorname{\shrink{Var}}^*%
	\IfValueT{#1}{\left[#1\right]}%
}
\NewDocumentCommand{\varsmpl}{g}{%
	\operatorname{\shrink{Var}}_{\rbindexem{smpl}}%
	\IfValueT{#1}{\left[#1\right]}%
}
\NewDocumentCommand{\varsmplinner}{g}{%
	\operatorname{\shrink{Var}}_{\rbindexem{smpl}}%
	\IfValueT{#1}{[#1]}%
}

\NewDocumentCommand{\varsmplstar}{g}{%
	\operatorname{\shrink{Var}}_{\rbindexem{smpl}}^*%
	\IfValueT{#1}{\left[#1\right]}%
}
\NewDocumentCommand{\varldp}{g}{%
	\operatorname{\shrink{Var}}_{\rbindexem{LDP}}%
	\IfValueT{#1}{\left[#1\right]}%
}
\newcommand{\simplevar}{\operatorname{V}}
\newcommand{\simplevarpldp}{\operatorname{V}_{\rbindexem{PFO}}}
\newcommand{\simplevarsmpl}{\operatorname{V}_{\rbindexem{smpl}}}

\newcommand{\simplevarldp}{\operatorname{V}_{\rbindexem{LDP}}}

\NewDocumentCommand{\varc}{g}{%
	\operatorname{\shrink{Varc}}%
	\IfValueT{#1}{\left[#1\right]}%
}
\NewDocumentCommand{\varcem}{g}{%
	\operatorname{\shrink{\em Varc}}%
	\IfValueT{#1}{\left[#1\right]}%
}

\NewDocumentCommand{\vara}{g}{%
	\operatorname{\overline{\shrink{Var}}}%
	\IfValueT{#1}{\left[#1\right]}%
}
\NewDocumentCommand{\varaem}{g}{%
	\operatorname{\overline{\shrinkem{Var}}}%
	\IfValueT{#1}{\left[#1\right]}%
}
\NewDocumentCommand{\varainner}{g}{%
	\operatorname{\overline{\shrink{Var}}}%
	\IfValueT{#1}{[#1]}%
}

\newcommand{\privacyDefinitionSimpleName}{$(\vectorfont{w},\vectorfont{\epsilon})$-EPLDP}
\newcommand{\privacyDefinitionTotalName}{$\vectorfont{w}$-Event $\vectorfont{\epsilon}$-Personalized Local Differential Privacy}
\newcommand{\privacyDefinitionOnlySimpleName}{EPLDP}

\newcommand{\pldpDefinitionSimpleName}{$\vectorfont{\epsilon}$-PLDP}
\newcommand{\pldpDefinitionTotalName}{$\vectorfont{\epsilon}$-Personalized Local Differential Privacy}
\newcommand{\pldpDefinitionOnlySimpleName}{PLDP}

\newcommand{\problemDefineSimpleName}{PPSDELS}
\newcommand{\problemDefineTotalName}{Personalized Private Streaming Data Estimation in Local Setting}

\newcommand{\specialUnbiasedMatrixBTotalName}{Empirical Fusion-Error Matrix}
\newcommand{\specialCorrectedMatrixCTotalName}{\basicClipping{}-Corrected Fusion-Error Matrix}

\newcommand{\basicSelection}{OPS}
\newcommand{\basicSelectionTotalName}{Optimal Population Selection}
\newcommand{\advancedSelection}{GA-OPS}
\newcommand{\advancedSelectionTotalName}{Group-Aware Optimal Population Selection}
\newcommand{\advancedEnhance}{BA-PUE}
\newcommand{\advancedEnhanceTotalName}{Bias-Aware Personalized Utility Enhancement}
\newcommand{\basicClipping}{PDEC}
\newcommand{\basicClippingTotalName}{Positive Definite Eigenvalue Clipping}

\newcommand{\advancedCorretedWeight}{PAA-CW}
\newcommand{\advancedCorretedWeightTotalName}{Partial-Absorption-Aware Corrected Weight}

\newcommand{\solutionA}{\ensuremath{\mathrm{PLPD}}}
\newcommand{\solutionATotalName}{Personalized Local Population Distribution}
\newcommand{\solutionB}{\ensuremath{\mathrm{PLPA}}}
\newcommand{\solutionBTotalName}{Personalized Local Population Absorption}

\newcommand{\solutionC}{\ensuremath{\mathrm{PLPD}^{+}}}

\newcommand{\solutionCTotalName}{Personalized Local Population Distribution Plus}
\newcommand{\solutionD}{\ensuremath{\mathrm{PLPA}^{+}}}
\newcommand{\solutionDTotalName}{Personalized Local Population Absorption Plus}
\newcommand{\solutionCMPC}{LPD}
\newcommand{\solutionCMPCTotalName}{Local Population Distribution}
\newcommand{\solutionCMPD}{LPA}
\newcommand{\solutionCMPDTotalName}{Local Population Absorption}

\newcommand{\solutionCMPG}{BAPUE-PLPD}
\newcommand{\solutionCMPH}{BAPUE-PLPA}
\newcommand{\solutionCMPI}{GAOPS-PLPD}
\newcommand{\solutionCMPJ}{GAOPS-PLPA}

\newcommand{\checkInDatasetName}{Foursquare}
\newcommand{\trajectoryDatasetName}{Taxi}
\newcommand{\tlnsDatasetName}{TLNS}
\newcommand{\sinDatasetName}{Sin}
\newcommand{\logDatasetName}{Log}

\newcommand{\outereqsize}[1]{{\scriptsize#1}}

\begin{document}

\title{PLDP-IDS: Personalized Local Differential Privacy for Infinite Data Stream
}

\author{
{Leilei Du{\small$~^{1}$}, Yikun Hu{\small$~^{1}$}, Xu Zhou{\small$~^{1}$}, Yang Cao{\small$~^{2}$}, Peng Cheng{\small$~^{3}$}, Wangze Ni{\small$~^{4}$}, Kenli Li{\small$~^{1}$}
}\\
\fontsize{10}{10}\itshape
$~^{1}$Hunan University, Changsha, China; 
$~^{2}$Institute of Science Tokyo, Tokyo, Japan; \\
$~^{3}$Tongji University, Shanghai, China;
$~^{4}$Zhejiang University, China
\fontsize{9}{9}\upshape\\
leileidu@hnu.edu.cn; yikunhu@hnu.edu.cn; zhxu@hnu.edu.cn; cao@c.titech.ac.jp; \\ 
cspcheng@tongji.edu.cn; niwangze@zju.edu.cn;
lkl@hnu.edu.cn
}

\maketitle

%

\begin{abstract}
	Existing $w$-event local differential privacy (LDP) mechanisms for streaming data impose a uniform privacy window and budget, forcing all users to adopt the most stringent privacy requirement.
	We study personalized continual frequency estimation under LDP, where each user specifies her own window size and privacy budget, and identify two challenges. First, a single
	sampling size cannot balance sampling and perturbation errors for privacy
	groups with heterogeneous requirements. Second, borrowing estimates from groups
	with larger budgets reduces variance but introduces bias when privacy
	preferences correlate with the data. We first propose Optimal Population
	Selection (OPS) and two OPS-guided mechanisms, PLPD and PLPA,
	which use one privacy-compatible sampling size for all users. Since one size cannot fit all groups and pooling their estimates is biased under privacy--data correlation, we further develop
	PLPD$^+$ and PLPA$^+$, which combine group-aware population scheduling with
	bias-aware cross-group fusion while preserving personalized $w$-event LDP.
	Experiments on real-world and synthetic streams show that, compared with the direct personalized extension of LDP-IDS, PLPA$^+$ reduces the average mean squared error (AMSE) by 52.3\% and 40.8\% on average, respectively, and remains 4.3--5.1$\times$ more accurate than PLPA under strong privacy--data correlation.
\end{abstract}

\begin{IEEEkeywords}
	local differential privacy, data stream, personalized differential privacy.
\end{IEEEkeywords}

\newcommand{\mycolor}{\color{black}}
\section{Introduction}
Mobile devices, wearable sensors, and location-based services continuously
generate user data streams that support real-time analysis such as traffic
monitoring, event detection, and mobility-aware recommendation~\cite{DBLP:conf/ccs/0001C0SC0LJ21,DBLP:journals/tkde/XueYHZW23}. A
fundamental task behind these applications is continual frequency estimation,
where a collector estimates the population-level distribution of user values
(e.g., the regions where drivers are located) at every time slot. Collecting
raw streams, however, can reveal sensitive individual patterns: consecutive
location reports may expose daily routines or repeated visits to hospitals,
religious facilities, or private residences. Local Differential Privacy
(LDP)~\cite{DBLP:conf/uss/WangBLJ17,DBLP:conf/stoc/BassilyS15} addresses this risk without a trusted curator by letting each
user perturb her data before reporting.


Unlike static data, consecutive reports from the same user are temporally
correlated, thus, protecting each event independently fails to hide patterns
spanning multiple slots, whereas protecting the entire stream exhausts the
privacy budget on infinite streams. $w$-event privacy~\cite{DBLP:journals/pvldb/KellarisPXP14} strikes a balance by
protecting any event sequence within a sliding window of length $w$, and
LDP-IDS~\cite{DBLP:conf/sigmod/RenSYYZX22} extends it to the local setting. Existing $w$-event LDP mechanisms,
however, impose a uniform window size and privacy budget on all users. In
practice, users differ in how long their behaviors remain correlated and in how
sensitive their data are, and thus naturally demand different
$(w_i,\epsilon_i)$. A uniform mechanism has to adopt the most stringent
requirement for everyone, as the following car-hailing example shows.

%

\begin{figure}[ht!]
  \centering
  \includegraphics[width=0.5\textwidth]{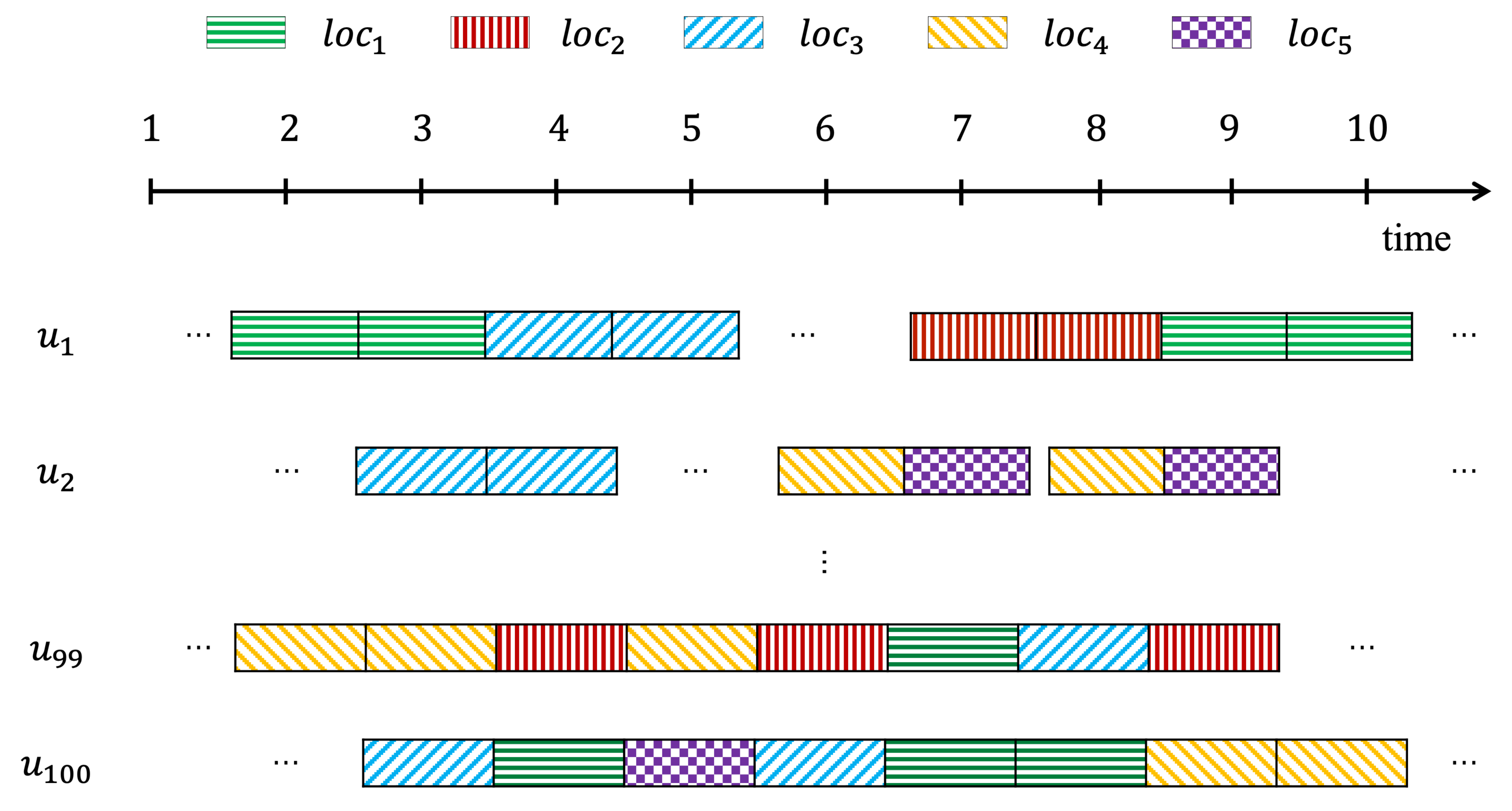}
  \caption{An example for personalized $w$-event}\label{fig:introduction_example}
\end{figure}

\begin{example}\label{exm:car-hailing}
	As illustrated in Figure~\ref{fig:introduction_example}, consider $n=10{,}000$ drivers over $d=5$ regions, where $9{,}800$ drivers require protection within $w=4$ window length and the remaining $200$ require $w=8$. A conventional uniform $w$-event LDP mechanism has to set $w=8$ for all users. With privacy budget $\epsilon=1$, its standard error is $\scriptmath{\shrink{SE} = \sqrt{\var_{\rbindexsmall{FO}}\left[\frac{\epsilon}{w}, n\right]/n^2} = \sqrt{\frac{d-2+e^{\epsilon/w}}{n(e^{\epsilon/w}-1)^2}} \approx 0.153}$, where $\shrink{Var}_{\rbindexsmall{FO}}[\epsilon,n]=n\cdot\frac{d-2+e^{\epsilon}}{(e^{\epsilon}-1)^2}$~\cite{DBLP:conf/uss/WangBLJ17}. However, if users are divided according to their personalized windows, using $w_1=4$ for the first group and $w_2=8$ for the second group, the standard error decreases to approximately $\scriptmath{\shrink{SE} = \sqrt{\left(\var_{\rbindexsmall{FO}}\left[\frac{\epsilon}{w_1}, n_1\right]+\var_{\rbindexsmall{FO}}\left[\frac{\epsilon}{w_2}, n_2\right]\right)/n^2} \approx 0.075}$. This example shows that a worst-case uniform window can cause substantial utility loss, motivating personalized continual LDP.
\end{example}

In this paper, we study personalized continual frequency estimation under $w$-event personalized
LDP, formalized as \privacyDefinitionSimpleName{}, where each user specifies her own window size $w_i$ and privacy budget
$\epsilon_i$, and the collector estimates the population-level frequency
distribution at every time slot.
Personalized temporal privacy has been studied under centralized DP~\cite{DBLP:journals/vldb/DuZCCLXL26}; however,
those mechanisms inject noise into raw streams held by a trusted curator and
thus do not apply to the local model.


\textbf{Existing methods and limitations.} A straightforward solution splits
each user's budget across the reports in her window, which quickly shrinks the
per-report budget under large $w$ and infinite streams. LDP-IDS~\cite{DBLP:conf/sigmod/RenSYYZX22} avoids such
fragmentation via population division: at each slot, only a sampled population
reports with a larger budget, while the others are recycled or approximated by
the latest release. However, LDP-IDS assumes a uniform $(w,\epsilon)$.
Applying it with the most stringent requirement incurs the utility loss shown
in Example~\ref{exm:car-hailing}.



\textbf{Challenges.} Directly extending LDP-IDS to heterogeneous requirements leaves two challenges unresolved:

\emph{Challenge~1: heterogeneous population scheduling.}
Population division couples sampling with privacy. A larger sampling size
reduces the sampling error but shortens the recycling period, forcing users
with long windows to split their budgets over more reports and thus increasing
their perturbation error; a smaller size does the opposite. Since this
trade-off differs across users with heterogeneous $(w_i,\epsilon_i)$ and
across groups of different sizes, a single sampling size can only strike one
compromise for all of them, over-sampling some groups and under-sampling
others.


\emph{Challenge~2: biased cross-group fusion.} Groups with small budgets suffer
high estimation variance. Borrowing estimates from groups with larger budgets
can reduce this variance. However, privacy preferences may correlate with the data:
users with more sensitive data tend to choose stronger protection. 
Thus, directly borrowing estimates from other groups can introduce bias. 
Moreover, when groups update asynchronously, the borrowed estimates may be stale. 
Compounding these issues, the true group distributions are unknown, deciding how much to borrow is
itself an estimation problem.


\textbf{Our approaches.} To address Challenge~1, we first derive the reporting
error of population division under personalized budgets as the sum of
sampling and perturbation errors, and propose \basicSelectionTotalName{}
(\basicSelection{}), which selects the privacy-compatible sampling size minimizing this
error. Built on \basicSelection{}, we propose two personalized population-division
mechanisms, \solutionATotalName{} (\solutionA{}) and \solutionBTotalName{} (\solutionB{}). 
Since a unified sampling size limits gains from personalization, we
further propose \advancedSelectionTotalName{} (\advancedSelection{}), which assigns each privacy group its
own optimal sampling size, recycling period, and effective budget, so that each group
follows its own error trade-off while satisfying its privacy requirement.

To address Challenge~2, we propose \advancedEnhanceTotalName{} (\advancedEnhance{}). For each group, \advancedEnhance{} fuses its own estimate with those of groups with weaker privacy requirements, using simplex weights that minimize a risk accounting for both variance and the bias due to distribution shift or staleness. Since this risk depends on the unknown group distributions, \advancedEnhance{} estimates it with an unbiased empirical risk matrix, clips its eigenvalues, and solves a small quadratic program for the weights. The fused group estimates are aggregated by population proportions. 
As a result, \advancedEnhance{} borrows heavily when groups are similar and lightly when the shift grows. Integrating \advancedSelection{} and \advancedEnhance{} yields the enhanced mechanisms \solutionC{} and \solutionD{}.

\textbf{Contributions.} In summary, we make the following contributions.
\begin{itemize}
	\item We formalize the problem of \problemDefineTotalName{} under a newly defined \privacyDefinitionSimpleName{}, where each user specifies her own window size and
	privacy budget in Section~\ref{predef-prodef}.
	\item We analyze the error of population-division under personalized LDP and propose \solutionA{} and \solutionB{}, guided by a new \basicSelection{} strategy that selects a privacy-compatible sampling size, together with their
	privacy and error analysis in Section~\ref{method:basic}.
	\item We propose \advancedSelection{} for group-specific scheduling and BA-PUE for
		bias-aware cross-group fusion, yielding \solutionC{} and \solutionD{}, and establish
		their privacy guarantees and error bounds in Section~\ref{method:enhanced}.
	\item We conduct experiments on two real and three synthetic datasets, which shows that \solutionD{}
	reduces the $\shrinkem{AMSE}$ of Direct-\solutionB{} by 52.3\% and 40.8\% on average, respectively in Section~\ref{experiments}.
	Besides, a controlled study further shows that \solutionD{} remains about 4.3--5.1$\times$ more accurate than \solutionB{} under strong privacy-data correlation.
\end{itemize}

\section{Related Work}
\subsection{Local Differential Privacy on Streaming Data}
Local Differential Privacy (LDP)\cite{DBLP:conf/stoc/BassilyS15} enables privacy-preserving data collection without a trusted curator and has been widely adopted in practice.
Existing studies on LDP for streaming data can be broadly categorized into event-level, user-level, and $w$-event privacy models.

Event-level LDP (event-LDP) mechanisms, such as RAPPOR~\cite{DBLP:conf/ccs/ErlingssonPK14} and ToPL~\cite{DBLP:conf/ccs/0001C0SC0LJ21}, perturb each record independently and do not account for temporal correlations.
User-level LDP methods~\cite{DBLP:journals/pvldb/BaoYXD21,DBLP:journals/tkde/XueYHZW23,DBLP:conf/infocom/YeHHAX23,DBLP:conf/icde/Mao00WH24} aim to protect entire user streams, but typically suffer from rapid
privacy budget consumption or limited adaptability in infinite streams.
To bridge the gap between event-level and user-level protection,
$w$-event privacy was introduced to protect event sequences within a sliding window of length $w$~\cite{DBLP:journals/pvldb/KellarisPXP14}.
Ren et al.~\cite{DBLP:conf/sigmod/RenSYYZX22} extend this notion to the local setting and propose LDP-IDS for infinite data streams. However, LDP-IDS assumes a uniform window size and privacy budget for all users, which limits its applicability in personalized settings.
SHieLD~\cite{DBLP:journals/jbd/ZolfaghariMJ26} supports personalized continual reporting via DDRM-style difference trees~\cite{DBLP:journals/tkde/XueYHZW23}, cross-level subsampling, and inverse-variance fusion. However, it provides only per-report $\epsilon$-LDP under independent records, rather than user-specific $w_i$-event privacy.


\subsection{Personalized Differential Privacy}
Personalized Differential Privacy (PDP), also known as Heterogeneous Differential Privacy (HDP)~\cite{DBLP:journals/jpc/AlagganGK16}, allows heterogeneous privacy requirements across users or data attributes. Existing PDP models can be broadly divided into item-grained and user-grained settings.


Item-grained approaches, such as One-Sided Differential Privacy (OSDP)~\cite{DBLP:conf/icde/KotsogiannisDHM20}, Pufferfish~\cite{DBLP:journals/tods/KiferM14} and Blowfish~\cite{DBLP:conf/sigmod/HeMD14}, relax protection for less sensitive attributes to improve utility. However, they rely on centralized mechanisms and policy specification, making them unsuitable for decentralized streaming settings.
User-grained PDP allows each user to specify an individual privacy budget.
Representative works include Personalized Differential Privacy (PDP)~\cite{DBLP:conf/icde/JorgensenYC15}, Adaptive Personalized Differential Privacy (AdaPDP)~\cite{DBLP:conf/infocom/NiuCWWLC21}, and Affine Differentially Private Mean Estimator (ADPM)~\cite{DBLP:journals/tit/ChaudhuriMC25}, which primarily focus on centralized or static data.
More recently, Chen et al.~\cite{DBLP:conf/icde/ChenLQKJ16} and He et al.~\cite{he2025personalized} propose Personalized Local Differential Privacy (PLDP) frameworks for count and range queries. 
Li et al.~\cite{DBLP:journals/tdsc/LiYCSL23} propose HDG-COE, which enables users to specify personalized indistinguishable areas for protection, achieving high utility.
However, existing PLDP methods do not consider temporal correlations in infinite data streams.

Beyond LDP and PLDP, several studies investigate streaming data privacy under centralized DP with a trusted curator, including $w$-event privacy~\cite{DBLP:journals/pvldb/KellarisPXP14}, $\ell$-trajectory privacy~\cite{DBLP:conf/mdm/CaoY15}, and $(\vectorfont{w},\vectorfont{\epsilon})$-personalized privacy~\cite{DBLP:journals/vldb/DuZCCLXL26}.
In contrast, we study personalized local differential privacy for infinite, correlated attribute streams without requiring any trusted curator.
Table~\ref{relatedwork_table} summarizes representative studies.

\begin{table}[t!]
	\caption{Summary for related work.}
	\label{relatedwork_table}\vspace{1ex}
	\centering
	\resizebox{\linewidth}{!}{
		\begin{tabular}{|ccc|c|c|c|}
			\hline
			\multicolumn{1}{|c|}{\textbf{Categories}} &
			\multicolumn{1}{c|}{\textbf{Model Types}} &
			\textbf{Methods} &
			\textbf{\begin{tabular}[c]{@{}c@{}}Infinite \&\\ correlated\end{tabular}} &
			\textbf{Trust-free} &
			\textbf{Personalized} \\ \hline
			
			\multicolumn{1}{|c|}{} &
			\multicolumn{1}{c|}{} &
			RAPPOR~\cite{DBLP:conf/ccs/ErlingssonPK14} &
			$\times$ & \checkmark & $\times$ \\ \cline{3-6}
			
			\multicolumn{1}{|c|}{} &
			\multicolumn{1}{c|}{} &
			ToPL~\cite{DBLP:conf/ccs/0001C0SC0LJ21} &
			$\times$ & \checkmark & $\times$ \\ \cline{3-6}
			
			\multicolumn{1}{|c|}{} &
			\multicolumn{1}{c|}{\multirow{-3}{*}{\begin{tabular}[c]{@{}c@{}}event-level \\ LDP\end{tabular}}} &
			SHieLD~\cite{DBLP:journals/jbd/ZolfaghariMJ26} &
			$\times$ & \checkmark & $\checkmark$ \\ \cline{2-6}
			
			\multicolumn{1}{|c|}{} &
			\multicolumn{1}{c|}{} &
			CGM~\cite{DBLP:journals/pvldb/BaoYXD21} &
			\checkmark & \checkmark & $\times$ \\ \cline{3-6}
			
			\multicolumn{1}{|c|}{} &
			\multicolumn{1}{c|}{} &
			DDRM~\cite{DBLP:journals/tkde/XueYHZW23} &
			\checkmark & \checkmark & $\times$ \\ \cline{3-6}
			
			\multicolumn{1}{|c|}{} &
			\multicolumn{1}{c|}{} &
			StaSwitch~\cite{DBLP:conf/infocom/YeHHAX23} &
			\checkmark & \checkmark & $\times$ \\ \cline{3-6}
			
			\multicolumn{1}{|c|}{} &
			\multicolumn{1}{c|}{\multirow{-4}{*}{\begin{tabular}[c]{@{}c@{}}user-level \\ LDP\end{tabular}}} &
			PrivShape~\cite{DBLP:conf/icde/Mao00WH24} &
			\checkmark & \checkmark & $\times$ \\ \cline{2-6}
			
			\multicolumn{1}{|c|}{\multirow{-8}{*}{\begin{tabular}[c]{@{}c@{}}LDP on \\ streaming \\ data\end{tabular}}} &
			\multicolumn{1}{c|}{$w$-event LDP} &
			LDP-IDS~\cite{DBLP:conf/sigmod/RenSYYZX22} &
			\checkmark & \checkmark & $\times$ \\ \hline
			
			\multicolumn{1}{|c|}{} &
			\multicolumn{1}{c|}{} &
			OSDP~\cite{DBLP:conf/icde/KotsogiannisDHM20} &
			$\times$ & $\times$ & \checkmark \\ \cline{3-6}
			
			\multicolumn{1}{|c|}{} &
			\multicolumn{1}{c|}{} &
			Pufferfish~\cite{DBLP:journals/tods/KiferM14,DBLP:journals/tit/NuradhaG23,DBLP:conf/icml/PierquinBTB24,DBLP:journals/tit/NuradhaGW24,DBLP:journals/tifs/Ding24} &
			\checkmark & $\times$ & \checkmark \\ \cline{3-6}
			
			\multicolumn{1}{|c|}{} &
			\multicolumn{1}{c|}{\multirow{-3}{*}{\begin{tabular}[c]{@{}c@{}}item-grained\\ privacy\end{tabular}}} &
			Blowfish~\cite{DBLP:conf/sigmod/HeMD14,DBLP:journals/pvldb/LiuKTDH21,DBLP:journals/tit/EdwardsRZZ22} &
			\checkmark & $\times$ & \checkmark \\ \cline{2-6}
			
			\multicolumn{1}{|c|}{} &
			\multicolumn{1}{c|}{} &
			SM~\cite{DBLP:conf/icde/JorgensenYC15} &
			$\times$ & $\times$ & \checkmark \\ \cline{3-6}
			
			\multicolumn{1}{|c|}{} &
			\multicolumn{1}{c|}{} &
			AdaPDP~\cite{DBLP:conf/infocom/NiuCWWLC21} &
			$\times$ & $\times$ & \checkmark \\ \cline{3-6}
			
			\multicolumn{1}{|c|}{} &
			\multicolumn{1}{c|}{} &
			ADPM~\cite{DBLP:journals/tit/ChaudhuriMC25} &
			$\times$ & $\times$ & \checkmark \\ \cline{3-6}
			
			\multicolumn{1}{|c|}{} &
			\multicolumn{1}{c|}{} &
			PCE~\cite{DBLP:conf/icde/ChenLQKJ16} &
			$\times$ & \checkmark & \checkmark \\ \cline{3-6}
			
			\multicolumn{1}{|c|}{} &
			\multicolumn{1}{c|}{} &
			HDG-COE~\cite{DBLP:journals/tdsc/LiYCSL23} &
			$\times$ & \checkmark & \checkmark \\ \cline{3-6}
			
			\multicolumn{1}{|c|}{\multirow{-9}{*}{\begin{tabular}[c]{@{}c@{}}Personalized \\ DP\end{tabular}}} &
			\multicolumn{1}{c|}{\multirow{-6}{*}{\begin{tabular}[c]{@{}c@{}}user-grained \\ privacy\end{tabular}}} &
			PLDP-MRQ~\cite{he2025personalized} &
			$\times$ & \checkmark & \checkmark \\ \hline
			
			\multicolumn{3}{|c|}{{\color[HTML]{000000} \textbf{Our methods}}} &
			{\color[HTML]{000000} \CheckmarkBold} &
			{\color[HTML]{000000} \CheckmarkBold} &
			{\color[HTML]{000000} \CheckmarkBold} \\ \hline
		\end{tabular}
	}
\end{table}

\section{Problem Settings}\label{predef-prodef}
In this section, we introduce the stream model, define \privacyDefinitionTotalName{} \big(\privacyDefinitionSimpleName{}\big), and formulate \problemDefineTotalName{} (\problemDefineSimpleName{}).
Table~\ref{tbl:notations} summarizes the common notations.
\begin{table}[t!]
	\caption{Notations.}
	\label{tbl:notations}
	\centering
	\resizebox{\linewidth}{!}{
		\begin{tabular}{c|c}
			\hline
			\textbf{Notations}        & \textbf{Description}                                       \\ \hline 
			$u_i$; $n$			& the $i$-th user; 	the total number of users		\\ 
			$d$             & the domain size of the attribute                                   \\ 
			$w_{i}$; $\epsilon_{i}$           & $u_i$'s privacy window size; $u_i$'s privacy budget                                      \\
			$PG_r$            & the $r$-th privacy group $(\widetilde{w}_r,\widetilde{\epsilon}_r, U_r)$                                 \\ 
			$m$, $K$             & the total number of privacy groups  grouped by privacy budgets or privacy requirements                      \\ 
			$\pi_r$           & the population proportion of privacy group $PG_r$                         \\  
			$\omega_j$										& the $j$-th value in the value domain \\
			$\vectorfont{f}$             & the true population-level frequency histogram at any time slot									\\
			$\vectorfont{f}_t$             & the true population-level frequency histogram at time slot $t$                                     \\
			$\widehat{\vectorfont{f}}_t$             & the estimated statistical population-level frequency histogram at time slot $t$                                     \\ 
			$\vectorfont{f}_{t}^{(r)}$				& the true frequency histogram of $PG_r$ at time slot $t$ 				\\			
			$\widehat{\vectorfont{f}}_{t}^{(r)}$				& the estimated frequency histogram of $PG_r$ at time slot $t$ 				\\	\hline			
		\end{tabular}
	}
\end{table}

\subsection{Data Streams}
%
We follow the standard streaming data model~\cite{DBLP:journals/tissec/ChanSS11,DBLP:journals/pvldb/KellarisPXP14}.
At each time slot $t$, users report attribute values from a finite domain $\Omega$ of size $d$, forming a database $D_t$.
The sequence $S=( D_1,D_2,\ldots)$ constitutes an infinite data stream.
The server aims to estimate the frequency distribution $\vectorfont{f}_t$ at each time slot under local privacy constraints.

\begin{definition}[$w$-Neighboring Stream Prefixes~\cite{DBLP:conf/sigmod/RenSYYZX22,DBLP:journals/pvldb/KellarisPXP14}]
		Let $w$ be a positive integer. For user $u_i$, two stream prefixes $S_{i,t}=(x_{i,1},\dots,x_{i,t})$ and $S'_{i,t}=(x'_{i,1},\dots,x'_{i,t})$ are $w$-neighboring, denoted by $S_{i,t}\sim_w S'_{i,t}$, if for all $\tau_1\le\tau_2\le t$ with $x_{i,\tau_1}\neq x'_{i,\tau_1}$ and $x_{i,\tau_2}\neq x'_{i,\tau_2}$, it holds that $\tau_2-\tau_1+1\le w$.
\end{definition}


\subsection{Privacy Definitions}\label{subsec:prodef}
Let $U=\{u_1,\ldots,u_n\}$ be the user set. Each user $u_i$
holds a personalized \textit{privacy requirement} $(w_i,\epsilon_i)$~\cite{DBLP:journals/vldb/DuZCCLXL26},
where $w_i$ specifies the sliding-window size and
$\epsilon_i$ is the privacy budget for that window.
In practice, personalized privacy requirements can be discretized into a finite set of window-budget pairs and encapsulated as different privacy levels (Table~\ref{tlb:privacy_level}), allowing users to choose privacy requirements without understanding $(w_i,\epsilon_i)$ while also ensuring that users  with identical requirements are assigned to the same privacy group.

\begin{table}
	\caption{Privacy Level Table}
	\label{tlb:privacy_level} 
	\centering
	\begin{tabular}{|c|c|c|c|c}
		\hline
		\diagbox{$w$}{$\epsilon$} & $1.5$ & $1.0$ & $0.5$ \\
		\hline
		$40$ & Level $(1,1)$ & Level $(1,2)$ & Level $(1,3)$ \\
		\hline
		$120$ & Level $(2,1)$ & Level $(2,2)$ & Level $(2,3)$ \\
		\hline
		$200$ & Level $(3,1)$ & Level $(3,2)$ & Level $(3,3)$ \\
		\hline
	\end{tabular}
\end{table}

\begin{definition}[\privacyDefinitionTotalName{}, \privacyDefinitionSimpleName{}]\label{def:privacy}
	Let $\vectorfont{\entity{R}}=\{\entity{R}_i\mid i\in[|U|]\}$ be a local randomizer set with which each user $u_i\in U$ perturbs her own stream $x_i=(x_{i,1},x_{i,2},\dots)$.
	$\vectorfont{\entity{R}}$ satisfies \privacyDefinitionSimpleName{} if $\forall u_i\in U$, $\forall t$, 
	$\forall S_{i,t},S'_{i,t}$ satisfying $S_{i,t}\sim_{w_i}S'_{i,t}$, and $\forall O\in\entity{O}$ of output sequences,
	\begin{equation}\notag
		\outereqsize{
			\begin{aligned}
				\Pr[\mathcal{R}_i(S_{i,t})\in O]\le e^{\epsilon_i}\Pr[\mathcal{R}_i(S'_{i,t})\in O].
			\end{aligned}
		}
	\end{equation}
\end{definition}
When $w_i=1$ for all users, the definition above reduces to
\pldpDefinitionTotalName{} (\pldpDefinitionSimpleName{})~\cite{he2025personalized}.
Additionally, when $(w_i,\epsilon_i)$ is a constant pair \big(i.e., $(w,\epsilon)$\big), it collapses to $w$-Event $\epsilon$-LDP~\cite{DBLP:journals/pvldb/KellarisPXP14,DBLP:conf/sigmod/RenSYYZX22}.
Following References~\cite{DBLP:conf/icde/JorgensenYC15,he2025personalized}, we treat $(w_i,\epsilon_i)$ as public metadata for scheduling; neighboring streams differ only in data values. 
Thus, the guarantee is conditional on the requirement, and leakage from requirements correlated with data, shared by all personalized DP models~\cite{DBLP:journals/vldb/DuZCCLXL26,DBLP:conf/icde/JorgensenYC15,he2025personalized}, is left to future work. 

To make the guarantee independent of the server’s honesty, each client locally records its report history and responds only if its total budget consumption within the latest $w_i$ slots, including the new report, does not exceed $\epsilon_i$; otherwise it declines. Since this check depends only on report times rather than data values, a server deviating from the sampling schedule can degrade utility but not privacy.

\begin{definition}[Privacy Group and Budget Group]\label{def:privacy-group}
		Given a unique privacy requirement set $(\widetilde{\vectorfont{w}},\widetilde{\vectorfont{\epsilon}})=\{(\widetilde{w}_j, \widetilde{\epsilon}_j)\; |\; \forall r\neq j, \widetilde{w}_r\neq\widetilde{w}_j \; or \; \widetilde{\epsilon}_r\neq\widetilde{\epsilon}_j\}$ with $|(\widetilde{\vectorfont{w}},\widetilde{\vectorfont{\epsilon}})|=K$ and a  user set $U\subseteq\entity{U}$ with $n$ users, where each user $u_i$ holds a privacy requirement $(w_i, \epsilon_i)\in(\widetilde{\vectorfont{w}},\widetilde{\vectorfont{\epsilon}})$, $U$ can be divided into $K$ groups $\{U_1,U_2,\dots U_K\}$ based on distinct privacy requirement.
		The  pair $\shrinkem{PG}_j=(\widetilde{w}_j, \widetilde{\epsilon}_j, U_j)$ is called the $j$-th privacy group of $U$.
		If all privacy groups share the same window size (i.e., $\widetilde{\vectorfont{w}}\equiv w\cdot\vectorfont{1}$), we refer to them as budget groups, and denote their number by $m$ instead of $K$.
\end{definition}

Based on Definition~\ref{def:privacy-group}, we can define privacy domination in Definition~\ref{def:privacy-dominate}, which means $PG_j$ requires no-weaker privacy requirement than $PG_i$ if $PG_j$ privacy-dominates $PG_i$.

\begin{definition}[Privacy Domination]\label{def:privacy-dominate}
	For any two Privacy Group $PG_i=(\widetilde{w}_i, \widetilde{\epsilon}_i, U_i)$ and $PG_j=(\widetilde{w}_j, \widetilde{\epsilon}_j, U_j)$, $PG_j$ privacy-dominates $PG_i$ denoted by $PG_i\preceq_p PG_j$, if $\widetilde{w}_i\leq \widetilde{w}_j$ and $\widetilde{\epsilon}_i\geq \widetilde{\epsilon}_j$.
\end{definition}

Let $\vectorfont{PG}$ be the total privacy groups. 
For any privacy group $PG_i\in\vectorfont{PG}$, we define its \textit{source} set as the set of privacy groups \textit{strictly} privacy-dominated by $PG_i$, denoted as $\entity{S}(i)=\{PG_r \,|\, \ PG_r\preceq_p PG_i, PG_r\neq PG_i, PG_r\in\vectorfont{PG}\}$, which is abbreviated as $\entity{S}_i$.

\subsection{Problem Definition}
Based on the above definitions, the server observes only locally perturbed reports and reconstructs an estimated frequency stream $(\widehat{\vectorfont{f}}_1, \widehat{\vectorfont{f}}_2, \ldots)$. The goal is to minimize estimation error while satisfying \privacyDefinitionSimpleName{}.
\begin{definition} (\problemDefineSimpleName{} Problem).\label{Problem_Def} 
	Given any positive integer $T\in\constvar{N}^+$, a user set $U=\{u_1, u_2,..., u_n\}\in\entity{U}$, where each $u_i$ holds a privacy requirement $(w_{i}, \epsilon_{i})$ and a series of data $x_{i,t}$ for $t\in[T]$, all the $x_{i,t}$ at time slot $t$ form $D_t$, and all the $D_t$ form a data stream prefix $S_T=\left(D_1, D_2, \ldots, D_T\right)$. \problemDefineSimpleName{} is to publish an obfuscated stream prefix frequency estimation $(\widehat{\vectorfont{f}}_1, \widehat{\vectorfont{f}}_2, \ldots, \widehat{\vectorfont{f}}_T)$ of $S_T$ while achieving \privacyDefinitionSimpleName{} with the error between $\widehat{\vectorfont{f}}$ and $\vectorfont{f}$ minimized, namely:
	\begin{equation}\notag
		\outereqsize{
			\begin{aligned}
				\min\;\; & \sum\limits_{t\in[T]}\simpleexpectation{ \left\|\widehat{\vectorfont{f}}_t-\vectorfont{f}_t\right\|_2^2} &\\
				\mathrm{s.t.}\;\; & \sum_{\tau=\max{(t-w_{i}+1,1)}}^{t}\epsilon_{i,\tau}\leq \epsilon_{i}, \;\;\;\; \forall u_i\in U, & 
			\end{aligned}
		}
	\end{equation}
	\noindent where $\epsilon_{i,\tau}$ denotes the privacy budget consumption of $u_i$ at time slot $\tau$.
\end{definition}

\section{OPS-Guided Personalized Population Division}\label{method:basic}
In this section, we study the case where privacy requirements are independent of user values. 
The dependent case is studied in Section~\ref{method:enhanced}.

Following the population-division paradigm~\cite{DBLP:conf/sigmod/RenSYYZX22,DBLP:journals/tdsc/0001LJ21,DBLP:conf/uss/WangBLJ17}, we first derive the reporting error under personalized LDP, then use it to select the sampling size, and finally instantiate two personalized population-division mechanisms, \solutionATotalName{} (\solutionA{}) and \solutionBTotalName{} (\solutionB{}). 

\subsection{Personalized Frequency Oracle and Reporting Error}
We use a GRR-based Personalized Frequency Oracle (PFO) that combines
the unbiased estimate $\widehat{f}_j^{(r)}$ of the $m$ budget groups, each
with $z_r$ sampled reporters.
It fuses the grouped estimates by the \textit{inverse average-variance weight}, which is a modification of the inverse-variance weight~\cite{he2025personalized}:


\begin{equation}\notag 
	\outereqsize{
		\begin{aligned}
			\forall \widetilde{f}_j\in\widetilde{\vectorfont{f}},\widetilde{f}_j = \sum_{r=1}^{m}\eta_{r}\cdot \widehat{f}_{j}^{(r)},\;\; \eta_r = \frac{1/\vara{\widehat{f}_{j}^{(r)};\vectorfont{z}}}{\sum_{r'=1}^{m}1/\vara{\widehat{f}_{j}^{(r')};\vectorfont{z}}},
		\end{aligned}
	}
\end{equation}
where
$\varainner{\widehat{f}_{j}^{(r)};\vectorfont{z}}=\frac{q_{r}(1-q_{r})}{z_r(p_{r}-q_{r})^2}+\frac{1-p_{r}-q_{r}}{z_r(p_{r}-q_{r})}\cdot\frac{1}{d}$
is an average approximating value of LDP unbiased variance~\cite{DBLP:conf/uss/WangBLJ17},
 $p_r=e^{\epsilon_r}/(e^{\epsilon_r}+d-1)$ and
$q_r=1/(e^{\epsilon_r}+d-1)$ are the truthful and fixed non-truthful
report probabilities, respectively.

Let $z_{\rbindexem{sum}}=\sum_{r\in[m]}z_r$, $a_{r}=\frac{q_{r}(1-q_{r})}{(p_{r}-q_{r})^2}$, and
$b_{r}=\frac{1-p_{r}-q_{r}}{(p_{r}-q_{r})}$.
The error consists of two components: (1) the PFO \textit{perturbation variance} $\simplevarpldp(\widetilde{\vectorfont{\epsilon}},\vectorfont{z}, d)=1/\sum_{r=1}^{m}z_r/\left(d a_{r}+b_{r}\right)$ introduced by personalized local randomization; and (2) the \textit{sampling variance} $\varsmplinner{\widehat{\vectorfont{f}}} = \frac{n-z_{\rbindexem{sum}}}{z_{\rbindexem{sum}}(n-1)}(1-\sum_{j=1}^{d}f_j^2)\leq \frac{n-z_{\rbindexem{sum}}}{z_{\rbindexem{sum}}(n-1)}$ caused by using only a subset of the population. 
Let $\simplevarsmpl\left(n,z_{\rbindexem{sum}}\right) = \frac{n-z_{\rbindexem{sum}}}{z_{\rbindexem{sum}}(n-1)}$. 
We use the distribution-free upper bound on the per-domain reporting error:
\begin{equation}\label{eq:finalavgvar}
	\outereqsize{
		\begin{aligned}
			\shrinkem{err}(\widetilde{\vectorfont{\epsilon}},n,\vectorfont{z}, d)
			&= \left(\simplevarsmpl\left(n,z_{\rbindexsmallem{sum}}\right) + \simplevarpldp(\widetilde{\vectorfont{\epsilon}},\vectorfont{z}, d)\right)/d \\
			&= \frac{n-z_{\rbindexsmallem{sum}}}{dz_{\rbindexsmallem{sum}}(n-1)} +
			1/\sum_{r=1}^{m}z_r/\left(a_{r}+b_{r}/d\right),
		\end{aligned}
	}
\end{equation}

Derivations and the optimal value of the weights are provided in 
Appendix~\ref{appendix:PFO-and-variances}.

The error in Eq.~\eqref{eq:finalavgvar} plays two roles below: it is the objective for selecting the sampling size and the threshold for deciding whether to publish.

\begin{algorithm}[!t]\small \DontPrintSemicolon
	\caption{\basicSelectionTotalName{} (\basicSelection{})}
	\label{alg:ops}
	\DontPrintSemicolon
	\KwIn{The sampling size requirement list $\vectorfont{z}=(z_1,\dots z_n)$; the privacy budget requirement list $\vectorfont{\epsilon}=(\epsilon_1,\dots\epsilon_n)$; the window size list $\vectorfont{w}=(w_1,\dots,w_n)$; the sampling ratio $\rho$; the value domain size $d$}
	\KwOut{The optimal sampling size $z^*$; the optimal recycle period $\tau^*$; the effective privacy budget vector $\vectorfont{\epsilon}'$}
	Initialize $\shrinkem{err}^*$ as the upper bound of error value; \\
	Set $\widetilde{\vectorfont{z}}\gets\shrink{Unique}(\vectorfont{z})$ as the unique list of $\vectorfont{z}$; \\ \label{uz}
	
	\For{$\widetilde{z}\in\widetilde{\vectorfont{z}}$}{ \label{traversal-begin}
		Set $\widetilde{\tau}\gets\max{\left(1,\left\lfloor \rho n/\widetilde{z} \right\rfloor\right)}$;\\ \label{alg:ops:tau}
		Initialize $\vectorfont{\epsilon}''\gets\emptyset$;\\
		\For{$i\in[n]$}{ \label{changebudgetstart}
			$\epsilon_{\rbindexem{eff},i}\gets\epsilon_{i} / \left\lceil w_i/\widetilde{\tau} \right\rceil$;\\ \label{changebudgetend}
			Add $\epsilon_{\rbindexem{eff},i}$ to $\vectorfont{\epsilon}''$;
		}
		Set $\widetilde{\vectorfont{\epsilon}}''\gets\shrink{Unique}(\vectorfont{\epsilon}'')$;\\ \label{recordstart}
		Calculate $\vectorfont{\alpha}\gets|\{i:\epsilon_{\rbindexem{eff},i}=\widetilde{\epsilon}_r\}|/n$;\\ \label{alg:alpha}
		Get the sampling size set $\vectorfont{z}'=\left\{z'_r\mid z'_r=\widetilde{z}\cdot \alpha_r\right\}$;\\ \label{alg:zprime}
		Calculate $\shrinkem{err}_{\rbindexem{tmp}}\gets\shrinkem{err}\left(\widetilde{\vectorfont{\epsilon}}'',n,\vectorfont{z}',d\right)$ by Eq.~\eqref{eq:finalavgvar}; \\ \label{alg:errtmp}
		\If{$\shrink{err}_{\rbindex{tmp}}<\shrink{err}^*$}{
			$\shrinkem{err}^*\gets\shrinkem{err}_{\rbindexem{tmp}}$; 
			$z^*\gets \widetilde{z}$; 
			$\tau^*\gets\widetilde{\tau}$;
			$\vectorfont{\epsilon}'\gets\vectorfont{\epsilon}''$; \label{recordend}
		}
	} \label{traversal-end}
	\Return{$z^*$, $\tau^*$, $\vectorfont{\epsilon}'$};
\end{algorithm}

\subsection{\basicSelectionTotalName{} (\basicSelection{})}
After knowing the error form, we can choose the optimal sampling size to minimize the error.
The process (\basicSelection{}) is shown in Algorithm~\ref{alg:ops}.
In this process, the sampling size $z$ affects the error in Eq.~\eqref{eq:finalavgvar} twice: directly through $\simplevarsmpl$, and indirectly through the recycling period $\widetilde{\tau}=\max\{1,\lfloor\rho n/z\rfloor\}$. 
It sets the effective budgets $\epsilon_{\rbindexem{eff},i}=\epsilon_i/\lceil w_i/\widetilde{\tau}\rceil$ entering $\simplevarpldp$. A large $z$ thus fragments the budgets of long-window users, whereas a conservative $z$ incurs unnecessary sampling error.
\basicSelection{} resolves this trade-off by the error in evaluating Eq.~\eqref{eq:finalavgvar} at the effective budgets:

%
%
%

\begin{equation}\notag
	\outereqsize{
		\begin{aligned}
			z^{*}=\arg\min_{\vectorfont{z}'\in \mathcal{Z}} 
			\shrinkem{err}(\widetilde{\vectorfont{\epsilon}}'',n,\vectorfont{z}',d),
		\end{aligned}
	}
\end{equation}
where $\widetilde{\vectorfont{\epsilon}}''$ is the distinct effective budget vector induced by the distinct sampling size vector $\widetilde{\vectorfont{z}}$ (Lines~\ref{alg:ops:tau}--\ref{recordstart}) and $\widetilde{\boldsymbol{z}}'$ are the corresponding group-wise sampled populations (Lines~\ref{alg:alpha}--\ref{alg:zprime}); \basicSelection{} keeps the candidate with minimum error (Lines~\ref{alg:errtmp}--\ref{recordend}).




\begin{algorithm}[!t]
	\caption{Unified \solutionA{}/\solutionB{} Stream Estimation}
	\label{alg:ops-guided-plpd-plpa}
	\KwIn{Available user set $U_A$; privacy requirements $(\boldsymbol w,\boldsymbol\epsilon)$;
		history $(\widehat{\vectorfont{f}}_{{1}},\ldots, \widehat{\vectorfont{f}}_{{t-1}})$; $\mathsf{mode}\in\{\mathrm{PLPD},\mathrm{PLPA}\}$}
	\KwOut{Current estimate $\widehat{\boldsymbol f}_t$}
	$\vectorfont{z}\gets\scriptmath{\left( \left\lfloor\frac{n}{2w_{{1}}}\right\rfloor,\left\lfloor\frac{n}{2w_{{2}}}\right\rfloor,\ldots,\left\lfloor\frac{n}{2w_{{n}}}\right\rfloor\right)}$;\tcp*[f]{once} \\ \label{alg:z-basic}
	$\left(z^*,\tau^*,\vectorfont{\epsilon}'\right)\gets\shrink{OPS}(\vectorfont{z},\vectorfont{\epsilon},\vectorfont{w}, \frac{1}{2},d)$ (Alg.~\ref{alg:ops}); \tcp*[f]{once} \\ \label{alg:bigZ} 
	
	Sample the dissimilarity population $U_{\rbindexem{dis},t}$ of size $z^*$ and set $U_A\gets U_A\backslash U_{\rbindexem{dis},t}$;\\
	$(\widehat{\vectorfont{f}}_{\rbindexem{dis},t}, \widehat{V}_{\rbindexem{dis},t}) \gets \operatorname{PFO}(D_{\rbindexem{dis},t};\vectorfont{\epsilon}')$;\\
	$\shrinkem{dis}\gets \frac{1}{d}\|\widehat{\vectorfont{f}}_{\rbindexem{dis},t} -\widehat{\vectorfont{f}}_{\rbindexem{pub},l}\|_2^2 -\frac{1}{d}\widehat{V}_{\rbindexem{dis},t}$;\\
	
	$(n_{\rbindexem{pp}},\shrinkem{status},\shrinkem{err}_{\rbindexem{pub}})
	\gets \operatorname{PubSchedule}(\mathsf{mode})$;\\
	
	\If{$\shrink{status}=\operatorname{reuse}$}{
		$\widehat{\vectorfont{f}}_{\rbindexem{pub},t}\gets
		\widehat{\vectorfont{f}}_{\rbindexem{pub},t-1}$;
	}\ElseIf{$\shrink{dis}>\shrink{err}_{\rbindex{pub}}$
		\KwSty{and} $n_{\rbindex{pp}}\geq u_{\rbindex{min}}$}{
			Sample $U_{\rbindexem{pub},t}$ of size $n_{\rbindexem{pp}}$ and set $U_A\gets U_A\backslash U_{\rbindexem{pub},t}$;\\
		$(\widehat{\vectorfont{f}}_{\rbindexem{pub},t},
		\widehat{V}_{\rbindexem{pub},t})\gets \operatorname{PFO}(D_{\rbindexem{pub},t};\vectorfont{\epsilon}')$;
	}\Else{
		$\widehat{\vectorfont{f}}_{\rbindexem{pub},t}\gets
		\widehat{\vectorfont{f}}_{\rbindexem{pub},t-1}$;
	}
	
	Recycle expired users at time slot $t-\tau^*+1$;\\
	\Return{$\widehat{\vectorfont{f}}_{\rbindex{pub},t}$}
\end{algorithm}

\subsection{OPS-Guided \solutionA{} and \solutionB{}}\label{subsec:basic_solutions}

\noindent
\textbf{Initialization.} Alg.~\ref{alg:ops-guided-plpd-plpa} invokes \basicSelection{} once (Lines~\ref{alg:z-basic}--\ref{alg:bigZ}) with the per-user requirements $z_i=\{\lfloor n/(2w_i)\rfloor\}$ and $\rho=1/2$, since each time slot spends half of the users on dissimilarity and half on publication~\cite{DBLP:conf/sigmod/RenSYYZX22}.
Each candidate yields $\widetilde\tau\approx w_i$. 
Thus, \basicSelection{} effectively selects which window sets the recycling period.

\noindent
\textbf{Publication Decision.}
Following Reference~\cite{DBLP:conf/sigmod/RenSYYZX22}, a new estimate is generated only when the private dissimilarity exceeds the potential publication error $\shrinkem{err}_{\rbindexem{pub}}$ (i.e., Eq.~\eqref{eq:finalavgvar}),  which is evaluated according to the candidate publication population; otherwise, the latest release $\widehat{\vectorfont{f}}_{\rbindexem{pub},l}$, published at slot $l$, is reused. Because the private dissimilarity depends on real ditribution, we use the variance-corrected dissimilarity instead:

%
\begin{equation}\notag
	\outereqsize{
		\begin{aligned}
			\shrinkem{dis} &= \frac{1}{d}\sum_{j=1}^{d}\left(\widehat{\vectorfont{f}}_{\rbindexem{dis},t}[j]-\widehat{\vectorfont{f}}_{\rbindexem{pub},l}[j]\right)^2-\frac{1}{d}\sum_{j=1}^{d}\varem{\widehat{\vectorfont{f}}_{\rbindexem{dis},t}[j]},
		\end{aligned}
	}
\end{equation}
which removes the estimation variance of the current private estimate (see
Appendix~\ref{subsec:DetailsForDiss}.
for the derivation).

\noindent
\textbf{Publication Schedule.}
Both \solutionA{} and \solutionB{} follow Alg.~\ref{alg:ops-guided-plpd-plpa} with $U_A$ initialized as $U$.
They share \basicSelection{}, the dissimilarity, and the publication test, and differ only in $\operatorname{PubSchedule}$, i.e., how the publication half of the users is spent. 
\solutionA{} gives each potential publication half of the publication users unused in the current recycling period: $n^{\operatorname{\solutionA{}}}_{\rbindexem{pp}}=\lfloor(n/2-\sum_{\tau=t-\tau^*+1}^{t-1}|U_{pub,\tau}|)/2\rfloor$. 
\solutionB{} absorbs the quota of skipped slots: a publication at slot $l$ nullifies the next $t_N=|U_{pub,l}|/z^*-1$ slots and the nullified slots reuse the publication at slot $l$; afterwards, $n^{\operatorname{\solutionB{}}}_{\rbindexem{pp}}=z^*\min\{t-l-t_N,\tau^*\}$. A publication occurs only if $dis>\shrinkem{err}_{\rbindexem{pub}}=\shrinkem{err}(\widetilde{\vectorfont{\epsilon}}_{\rbindexem{pub}},n,\vectorfont{z}_{\rbindexem{pp}},d)$ and $n_{\rbindexem{pp}}$ is no less than the population size lower bound (i.e., $n_{\rbindexem{pp}}\ge u_{\rbindexem{min}}$).

\subsection{Analysis}
\noindent
\textbf{Time Complexity Analysis.} 
Let $m_{\rbindexem{eff}}=\left|\operatorname{Unique}\left(\vectorfont{\epsilon}'\right)\right|\leq n$ denote
the number of effective privacy budget groups after \basicSelection{}.
Then we have Theorem~\ref{thm:timecost_enhanced} as follows.
\begin{theorem}\label{thm:timecost_enhanced}
	The time complexity of both \solutionA{} and \solutionB{} is $O\left(n+d\cdot m_{\rbindex{eff}}\right)$.
\end{theorem}
\begin{proof}[Proof Sketch]
	\basicSelection{} is executed once at initialization. 
	At each slot, sampling and perturbation take $O(n)$ time and PFO estimation takes $O(d\,m_{\rbindexem{eff}})$ time.
	Therefore, the total complexity is $O(n+d\,m_{\rbindexem{eff}})$.
	The proof details are shown in  Appendix~\ref{Proof:timecost_enhanced}.
\end{proof}

\noindent
\textbf{Privacy Analysis.} Theorem~\ref{thm:PLP+PrivacyAnalysis} gives the privacy guarantee for \solutionA{} and \solutionB{}.

\begin{theorem}\label{thm:PLP+PrivacyAnalysis}
	\solutionA{} and \solutionB{} satisfy \privacyDefinitionSimpleName{}.
\end{theorem}
\begin{proof}
	Enforced by the client-side check, under \basicSelection{}, each user participates at most once in every $\tau^{*}$ consecutive slots. 
	Hence, a $w_i$-window contains at most $c_i=\lceil w_i/\tau^{*}\rceil$ reports.
	Each report uses $\epsilon_i'=\epsilon_i/c_i$. 
	Thus, sequential composition gives a total privacy loss no larger than $\epsilon_i$.
	Dissimilarity computation, reuse, absorption, aggregation, and population selection are post-processing, which do not consume privacy budgets.
	Therefore, \solutionA{} and \solutionB{} satisfy \privacyDefinitionSimpleName{}.
	The proof details are shown in Appendix~\ref{Proof:PLP+PrivacyAnalysis}.
\end{proof}

\noindent
\textbf{Utility Analysis.}
Let $\tau^{*}$ denote the optimal recycling period determined by \basicSelection{}.
For analytical simplicity, we assume that at most $s<\tau^{*}$ new publications occur within a $\tau^{*}$-length window and that these publications are uniformly distributed. 
Let  $\widetilde{\vectorfont{\epsilon}}'=\operatorname{Unique}(\vectorfont{\epsilon}')$ denote the distinct effective privacy budget obtained after \basicSelection{}, and let  $\widetilde{\epsilon}_{\rbindexem{min}}'=\min(\widetilde{\vectorfont{\epsilon}}')$ denote the minimum effective privacy budget. 
Let $A^{*}=\frac{d-1}{n}\cdot\frac{2e^{\widetilde{\epsilon}_{\rbindexsmallem{min}}'}+d-2}{(e^{\widetilde{\epsilon}_{\rbindexsmallem{min}}'}-1)^2}$ and $R_{\rbindexem{opt}}=\frac{e^{\widetilde{\epsilon}_{\rbindexsmallem{min}}'}+d-2}{e^{\widetilde{\epsilon}_{\rbindexsmallem{min}}'}-1}$.
We analyze the expected $2$-norm error of \solutionA{} and \solutionB{} as follows.

\begin{theorem}\label{thm:PLPD+UtilityAnalysis}
	The error upper bound of per time slot in \solutionA{} is
	$\scriptmath{3R_{\rbindex{opt}}\sqrt{\frac{nd}{z^{*}} \left(\frac1{n-1}+A^{*}\right) - \frac{d}{n-1}}+ \frac{4(2^{s}-1)}{s}\left(\frac{1}{n-1}+A^{*}\right)-\frac{1}{n-1}}$.
\end{theorem}
\begin{proof}
	Please refer to Appendix~\ref{Proof:PLP+UtilityAnalysis}.
\end{proof}
\begin{theorem}\label{thm:PLPA+UtilityAnalysis}
	The error upper bound of per time slot in \solutionB{} is $\scriptmath{3R_{\rbindexsmall{opt}}\sqrt{\frac{nd}{z^{*}} (\frac1{n-1}+A^{*}) - \frac{d}{n-1}} }$  $\scriptmath{+ \frac{nC}{z^{*}}(\frac{1}{n-1}+A^{*})	H_{B+1}}$ $\scriptmath{- \frac{C(B+1)}{n-1}}$ $\scriptmath{+ 	BC\cdot\shrink{err}_{\rbindex{nlf}}}$, where  $\scriptmath{B=\frac{\tau^{*}-s}{2s}}$, $\scriptmath{C=\frac{1}{2B+1}}$, $\scriptmath{H_{B+1}=\sum_{j=1}^{B+1}\frac{1}{j}}$ and $\scriptmath{\shrink{err}_{\rbindex{nlf}}}$ denotes the average approximation error of a nullified time slot.
\end{theorem}
\begin{proof}
	Please refer to Appendix~\ref{Proof:PLP+UtilityAnalysis}.
\end{proof}

\section{Enhanced Approaches}\label{method:enhanced}

\solutionA{} and \solutionB{} still have two limitations.
First, although \basicSelection{} avoids imposing the stringent privacy requirement on all users and minimizes the reporting-error bound, a single sampling size limits utility gains from personalization. 
Second, high-privacy groups suffer high estimation variance, yet cross-group borrowing risks estimation bias when the privacy–data independence assumed by \solutionA{} and \solutionB{} fails.
To address these limitations, we develop \advancedSelection{} for group-specific scheduling and \advancedEnhance{} for bias-aware cross-group estimation, yielding the enhanced mechanisms \solutionC{} and \solutionD{}.

\subsection{\advancedSelectionTotalName{} (\advancedSelection{})}

\begin{algorithm}[!t]\small
	\caption{\advancedSelectionTotalName{}  (\advancedSelection{})}
	\label{alg:adv_ops}
	\KwIn{Privacy groups
		$\vectorfont{PGT}=\{(\widetilde w_r,\widetilde\epsilon_r,n_r)\}_{r=1}^{K}$;
		sampling ratio $\rho$; domain size $d$}
	\KwOut{Sampling-size vector $\vectorfont z^*$;
		recycle-period vector $\vectorfont\tau^*$;
		effective-budget vector $\vectorfont\epsilon'$}
	\For{$r\in[K]$}{
		$N_r\gets\lfloor\rho n_r\rfloor$;\\
		Construct $\mathcal Z_r\gets\left\{\left\lfloor\omega N_{r}/\widetilde{w}_r\right\rfloor\mid \omega\in[\widetilde{w}_r]\right\}\cap[2,N_{r}]$;\\
		\For{$z_r\in\mathcal Z_r$}{
			$\tau_r\gets
			\max\{1,\lfloor N_r/z_r\rfloor\}$;
			$c_r\gets
			\lceil\widetilde w_r/\tau_r\rceil$;
			$\epsilon'_r\gets
			\widetilde\epsilon_r/c_r$;\\
			Evaluate $err_r(z_r)$ and retain the minimizing candidate;
		}
	}
	\Return{$\vectorfont z^*,\vectorfont\tau^*,\vectorfont\epsilon'$};
\end{algorithm}


For each privacy group $PG_r$, \advancedSelection{} reserves
$N_r=\lfloor\rho n_r\rfloor\ge 2$ users for dissimilarity and searches the candidate
sampling sizes specified in Alg.~\ref{alg:adv_ops}.
Each candidate $z_r$ induces a feasible recycling period and
effective per-report privacy budget, and is evaluated by the
average per-domain error
$\shrinkem{err}_{r}(z_r)=\frac{1}{d}\left(
\simplevarsmpl(N_r,z_r)+
\simplevarldp(\epsilon'_r(z_r),z_r,d)\right)$.
The minimum-error candidate determines
$(z_r^*,\tau_r^*,\epsilon'_r)$ for group each $PG_r$.

Although \advancedSelection{} improves population scheduling, it does not directly address estimation bias arising from heterogeneous group distributions. We therefore introduce  \advancedEnhance{} to improve accuracy through bias-aware cross-group borrowing.

\subsection{\advancedEnhanceTotalName{} (\advancedEnhance{})}\label{subsec:ba-pue}
Cross-group borrowing can reduce the estimation variance of high-privacy groups~\cite{he2025personalized,DBLP:journals/pvldb/LiuLXLM21}. 
However, it may introduce bias due to distribution shift or staleness. 
To balance these effects, we propose \advancedEnhance{}, which formulates a target-specific bias–variance risk and employs \advancedCorretedWeightTotalName{} (\advancedCorretedWeight{}) to estimate the risk and optimize fusion weights. 
The fused group estimates are then aggregated by population proportions.



\noindent
\textbf{Target-Specific Bias--Variance Objective.}
For each target group $PG_H$, \advancedEnhance{} combines its own
candidate estimate with the estimates from its source set
$\entity{S}_H$. Let
$\entity{C}_H=\entity{S}_H\cup\{PG_H\}$.
Because a candidate may represent a current or historical distribution,
let $\vectorfont{g}_t^{(r)}$ denote the distribution represented by the
candidate of $PG_r$ at time $t$. Each candidate provides an unbiased
estimate
$\widehat{\vectorfont{g}}_{\rbindexem{risk},t}^{(r)}
=\vectorfont{g}_t^{(r)}
+\vectorfont{e}_{\rbindexem{risk},r,t}$, where
$\simpleexpectation{\vectorfont{e}_{\rbindexem{risk},r,t}}
=\vectorfont{0}$ and
$V_{\rbindexem{risk},r,t}
=\simpleexpectation{\|\vectorfont{e}_{\rbindexem{risk},r,t}\|_2^2}$.

For simplex weights
$\vectorfont{\lambda}_{H,t}\succeq\vectorfont{0}$ with
$\vectorfont{1}^{\top}\vectorfont{\lambda}_{H,t}=1$, the enhanced
estimate is
\begin{equation}\label{eq:linear-combine}
	\outereqsize{
		\begin{aligned}
			\widetilde{\vectorfont{f}}_t^{(H)}= \sum\nolimits_{PG_r\in\entity{C}_H}\lambda_{H,r,t}\cdot\widehat{\vectorfont{g}}_{\rbindexsmallem{risk},t}^{(r)}.
		\end{aligned}
	}
\end{equation} 
Let
$\vectorfont{\xi}_{r\mid H,t}
=\vectorfont{g}_t^{(r)}-\vectorfont{f}_t^{(H)}$
denote the distribution shift from candidate $PG_r$ to target $PG_H$.
Assuming zero-mean and mutually uncorrelated candidate errors, the
target-specific fusion risk
$R_H(\vectorfont{\lambda}_{H,t})
=\simpleexpectationinner{\|\widetilde{\vectorfont{f}}_t^{(H)}
	-\vectorfont{f}_t^{(H)}\|_2^2}$ is
\begin{equation}\notag
	\outereqsize{
		\begin{aligned}
			R_H(\vectorfont{\lambda}_{H,t}) &= \left\| \sum\nolimits_{PG_r\in\entity{C}_H}\lambda_{H,r,t}\vectorfont{\xi}_{r\mid H,t}\right\|_2^2+\sum\nolimits_{PG_r\in\entity{C}_H}\lambda_{H,r,t}^2 V_{\rbindexsmallem{risk},r,t}\\
			&= \vectorfont{\lambda}_{H,t}^{\top}\vectorfont{M}_{H,t}\vectorfont{\lambda}_{H,t},
		\end{aligned}
	}
\end{equation}
where
\begin{equation}\label{eq:fusion-error-matrix}
	\outereqsize{
		\begin{aligned}
			\vectorfont{M}_{H,t}(r,s)=\left\langle \vectorfont{\xi}_{r\mid H,t},\vectorfont{\xi}_{s\mid H,t} \right\rangle+V_{\rbindexsmallem{risk},r,t}\delta_{r,s}
		\end{aligned}
	}
\end{equation}
and $\delta_{r,s}=1$ if $r=s$ and $\delta_{r,s}=0$ otherwise.

Our objective is to find the optimal $\vectorfont{\lambda}_{H,t}$ for each target group $PG_H\in\vectorfont{PG}$ that minimizes $R_H(\vectorfont{\lambda}_H)$.
Since each target group may exhibit different distribution shifts and estimation variance, its optimal fusion weights need to be determined separately.

\noindent
\textbf{Empirical \advancedCorretedWeight{} Solver.}
The ideal matrix $\vectorfont{M}_{H,t}$ is unobservable because the real
group distributions are unknown.
Thus, we need to construct an unbiased empirical risk matrix $\widehat{\vectorfont{M}}_{H,t}$ to calculate $\vectorfont{\lambda}_{H,t}$.
To get $\widehat{\vectorfont{M}}_{H,t}$, we need (1) auxiliary group estimates and (2) their unbiased variances.

For any group $PG_r$, we use the following unbiased estimate for the group-level variance $V_r$:
\begin{equation} \label{thm:precise_var}  
	\outereqsize{
		\begin{aligned}
				\widehat{V}_{r}=\left(\theta_{r}\left(1-\left\| \widehat{\vectorfont{f}}^{(r)} \right\|_2^2\right) + V_{\rbindexsmallem{LDP,r}}\right)/\left(1-\theta_{r}\right),
		\end{aligned}
	}
\end{equation}
 where ${\theta_{r} = \frac{N_{r}-z_{r}}{z_{r}\left(N_{r}-1\right)}}$ and ${V_{\rbindexem{LDP,r}} = V_{\rbindexem{LDP}}\left(\epsilon_{r}',z_{r},d\right)}$.

For auxiliary group estimates, we use $\widehat{\vectorfont{f}}_{\rbindexem{est},t}^{(r)}=\vectorfont{f}_{\rbindexem{est},t}^{(r)}+\vectorfont{e}_{\rbindexem{est},t}$ and
$\widehat{\vectorfont{f}}_{\rbindexem{tar},t}^{(H)}=\vectorfont{f}_{\rbindexem{tar},t}^{(H)}+\vectorfont{e}_{\rbindexem{tar},t}$ with unbiased variance
estimates $\widehat{V}_{\rbindexem{est},r,t}$ and
$\widehat{V}_{\rbindexem{tar},H,t}$, which are obtained by Eq.~\eqref{thm:precise_var}.
The process (\advancedCorretedWeight{}) is shown in Algorithm~\ref{alg:pa-gcw}.

For
$PG_r\in\entity{C}_H$, define $\widehat{\vectorfont{\xi}}_{r\mid H,t}=\widehat{\vectorfont{f}}_{\rbindexem{est},t}^{(r)}-\widehat{\vectorfont{f}}_{\rbindexem{tar},t}^{(H)}$.
Under mutually uncorrelated estimation errors between different groups, we have
\begin{equation}\notag  
	\outereqsize{
		\begin{aligned}
			&\simpleexpectation{\left\langle \widehat{\vectorfont{\xi}}_{r\mid H,t},\widehat{\vectorfont{\xi}}_{s\mid H,t}\right\rangle} \\
			=& \left\langle \vectorfont{\xi}_{r\mid H,t},\vectorfont{\xi}_{s\mid H,t}\right\rangle+V_{\rbindexsmallem{tar},H,t}+V_{\rbindexsmallem{est},r,t}\delta_{r,s}-E_{r,H}-E_{s,H},
		\end{aligned}
	}
\end{equation} 
where $E_{r,s}=\simpleexpectationinner{\langle \vectorfont{e}_{\rbindexem{est},t}^{(r)} , \vectorfont{e}_{\rbindexem{tar},t}^{(s)} \rangle}$.
\advancedCorretedWeight{} distinguishes two cases based on whether the auxiliary estimate of the target group coincides with its target estimate (EQ) or has uncorrelated estimation errors (ND).

For \textbf{case}~(EQ), since $\widehat{\vectorfont{f}}_{\rbindexem{est},t}^{(H)}=\widehat{\vectorfont{f}}_{\rbindexem{tar},t}^{(H)}$, for $r,s\neq H$, we have 
\begin{equation}  \label{eq:fusion-error-matrix-exp-special}
	\outereqsize{
		\begin{aligned}
			\simpleexpectation{\left\langle \widehat{\vectorfont{\xi}}_{r\mid H,t},\widehat{\vectorfont{\xi}}_{s\mid H,t}\right\rangle}
			= \left\langle \vectorfont{\xi}_{r\mid H,t},\vectorfont{\xi}_{s\mid H,t}\right\rangle+V_{\rbindexsmallem{est},H,t}+V_{\rbindexsmallem{est},r,t}\delta_{r,s};
		\end{aligned}
	}
\end{equation} 
otherwise, $\simpleexpectationinner{\langle \widehat{\vectorfont{\xi}}_{r\mid H,t},\widehat{\vectorfont{\xi}}_{s\mid H,t}\rangle}=0$.
Then for $\vectorfont{M}_{H,t}$ with $r,s\neq H$, using unbiased estimator  $\widehat{V}_{\rbindexem{est},r,t}$, according to Eq.~\eqref{eq:fusion-error-matrix} and~\eqref{eq:fusion-error-matrix-exp-special}, we have
\begin{equation} \notag
	\outereqsize{
		\begin{aligned}
			\vectorfont{M}_{H,t}(r,s) &=\left\langle \vectorfont{\xi}_{r\mid H,t},\vectorfont{\xi}_{s\mid H,t} \right\rangle + V_{\rbindexsmallem{risk},r,t}\delta_{r,s}\\
			&= \simpleexpectation{\left\langle \widehat{\vectorfont{\xi}}_{r\mid H,t},\widehat{\vectorfont{\xi}}_{s\mid H,t}\right\rangle} - V_{\rbindexsmallem{est},H,t}- V_{\rbindexsmallem{est},r,t}\delta_{r,s} + V_{\rbindexsmallem{risk},r,t}\delta_{r,s}\\
			&=\simpleexpectation{\left\langle \widehat{\vectorfont{\xi}}_{r\mid H,t},\widehat{\vectorfont{\xi}}_{s\mid H,t}\right\rangle - \widehat{V}_{\rbindexsmallem{est},H,t} + \left(V_{\rbindexsmallem{risk},r,t}- \widehat{V}_{\rbindexsmallem{est},r,t}\right)\delta_{r,s}}.
		\end{aligned}
	}
\end{equation} 
Thus, \advancedCorretedWeight{} can construct $\widehat{\vectorfont{M}}_{H,t}$'s unbiased estimation matrix
\begin{equation}\label{eq:risk-block}
	\outereqsize{
		\begin{aligned}
			\widehat{\vectorfont{M}}_{H,t}(r,s) = 
			\begin{bmatrix}
				V_{\rbindexsmallem{risk},H,t}& \vectorfont0^\top\\
				\vectorfont 0 & \widehat{\vectorfont{A}}_{H,t}
			\end{bmatrix}
		\end{aligned}
	}
\end{equation} 
where 
\begin{equation}\notag
	\outereqsize{
		\begin{aligned}
			\widehat{A}_{H,t}(r,s) = \left\langle \widehat{\vectorfont{\xi}}_{r\mid H,t},\widehat{\vectorfont{\xi}}_{s\mid H,t}\right\rangle - \widehat{V}_{\rbindexsmallem{est},H,t} + \left(V_{\rbindexsmallem{risk},r,t}- \widehat{V}_{\rbindexsmallem{est},r,t}\right)\delta_{r,s}.
		\end{aligned}
	}
\end{equation}

For \textbf{case}~(ND), we have $E_{r,H}=E_{s,H}=0$, thus, 
\begin{equation}  \label{eq:fusion-error-matrix-exp-total-general}
	\outereqsize{
		\begin{aligned}
			\simpleexpectation{\left\langle \widehat{\vectorfont{\xi}}_{r\mid H,t},\widehat{\vectorfont{\xi}}_{s\mid H,t}\right\rangle} 
			= \left\langle \vectorfont{\xi}_{r\mid H,t},\vectorfont{\xi}_{s\mid H,t}\right\rangle+V_{\rbindexsmallem{tar},H,t}+V_{\rbindexsmallem{est},r,t}\delta_{r,s}.
		\end{aligned}
	}
\end{equation} 

Then for $\vectorfont{M}_{H,t}$, using unbiased estimator $\widehat{V}_{\rbindexem{tar},H,t}$ and $\widehat{V}_{\rbindexem{est},r,t}$, according to Eq.~\eqref{eq:fusion-error-matrix} and~\eqref{eq:fusion-error-matrix-exp-total-general}, we have
\begin{equation} \notag
	\outereqsize{
		\begin{aligned}
			\vectorfont{M}_{H,t}(r,s) &=\left\langle \vectorfont{\xi}_{r\mid H,t},\vectorfont{\xi}_{s\mid H,t} \right\rangle + V_{\rbindexsmallem{risk},r,t}\delta_{r,s}\\
			&= \simpleexpectation{\left\langle \widehat{\vectorfont{\xi}}_{r\mid H,t},\widehat{\vectorfont{\xi}}_{s\mid H,t}\right\rangle} - V_{\rbindexsmallem{tar},H,t} - V_{\rbindexsmallem{est},r,t}\delta_{r,s} + V_{\rbindexsmallem{risk},r,t}\delta_{r,s}\\
			&=\simpleexpectation{\left\langle \widehat{\vectorfont{\xi}}_{r\mid H,t},\widehat{\vectorfont{\xi}}_{s\mid H,t}\right\rangle - \widehat{V}_{\rbindexsmallem{tar},H,t} + \left(V_{\rbindexsmallem{risk},r,t}- \widehat{V}_{\rbindexsmallem{est},r,t}\right)\delta_{r,s}}.
		\end{aligned}
	}
\end{equation} 
Thus, \advancedCorretedWeight{} can construct unbiased matrix
\begin{equation} \label{eq:block-general}
		\outereqsize{
			\begin{aligned}
				\widehat{\vectorfont{M}}_{H,t}(r,s) = \left\langle \widehat{\vectorfont{\xi}}_{r\mid H,t},\widehat{\vectorfont{\xi}}_{s\mid H,t}\right\rangle - \widehat{V}_{\rbindexsmallem{tar},H,t} + \left(V_{\rbindexsmallem{risk},r,t}- \widehat{V}_{\rbindexsmallem{est},r,t}\right)\delta_{r,s}.
			\end{aligned}
		}
\end{equation} 

Finite-sample fluctuations may make
$\widehat{\vectorfont{M}}_{H,t}$ indefinite. \advancedCorretedWeight{} therefore clips its
eigenvalues at a fixed threshold $\mu>0$: $\widetilde{\vectorfont{M}}_{H,t}=\vectorfont{Q}\operatorname{diag}\!\left(\max\{\nu_j,\mu\}\right)
\vectorfont{Q}^{\top}
\succeq \mu\vectorfont{I}$.
We denote this process as \basicClippingTotalName{} (\basicClipping{}).
\advancedCorretedWeight{} invokes \basicClipping{} and then get the fusion weights by solving $\vectorfont{\lambda}_{H,t}^{*}=\arg\min_{\vectorfont{\lambda}\succeq\vectorfont{0},\,\vectorfont{1}^{\top}\vectorfont{\lambda}=1}
\vectorfont{\lambda}^{\top}
\widetilde{\vectorfont{M}}_{H,t}
\vectorfont{\lambda}$.

\begin{algorithm}[!t]\small \DontPrintSemicolon
	\caption{\advancedCorretedWeightTotalName{} (\advancedCorretedWeight{})}
	\label{alg:pa-gcw}
	\KwIn{
		$\widehat{\vectorfont{F}}_{\rbindexem{est}}$; $\widehat{\vectorfont{V}}_{\rbindexem{est}}$; $\widehat{\vectorfont{F}}_{\rbindexem{tar}}$; $\widehat{\vectorfont{V}}_{\rbindexem{tar}}$; $\vectorfont{V}_{\rbindexem{risk}}$;
		total groups $\vectorfont{PG}$;
		source sets $\vectorfont{\entity{S}}$;
		clipping threshold $\mu$;
		auxiliary status set $\vectorfont{ST}=\{ST_r\mid PG_r\in\vectorfont{PG}, ST_R\in\{\operatorname{EQ}, \operatorname{ND}\}\}$ for $\vectorfont{PG}$
	}
	\KwOut{Fusion weights $\vectorfont{\Lambda}^{*}$}

	\For{$PG_H\in\vectorfont{PG}$}{ 
		Set $\entity{C}_H\gets\{PG_H\}\cup \entity{S}_H$;\\
		\If{$ST_H=\operatorname{EQ}$}{
			Construct blocked $\widehat{M}_H$ by Eq.~\eqref{eq:risk-block};\\
		}\Else{
			Construct general $\widehat{M}_H$ by Eq.~\eqref{eq:block-general};\\
		}
		
		$\widetilde{\vectorfont{M}}_{H}\gets\operatorname{\basicClipping{}}(\widehat{\vectorfont{M}}_{H},\mu)$;\\ 	
		Obtain $\vectorfont{\lambda}_{H}^{*}\gets\arg\min_{\vectorfont{\lambda}\succeq 0,\vectorfont{1}^\top\vectorfont{\lambda}=1}
		\vectorfont{\lambda}^\top\widetilde{\vectorfont{M}}_{H}\vectorfont{\lambda}$;\\	
		
		Add $\vectorfont{\lambda}_{H}^{*}$ to
		$\vectorfont{\Lambda}^{*}$;\\
	}
	
	\Return{$\vectorfont{\Lambda}^{*}$}	
\end{algorithm}


\noindent
\textbf{Population-Level Aggregation.}
Let $\pi_H=|\vectorfont{U}_H|/n$ denote the population proportion of
$PG_H$. Because privacy groups may have different distributions, \advancedEnhance{}
aggregates the fused group estimates using $\pi_H$ rather than
inverse-variance weights of PFO. With
$\gamma_{r,t}=\sum_{H=1}^{K}\pi_H\lambda_{H,r,t}^{*}$, the population
estimate is
\begin{equation}\notag 
	\outereqsize{
		\begin{aligned}
			\widetilde{\vectorfont{f}}_{t} = \sum_{H=1}^{K}\pi_H\cdot\widetilde{\vectorfont{f}}_t^{(H)}
			= \sum_{r=1}^{K}\left(\sum_{H=1}^{K}\pi_H\lambda_{H,r,t}^*\right) \widehat{\vectorfont{g}}_{\rbindexsmallem{risk},t}^{(r)} 
			= \sum_{r=1}^{K} \gamma_{r,t}\cdot\widehat{\vectorfont{g}}_{\rbindexsmallem{risk},t}^{(r)}.
		\end{aligned}
	}
\end{equation}

For
$\vectorfont{f}_t=\sum_{r=1}^{K}\pi_r\vectorfont{f}_t^{(r)}$,
the average per-domain mean-squared error is
\begin{equation}\label{eq:original-corrent-error}
	\outereqsize{
		\begin{aligned}
			\shrinkem{err}_t^+ 
			=& \frac{1}{d}\simpleexpectation{\left\| \widetilde{\vectorfont{f}}_t-\vectorfont{f}_t \right\|_2^2}\\
			=& \frac{1}{d}\left(\sum_{r=1}^{K}\gamma_{r,t}^2V_{\rbindexem{risk},r,t}+\left\| \sum_{r=1}^{K}\gamma_{r,t}\cdot\vectorfont{g}_{t}^{(r)}-\sum_{r=1}^{K}\pi_r\vectorfont{f}_t^{(r)} \right\|_2^2\right).\\
		\end{aligned}
	}
\end{equation}

\subsection{Personalized Local Population Enhanced Solutions}\label{subsec:plpd+plpa+}

\begin{table}[t]
	\centering
	\small
	\caption{Inputs supplied to PAA-CW under different publication modes.}
	\label{tab:paacw-inputs}
\resizebox{\linewidth}{!}{
	\begin{tabular}{c|ccccc}
		\hline
		\multirow{2}{*}{\textbf{Cases}} & \multicolumn{5}{c}{$PG_r\in\vectorfont{PG}_t^{(A)}$}                                                                                                                                                                                                                                                                            \\ \cline{2-6} 
		& \multicolumn{1}{c|}{$\widehat{\vectorfont{f}}_{\rbindexem{est}}^{(r)}$}     & \multicolumn{1}{c|}{$\widehat{\vectorfont{f}}_{\rbindexem{tar}}^{(r)}$}   & \multicolumn{1}{c|}{$\widehat{V}_{\rbindexem{est},r}$}     & \multicolumn{1}{c|}{$\widehat{V}_{\rbindexem{tar},r}$}   & $V_{\rbindexem{risk},r}$                   \\ \hline
		\solutionC{}-Pre       & \multicolumn{2}{c|}{$\widehat{\vectorfont{f}}_{\rbindexem{dis},t}^{(r)}$}                                                                               & \multicolumn{2}{c|}{$\widehat{V}_{\rbindexem{dis},r,t}$}                                                              & $V_{\rbindexem{pub},r,t}^{\operatorname{Pre}}$ \\ \hline
		\solutionC{}-Post      & \multicolumn{2}{c|}{$\widehat{\vectorfont{f}}_{\rbindexem{pub},t}^{(r)}$}                                                                               & \multicolumn{2}{c|}{$\widehat{V}_{\rbindexem{pub},r,t}$}                                                              & $\widehat{V}_{\rbindexem{pub},r,t}$                      \\ \hline
		\solutionD{}-Pre       & \multicolumn{2}{c|}{$\widehat{\vectorfont{f}}_{\rbindexem{dis},t}^{(r)}$}                                                                               & \multicolumn{2}{c|}{$\widehat{V}_{\rbindexem{dis},r,t}$}                                                              & $V_{\rbindexem{pub},r,t}^{\operatorname{Pre}}$ \\ \hline
		\solutionD{}-Post      & \multicolumn{2}{c|}{$\widehat{\vectorfont{f}}_{\rbindexem{pub},t}^{(r)}$}                                                                               & \multicolumn{2}{c|}{$\widehat{V}_{\rbindexem{pub},r,t}$}                                                              & $\widehat{V}_{\rbindexem{pub},r,t}$                      \\ \hline
		\multirow{2}{*}{\textbf{Cases}} & \multicolumn{5}{c}{$PG_r\in\vectorfont{PG}_t^{(R)}$}                                                                                                                                                                                                                                                                            \\ \cline{2-6} 
		& \multicolumn{1}{c|}{$\widehat{\vectorfont{f}}_{\rbindexem{est}}^{(r)}$}     & \multicolumn{1}{c|}{$\widehat{\vectorfont{f}}_{\rbindexem{tar},r}$}       & \multicolumn{1}{c|}{$\widehat{V}_{\rbindexem{est},r}$}     & \multicolumn{1}{c|}{$\widehat{V}_{\rbindexem{tar},r}$}   & $V_{\rbindexem{risk},r}$                       \\ \hline
		\solutionD{}-Pre       & \multicolumn{1}{c|}{$\widehat{\vectorfont{f}}_{\rbindexem{pub},l_r}^{(r)}$} & \multicolumn{1}{c|}{$\widehat{\vectorfont{f}}_{\rbindexem{dis},t}^{(r)}$} & \multicolumn{1}{l|}{$\widehat{V}_{\rbindexem{pub},r,l_r}$} & \multicolumn{1}{c|}{$\widehat{V}_{\rbindexem{dis},r,t}$} & $\widehat{V}_{\rbindexem{pub},r,l_r}$                    \\ \hline
		\solutionD{}-Post      & \multicolumn{1}{c|}{$\widehat{\vectorfont{f}}_{\rbindexem{pub},l_r}^{(r)}$} & \multicolumn{1}{c|}{$\widehat{\vectorfont{f}}_{\rbindexem{dis},t}^{(r)}$} & \multicolumn{1}{l|}{$\widehat{V}_{\rbindexem{pub},r,l_r}$} & \multicolumn{1}{c|}{$\widehat{V}_{\rbindexem{dis},r,t}$} & $\widehat{V}_{\rbindexem{pub},r,l_r}$                    \\ \hline
	\end{tabular}
}
\end{table}

Both \solutionC{} and \solutionD{} use a variance-corrected plug-in estimate to decide
whether a new population-level publication is necessary. 
The ideal risk $\shrinkem{err}_t^+$ in Eq.~\eqref{eq:original-corrent-error} contains unknown current and historical group distributions. 
We first derive the observable risk for general mixed current--historical candidates used by \solutionD{} and then discuss the current candidates used by \solutionC{}.

\noindent
\textbf{Risk Estimation.} We divide the current privacy groups $\vectorfont{PG}_t$ as two parts: (1) the \textit{active groups} $\vectorfont{PG}_t^{(A)}$; and (2) the \textit{reused groups} $\vectorfont{PG}_t^{(R)}$. 
Then $\shrinkem{err}_t^+$ can be further expressed as 
\begin{equation}\notag
	\outereqsize{
		\begin{aligned}
			\shrinkem{err}_t^+ 
			=& \frac{1}{d}\Bigg(\sum_{r=1}^{K}\gamma_{r,t}^2V_{\rbindexsmallem{risk},r,t}+\Bigg\| \sum_{PG_r\in\vectorfont{PG}_t^{(A)}}\left(\gamma_{r,t}-\pi_{r}\right)\vectorfont{f}_t^{(r)} \\
			&+ \sum\nolimits_{PG_r\in\vectorfont{PG}_t^{(R)}}\left(\gamma_{r,t}\vectorfont{f}_{l_r}^{(r)}-\pi_r\vectorfont{f}_t^{(r)}\right) \Bigg\|_2^2\Bigg).\\
		\end{aligned}
	}
\end{equation}
Let $B(\vectorfont{f}_t,\vectorfont{f}_{l_r})=\sum_{PG_r\in\vectorfont{PG}_t^{(A)}}\left(\gamma_{r,t}-\pi_{r}\right)\vectorfont{f}_t^{(r)} + \sum_{PG_r\in\vectorfont{PG}_t^{(R)}}(\gamma_{r,t}\vectorfont{f}_{l_r}^{(r)}-\pi_{r}\vectorfont{f}_t^{(r)})$.
Let $\vectorfont{H}_t=\{\vectorfont{h}_t^{(r)}\mid r\in[K]\}$, 
where $\vectorfont{h}_t^{(r)}$ is an unbiased proxy for $\vectorfont{f}_t^{(r)}$; \solutionC{} and \solutionD{} instantiate it with the group-level dissimilarity estimate $\widehat{\vectorfont{f}}_{\rbindexem{dis},t}$ (Table~\ref{tab:paacw-inputs}), whose variance is estimated by Theorem~\ref{thm:precise_var}.
Let $\widehat{\vectorfont{F}}_{l,t}=\{\widehat{\vectorfont{f}}_{l_r}^{(r)}\mid PG_r\in\vectorfont{PG}_t^{(R)}\}$ contain the latest historical estimates of reused groups.
For $B(\vectorfont{f}_t,\vectorfont{f}_{l_r})$, we have 
\begin{equation}
	\outereqsize{
		\begin{aligned}\notag 
			&\left\| B\left(\vectorfont{f}_t,\vectorfont{f}_{l_r}\right)  \right\|_2^2 \\
			=& \simpleexpectation{\left\| B\left(\vectorfont{H}_t,\widehat{\vectorfont{F}}_{l,t}\right) \right\|_2^2} 
			- \sum\nolimits_{PG_r\in\vectorfont{PG}_t^{(A)}} \left(\gamma_{r,t}-\pi_r\right)^2\var{\vectorfont{h}_t^{(r)}} \\
			&- \sum\nolimits_{PG_r\in\vectorfont{PG}_t^{(R)}}\left(\pi_r^2\var{\vectorfont{h}_t^{(r)}}+\gamma_{r,t}^2\var{\widehat{\vectorfont{f}}_{l_r}^{(r)}}\right). 
		\end{aligned}
	}
\end{equation}
By setting  the variance-corrected squared-bias estimate as:
\begin{equation}\notag
	\outereqsize{
		\begin{aligned}
			\widehat{B}^2_{t} =& \max\Big\{0,\left\| B\left(\vectorfont{H}_t,\widehat{\vectorfont{F}}_{l,t}\right) \right\|_2^2-\sum\nolimits_{r\in\vectorfont{PG}_t^{(A)}}\left(\gamma_{r,t}-\pi_r\right)^2\var{\vectorfont{h}_{t}^{(r)}}\\
			&-\sum\nolimits_{r\in\vectorfont{PG}_t^{(R)}}\left(\pi_r^2\var{\vectorfont{h}_{t}^{(r)}}+\gamma_{r,t}^2\var{\widehat{\vectorfont{f}}_{l_r}^{(r)}}\right)\Big\}. 
		\end{aligned}
	}
\end{equation}
We can get the corrected error:
\begin{equation}\label{eq:final_corrected_time_error}
	\outereqsize{
		\begin{aligned}
			\widetilde{err}^{+}\left(\vectorfont{H}_t,\vectorfont{V}_{\rbindexsmallem{risk}}\right) = \frac{1}{d}\left(\sum\limits_{r\in\vectorfont{PG}_t^{(A)}}\gamma_{r,t}^2 V_{\rbindexsmallem{risk},r}^{(A)}
			+\sum\limits_{r\in\vectorfont{PG}_t^{(R)}}\gamma_{r,t}^2 V_{\rbindexsmallem{risk},r}^{(R)}+\widehat{B}^2_{t}\right),
		\end{aligned}
	}
\end{equation}
where $\vectorfont{V}_{\rbindexsmallem{risk}}=\vectorfont{V}_{\rbindexsmallem{risk}}^{(A)}\cup\vectorfont{V}_{\rbindexsmallem{risk}}^{(R)}$ is composed of the active variance set and reused variance set. 
Eq.~\eqref{eq:final_corrected_time_error} gives a variance-corrected plug-in estimate for mixed current-and-historical publication candidates; it is not claimed to be strictly unbiased because the fusion weights and bias estimate depend on noisy data.

When $\vectorfont{g}_t=\vectorfont{f}_t$, $B\left(\vectorfont{f}_t,\vectorfont{f}_{l_r}\right)$ degrade into 
$B\left(\vectorfont{f}_t\right)=\sum_{r=1}^{K}\left(\gamma_{r,t}-\pi_{r}\right)\vectorfont{f}_t^{(r)}$.
Consequently, 
\begin{equation}\label{eq:final_corrected_error}
	\outereqsize{
		\begin{aligned}
			\widetilde{err}^+(\vectorfont{H}_t,\vectorfont{V}_{\rbindexsmallem{risk}}) &= \frac{1}{d}\left(\sum_{r=1}^{K}\gamma_{r,t}^2 V_{\rbindexsmallem{risk},r}+\widehat{B}_t^2\right),
		\end{aligned}
	}
\end{equation}
where 
\begin{equation}\notag
	\outereqsize{
		\begin{aligned}
			\widehat{B}_t^2 &= \max\left\{0,\left\|\sum_{r=1}^{K}(\gamma_{r,t}-\pi_r)\vectorfont{h}^{(r)}\right\|_2^2-\sum_{r=1}^{K}(\gamma_{r,t}-\pi_r)^2\var{\vectorfont{h}^{(r)}}\right\}.
		\end{aligned}
	}
\end{equation}

Both \solutionC{} and \solutionD{} combine
\advancedSelection{} and \advancedEnhance{} with personalized
population division. They share the same dissimilarity-estimation
stage and the same \advancedCorretedWeight{} interface but differ in
their publication schedules: \solutionC{} updates all privacy groups
synchronously, whereas \solutionD{} updates active groups and reuses
the latest publications for the remaining groups.

\begin{figure*}[!t]\vspace{-2ex}
	\centering
	\includegraphics[width=1\textwidth]{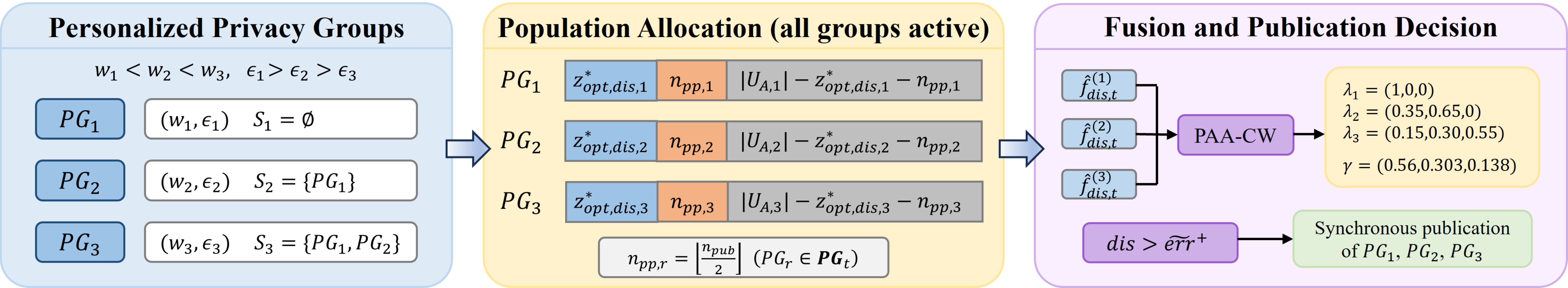}
	\caption{A \solutionC{}-Pre example: all groups are active and synchronously updated.}\label{fig:plpdplus-example}
	\vspace{-2ex}
\end{figure*}

\begin{algorithm}[!t]\small \DontPrintSemicolon
	\caption{\solutionC{} Stream Estimation}
	\label{alg:PLPD+}
	\DontPrintSemicolon
	\KwIn{Available user set ${U_A}$; privacy requirements ${(\vectorfont{w},\vectorfont{\epsilon})}$; 
		history ${(\widehat{\vectorfont{f}}_{{1}},\ldots, \widehat{\vectorfont{f}}_{{t-1}})}$; 
		${\mathsf{mode}_{\rbindexem{fsn}}\in\{\operatorname{Pre},\operatorname{Post}\}}$} 
	\KwOut{${\widehat{\vectorfont{f}}_t}$}
	Initialize $\widehat{\vectorfont{F}}_{\rbindexem{pub},t},\vectorfont{V}_{\rbindexem{pub},t}^{\operatorname{Pre}},\widehat{\vectorfont{V}}_{\rbindexem{pub},t},\widehat{\vectorfont{F}}_{\rbindexem{dis},t},\widehat{\vectorfont{V}}_{\rbindexem{dis},t}$ as empty;\\ \label{alg:plpd+init_start}
	Construct $\vectorfont{PG}$, $\vectorfont{PGT}$, and ${\vectorfont{\entity{S}}\gets\left\{\entity{S}_r\mid r\in[K]\right\}}$; \tcp*[f]{once} \\ \label{alg:pg} \label{alg:source}\label{alg:plpd+init_end}
	\tcp{sub-mechanism $\entity{M}_{dis,t}$}
	${(\vectorfont{z}^*_{\rbindexem{opt,dis}},\vectorfont{\tau}^*, \vectorfont{\epsilon}')\gets\operatorname{\advancedSelection}(\vectorfont{PGT},\frac{1}{2},d)}$ (Alg.~\ref{alg:adv_ops}); \tcp*[f]{once}\\ \label{alg:apply_ops}\label{alg:plpd+dis_start} 
	\For{${PG_r\in\vectorfont{PG}}$}{
		Sample ${U_{\rbindexem{dis},r,t}}$ of size ${z^*_{\rbindexem{opt,dis,r}}}$ from ${U_{A,r}}$;\\ \label{alg:sampling+}
		${U_{A,r}\gets U_{A,r}\backslash U_{\rbindexem{dis},r,t}}$; \\ 
		${\widehat{\vectorfont{f}}^{(r)}_{\rbindexem{dis},t}\gets\shrink{FO}(D_{\rbindexem{dis},r,t},\epsilon'_{r})}$, and
		compute ${\widehat{\vectorfont{f}}^{(r)}_{\rbindexem{dis},t}}$'s unbiased vector-level variance estimate ${\widehat{V}_{\rbindexem{dis},r,t}}$ by Eq.~\ref{thm:precise_var}; \\ \label{alg:PFO.P+}
		Add ${\widehat{\vectorfont{f}}^{(r)}_{\rbindexem{dis},t}}$ to ${\widehat{\vectorfont{F}}_{\rbindexem{dis},t}}$ and ${\widehat{V}_{\rbindexem{dis},r,t}}$ to  ${\widehat{\vectorfont{V}}_{\rbindexem{dis},t}}$;
	}
	Aggregate ${\widehat{\vectorfont{f}}_{\rbindexem{dis},t}\gets\sum_{r\in[K]}\pi_r\widehat{\vectorfont{f}}^{(r)}_{\rbindexem{dis},t}}$;\\ \label{alg:estimate_group_combine}
	${\shrinkem{dis}\gets \frac{1}{d}\sum_{j=1}^{d}\left(\widehat{\vectorfont{f}}_{\rbindexem{dis},t}[j]-\widehat{\vectorfont{f}}_l[j]\right)^2-\frac{1}{d}\sum_{r=1}^{K}\pi_r^2 \widehat{V}_{\rbindexem{dis},r,t}}$;\\ \label{alg:dis+} \label{alg:plpd+dis_end}
	\tcp{sub-mechanism $\entity{M}_{pub,t}$}
	\For{${PG_r\in\vectorfont{PG}}$}{
		Compute ${n_{\rbindexem{pub},r}\gets \left\lfloor n_r/2\right\rfloor-\sum_{\tau=t-\tau^*_{r}+1}^{t-1}|U_{{\rbindexem{pub},r,\ensuremath{\tau}}}|}$;\\ \label{alg:repop:start+} 
		Set ${n_{\rbindexem{pp},r}\gets \left\lfloor n_{\rbindexem{pub,r}}/2\right\rfloor}$, and $ST_r\gets\operatorname{EQ}$;\\ \label{alg:group_half_remain}
		${V_{\rbindexem{pub},r,t}^{\operatorname{Pre}}\gets V_{\rbindexem{smpl}}\left(\left\lfloor n_r/2\right\rfloor,n_{\rbindexem{pp},r}\right)+V_{\rbindexem{LDP}}\left(\epsilon'_r,n_{\rbindexem{pp},r},d\right)}$;\\ \label{alg:pre:end} \label{alg:plpd+:v_pub_pre}
		Add ${V_{\rbindexem{pub},r,t}^{\operatorname{Pre}}}$ to ${\vectorfont{V}_{\rbindexem{pub},t}^{\operatorname{Pre}}}$; \\
		\label{alg:plpd+:v_pub_pre_add}
	}
	Construct Pre inputs according to Table~\ref{tab:paacw-inputs};\\
	$\vectorfont{\Lambda}^{\operatorname{Pre}}_{t}\gets\operatorname{\advancedCorretedWeight{}}\left(\text{Pre inputs}\right)$ (Alg.~\ref{alg:pa-gcw});\\ \label{alg:pre_weight}
	Compute publication error ${\widetilde{\shrinkem{err}}^{+}}(\widehat{\vectorfont{F}}_{\rbindexem{dis},t},\vectorfont{V}_{\rbindexem{pub},t}^{\operatorname{Pre}})$ using Eq.~\eqref{eq:final_corrected_error}; \\ \label{alg:repop:end+}
	
	\If{${\shrink{dis}>\widetilde{\shrink{err}}^{+}}$ and ${\min_{r\in[K]}n_{\rbindex{pp},r}\geq u_{\rbindex{min}}}$}{\label{line:diserrjudge_start+}
		\For{${PG_r\in\vectorfont{PG}}$}{\label{alg:begin_pub_f}
			Sample ${U_{\rbindexem{pub},r,t}}$ of size ${n_{\rbindexem{pp},r}}$ from  ${U_{A,r}}$;\\ \label{alg:sampleinadvance}
			${U_{A,r}\gets U_{A,r}\backslash U_{\rbindexem{pub},r,t}}$; \\ \label{alg:smpl+}
			${\widehat{\vectorfont{f}}^{(r)}_{\rbindexem{pub},t}\gets\shrink{FO}(D_{\rbindexem{pub},r,t},\epsilon'_{r})}$ with its unbiased ${\widehat{V}_{\rbindexem{pub},r,t}}$; \\
			Add ${\widehat{\vectorfont{f}}^{(r)}_{\rbindexem{pub},t}}$ to ${\widehat{\vectorfont{F}}_{\rbindexem{pub},t}}$ and ${\widehat{V}_{\rbindexem{pub},r,t}}$ to ${\widehat{\vectorfont{V}}_{\rbindexem{pub},t}}$;
			\label{alg:end_pub_f}
		}
		\If{${\mathsf{mode}_{\rbindex{fsn}}=\operatorname{Post}}$}{\label{alg:plpd+:mode_start} 
			Construct Post inputs according to Table~\ref{tab:paacw-inputs};\\
			$\vectorfont{\Lambda}^{\operatorname{Post}}_{t}\gets\operatorname{\advancedCorretedWeight{}}\left(\text{Post inputs}\right)$ (Alg.~\ref{alg:pa-gcw});\\ \label{alg:plpd+post_start}
			Set ${\vectorfont{\Lambda}^{(\rbindexem{fnl})}_{t}\gets\vectorfont{\Lambda}^{\operatorname{Post}}_{t}}$;\label{alg:plpd+post_end}
		} \Else{
			Set ${\vectorfont{\Lambda}^{(\rbindexem{fnl})}_{t}\gets\vectorfont{\Lambda}^{\operatorname{Pre}}}$;\label{alg:plpd+pre}
		} \label{alg:plpd+:mode_end}
		Fuse $\widetilde{\vectorfont{F}}_{\rbindexem{pub},t}$ by Eq.~\eqref{eq:linear-combine} with $\vectorfont{\Lambda}^{(\rbindexem{fnl})}_t$;\\
		\label{alg:ba-pue}
		Aggregate $\widehat{\vectorfont{f}}_{\rbindexem{pub},t}\gets\Pi_{\Delta}\big(\sum_{r\in[K]}\pi_r\widetilde{\vectorfont{f}}^{(r)}_{\rbindexem{pub},t}\big)$;\\ \label{alg:final_new_estimates}
	} \Else { \label{alg:plpd+:reuse_start}
		Set $\widehat{\vectorfont{f}}_{\rbindexem{pub},t}\leftarrow\widehat{\vectorfont{f}}_{\rbindexem{pub},t-1}$;\\ \label{alg:approximate} \label{alg:plpd+:reuse_end}
	}\label{line:diserrjudge_end+}
	\For{${PG_r\in\vectorfont{PG}}$}{ \label{alg:begin-recycle-users}
		\If{${t\geq \tau^*_{r}}$}{
			${U_{A,r}\gets U_{A,r}\cup U_{\rbindexem{dis},r,t-\tau^*_{r}+1}\cup U_{\rbindexem{pub},r,t-\tau^*_{r}+1}}$; \\ \label{alg:recycle}
		}
	}\label{alg:end-recycle-users}
	\Return{${\widehat{\vectorfont{f}}_{\rbindex{pub},t}}$.}
\end{algorithm}

\noindent
\textbf{\solutionCTotalName{}: Synchronous Group Updates.}
In \solutionC{}, the available user set $U_A$ is initialized as total user set $U$. 
All groups provide current candidates, i.e., $\vectorfont{g}_t^{(r)}=\vectorfont{f}_t^{(r)}$, eliminating temporal-staleness bias. At each slot, \advancedSelection{} determines group-specific reporting populations, and \advancedCorretedWeight{} computes fusion weights using the inputs in Table~\ref{tab:paacw-inputs}. A synchronous publication occurs only when $\shrinkem{dis}>\widetilde{\shrinkem{err}}^{+}$ in Eq.~\eqref{eq:final_corrected_error} and all groups satisfy the minimum-population requirement. Pre retains the planned fusion weights, whereas Post recomputes them from the realized publication reports. Alg.~\ref{alg:PLPD+} summarizes the procedure, where $\Pi_\Delta$ is the Euclidean projection onto the probability simplex.

\begin{figure*}[!t]
	\centering
	\includegraphics[width=1\textwidth]{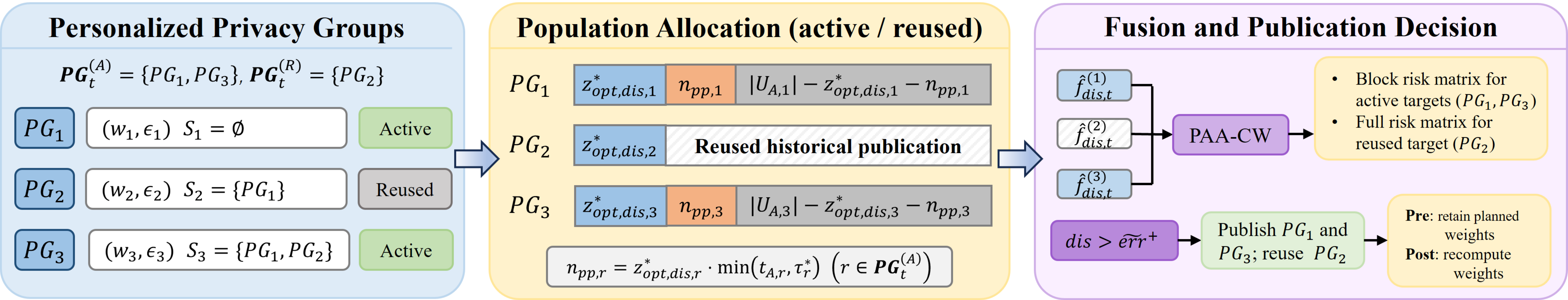}
	\caption{A \solutionD{} example: active groups form candidate publication populations, whereas groups within their nullified intervals reuse historical publications; new reports are generated only if the publication test is satisfied.}\label{fig:plpaplus-example}\vspace{-2ex}
\end{figure*}

\begin{example}\label{exmpl:solutionC}
	Consider three privacy groups satisfying
	$w_1<w_2<w_3,\qquad \epsilon_1>\epsilon_2>\epsilon_3$,
	so that $PG_1\preceq_p PG_2\preceq_p PG_3$, with
	$S_1=\emptyset$, $S_2=\{PG_1\}$, and
	$S_3=\{PG_1,PG_2\}$.
	Let $d=4$ and $\pi=(0.4,0.35,0.25)$. 
	\advancedSelection{} independently determines $(z_{\rbindex{opt},\rbindex{dis},r}^*,\tau_r^*,\epsilon_r')$ for each group, 
	yielding the group-specific population allocation illustrated in Figure~\ref{fig:plpdplus-example}. 
	In \solutionC{}-$\operatorname{Pre}$, \advancedCorretedWeight{} gives
	$\lambda_1=(1,0,0)$,
	$\lambda_2=(0.35,0.65,0)$, and
	$\lambda_3=(0.15,0.30,0.55)$,
	yielding $\gamma=(0.56,0.303,0.138)$.
	If $dis>\widetilde{err}^{+}$, \solutionC{} triggers a synchronous publication for all groups. 	
\end{example}

\noindent
\textbf{\solutionDTotalName{}: Partial Group Updates with Historical Reuse.}
Let $l_r$ denote the latest publication time of group $PG_r$, and let $\widehat{\vectorfont{f}}_{\rbindexem{pub},l_r}^{(r)}$ and $U_{\rbindexem{pub},l_r}$ denote its latest group-level publication and reporting users, respectively. These quantities are maintained as stream state.
\solutionD{} improves population utilization by updating privacy groups independently. At each time slot, groups whose nullified intervals have not expired reuse their latest publications, whereas active groups absorb the populations accumulated since their previous publications to form new candidate publication populations. 
By setting auxiliary estimates and variances in Table~\ref{tab:paacw-inputs}, \advancedCorretedWeight{} therefore operates on mixed current–historical candidates: Pre uses current dissimilarity estimates for active groups and historical publications for reused groups, while Post replaces the active inputs with their realized publication estimates. Alg.~\ref{alg:PLPA+} summarizes the complete procedure.

\begin{algorithm}[!t]\small \DontPrintSemicolon
	\caption{\solutionD{} Stream Estimation}
	\label{alg:PLPA+}
	\DontPrintSemicolon
	\KwIn{Available user set ${U_A}$; privacy requirements ${(\vectorfont{w},\vectorfont{\epsilon})}$;
	history ${(\widehat{\vectorfont{f}}_{{1}},\ldots, \widehat{\vectorfont{f}}_{{t-1}})}$;  
	${\mathsf{mode_{\rbindexem{fsn}}}\in\{\operatorname{Pre},\operatorname{Post}\}}$} 
	\KwOut{${\widehat{\vectorfont{f}}_t}$}
	The same as Lines~\ref{alg:plpd+init_start}-\ref{alg:plpd+dis_end} in Alg.~\ref{alg:PLPD+};\\ \label{alg:plpa+:repeat-1}\label{alg:plpa+:repeat-2}
	\tcp{sub-mechanism ${\entity{M}_{pub,t}}$}
	Initialize ${\vectorfont{PG}_t^{(A)}\gets\emptyset}$;\\
	\For{${PG_r\in\vectorfont{PG}}$}{
		Calculate ${t_{N,r}=\frac{\big|U_{\rbindexem{pub},r,l_r}\big|}{z^*_{\rbindexem{opt,dis},r}}-1}$;\\ \label{alg:nullifiedrightbound}
		\If{${t-l_r\leq t_{N,r}}$}{ \label{line:judgenullified+}
			Set $ST_r\gets\operatorname{ND}$;\\
			Add ${\widehat{\vectorfont{f}}^{(r)}_{\rbindexem{pub},l_r}}$ to ${\widehat{\vectorfont{F}}_{\rbindexem{pub},t}}$,
			${\widehat{V}_{\rbindexem{pub,r,l}_r}}$ to ${\vectorfont{V}_{\rbindexem{pub},t}^{\operatorname{Pre}}}$ and ${\widehat{\vectorfont{V}}_{\rbindexem{pub},t}}$;\\ \label{alg:plpa+reused}
		} \Else{
			Calculate ${t_{A,r}=t-l_r-t_{N,r}}$;\\ \label{line:canbeabsorbed+}
			Set ${n_{\rbindexem{pp},r}\gets z_{\rbindexem{opt,dis,r}}^*\cdot\min(t_{A,r},\tau^*_{r})}$;\\ \label{line:absorbed+}
			${V_{\rbindexem{pub},r,t}^{\operatorname{Pre}}\gets V_{\rbindexem{smpl}}\left(\left\lfloor\frac{n_r}{2}\right\rfloor,n_{\rbindexem{pp},r}\right)+V_{\rbindexem{LDP}}\left(\epsilon'_r,n_{\rbindexem{pp},r},d\right)}$;\\
			Set $ST_r\gets\operatorname{EQ}$;\\
			Add ${V_{\rbindexem{pub},r,t}^{\operatorname{Pre}}}$ to ${\vectorfont{V}_{\rbindexem{pub},t}^{\operatorname{Pre}}}$, ${PG_r}$ to ${\vectorfont{PG}_t^{(A)}}$;\\ \label{alg:plpa+pga}
		}
	}
	Set $\widehat{\vectorfont{f}}_{\rbindexem{pub},t}\leftarrow\widehat{\vectorfont{f}}_{\rbindexem{pub},t-1}$;\\ \label{alg:plpa+allnullified}
	\If{${\vectorfont{PG}_t^{(A)}\neq\emptyset}$}{
		Construct Pre inputs according to Table~\ref{tab:paacw-inputs};\\
		$\vectorfont{\Lambda}^{\operatorname{Pre}}_{t}\gets\operatorname{\advancedCorretedWeight{}}\left(\text{Pre inputs}\right)$ (Alg.~\ref{alg:pa-gcw});  \\ \label{alg:pre_weight_plpa+}
		Compute $\widetilde{\shrinkem{err}}^{+}(\widehat{\vectorfont{F}}_{\rbindexem{dis},t},\vectorfont{V}_{\rbindexem{pub},t}^{\operatorname{Pre}})$ using Eq.~\eqref{eq:final_corrected_time_error}; \\ \label{alg:repop_plpa:end+}
		\If{$\shrink{dis}>\widetilde{\shrink{err}}^{+}$ and $\min_{PG_r\in\vectorfont{PG}_t^{(A)}}n_{\rbindex{pp},r}\geq u_{\rbindex{min}}$}{\label{line:diserrjudge_plpa+_start}
			\For{${PG_r\in\vectorfont{PG}_t^{(A)}}$}{
				The same as Lines~\ref{alg:sampleinadvance}--\ref{alg:end_pub_f} in Alg.~\ref{alg:PLPD+};\\
			}
			The same as Lines~\ref{alg:plpd+:mode_start}--\ref{alg:final_new_estimates} in Alg.~\ref{alg:PLPD+};
		} \label{line:diserrjudge_plpa+_end}
	}
	Recycle users (The same as Lines~\ref{alg:begin-recycle-users}--\ref{alg:end-recycle-users} in Alg.~\ref{alg:PLPD+});\\
	\Return{${\widehat{\vectorfont{f}}_{\rbindex{pub},t}}$}.
\end{algorithm}


\begin{example}\label{exmpl:solutionD}
Continue with the three groups in Example~\ref{exmpl:solutionC}. 
Suppose that at time $t$, $PG_2$ remains in its nullified interval, whereas the intervals of $PG_1$ and $PG_3$ have expired; 
hence $PG_t^{(A)}=\{PG_1,PG_3\}$ and $PG_t^{(R)}=\{PG_2\}$, as illustrated in Figure~\ref{fig:plpaplus-example}. 
Each active group absorbs the population accumulated since its latest publication and forms a potential publication population $n_{\rbindex{pp},r}=z_{\rbindex{opt},\rbindex{dis},r}^* \cdot\min\{t_{A,r},\tau_r^*\}$,
whereas $PG_2$ generates no new publication and reuses $\widehat{\vectorfont{f}}_{\rbindex{pub},l_2}^{(2)}$.

In Pre, the active groups use their current dissimilarity estimates for \advancedCorretedWeight{}. 
For the reused $PG_2$, $\widehat{\vectorfont{f}}_{\rbindex{dis},t}^{(2)}$ represents its current target distribution while $\widehat{\vectorfont{f}}_{\rbindex{pub},l_2}^{(2)}$ remains the publication candidate; their mismatch is handled by the full risk matrix to account for temporal-staleness bias. If $dis>\widetilde{err}^{+}$ and the active groups satisfy $n_{\rbindex{pp},r}\ge u_{\rbindex{min}}$, $PG_1$ and $PG_3$ publish new estimates while $PG_2$ is reused. Pre retains the planned weights, whereas Post recomputes them from the realized publication estimates.

\end{example}

\subsection{Analysis}\label{subsec:analysis}

\noindent
\textbf{Time Complexity Analysis.}
Let $d$ denote the domain size, $K$ the
number of privacy groups, and $n$ the total number of users. 
For each target privacy group $PG_H$, let $q_H=|\entity{C}_H|=|\entity{S}_H|+1$, and define $q_{\rbindexem{max}}=\max_{H\in[K]}q_H\leq K$.
Let $I_{\rbindexem{QP}}$ denote the maximum number of iterations required by the active-set simplex-QP solver.
\begin{theorem}\label{thm:advanced_time_cost_analysis}
	The per-time-slot time complexity of both \solutionC{} and \solutionD{} is $O\scriptmath{\left(n+d\sum_{H=1}^{K}q_H^2+I_{\rbindex{QP}}\sum_{H=1}^{K}q_H^3\right)}$, which is upper bounded by
	$O\scriptmath{\left(n+dKq_{\rbindex{max}}^2+I_{\rbindex{QP}}Kq_{\rbindex{max}}^3\right)}$.
	In the worst case where $q_{\rbindex{max}}=K$, the complexity becomes
	$O\scriptmath{\left(n+dK^3+I_{\rbindex{QP}}K^4\right)}$.
\end{theorem}
\begin{proof}
	Please refer to Appendix~\ref{Proof:advanced_time_cost}.
\end{proof}

\noindent
\textbf{Privacy Analysis.} Theorem~\ref{thm:plus:privacy} gives the privacy guarantee for \solutionC{} and \solutionD{}.

\begin{theorem}\label{thm:plus:privacy}
	 \solutionC{} and \solutionD{} satisfy $(\vectorfont{w},\vectorfont{\epsilon})$-EPLDP.
\end{theorem}

\begin{proof}(sketch)
	\advancedSelection{} ensures that each user in $U_r$ reports at
	most once in any $\tau^*_r$ consecutive slots with budget
	$\epsilon'_r=\tilde\epsilon_r/\lceil\tilde w_r/\tau^*_r\rceil$, and all
	cross-group operations are post-processing and incur no additional privacy loss.
	Therefore, both solutions satisfy $(\vectorfont{w},\vectorfont{\epsilon})$-EPLDP.
	The proof details are shown in Appendix~\ref{Proof:plus:privacy}.
\end{proof}

\noindent
\textbf{Utility Analysis.}
We analyze \solutionC{} and \solutionD{} in the $\operatorname{Pre}$ mode, where fusion weights are determined before the current publication reports are generated. We assume that dissimilarity-estimation noise is independent of subsequent publication-estimation noise. For each privacy group $PG_r$, assume that $1\leq s_r<\tau_r^*$ new publications occur in each $\tau_r^*$-length recycling cycle, are uniformly distributed, and satisfy the population lower-bound condition in Alg.~\ref{alg:PLPD+}; integer rounding is omitted.

Let $\Xi\triangleq\sup_t\max_{r,s}\|\vectorfont f^{(r)}_t-\vectorfont f^{(s)}_t\|_2$ be the
maximum cross-group shift. For reused groups, let
$\Gamma_t^2=\sum_{r\in\vectorfont{PG}^{(R)}_t}\|\vectorfont f^{(r)}_{l_r}-\vectorfont f^{(r)}_t\|_2^2$
be the temporal staleness and
$V_{\rbindexem{hist},t}=\sum_{r\in\vectorfont{PG}^{(R)}_t}V_{\rbindexem{pub},r,l_r}$.

Let $\overline{V}_{\rbindexem{pub},r}=\scriptmath{\frac{2(2^{s_r}-1)}{s_r}(\frac{1}{\lfloor n_r/2\rfloor-1}
	+\frac{d-1}{\lfloor n_r/2\rfloor}\cdot\frac{2e^{\epsilon'_r}+d-2}{(e^{\epsilon'_r}-1)^2})
	-\frac{1}{\lfloor n_r/2\rfloor-1}}$,
$S_{\rbindexem{dis}}={\sum_r\overline V_{\rbindexem{dis},r}}$,
$D_{\rbindexem{dis}}=\scriptmath{3R_{\rbindexsmallem{grp}}\sqrt{d\sum_r\pi_r^2\overline V_{\rbindexsmallem{dis},r}}+\sqrt{\sum_r\pi_r^4Q_r^2}}$,
where $R_{\rbindexem{grp}}={\sum_{r=1}^{K}\pi_rR_r}$, 
$Q_r=\scriptmath{\frac{\theta_{\rbindexsmallem{dis},r}R_r\sqrt{d\left(1+\overline{V}_{\rbindexsmallem{dis},r}\right)}}{1-\theta_{\rbindexsmallem{dis},r}}}$, 
$R_r=\scriptmath{\frac{e^{\epsilon'_r}+d-2}{e^{\epsilon'_r}-1}}$,
$\overline{V}_{\rbindexem{dis},r}=\scriptmath{\widetilde{err}_{\rbindexsmallem{FO}}\left(\left\lfloor n_r/2\right\rfloor,z_{\rbindexsmallem{opt},\rbindexsmallem{dis},r}^{*},\epsilon'_r\right)}$,
and $\theta_{\rbindexem{dis},r}=\scriptmath{\frac{\left\lfloor n_r/2\right\rfloor-z_{\rbindexsmallem{opt},\rbindexsmallem{dis},r}^{*}}{z_{\rbindexsmallem{opt},\rbindexsmallem{dis},r}^{*}\left(\left\lfloor n_r/2\right\rfloor-1\right)}}$.


\begin{theorem}\label{thm:plpd-plus-utility}
	Assume the historical estimates of reused groups are unbiased. For any slot
	$t$ with $\vectorfont{PG}^{(A)}_t\neq\emptyset$ and
	$\min_{PG_r\in\vectorfont{PG}^{(A)}_t}n_{pp,r}\ge u_{min}$, \solutionD{} in Pre mode
	satisfies
	\begin{equation}\notag
		\outereqsize{
			\begin{aligned}
				\shrink{err}_t\le\sum\nolimits_{r\in\vectorfont{PG}^{(A)}_t}V^{\operatorname{Pre}}_{\rbindexsmall{pub},r,t}
				+\beta_t^2+2\sqrt2\,\beta_t\sqrt{S_{\rbindexsmall{dis}}}+V_{\rbindexsmall{hist},t}+D,
			\end{aligned}
		}
	\end{equation}
	where $\beta_t=\Xi+\sqrt{\Gamma_t^2+V_{\rbindex{hist},t}}$, and
	$D=3S_{\rbindex{dis}}+\sum_r Q_r+D_{\rbindex{dis}}$ collects dissimilarity-estimation terms that
	vanish as group sizes grow. For PLPD$^+$, all groups are active, thus,
	$\Gamma_t=V_{\rbindex{hist},t}=0$, and the time-averaged error satisfies
	$\overline{err}\le\sum_r\overline V_{\rbindex{pub},r}+\Xi^2+2\sqrt2\,\Xi\sqrt{S_{\rbindex{dis}}}+D$.
\end{theorem}
\begin{proof}
	Please refer to Appendix~\ref{proof:plpa-plus-utility}.
\end{proof}

\textit{Remark.} The bounds separate variance from fusion bias: if all privacy groups share the same current distribution, $\Xi=0$; otherwise the bias is governed by $\Xi$
and $\Gamma_t$. They hold for Pre, where the fusion weights are independent of
the publication noise; Post is evaluated empirically in Section~\ref{experiments}.

\section{Experiments}\label{experiments}
We evaluate utility and runtime on two real and three synthetic datasets, and examine the contributions of \advancedSelection{} and \advancedEnhance{} through ablation and privacy–data correlation studies.


\subsection{Datasets}
\noindent
\textbf{Real Datasets.} 
We employ \textit{\trajectoryDatasetName{}} and \textit{\checkInDatasetName} to evaluate the performance of all methods.

\textit{\trajectoryDatasetName{}.} 
This dataset records real-time trajectories of $10{,}357$ taxis in Beijing from February $2$ to February $8$, 2008.
After cleaning, $14{,}859{,}377$ valid records from $n=10{,}269$ taxis are retained, and locations are mapped into $d=5$ areas.  
We sample the trajectories every minute, resulting in $T=8{,}889$ time slots.

\textit{\checkInDatasetName{}.} 
This dataset consists of $33{,}278{,}683$ Foursquare check-ins from $266{,}909$ users collected between April 2012 and September 2013.
We map all venues into $d=5$ discrete categories. Using $100$-minute time slots, we obtain $n=13{,}216$ user streams over $T=7{,}649$ time slots.

\noindent
\textbf{Synthetic Datasets.}
We generate three synthetic categorical streams
with $n=10{,}000$ users, $T=10{,}000$ time slots, and $d=5$. At each time slot
$t$, a generation function produces a probability $p_t\in[0,1]$ for a
designated status, and the remaining mass $1-p_t$ is spread uniformly over the
other $d-1$ statuses.

\textit{\tlnsDatasetName{}.}
The probability evolves according to a clipped random walk:
$p_t=\min\{1,\max\{0,p_{t-1}+\eta_t\}\}$, $\eta_t\sim\mathcal{N}(0,Q)$, where $p_0=0.05$ and $\sqrt{Q}=0.0025$.

\textit{\sinDatasetName{}.}
The probability follows a sinusoidal pattern: $p_t=A\sin(\omega t)+h$, where $A=0.05$, $\omega=0.01$, and $h=0.5$.

\textit{\logDatasetName{}.}
The probability follows a logistic growth pattern: $p_t=\frac{A}{1+e^{-bt}}$, where $A=0.7$ and $b=0.02$.

\begin{figure}[h]\centering
	\subfigure{
		\scalebox{0.33}[0.33]{\includegraphics{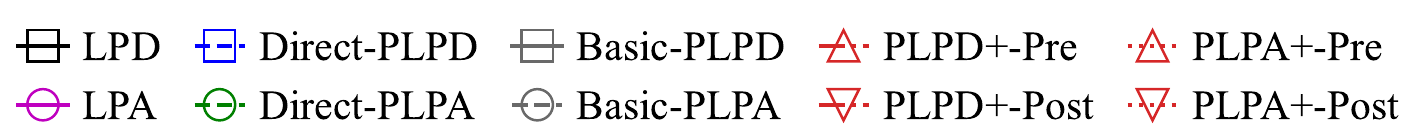}}}\hfill\\
	\addtocounter{subfigure}{-1}\vspace{-2ex}
	\subfigure[][{\small \trajectoryDatasetName{}}]{
		\scalebox{0.295}[0.295]{\includegraphics{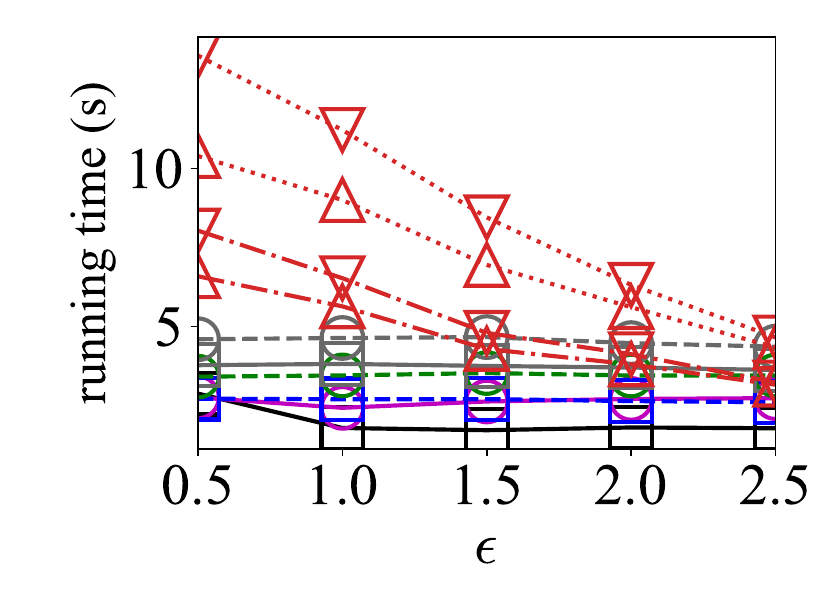}}
		\label{subfig:trajectory_budget_change_running_time}}\hfill
	\subfigure[][{\small \tlnsDatasetName{}}]{
		\scalebox{0.295}[0.295]{\includegraphics{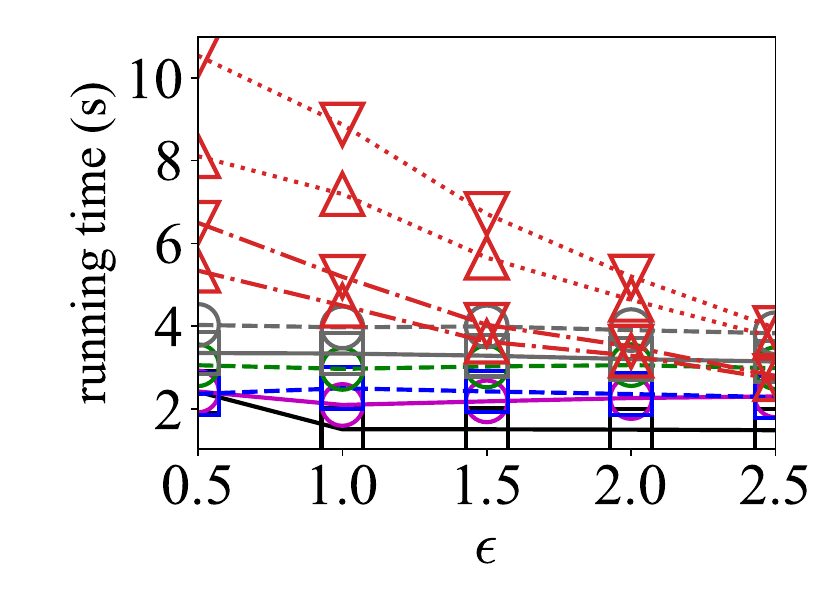}}
		\label{subfig:tlns_budget_change_running_time}}\hfill 
	\subfigure[][{\small \trajectoryDatasetName{}}]{
		\scalebox{0.295}[0.295]{\includegraphics{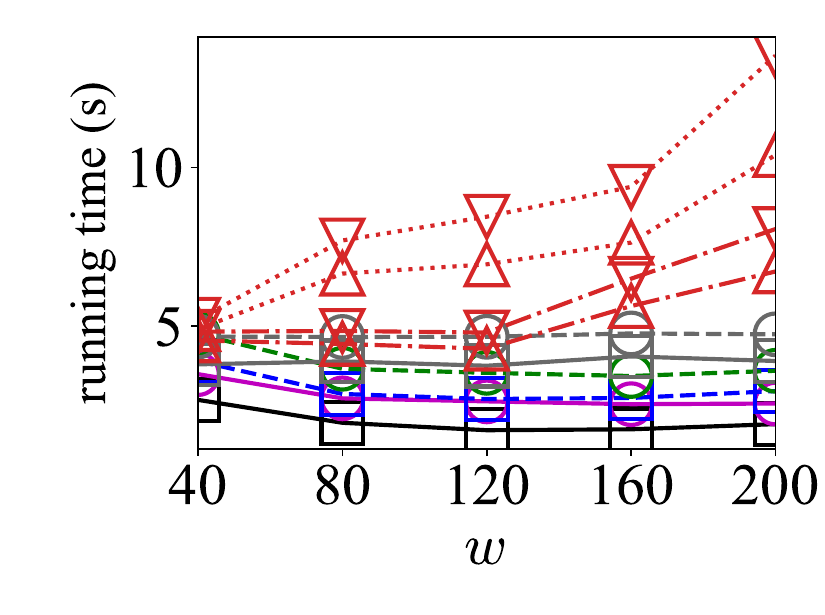}}
		\label{subfig:trajectory_window_size_change_running_time}}\hfill
	\subfigure[][{\small \tlnsDatasetName{}}]{
		\scalebox{0.295}[0.295]{\includegraphics{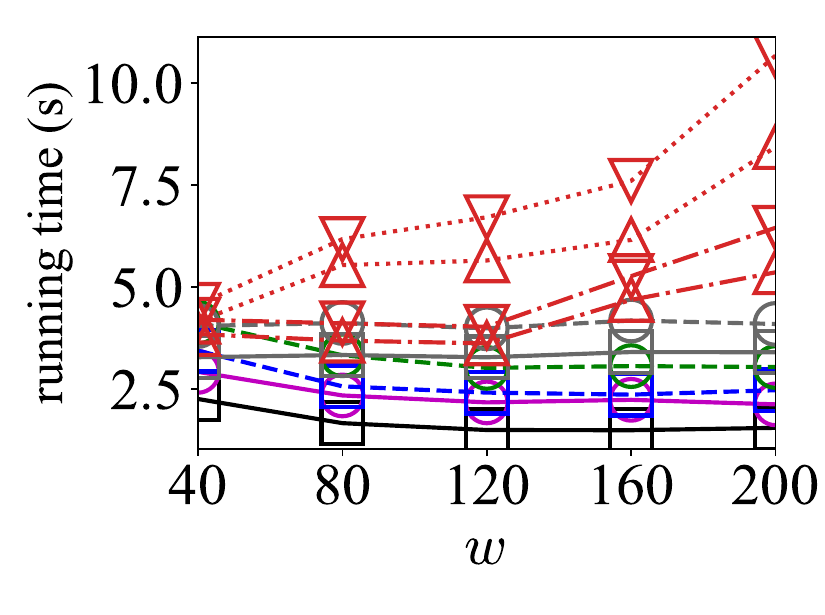}}
		\label{subfig:tlns_window_size_change_running_time}}\hfill 
	\caption{\small The running time with $\epsilon$ or $w$ varied.}
	\label{fig:alter_e_w_running_time}
\end{figure}


\begin{figure*}[!t]\centering
	\subfigure{
		\scalebox{0.33}[0.33]{\includegraphics{figures/legend_main.pdf}}}\hfill\\
	\addtocounter{subfigure}{-1}\vspace{-2ex}
	\subfigure[][{\small \trajectoryDatasetName{}, $w=120$}]{
		\scalebox{0.238}[0.238]{\includegraphics{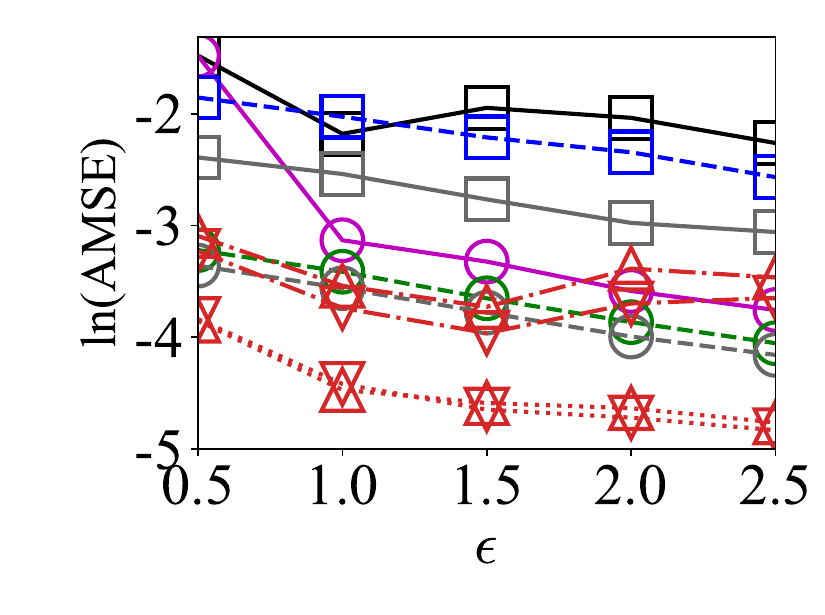}}
		\label{subfig:trajectory_budget_change_AMSE_main_ldp}}\hfill
	\subfigure[][{\small \checkInDatasetName{}, $w=120$}]{
		\scalebox{0.238}[0.238]{\includegraphics{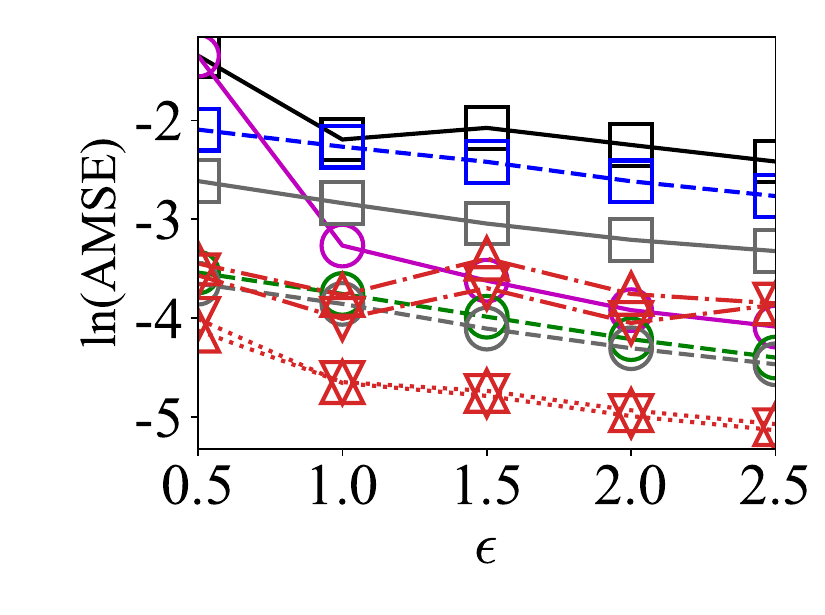}}
		\label{subfig:check_in_budget_change_AMSE_main_ldp}}\hfill	
	\subfigure[][{\small \tlnsDatasetName{}, $w=120$}]{
		\scalebox{0.238}[0.238]{\includegraphics{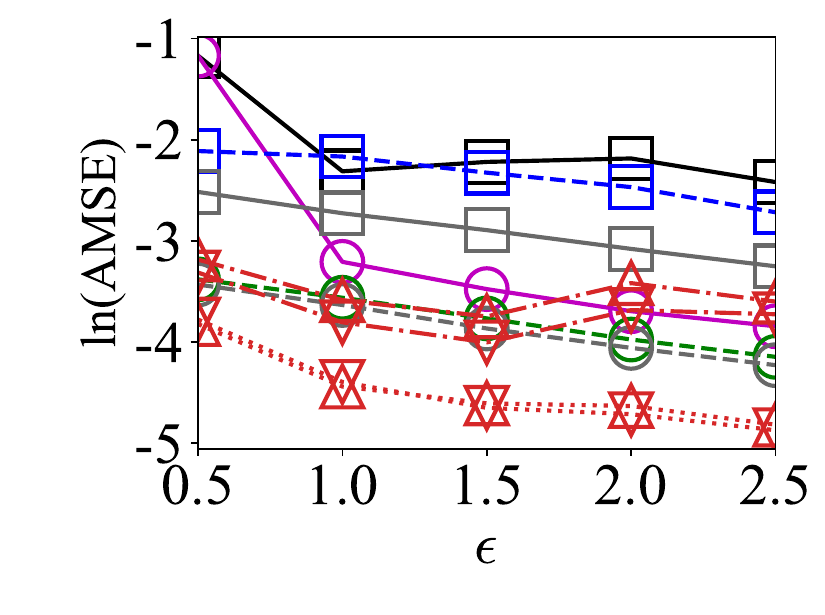}}
		\label{subfig:tlns_budget_change_AMSE_main_ldp}}\hfill 
	\subfigure[][{\small \sinDatasetName{}, $w=120$}]{
		\scalebox{0.238}[0.238]{\includegraphics{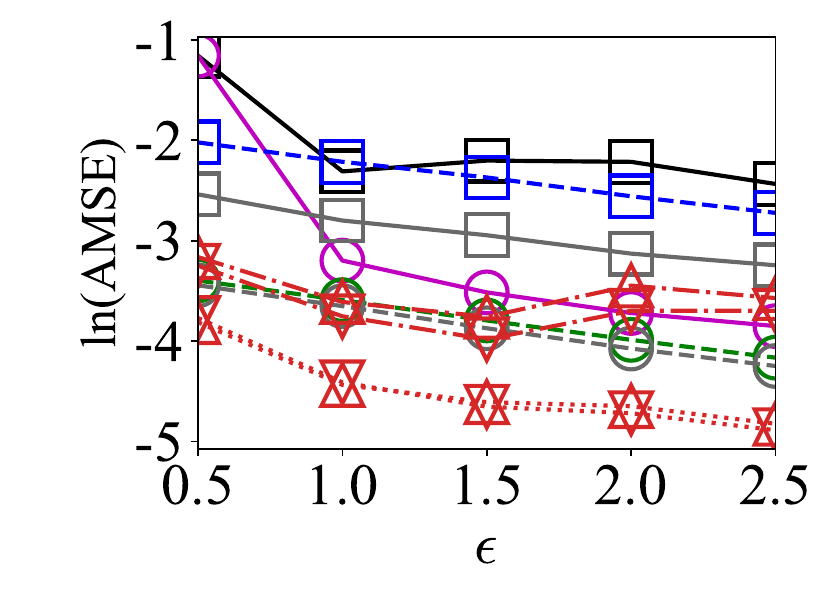}}
		\label{subfig:sin_budget_change_AMSE_main_ldp}}\hfill 
	\subfigure[][{\small \logDatasetName{}, $w=120$}]{
		\scalebox{0.238}[0.238]{\includegraphics{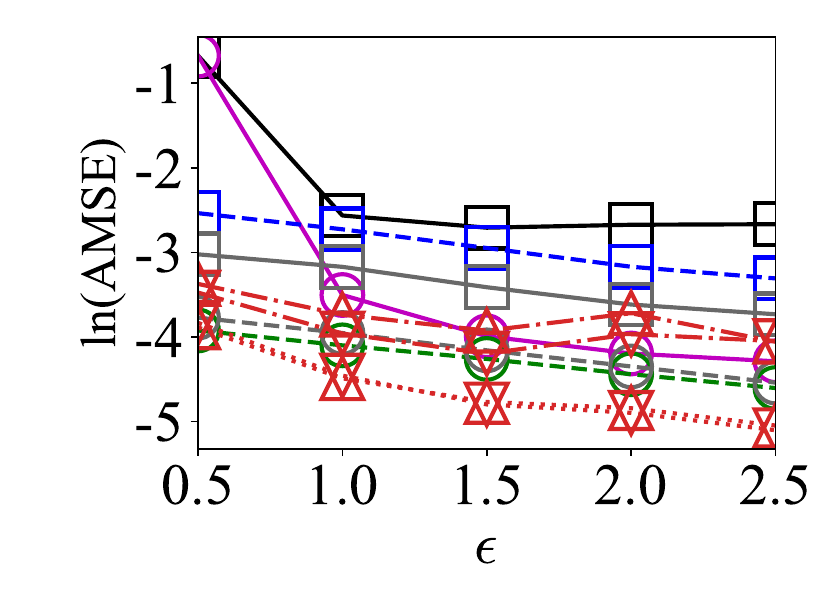}}
		\label{subfig:log_budget_change_AMSE_main_ldp}}\hfill \\
	\subfigure[][{\small \trajectoryDatasetName{}, $\epsilon=1.5$}]{
		\scalebox{0.238}[0.238]{\includegraphics{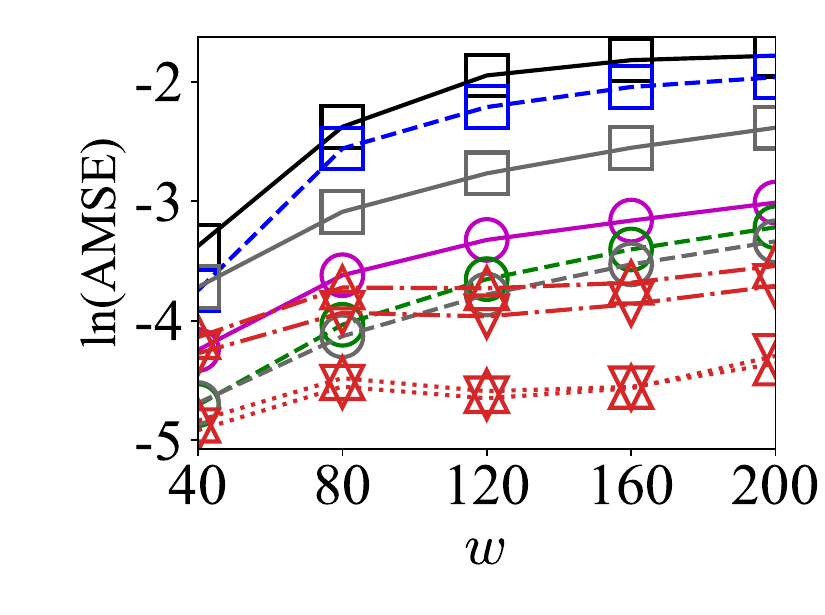}}
		\label{subfig:trajectory_window_size_change_AMSE_main_ldp}}\hfill
	\subfigure[][{\small \checkInDatasetName{}, $\epsilon=1.5$}]{
		\scalebox{0.238}[0.238]{\includegraphics{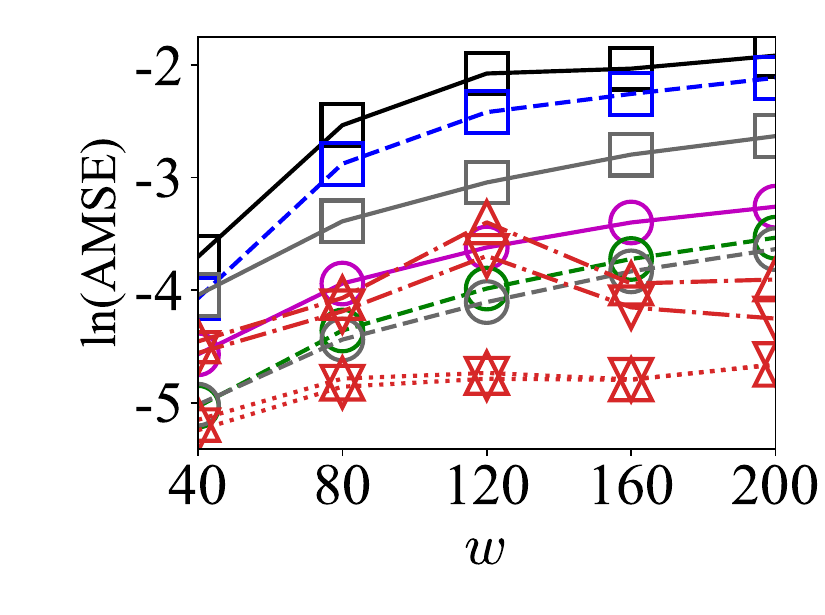}}
		\label{subfig:check_in_window_size_change_AMSE_main_ldp}}\hfill	
	\subfigure[][{\small \tlnsDatasetName{}, $\epsilon=1.5$}]{
		\scalebox{0.238}[0.238]{\includegraphics{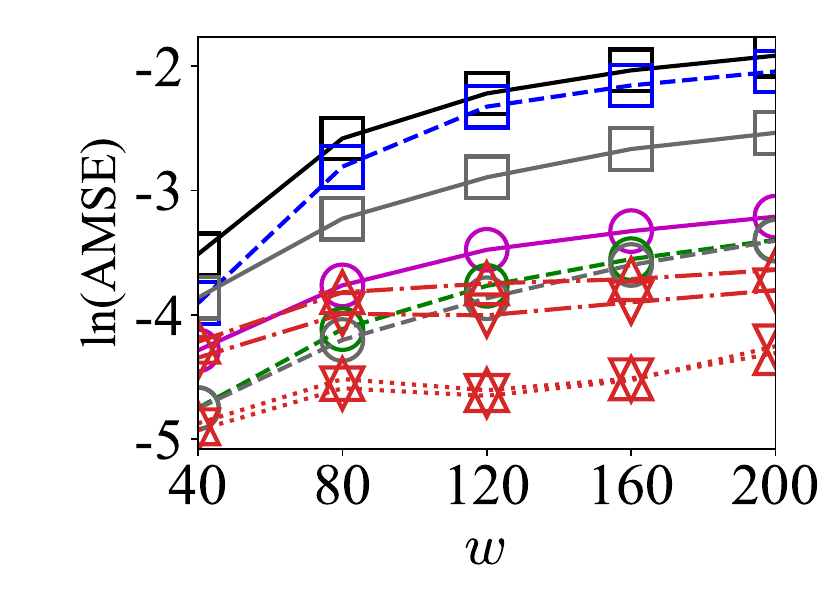}}
		\label{subfig:tlns_window_size_change_AMSE_main_ldp}}\hfill 
	\subfigure[][{\small \sinDatasetName{}, $\epsilon=1.5$}]{
		\scalebox{0.238}[0.238]{\includegraphics{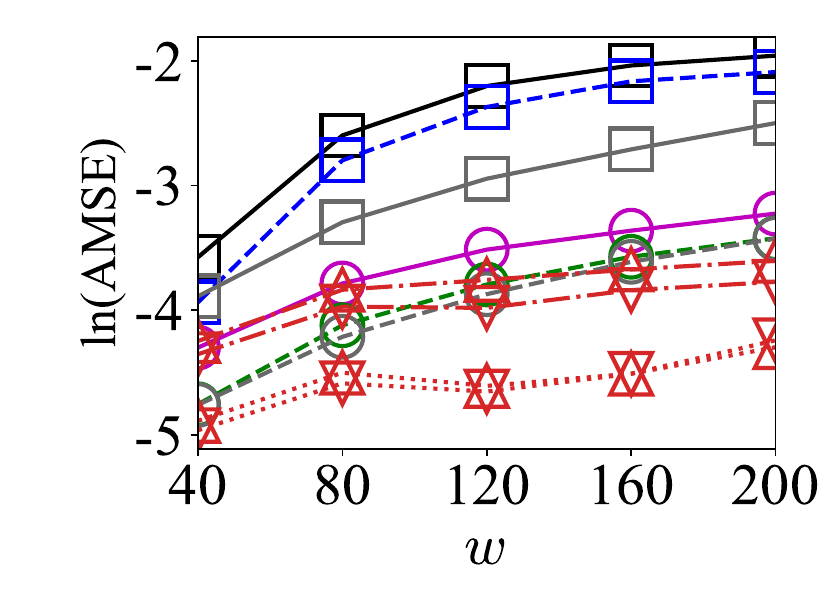}}
		\label{subfig:sin_window_size_change_AMSE_main_ldp}}\hfill 
	\subfigure[][{\small \logDatasetName{}, $\epsilon=1.5$}]{
		\scalebox{0.238}[0.238]{\includegraphics{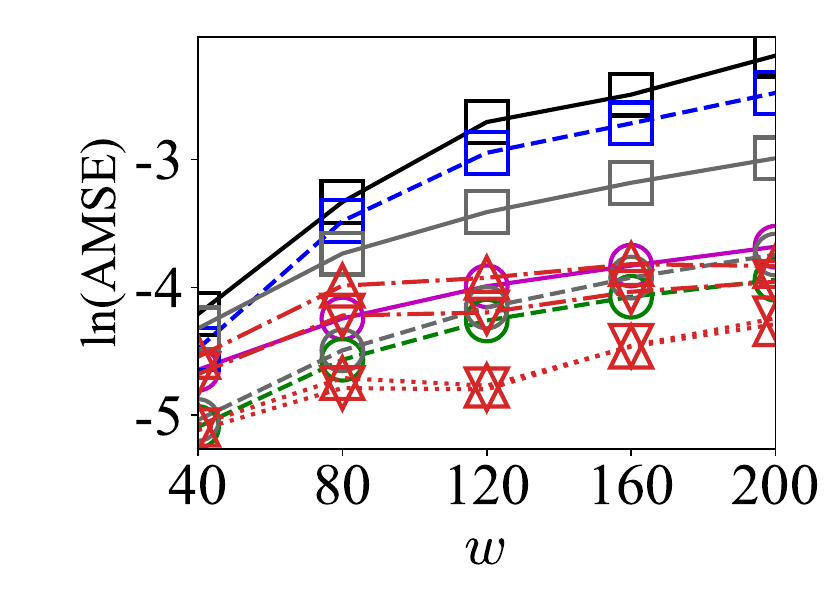}}
		\label{subfig:log_window_size_change_AMSE_main_ldp}}\hfill 
	\caption{\small $\shrinkem{AMSE}$ with different $\epsilon$ or $w$.}
	\label{fig:alter_e_w_AMSE_main_ldp}\vspace{-2ex}
\end{figure*}


\subsection{Experiment Setup}
\noindent
\textbf{Compared Algorithms.}
We compare ten mechanisms.
\solutionCMPCTotalName{} (\solutionCMPC{}) and \solutionCMPDTotalName{} (\solutionCMPD{}) are the non-personalized population-distribution and population-absorption mechanisms, respectively~\cite{DBLP:conf/sigmod/RenSYYZX22}.
Direct-\solutionA{} and Direct-\solutionB{} directly extend these two mechanisms to users with heterogeneous privacy budgets and window sizes, without applying \basicSelection{}.
Basic-\solutionA{} (\solutionA{}) and Basic-\solutionB{} (\solutionB{}) further employ \basicSelection{} to determine a unified privacy-compatible sampling size.
Our enhanced mechanisms, \solutionC{} and \solutionD{}, incorporate both \advancedSelection{} and \advancedEnhance{}.
For each enhanced mechanism, we evaluate two fusion modes: the Pre mode determines the fusion weights before the current publication estimates are observed, whereas the Post mode recomputes the weights after the current publication estimates become available.
Accordingly, the four enhanced variants are denoted by \solutionC{}-Pre, \solutionC{}-Post, \solutionD{}-Pre, and \solutionD{}-Post.


\begin{table}[!t]
	\begin{center}
		{\small  \vspace{1ex}
			\caption{\small Experimental settings with default values in bold.} \label{tab:settings}
			\begin{tabular}{l|l}\hline
				{\bf Parameters} & {\bf Values} \\ \hline
				global privacy budget $\epsilon$
				& $0.5,1.0,\mathbf{1.5},2.0,2.5$ \\
				global window size $w$
				& $40,80,\mathbf{120},160,200$ \\
				personalized privacy budget $\epsilon_i$
				& $\epsilon,\ldots,2.0,2.5$ \\
				personalized window size $w_i$
				& $40,80,\ldots,w$ \\
				population lower bound $u_{\rbindexem{min}}$
				& $10$ \\
				eigenvalues' clipping lower bound $\mu$
				& $10^{-6}$ \\
				\hline
			\end{tabular}
		}
	\end{center}
\end{table}

\noindent
\textbf{Parameter Settings.}
We vary $\epsilon$ from $0.5$ to $2.5$ and $w$ from $40$ to $200$. When varying $\epsilon$, we fix $w=120$;
when varying $w$, we fix $\epsilon=1.5$. For personalized mechanisms,
$\epsilon_i$ is uniformly sampled from candidate values no smaller than
$\epsilon$, while $w_i$ is sampled from values no larger than $w$~\cite{DBLP:journals/vldb/DuZCCLXL26}, independently of user data. We set the population lower bound as $u_{\rbindexem{min}}=10$, and eigenvalues' clipping lower bound in \basicClipping{} as $\mu=10^{-6}$.
The summary of parameter settings is shown in Table~\ref{tab:settings}.


All experiments are conducted in Java on an Intel(R) Xeon(R) Silver 4210R CPU at 2.4 GHz with 128 GB of RAM.
Each experiment is repeated ten times using different random seeds ranging from $0$ to $9$, and we report the average results.

\noindent
\textbf{Performance Metrics.} 
We evaluate computational efficiency using running time and measure data utility using the average mean squared error ($\shrinkem{AMSE}$):
\begin{equation}\notag
	\outereqsize{
		\begin{aligned}
			\shrinkem{AMSE}
			=
			\frac{1}{T}
			\sum\nolimits_{\tau=1}^{T}
			\left\|
			\widehat{\vectorfont{f}}_{\tau}
			-
			\vectorfont{f}_{\tau}
			\right\|_2^2
			=
			\frac{1}{T}
			\sum\nolimits_{\tau=1}^{T}
			\sum\nolimits_{j=1}^{d}
			\left(
			\hat{f}_{j,\tau}-f_{j,\tau}
			\right)^2.
		\end{aligned}
	}
\end{equation}
We also use average Jensen--Shannon divergence ($\shrinkem{AJSD}$) to measure the data utility.
Its definition and experiment results are shown in 
Appendix~\ref{appendix:extra-AJSD-result}.
Lower values indicate better data utility.
For readability, the utility figures report $\ln(\shrinkem{AMSE})$ and $\ln(\shrinkem{AJSD})$.

\subsection{Time Cost Analysis}
Figure~\ref{fig:alter_e_w_running_time} summarizes the running time under varying $\epsilon$ and $w$. 
The enhanced variants are slower because \advancedSelection{} and \advancedEnhance{} introduce group-specific scheduling and fusion, with Post further recomputing the fusion weights. 
Increasing $\epsilon$ reduces the number of possible privacy groups and therefore lowers the overhead of the enhanced variants. In contrast, increasing $w$ enlarges the candidate
set of personalized window sizes and increases the number of groups.
The time cost under the other datasets holds the same trend and is shown in 
Appendix~\ref{appendix:experiment:timecost-extra}.

\subsection{Overall Utility Analysis}\label{subsec:overall-utility-analysis}
Figures~\ref{subfig:trajectory_budget_change_AMSE_main_ldp}--\ref{subfig:log_budget_change_AMSE_main_ldp} show that $\ln(\shrinkem{AMSE})$ generally decreases as $\epsilon$ increases, since larger personalized budgets reduce perturbation noise. Across matched variants, population absorption consistently improves over population distribution by accumulating unused publication users, while personalized mechanisms outperform their non-personalized counterparts by exploiting heterogeneous $(w_i,\epsilon_i)$.
\solutionD{}-Post reduces the $\shrinkem{AMSE}$ of Direct-\solutionB{} by $47.2\%$ on average over the five datasets,
demonstrating the benefits of \advancedSelection{} and \advancedEnhance{}; \solutionD{}-Post achieves the lowest or nearly lowest error on all five datasets.


Figures~\ref{subfig:trajectory_window_size_change_AMSE_main_ldp}--\ref{subfig:log_window_size_change_AMSE_main_ldp} show the complementary trend with $w$: larger windows reduce the available reporting population and generally increase $\ln(\shrinkem{AMSE})$. 
\solutionD{}-Post reduces the $\shrinkem{AMSE}$ of Direct-\solutionB{} by $43.5\%$ on average.
The relative ranking of the methods remains consistent with the $\epsilon$-varying results, with \solutionD{}-Post again providing the best or nearly best utility. 

Averaged over all $\epsilon$ and $w$ varying settings, \solutionD{}-Post reduces
the AMSE of Direct-PLPA by $52.3\%$ on the two real datasets and by $40.8\%$ on
the three synthetic datasets.

\subsection{Ablation Study}
To isolate the two components, we compare the basic methods with \advancedSelection{}-only
variants, \advancedEnhance{}-only variants, and the enhanced methods; the
latter two are evaluated in both Pre and Post modes.

\begin{figure}[t]\centering
	\subfigure{
		\scalebox{0.3}[0.3]{\includegraphics{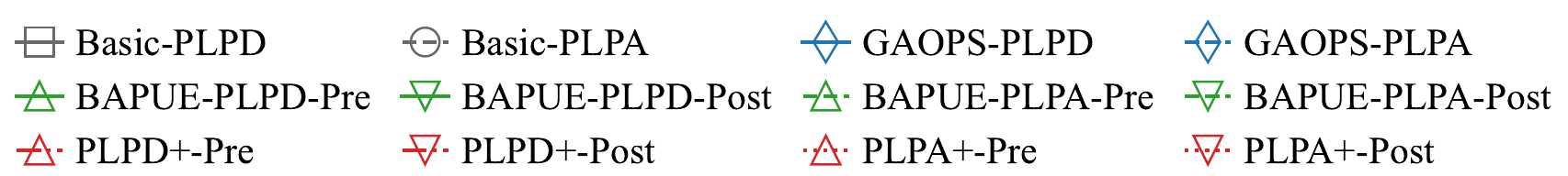}}}\hfill\\
	\addtocounter{subfigure}{-1}
	\subfigure[][{\small \trajectoryDatasetName{}, $w=120$}]{
		\scalebox{0.295}[0.295]{\includegraphics{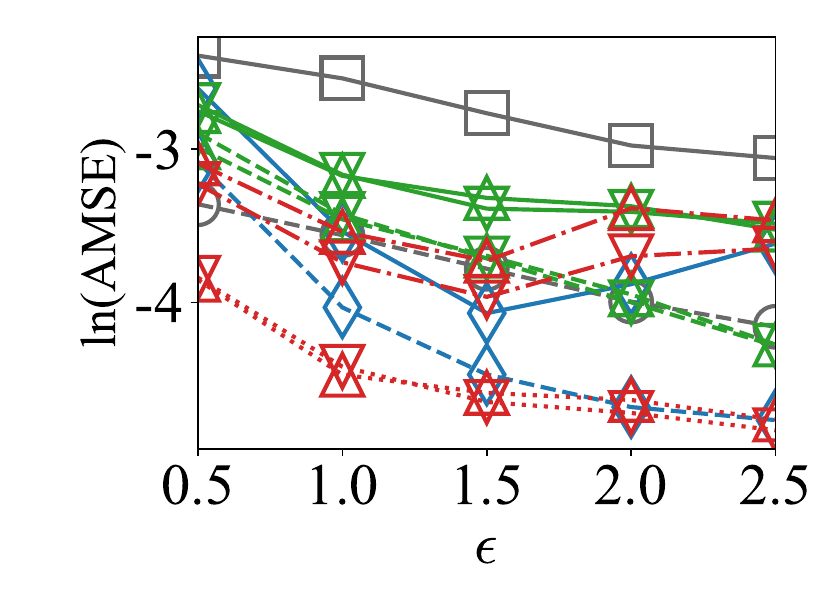}}
		\label{subfig:trajectory_budget_change_AMSE_ablation_ldp}}\hfill
	\subfigure[][{\small \tlnsDatasetName{}, $w=120$}]{
		\scalebox{0.295}[0.295]{\includegraphics{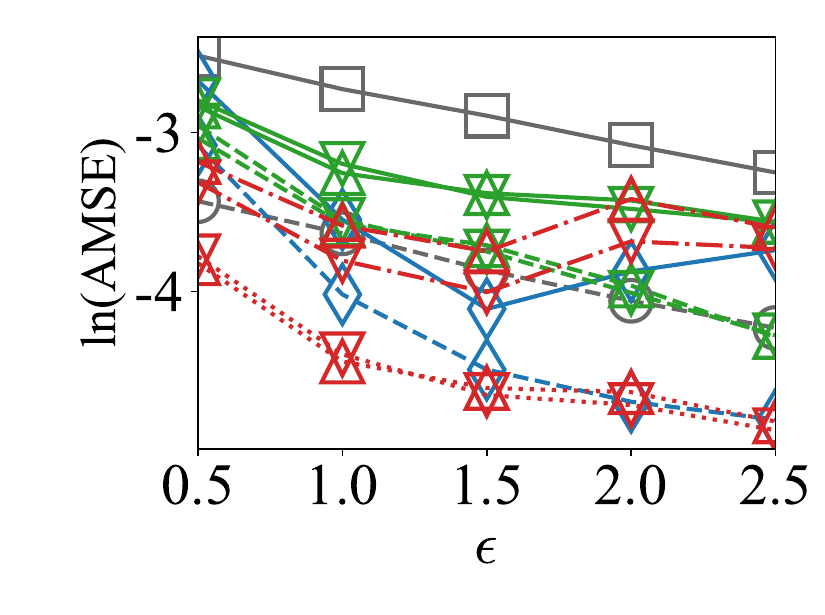}}
		\label{subfig:tlns_budget_change_AMSE_ablation_ldp}}\hfill  \\
	\subfigure[][{\small \trajectoryDatasetName{}, $\epsilon=1.5$}]{
		\scalebox{0.295}[0.295]{\includegraphics{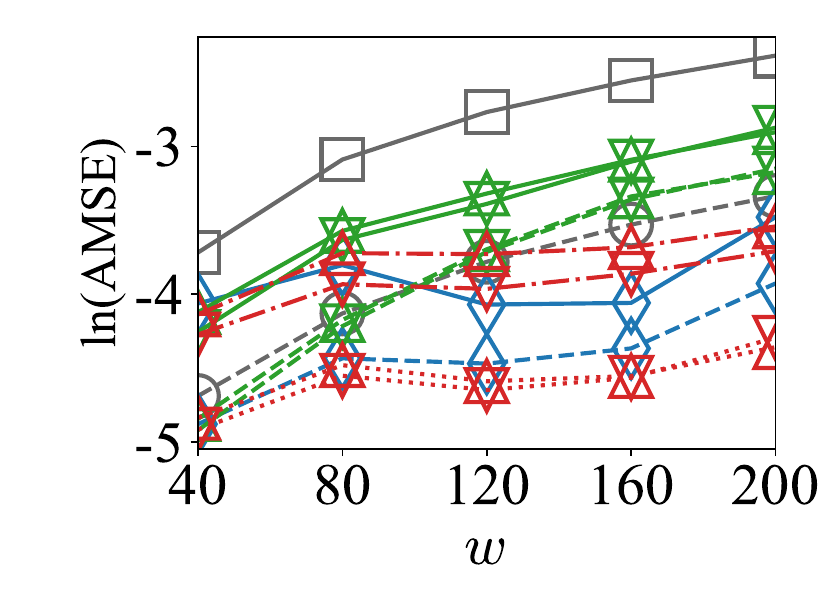}}
		\label{subfig:trajectory_window_size_change_AMSE_ablation_ldp}}\hfill
	\subfigure[][{\small \tlnsDatasetName{}, $\epsilon=1.5$}]{
		\scalebox{0.295}[0.295]{\includegraphics{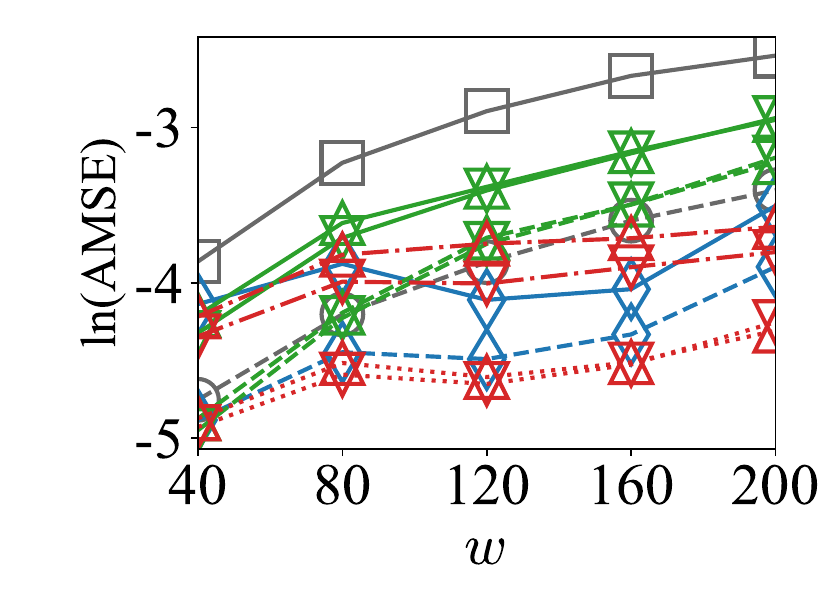}}
		\label{subfig:tlns_window_size_change_AMSE_ablation_ldp}}\hfill 
	\caption{\small The ablation study of $\shrinkem{AMSE}$ with different $\epsilon$ or $w$.}
	\label{fig:alter_e_w_AMSE_ablation_ldp}
\end{figure}


Figure~\ref{fig:alter_e_w_AMSE_ablation_ldp} compares the Basic, \advancedSelection{}-only, \advancedEnhance{}-only, and complete variants. \advancedSelection{} yields gains under heterogeneous windows, particularly for population absorption, confirming the benefit of group-specific scheduling. \advancedEnhance{} provides additional improvements when variance reduction outweighs distribution-shift bias. Combining both components gives the best overall performance. The Post methods generally outperform the Pre ones because Post methods' fusion weights use the realized publication estimates.

\subsection{Privacy--Data Correlation Analysis}\label{subsec:privacy-data-correlation}
We build a Sin variant ($n{=}120{,}000$, $T{=}1{,}000$, $d{=}5$) with four
equal-size groups ($\epsilon_i\in\{0.5,1,1.5,2.5\}$, $w_i{=}40$), and shift
group $r$'s designated-status probability to $p_t+\sigma s_r$ with
$\vectorfont s=(1,\frac13,-\frac13,-1)$. Since $\sum_r s_r=0$, varying $\sigma$ changes only group-level distributions, leaving the population distribution unchanged. 
As Table~\ref{tbl:data-correlation} shows, \solutionB{} is the most accurate at $\sigma=0$, where its inverse-variance weight is efficient, but degrades $23\times$ at $\sigma=0.2$ since it borrows regardless of the shift. \solutionCMPJ{} never borrows and stays flat, but is $3.2\times$ worse than PLPA at $\sigma=0$ because the noisy $\epsilon=0.5$ group enters with its full proportion. \advancedEnhance{} lies in between: it reduces the error of \solutionCMPJ{} by $2.5\times$ at $\sigma=0$, at a $1.2$--$1.3\times$ cost relative to \solutionB{} due to weights estimated from noisy data, and remains $4.3$--$5.1\times$ more accurate than \solutionB{} at $\sigma=0.2$. Beyond $\sigma\approx0.1$, \solutionCMPJ{} is preferable, since the empirical risk matrix still assigns weight to shifted sources.

\begin{table}[t]
	\centering
	\caption{AMSE ($\times10^{-3}$) under positive privacy--data correlation (Sin).}\label{tbl:data-correlation}\vspace{-2ex}
	\resizebox{\linewidth}{!}{
	\begin{tabular}{c|ccccc}
		\toprule
		$\sigma$ & 0 & 0.05 & 0.10 & 0.15 & 0.20\\
		\midrule
		\solutionB{}     & 1.42 & 3.39 & 9.26 & 19.01 & 32.67\\
		\solutionCMPJ{}    & 4.48 & 4.52 & 4.45 & 4.33 & 4.44\\
		\solutionD{}-Pre   & 1.74 & 2.80 & 4.78 & 6.74 & 7.59\\
		\solutionD{}-Post  & 1.89 & 2.72 & 4.12 & 5.71 & 6.45\\
		\bottomrule
	\end{tabular}
	}
\end{table}

\section{Conclusion}\label{conclusion}
We studied personalized continual frequency estimation under heterogeneous
privacy budgets and window sizes, and proposed \solutionA{} and \solutionB{} with \basicSelection{} and
their enhanced variants \solutionC{} and \solutionD{} with \advancedSelection{} and \advancedEnhance{}.
Experiments on two real and three synthetic datasets demonstrate their
effectiveness and confirm that \advancedEnhance{} limits the bias induced by
privacy--data correlation.

\newpage
\section{AI-Generated Content Acknowledgment}
No content of this paper is generated by Generative AI tools and technologies, such as ChatGPT.

\bgroup\small
\bibliographystyle{IEEEtran}
\bibliography{reference}

@inproceedings{DBLP:conf/ndss/0001LLSL20,
  author    = {Tianhao Wang and
               Milan Lopuha{\"{a}}{-}Zwakenberg and
               Zitao Li and
               Boris Skoric and
               Ninghui Li},
  title     = {Locally Differentially Private Frequency Estimation with Consistency},
  booktitle = {27th Annual Network and Distributed System Security Symposium, {NDSS}
               2020, San Diego, California, USA, February 23-26, 2020},
  publisher = {The Internet Society},
  year      = {2020}
}

@inproceedings{DBLP:conf/stoc/BassilyS15,
  author    = {Raef Bassily and
               Adam D. Smith},
  editor    = {Rocco A. Servedio and
               Ronitt Rubinfeld},
  title     = {Local, Private, Efficient Protocols for Succinct Histograms},
  booktitle = {Proceedings of the Forty-Seventh Annual {ACM} on Symposium on Theory
               of Computing, {STOC} 2015, Portland, OR, USA, June 14-17, 2015},
  pages     = {127--135},
  publisher = {{ACM}},
  year      = {2015}
}

@inproceedings{DBLP:conf/ccs/ErlingssonPK14,
  author    = {{\'{U}}lfar Erlingsson and
               Vasyl Pihur and
               Aleksandra Korolova},
  editor    = {Gail{-}Joon Ahn and
               Moti Yung and
               Ninghui Li},
  title     = {{RAPPOR:} Randomized Aggregatable Privacy-Preserving Ordinal Response},
  booktitle = {Proceedings of the 2014 {ACM} {SIGSAC} Conference on Computer and
               Communications Security, Scottsdale, AZ, USA, November 3-7, 2014},
  pages     = {1054--1067},
  publisher = {{ACM}},
  year      = {2014}
}

@inproceedings{DBLP:conf/uss/WangBLJ17,
  author    = {Tianhao Wang and
               Jeremiah Blocki and
               Ninghui Li and
               Somesh Jha},
  editor    = {Engin Kirda and
               Thomas Ristenpart},
  title     = {Locally Differentially Private Protocols for Frequency Estimation},
  booktitle = {26th {USENIX} Security Symposium, {USENIX} Security 2017, Vancouver,
               BC, Canada, August 16-18, 2017},
  pages     = {729--745},
  publisher = {{USENIX} Association},
  year      = {2017}
}

@inproceedings{DBLP:conf/sigmod/RenSYYZX22,
  author    = {Xuebin Ren and
               Liang Shi and
               Weiren Yu and
               Shusen Yang and
               Cong Zhao and
               Zongben Xu},
  editor    = {Zachary Ives and
               Angela Bonifati and
               Amr El Abbadi},
  title     = {{LDP-IDS:} Local Differential Privacy for Infinite Data Streams},
  booktitle = {{SIGMOD} '22: International Conference on Management of Data, Philadelphia,
               PA, USA, June 12 - 17, 2022},
  pages     = {1064--1077},
  publisher = {{ACM}},
  year      = {2022}
}

@article{DBLP:journals/vldb/DuZCCLXL26,
	author       = {Leilei Du and
	Xu Zhou and
	Peng Cheng and
	Lei Chen and
	Xuemin Lin and
	Wei Xi and
	Kenli Li},
	title        = {Personalized w-event privacy for infinite stream estimation},
	journal      = {{VLDB} J.},
	volume       = {35},
	number       = {5},
	pages        = {51},
	year         = {2026}
}

@article{he2025personalized,
	title={Personalized Local Differential Privacy for Multi-dimensional Range Queries over Mobile User Data},
	author={He, Yuanyuan and Wang, Meiqi and Deng, Xianjun and Yang, Peng and Xue, Qiao and Yang, Laurence T},
	journal={IEEE Transactions on Mobile Computing},
	year={2025},
	publisher={IEEE}
}

@article{DBLP:journals/tissec/ChanSS11,
	author       = {T.{-}H. Hubert Chan and
	Elaine Shi and
	Dawn Song},
	title        = {Private and Continual Release of Statistics},
	journal      = {{ACM} Trans. Inf. Syst. Secur.},
	volume       = {14},
	number       = {3},
	pages        = {26:1--26:24},
	year         = {2011}
}

@article{DBLP:journals/pvldb/KellarisPXP14,
	author       = {Georgios Kellaris and
	Stavros Papadopoulos and
	Xiaokui Xiao and
	Dimitris Papadias},
	title        = {Differentially Private Event Sequences over Infinite Streams},
	journal      = {Proc. {VLDB} Endow.},
	volume       = {7},
	number       = {12},
	pages        = {1155--1166},
	year         = {2014}
}

@inproceedings{DBLP:conf/ccs/0001C0SC0LJ21,
	author       = {Tianhao Wang and
	Joann Qiongna Chen and
	Zhikun Zhang and
	Dong Su and
	Yueqiang Cheng and
	Zhou Li and
	Ninghui Li and
	Somesh Jha},
	title        = {Continuous Release of Data Streams under both Centralized and Local
	Differential Privacy},
	booktitle    = {{CCS} '21: 2021 {ACM} {SIGSAC} Conference on Computer and Communications
	Security, Virtual Event, Republic of Korea, November 15 - 19, 2021},
	pages        = {1237--1253},
	year         = {2021}
}

@article{DBLP:journals/pvldb/BaoYXD21,
	author       = {Ergute Bao and
	Yin Yang and
	Xiaokui Xiao and
	Bolin Ding},
	title        = {{CGM:} An Enhanced Mechanism for Streaming Data Collection with Local Differential Privacy},
	journal      = {Proc. {VLDB} Endow.},
	volume       = {14},
	number       = {11},
	pages        = {2258--2270},
	year         = {2021}
}

@article{DBLP:journals/tkde/XueYHZW23,
	author       = {Qiao Xue and
	Qingqing Ye and
	Haibo Hu and
	Youwen Zhu and
	Jian Wang},
	title        = {{DDRM:} {A} Continual Frequency Estimation Mechanism With Local Differential
	Privacy},
	journal      = {{IEEE} Trans. Knowl. Data Eng.},
	volume       = {35},
	number       = {7},
	pages        = {6784--6797},
	year         = {2023}
}

@inproceedings{DBLP:conf/infocom/YeHHAX23,
	author       = {Qingqing Ye and
	Haibo Hu and
	Kai Huang and
	Man Ho Au and
	Qiao Xue},
	title        = {Stateful Switch: Optimized Time Series Release with Local Differential
	Privacy},
	booktitle    = {{IEEE} {INFOCOM} 2023 - {IEEE} Conference on Computer Communications,
	New York City, NY, USA, May 17-20, 2023},
	pages        = {1--10},
	year         = {2023}
}

@article{DBLP:journals/jpc/AlagganGK16,
	author       = {Mohammad Alaggan and
	S{\'{e}}bastien Gambs and
	Anne{-}Marie Kermarrec},
	title        = {Heterogeneous Differential Privacy},
	journal      = {J. Priv. Confidentiality},
	volume       = {7},
	number       = {2},
	year         = {2016}
}

@inproceedings{DBLP:conf/icde/KotsogiannisDHM20,
	author       = {Ios Kotsogiannis and
	Stelios Doudalis and
	Samuel Haney and
	Ashwin Machanavajjhala and
	Sharad Mehrotra},
	title        = {One-sided Differential Privacy},
	booktitle    = {36th {IEEE} International Conference on Data Engineering, {ICDE} 2020,
	Dallas, TX, USA, April 20-24, 2020},
	pages        = {493--504},
	publisher    = {{IEEE}},
	year         = {2020}
}

@article{DBLP:journals/tods/KiferM14,
	author       = {Daniel Kifer and
	Ashwin Machanavajjhala},
	title        = {Pufferfish: {A} framework for mathematical privacy definitions},
	journal      = {{ACM} Trans. Database Syst.},
	volume       = {39},
	number       = {1},
	pages        = {3:1--3:36},
	year         = {2014}
}

@article{DBLP:journals/tit/NuradhaG23,
	author       = {Theshani Nuradha and
	Ziv Goldfeld},
	title        = {Pufferfish Privacy: An Information-Theoretic Study},
	journal      = {{IEEE} Trans. Inf. Theory},
	volume       = {69},
	number       = {11},
	pages        = {7336--7356},
	year         = {2023}
}

@inproceedings{DBLP:conf/icml/PierquinBTB24,
	author       = {Cl{\'{e}}ment Pierquin and
	Aur{\'{e}}lien Bellet and
	Marc Tommasi and
	Matthieu Boussard},
	title        = {R{\'{e}}nyi Pufferfish Privacy: General Additive Noise Mechanisms
	and Privacy Amplification by Iteration via Shift Reduction Lemmas},
	booktitle    = {Forty-first International Conference on Machine Learning, {ICML} 2024,
	Vienna, Austria, July 21-27, 2024},
	publisher    = {OpenReview.net},
	year         = {2024}
}

@article{DBLP:journals/tit/NuradhaGW24,
	author       = {Theshani Nuradha and
	Ziv Goldfeld and
	Mark M. Wilde},
	title        = {Quantum Pufferfish Privacy: {A} Flexible Privacy Framework for Quantum
	Systems},
	journal      = {{IEEE} Trans. Inf. Theory},
	volume       = {70},
	number       = {8},
	pages        = {5731--5762},
	year         = {2024}
}

@article{DBLP:journals/tifs/Ding24,
	author       = {Ni Ding},
	title        = {Approximation of Pufferfish Privacy for Gaussian Priors},
	journal      = {{IEEE} Trans. Inf. Forensics Secur.},
	volume       = {19},
	pages        = {5630--5640},
	year         = {2024}
}

@inproceedings{DBLP:conf/sigmod/HeMD14,
	author       = {Xi He and
	Ashwin Machanavajjhala and
	Bolin Ding},
	editor       = {Curtis E. Dyreson and
	Feifei Li and
	M. Tamer {\"{O}}zsu},
	title        = {Blowfish privacy: tuning privacy-utility trade-offs using policies},
	booktitle    = {International Conference on Management of Data, {SIGMOD} 2014, Snowbird,
	UT, USA, June 22-27, 2014},
	pages        = {1447--1458},
	publisher    = {{ACM}},
	year         = {2014}
}

@article{DBLP:journals/pvldb/LiuKTDH21,
	author       = {Jiaxiang Liu and
	Karl Knopf and
	Yiqing Tan and
	Bolin Ding and
	Xi He},
	title        = {Catch a Blowfish Alive: {A} Demonstration of Policy-Aware Differential
	Privacy for Interactive Data Exploration},
	journal      = {Proc. {VLDB} Endow.},
	volume       = {14},
	number       = {12},
	pages        = {2859--2862},
	year         = {2021}
}

@article{DBLP:journals/tit/EdwardsRZZ22,
	author       = {Tobias Edwards and
	Benjamin I. P. Rubinstein and
	Zuhe Zhang and
	Sanming Zhou},
	title        = {A Graph Symmetrization Bound on Channel Information Leakage Under
	Blowfish Privacy},
	journal      = {{IEEE} Trans. Inf. Theory},
	volume       = {68},
	number       = {1},
	pages        = {538--548},
	year         = {2022}
}

@inproceedings{DBLP:conf/icde/JorgensenYC15,
	author       = {Zach Jorgensen and
	Ting Yu and
	Graham Cormode},
	editor       = {Johannes Gehrke and
	Wolfgang Lehner and
	Kyuseok Shim and
	Sang Kyun Cha and
	Guy M. Lohman},
	title        = {Conservative or liberal? Personalized differential privacy},
	booktitle    = {31st {IEEE} International Conference on Data Engineering, {ICDE} 2015,
	Seoul, South Korea, April 13-17, 2015},
	pages        = {1023--1034},
	publisher    = {{IEEE} Computer Society},
	year         = {2015}
}

@inproceedings{DBLP:conf/infocom/NiuCWWLC21,
	author       = {Ben Niu and
	Yahong Chen and
	Boyang Wang and
	Zhibo Wang and
	Fenghua Li and
	Jin Cao},
	title        = {AdaPDP: Adaptive Personalized Differential Privacy},
	booktitle    = {40th {IEEE} Conference on Computer Communications, {INFOCOM} 2021,
	Vancouver, BC, Canada, May 10-13, 2021},
	pages        = {1--10},
	publisher    = {{IEEE}},
	year         = {2021}
}

@article{DBLP:journals/tit/ChaudhuriMC25,
	author       = {Syomantak Chaudhuri and
	Konstantin Miagkov and
	Thomas A. Courtade},
	title        = {Mean Estimation Under Heterogeneous Privacy Demands},
	journal      = {{IEEE} Trans. Inf. Theory},
	volume       = {71},
	number       = {2},
	pages        = {1362--1375},
	year         = {2025}
}

@article{DBLP:journals/tdsc/LiYCSL23,
	author       = {Xiaoguang Li and
	Haonan Yan and
	Zelei Cheng and
	Wenhai Sun and
	Hui Li},
	title        = {Protecting Regression Models With Personalized Local Differential
	Privacy},
	journal      = {{IEEE} Trans. Dependable Secur. Comput.},
	volume       = {20},
	number       = {2},
	pages        = {960--974},
	year         = {2023}
}

@inproceedings{DBLP:conf/icde/ChenLQKJ16,
	author       = {Rui Chen and
	Haoran Li and
	A. Kai Qin and
	Shiva Prasad Kasiviswanathan and
	Hongxia Jin},
	title        = {Private spatial data aggregation in the local setting},
	booktitle    = {32nd {IEEE} International Conference on Data Engineering, {ICDE} 2016,
	Helsinki, Finland, May 16-20, 2016},
	pages        = {289--300},
	publisher    = {{IEEE} Computer Society},
	year         = {2016}
}

@inproceedings{DBLP:conf/icde/Mao00WH24,
	author       = {Yulian Mao and
	Qingqing Ye and
	Haibo Hu and
	Qi Wang and
	Kai Huang},
	title        = {PrivShape: Extracting Shapes in Time Series Under User-Level Local
	Differential Privacy},
	booktitle    = {40th {IEEE} International Conference on Data Engineering, {ICDE} 2024,
	Utrecht, The Netherlands, May 13-16, 2024},
	pages        = {1739--1751},
	publisher    = {{IEEE}},
	year         = {2024}
}

@article{DBLP:journals/tdsc/0001LJ21,
	author       = {Tianhao Wang and
	Ninghui Li and
	Somesh Jha},
	title        = {Locally Differentially Private Heavy Hitter Identification},
	journal      = {{IEEE} Trans. Dependable Secur. Comput.},
	volume       = {18},
	number       = {2},
	pages        = {982--993},
	year         = {2021}
}

@article{kemp1956generalized,
	title={Generalized hypergeometric distributions},
	author={Kemp, C David and Kemp, Adrienne W},
	journal={Journal of the Royal Statistical Society Series B: Statistical Methodology},
	volume={18},
	number={2},
	pages={202--211},
	year={1956}
}

@article{DBLP:journals/pvldb/LiuLXLM21,
	author       = {Junxu Liu and
	Jian Lou and
	Li Xiong and
	Jinfei Liu and
	Xiaofeng Meng},
	title        = {Projected Federated Averaging with Heterogeneous Differential Privacy},
	journal      = {Proc. {VLDB} Endow.},
	volume       = {15},
	number       = {4},
	pages        = {828--840},
	year         = {2021}
}

@article{DBLP:journals/tit/Lin91,
	author       = {Jianhua Lin},
	title        = {Divergence measures based on the Shannon entropy},
	journal      = {{IEEE} Trans. Inf. Theory},
	volume       = {37},
	number       = {1},
	pages        = {145--151},
	year         = {1991}
}

@inproceedings{DBLP:conf/mdm/CaoY15,
	author       = {Yang Cao and
	Masatoshi Yoshikawa},
	editor       = {Christian S. Jensen and
	Xing Xie and
	Vladimir Zadorozhny and
	Sanjay Madria and
	Evaggelia Pitoura and
	Baihua Zheng and
	Chi{-}Yin Chow},
	title        = {Differentially Private Real-Time Data Release over Infinite Trajectory
	Streams},
	booktitle    = {16th {IEEE} International Conference on Mobile Data Management, {MDM}
	2015, Pittsburgh, PA, USA, June 15-18, 2015 - Volume 2},
	pages        = {68--73},
	publisher    = {{IEEE} Computer Society},
	year         = {2015}
}

@article{DBLP:journals/jbd/ZolfaghariMJ26,
	author       = {Majid Zolfaghari and
	Ahmad Mohammadi and
	Rasool Jalili},
	title        = {{SHieLD}: a personalized local differential privacy framework for continual
	reports},
	journal      = {J. Big Data},
	volume       = {13},
	number       = {1},
	pages        = {144},
	year         = {2026}
}
\egroup

\section{Appendix}
\subsection{Personalized Frequency Oracle}
\subsubsection{GPRR-based PFO and Unbiasedness}\label{appendix:PFO-and-variances}

Similar to Frequency Oracle ($\shrink{FO}$)~\cite{DBLP:conf/ndss/0001LLSL20} in the standard LDP model, frequency estimation under PLDP can also be encapsulated into a standard protocol. 
Here, we refer to this as the Personalized Frequency Oracle ($\shrink{PFO}$).

$\shrink{PFO}$ consists of three functions: the \textit{\textbf{P}erturbation} function $\shrink{PFO.P}$, the \textit{\textbf{A}ggregate} function $\shrink{PFO.A}$ and the \textit{\textbf{E}stimation} function $\shrink{PFO.E}$. 
The function $\shrink{PFO.P}$ randomizes the raw data into obfuscated data using the PLDP algorithm.
$\shrink{PFO.A}$ calculates the aggregate statistic for different groups of the obfuscated data.
$\shrink{PFO.E}$ is responsible for estimating the distribution of raw data.

We adopt Generalized Personalized Random Response (\shrink{GPRR}) as the personalized frequency oracle due to its simplicity and unbiasedness.
Each user perturbs his/her value using GPRR with personalized privacy budget $\epsilon_i$:
\begin{equation}\notag
	\outereqsize{
		\begin{aligned}
			\eqbold{PFO.P:}\;\; \forall y_i\in \Omega, \Pr[M(x_i)=y_i]=
			\begin{cases}
				p_i=\dfrac{e^{\epsilon_i}}{e^{\epsilon_i}+d-1}, &\textrm{if } y_i = x_i, \\[6pt]
				q_i=\dfrac{1}{e^{\epsilon_i}+d-1}, &\textrm{otherwise.}
			\end{cases}
		\end{aligned}
	}
\end{equation}

After receiving the obfuscated reports $\vectorfont{y}_r$ from all users across different groups divided by distinct $\widetilde{\epsilon}_r\in\widetilde{\vectorfont{\epsilon}}$, the server computes the obfuscated statistic of each domain value $\omega_j\in\Omega$ for every  group $G_r$:
\begin{equation}\notag
	\outereqsize{
		\begin{aligned}
			\eqbold{PFO.A:}\;\;\forall r\in[m], j\in[d], h_{r,j}=\frac{\shrink{Count}\left(y_{r,j}\right)}{z_r},
		\end{aligned}
	}
\end{equation}
where $z_r\in\vectorfont{z}$ is the number of users requiring $\widetilde{\epsilon}_r$.
Thus, the obfuscated frequency list for each group $G_r$ can be calculated as $\vectorfont{h}_{r}=( h_{r,1},h_{r,2},$
$\dots, h_{r,d}) $.
Subsequently, the server estimates the overall frequency distribution as:
\begin{equation}\label{eq:estimation-2}
	\outereqsize{
		\begin{aligned}
			\eqbold{PFO.E:}\;\; \forall \widehat{f}_j\in\widehat{\vectorfont{f}},\widehat{f_j} = \sum_{r=1}^{m}\eta_{r}\cdot \widehat{f}_{r,j} = \sum_{r=1}^{m}\eta_{r}\cdot\frac{h_{{r,j}}-\widetilde{q}_{r}}{\widetilde{p}_{r}-\widetilde{q}_{r}},
		\end{aligned}
	}
\end{equation}
where $\eta_{r}=\frac{1/\var{\widehat{f}_{r,j}}}{\sum_{r'=1}^{m}1/\var{\widehat{f}_{r',j}}}$ denotes the \textit{inverse-variance weight } of group $G_r$, and $\var{\widehat{f}_{r,j}}$ represents the variance under $\epsilon_{r}$-LDP~\cite{he2025personalized} with
\begin{equation}\label{eq:grr_var}
	\outereqsize{
		\begin{aligned}
			\var{\widehat{f}_{{r,j}}}=\frac{\widetilde{q}_{{r}}(1-\widetilde{q}_{{r}})}{z_r(\widetilde{p}_{{r}}-\widetilde{q}_{{r}})^2}+\frac{1-\widetilde{p}_{{r}}-\widetilde{q}_{{r}}}{z_r(\widetilde{p}_{{r}}-\widetilde{q}_{{r}})}\cdot f_{{j}},
		\end{aligned}
	}
\end{equation}
where $\widehat{p}_r=\frac{e^{\widehat{\epsilon}_r}}{e^{\widehat{\epsilon}_r}+d-1}$ and $\widehat{q}_r=\frac{1}{e^{\widehat{\epsilon}_r}+d-1}$.

Eq.~\eqref{eq:grr_var} shows that the variance depends on the true frequency $f_{{j}}$. 
However, since $f_{j}$ is unknown, this definition is impractical for direct computation.
To address this issue, we replace $\var{\cdot}$ with the average variance $\vara{\cdot}$, and redefine the weight as $\eta_{{r}}=\frac{1/\vara{\widehat{f}_{{r,j}}}}{\sum_{r'=1}^{m}1/\vara{\widehat{f}_{{r',j}}}}$, where  

\begin{equation}\notag
	\outereqsize{
		\begin{aligned}
			\vara{\widehat{f}_{r,j}}=\frac{\widetilde{q}_{{r}}(1-\widetilde{q}_{{r}})}{z_r(\widetilde{p}_{{r}}-\widetilde{q}_{{r}})^2}+\frac{1-\widetilde{p}_{{r}}-\widetilde{q}_{{r}}}{z_r(\widetilde{p}_{{r}}-\widetilde{q}_{{r}})}\cdot\frac{1}{d}.
		\end{aligned}
	}
\end{equation}


Based on the modified weights, we can calculate the total variance in \pldpDefinitionOnlySimpleName{} as:
\begin{equation}\label{eq:vartotalupperboundsimple}
	\outereqsize{
		\begin{aligned}
			\var{\widehat{\vectorfont{f}}}
			=1/\sum_{r=1}^{m}z_r/\left(da_{r}+b_{r}\right),
		\end{aligned}
	}
\end{equation}
where $a_{r}=\frac{\widetilde{q}_{{r}}(1-\widetilde{q}_{{r}})}{(\widetilde{p}_{{r}}-\widetilde{q}_{{r}})^2}$ and
$b_{r}=\frac{1-\widetilde{p}_{{r}}-\widetilde{q}_{{r}}}{(\widetilde{p}_{{r}}-\widetilde{q}_{{r}})}$.
For more details of the modified weights and the calculation process for the total variance, please refer to Theorem~\ref{Var:PLDP}.
We abbreviate the total variance in Eq.~\eqref{eq:vartotalupperboundsimple} as $\simplevarpldp(\widetilde{\vectorfont{\epsilon}},\vectorfont{z},d)$.

The inverse-variance aggregation in PFO.E treats the budget-stratified estimates as estimators of a common underlying distribution, which corresponds to the conventional setting where privacy requirements are independent of users' data distributions. When privacy requirements are correlated with data distributions, different privacy groups may represent different underlying distributions, and inverse-variance aggregation no longer preserves the population composition. We address this issue in Section~\ref{method:enhanced} by estimating privacy groups separately and aggregating them according to their population proportions.

\begin{theorem}\label{Aggregation:PLDP}
	Suppose that the privacy-budget assignment is independent of users' data values, such that all budget groups share the same underlying distribution $\vectorfont{f}$.
	Then GPRR satisfies $\vectorfont{\epsilon}$-PLDP and $\widehat{f}_{{j}} = \sum_{b=1}^{m}\eta_{{b}}\cdot\frac{h_{{b,j}}-\widetilde{q}_{{b}}}{\widetilde{p}_{{b}}-\widetilde{q}_{{b}}}$ is an unbiased aggregation of $f_j$ for all $\omega_j\in\Omega$. 
\end{theorem}
\begin{proof}
	For any two input $x_i,x_i'\in\Omega$ of $u_i$ with the output $y_i\in\Omega$, we have
	\begin{equation}
		\outereqsize{
			\begin{aligned}
				\dfrac{\Pr[M(x_i)=y_i]}{\Pr[M(x_i')=y_i]} \leq \dfrac{p_i}{q_i} = e^{\epsilon_i}.
			\end{aligned}
		}
	\end{equation}
	Thus, GPRR is $\vectorfont{\epsilon}$-PLDP.
	
	For any expectation $\widehat{f}_{j}\in\widehat{\vectorfont{f}}$ of $f_{j}\in\vectorfont{f}$, we have
	\begin{equation}
		\outereqsize{
			\begin{aligned}\notag
				\simpleexpectation{\widehat{f}_j} &= \simpleexpectation\left[\sum_{b=1}^{m}\eta_b\cdot\frac{h_{b,j}-\widetilde{q}_{b}}{\widetilde{p}_{b}-\widetilde{q}_b}\right] \\
				&= \simpleexpectation\left[\sum_{b=1}^{m}\eta_{b}\cdot\frac{f_{j}\cdot\widetilde{p}_{b}+(1-f_{j})\widetilde{q}_{b}-\widetilde{q}_{b}}{\widetilde{p}_{b}-\widetilde{q}_b}\right] \\
				&= \simpleexpectation\left[f_{j}\cdot\sum_{b=1}^{m}\eta_{b}\right]\\
				&= f_j.
			\end{aligned}
		}
	\end{equation}
	Thus, $\widehat{f}_j$ is an unbiased estimation of $f_j$.
\end{proof}

\subsubsection{Variance of PFO}\label{subsec:modifiedPFO}

\begin{theorem}\label{Var:PLDP}
	Under the data-independent privacy-budget assignment assumed in
	Theorem~\ref{Aggregation:PLDP}, consider the GPRR-based PFO in
	Eq.~\eqref{eq:estimation-2}. For each privacy-budget group $G_b$,
	let $\widetilde{p}_b$ and $\widetilde{q}_b$ satisfy
	$\frac{1}{d}<\widetilde{p}_b\leq1$ and
	$\widetilde{q}_b=\frac{1-\widetilde{p}_b}{d-1}$.
	Then the variance of the $j$-th frequency estimate
	$\widehat{f}_j\in\widehat{\vectorfont{f}}$ is
	\begin{equation}\label{eq:varpart}
		\outereqsize{
			\begin{aligned}
				\varem{\widehat{f}_{{j}}}=
				\begin{cases}
					V_{{A,j}}=\frac{1}{\sum\limits_{i=1}^{m}z_i/\left(a_{i}+b_{i}\cdot f_{{j}}\right)}, &\shrinkem{if } \eta_b=\frac{1/\varem{\widehat{f}_{{i,j}}}}{\sum\limits_{i'=1}^{m}1/\varem{\widehat{f}_{{i',j}}}}, \\
					V_{{B,j}}=\frac{\sum\limits_{i=1}^{m}\frac{z_i(a_{i}+b_{i}\cdot f_{j})}{\left(a_{i}+b_{i}/d\right)^2}}{\left(\sum\limits_{i=1}^{m}z_i/\left(a_{i}+b_{i}/d\right)\right)^2}, &\shrinkem{if } \eta_i=\frac{1/\varaem{\widehat{f}_{{i,j}}}}{\sum\limits_{i'=1}^{m}1/\varaem{\widehat{f}_{{i',j}}}}; \\
				\end{cases}
			\end{aligned}
		}
	\end{equation} 
	Moreover, the total variance under the exact inverse-variance weights  $\sum_{j=1}^{d}V_{{A,j}}$
	is upper-bounded by the total variance under the average-variance
	weights $\sum_{j=1}^{d}V_{{B,j}}$ , which satisfies
	\begin{equation}\label{eq:vartotalupperbound}
		\outereqsize{
			\begin{aligned}
				\sum_{j=1}^{d}V_{{A,j}}
				\leq
				\frac{1}{
					\sum_{i=1}^{m}
					z_i/\left(da_{i}+b_{i}\right)
				}
				=
				\sum_{j=1}^{d}V_{{B,j}}.
			\end{aligned}
		}
	\end{equation}
	where $a_{i}=\frac{\widetilde{q}_{i}(1-\widetilde{q}_{i})}{(\widetilde{p}_{i}-\widetilde{q}_{i})^2}$ and
	$b_{i}=\frac{1-\widetilde{p}_{i}-\widetilde{q}_{i}}{(\widetilde{p}_{i}-\widetilde{q}_{i})}$.
\end{theorem}
\begin{proof}
	For a fixed privacy-budget group $G_b$ and domain value $\omega_j$,
	$h_{i,j}$ is the average of
	$z_i$ independent Bernoulli random variables.
	Since the privacy-budget assignment is independent of users' data
	values, $f_{j}z_i$ of these variables have
	parameter $\widetilde{p}_i$, while
	$(1-f_{j})z_i$ of them have parameter
	$\widetilde{q}_i$.
	Consequently,
	$\var{\widehat{f}_{i,j}}
	=(a_i
	+b_i f_{j})/z_i$,
	and
	$\vara{\widehat{f}_{i,j}}
	=(a_i
	+b_i/d)/z_i$.
	Thus, if setting $\eta_i=\frac{1/\var{\widehat{f}_{i,j}}}{\sum_{i'=1}^{m}1/\var{\widehat{f}_{i',j}}}$, then
	\begin{equation}\label{eq:proof:normal}
		\outereqsize{
			\begin{aligned}
				\var{\widehat{f}_j}
				=& \sum_{i=1}^{m}\eta_i^2\cdot\var{\widehat{f}_{i,j}}\\
				=& \sum_{i=1}^{m}\left(\frac{1/\var{\widehat{f}_{i,j}}}{\sum_{k=1}^{m}1/\var{\widehat{f}_{k,j}}}\right)^2\cdot\var{\widehat{f}_{i,j}}\\
				=& \frac{\sum\limits_{i=1}^{m}1/\var{\widehat{f}_{i,j}}}{\left(\sum\limits_{k=1}^{m}1/\var{\widehat{f}_{k,j}}\right)^2}\\
				=& \frac{1}{\sum\limits_{i=1}^{m}1/\var{\widehat{f}_{i,j}}}\\
				=& 1/\sum_{i=1}^{m}z_i/\left(a_{i}+b_{i}\cdot f_{j}\right);
			\end{aligned}
		}
	\end{equation}
	If setting $\eta_i=\frac{1/\vara{\widehat{f}_{i,j}}}{\sum_{i'=1}^{m}1/\vara{\widehat{f}_{i',j}}}$, then
	\begin{equation}\label{eq:proof:average}
		\outereqsize{
			\begin{aligned}
				\var{\widehat{f}_j}
				=& \sum_{i=1}^{m}\eta_i^2\cdot\var{\widehat{f}_{i,j}}\\
				=& \sum_{i=1}^{m}\left(\frac{1/\vara{\widehat{f}_{i,j}}}{\sum_{k=1}^{m}1/\vara{\widehat{f}_{k,j}}}\right)^2\cdot\var{\widehat{f}_{i,j}}\\
				=& \frac{\sum\limits_{i=1}^{m}\var{\widehat{f}_{i,j}}/\left(\vara{\widehat{f}_{i,j}}\right)^2}{\left(\sum\limits_{k=1}^{m}1/\vara{\widehat{f}_{k,j}}\right)^2}\\
				=&
				\frac{
					\sum\limits_{i=1}^{m}z_i
					\left(a_i+b_i f_{j}\right) / \left(a_i+b_i/d\right)^2
				}{
					\left(
					\sum\limits_{i=1}^{m}
					z_i/\left(a_i+b_i/d\right)
					\right)^2
				}.
			\end{aligned}
		}
	\end{equation}
	For Eq.~\eqref{eq:proof:normal}, define
	\begin{equation}\notag
		\outereqsize{
			\begin{aligned}
				\phi(x)
				=
				\frac{1}{
					\sum\limits_{i=1}^{m}
					z_i/\left(
					a_{i}
					+b_{i}x
					\right)
				}.
			\end{aligned}
		}
	\end{equation}
	Since $b_{i}\geq0$ and
	$a_{i}+b_{i}x>0$ for
	$x\in[0,1]$, $\phi(x)$ is a positive multiple of the harmonic
	mean of positive affine functions and is therefore concave.
	Using $\sum_{j=1}^{d}f_{j}=1$ and Jensen's inequality,
	we obtain
	\begin{equation}\notag
		\outereqsize{
			\begin{aligned}
				\sum_{j=1}^{d}V_{A,j}
				&=
				\sum_{j=1}^{d}\phi(f_{j})\\
				&\leq
				d\phi\left(\frac{1}{d}\right)\\
				&=
				\frac{1}{
					\sum\limits_{i=1}^{m}
					z_i/\left(
					da_{i}
					+b_{i}
					\right)
				}.
			\end{aligned}
		}
	\end{equation}
	Equality is attained when
	$f_{1}=\cdots=f_{d}=1/d$.

	For Eq.~\eqref{eq:proof:average}, summing over all
	$j\in[d]$ and using $\sum_{j=1}^{d}f_j=1$, we obtain
	\begin{equation}\notag
		\outereqsize{
			\begin{aligned}
				\sum_{j=1}^{d}V_{B,j}
				&=
				\frac{
					\sum\limits_{i=1}^{m}
					\frac{
						z_i(da_{i}
						+b_{i})
					}{
						\left(
						a_{i}
						+b_{i}/d
						\right)^2
					}
				}{
					\left(
					\sum\limits_{i=1}^{m}
					\frac{z_i}{
						a_{i}
						+b_{i}/d
					}
					\right)^2
				}\\
				&=
				\frac{
					d\sum\limits_{i=1}^{m}
					\frac{z_i}{
						a_{i}
						+b_{i}/d
					}
				}{
					\left(
					\sum\limits_{i=1}^{m}
					\frac{z_i}{
						a_{i}
						+b_{i}/d
					}
					\right)^2
				}\\
				&=
				\frac{1}{
					\sum\limits_{i=1}^{m}
					z_i/\left(
					da_{i}
					+b_{i}
					\right)
				}.
			\end{aligned}
		}
	\end{equation}
\end{proof}

\subsubsection{Sampling Variance and Reporting Error}
\noindent
\textbf{Reporting Errors.} 
Similar to population division methods under LDP~\cite{DBLP:journals/tdsc/0001LJ21}, population division under PLDP can also be decomposed into two components: the \textit{variance due to PLDP perturbation} and the \textit{variance due to sampling}.
The variance  for \pldpDefinitionOnlySimpleName{} perturbation, denoted as $\simplevarpldp(\widetilde{\vectorfont{\epsilon}},\vectorfont{g},n,d)$, has been shown in Eq.~\eqref{eq:vartotalupperboundsimple}.

\textit{Variance for Sampling.} 
Given a sample of users $U$ with size $z$, let $X_j$ denote the number of occurrences of $\omega_j\in\Omega$ within this sample.
Then $X_j$ follows a Hypergeometric Distribution~\cite{kemp1956generalized}, and its variance can be expressed as
\begin{equation}
	\outereqsize{
		\begin{aligned}\notag
			\varsmpl{X_j}=z\cdot f_j\cdot (1-f_j)\cdot \frac{n-z}{n-1}.
		\end{aligned}
	}
\end{equation}
Accordingly, the variance of estimated frequency for $\Omega_j$  is:
\begin{equation}\notag
	\outereqsize{
		\begin{aligned}
			\varsmpl{\widehat{f}_j}=\frac{1}{z^2}\var{X_j}=\frac{f_j\cdot (1-f_j)}{z}\cdot \frac{n-z}{n-1}.
		\end{aligned}
	}
\end{equation}
Thus, the variance of the entire frequency estimation is given by:
\begin{equation}\label{eq:samplevar}
	\outereqsize{
		\begin{aligned}
			\varsmpl{\widehat{\vectorfont{f}}} = \sum_{j=1}^{d}\frac{f_j\cdot (1-f_j)}{z}\cdot \frac{n-z}{n-1} = \frac{n-z}{z(n-1)}\left(1-\sum_{j=1}^{d}f_j^2\right).
		\end{aligned}
	}
\end{equation}
Since $\sum_{j=1}^{d}f_j^2\geq 0$, Eq.~\eqref{eq:samplevar} is upper bounded by:
\begin{equation}\label{eq:samplevarstar}
	\outereqsize{
		\begin{aligned}
			\varsmplstar{\widehat{\vectorfont{f}}} &= \frac{n-z}{z(n-1)}.
		\end{aligned}
	}
\end{equation}
We abbreviate Eq.~\eqref{eq:samplevarstar} as $\simplevarsmpl\left(n,z\right)$.

\textit{Reporting Errors for population division under PLDP.} 
The total variance combining both sampling and perturbation effects is expressed as:
\begin{equation}\label{eq:finalvar-2}
	\outereqsize{
		\begin{aligned}
			&\simplevar(\widetilde{\vectorfont{\epsilon}},n,\vectorfont{z}, d)\\ 
			=& \simplevarsmpl\left(n,z_{\rbindexsmallem{sum}}\right) + \simplevarpldp(\widetilde{\vectorfont{\epsilon}},\vectorfont{z}, d) \\
			=& \frac{n-z_{\rbindexmuchsmallem{sum}}}{z_{\rbindexmuchsmallem{sum}}(n-1)} +
			1/\sum_{r=1}^{m}z_{r}/\left(da_{r}+b_{r}\right),
		\end{aligned}
	}
\end{equation} 
We defined the reporting error as the average variance across all values in the output domain, given by:
\begin{equation}\notag 
	\outereqsize{
		\begin{aligned}
			&\shrinkem{err}(\widetilde{\vectorfont{\epsilon}},n,\vectorfont{z}, d)\\ 
			=& \frac{1}{d}\left(\simplevarsmpl\left(n,z_{\rbindexmuchsmallem{sum}}\right) + \simplevarpldp(\widetilde{\vectorfont{\epsilon}},\vectorfont{z}, d)\right) \\
			=& \frac{n-z_{\rbindexmuchsmallem{sum}}}{dz_{\rbindexmuchsmallem{sum}}(n-1)} + 
			1/\sum_{r=1}^{m}dz_r/\left(da_{r}+b_{r}\right),
		\end{aligned}
	}
\end{equation} 
which coincides with Eq.~\eqref{eq:finalavgvar}.

\subsection{Private Dissimilarity}\label{subsec:DetailsForDiss}
The ideal method of strategy determination is using non-private dissimilarity $\shrinkem{dis}^*$ for comparison. 
The expression of non-private dissimilarity is defined as the mean square error between the true frequency $\vectorfont{f}_t$ and the latest publication $\widetilde{\vectorfont{f}}_l$:
\begin{equation}\label{dis:non-private}
	\outereqsize{
		\shrinkem{dis}^* = \frac{1}{d}\sum_{j=1}^{d}\left(\vectorfont{f}_t[j]-\widehat{\vectorfont{f}}_l[j]\right)^2,
	}
\end{equation}
where $\widehat{\vectorfont{f}}_l$ represents the previous frequency publication that consumes population resources.
However, $\shrinkem{dis}^*$ as an intermediate result leads to privacy leakage. 
To overcome this limitation, we use a private dissimilarity $\shrinkem{dis}$ instead.
The design of $\shrinkem{dis}$ has two requirements: (1) $\shrinkem{dis}$ needs to be private, meaning that given a privacy requirement $\vectorfont{\epsilon}$ for $\shrinkem{dis}$ calculation, it needs to achieve $\vectorfont{\epsilon}$-PLDP; (2) $\shrinkem{dis}$ needs to be an unbiased estimation of the non-private dissimilarity $\shrinkem{dis}^*$, ensuring accurate results. 
\begin{theorem}\label{Thm:Dis}
	Let $\widehat{\vectorfont{f}}_{\rbindexem{dis},t}$ denote the unbiased frequency estimation of $\vectorfont{f}_t$ from an $\vectorfont{\epsilon}$-PLDP frequency estimation mechanism $\entity{M}_{\rbindexem{dis},t}$ at current time slot $t$. Then the following dissimilarity measure
	\begin{equation}\label{eq:dis}
		\outereqsize{
			\begin{aligned}
				\shrink{dis} &= \frac{1}{d}\sum_{j=1}^{d}\left(\widehat{\vectorfont{f}}_{\rbindexem{dis},t}[j]-\widehat{\vectorfont{f}}_l[j]\right)^2-\frac{1}{d}\sum_{j=1}^{d}\varem{\widehat{\vectorfont{f}}_{\rbindexem{dis},t}[j]}
			\end{aligned}
		}
	\end{equation}
	is $\vectorfont{\epsilon}$-PLDP and an unbiased estimation of $\shrink{dis}^*$ in Eq.~\eqref{dis:non-private}.
\end{theorem}
\begin{proof}
	We first prove (1) $dis$ in Eq.~\eqref{eq:dis} is $\vectorfont{\epsilon}$-PLDP (\textit{Privacy Proof}) and then prove 
	(2) $dis$ in Eq.~\eqref{eq:dis} is an unbiased estimation of $dis^*$ in Eq.~\eqref{dis:non-private} (\textit{Unbiasedness Proof}).
	
	\textit{Privacy Proof.} The calculation of $dis$ is only related to $\left\{y_1,\dots,y_n\right\}$ of all users, which is published utilizing GRR. Thus, According to Theorem~\ref{Aggregation:PLDP}, we get $dis$ is $\vectorfont{\epsilon}$-PLDP.

	\textit{Unbiasedness Proof.}
	Since $\widehat{\vectorfont{f}}_{\rbindexem{dis},t}$ is an unbiased estimation of $\vectorfont{f}_{t}$, for any $\widehat{\vectorfont{f}}_{\rbindexem{dis},t}[j]\in\widehat{\vectorfont{f}}_{\rbindexem{dis},t}$ and $\vectorfont{f}_{t}[j]\in\vectorfont{f}_{t}$, we have
	\begin{equation}\notag
		\outereqsize{
			\begin{aligned}
				\simpleexpectation{\widehat{\vectorfont{f}}_{\rbindexem{dis},t}[j]} &= \vectorfont{f}_{t}[j].
			\end{aligned}
		}
	\end{equation}
	Besides, for any random variable $X$ and constant $C$, according to $\var{X}=\simpleexpectation{(X-\simpleexpectation{X})^2}=\simpleexpectation{X^2}-\simpleexpectation{X}^2$, we have
	\begin{equation}\notag
		\outereqsize{
			\begin{aligned}
				\var{X} &= \simpleexpectation{(X-\simpleexpectation{X})^2}\\
				&= \simpleexpectation{((X-C)-(\simpleexpectation{X}-C))^2}\\
				&= \simpleexpectation{(X-C)^2}-\left(\simpleexpectation{X}-C\right)^2.
			\end{aligned}
		}
	\end{equation}
	Since last reporting value $\widehat{\vectorfont{f}}_l$ has been a constant for the current value $\widehat{\vectorfont{f}}_{\rbindexem{dis},t}$, thus, we have
	\begin{equation}\notag
		\outereqsize{
			\begin{aligned}
				\var{\widehat{\vectorfont{f}}_{\rbindexem{dis},t}[j]} &= \simpleexpectation{\left(\widehat{\vectorfont{f}}_{\rbindexem{dis},t}[j]-\widehat{\vectorfont{f}}_l[j]\right)^2}-\left(\vectorfont{f}_{t}[j]-\widehat{\vectorfont{f}}_l[j]\right)^2.
			\end{aligned}
		}
	\end{equation}
	Therefore, 
	\begin{equation}\notag
		\outereqsize{
			\begin{aligned}
				\simpleexpectation{\left(\widehat{\vectorfont{f}}_{\rbindexem{dis},t}[j]-\widehat{\vectorfont{f}}_l[j]\right)^2} &= \left(\vectorfont{f}_{t}[j]-\widehat{\vectorfont{f}}_l[j]\right)^2 + \var{\widehat{\vectorfont{f}}_{\rbindexem{dis},t}[j]}.
			\end{aligned}
		}
	\end{equation}
	Hence, the expectation of $dis$ in Eq.~\eqref{eq:dis} satisfies
	\begin{equation}\notag
		\outereqsize{
			\begin{aligned}
				&\simpleexpectation{dis}\\  =&\simpleexpectation{\frac{1}{d}\sum_{j=1}^{d}\left(\widehat{\vectorfont{f}}_{\rbindexem{dis},t}[j]-\widehat{\vectorfont{f}}_l[j]\right)^2-\frac{1}{d}\sum_{j=1}^{d}\var{\widehat{\vectorfont{f}}_{\rbindexem{dis},t}}} \\
				=& \frac{1}{d}\sum_{j=1}^{d}\simpleexpectation{\left(\widehat{\vectorfont{f}}_{\rbindexem{dis},t}[j]-\widehat{\vectorfont{f}}_l[j]\right)^2}-\frac{1}{d}\sum_{j=1}^{d}\var{\widehat{\vectorfont{f}}_{\rbindexem{dis},t}} \\
				=& \frac{1}{d}\sum_{j=1}^{d}\left(\left(\vectorfont{f}_{t}[j]-\widehat{\vectorfont{f}}_l[j]\right)^2 + \var{\widehat{\vectorfont{f}}_{\rbindexem{dis},t}[j]}\right)-\frac{1}{d}\sum_{j=1}^{d}\var{\widehat{\vectorfont{f}}_{\rbindexem{dis},t}} \\
				=& \frac{1}{d}\sum_{j=1}^{d}\left(\vectorfont{f}_{t}[j]-\widehat{\vectorfont{f}}_l[j]\right)^2\\
				=& dis^*.
			\end{aligned}
		}
	\end{equation}
\end{proof}


\subsection{Analysis of \advancedEnhance{}}
\subsubsection{Proof of Theorem~\ref{thm:precise_var}}\label{proof:Precise_var}
\begin{proof}
		By Eq.~\eqref{eq:samplevar}, the vector-level variance introduced by sampling $z_{r}$ users without replacement from a finite population of size $N_{r}$ is
		\begin{equation}\notag
			\outereqsize{
				\begin{aligned}
					V_{\rbindexem{smpl,r}}=\theta_{r}\left(1-\left\| \vectorfont{f}^{(r)} \right\|_2^2\right),
				\end{aligned}
			}
		\end{equation}
		where $\theta_{r} = \frac{N_{r}-z_{r}}{z_{r}\left(N_{r}-1\right)}$.
		By the law of the total variance, the sampling variance and the conditional $\operatorname{GPRR}$ perturbation variance $V_{\rbindexem{LDP,r}}$ contribute additively to the total vector-level variance. Namely, the total variance of $\widehat{\vectorfont{f}}^{(r)}$ is
		\begin{equation}\notag
			\outereqsize{
				\begin{aligned}
					V_{r} &= \theta_{r}\left(1-\left\| \vectorfont{f}^{(r)} \right\|_2^2\right) + V_{\rbindexem{LDP,r}}.
				\end{aligned}
			}
		\end{equation}
		Since $\widehat{\vectorfont{f}}^{(r)}$ is an unbiased estimate of $\vectorfont{f}^{(r)}$, we have 
		\begin{equation}\notag
			\outereqsize{
				\begin{aligned}
					\simpleexpectation{\left\| \widehat{\vectorfont{f}}^{(r)} \right\|_2^2} &= \sum_{j=1}^{d}\simpleexpectation{\widehat{\vectorfont{f}}^{(r)}[j]^2} \\
					&= \sum_{j=1}^{d}\left(\vectorfont{f}^{(r)}[j]^2+\operatorname{Var}\left[\widehat{\vectorfont{f}}^{(r)}[j]\right]\right)\\
					&= \left\| \vectorfont{f}^{(r)} \right\|_2^2 + V_{r}.
				\end{aligned}
			}
		\end{equation}
		Since $2\leq z_{r}\leq N_{r}$, thus, $0\leq \theta_{r}<1$.
		Taking the expectation of $\widehat{V}_{r}$ gives
		\begin{equation}\notag
			\outereqsize{
				\begin{aligned}
					\simpleexpectation{\widehat{V}_{r}} &=\simpleexpectation{\frac{ \theta_{r}\left(1-\left\| \widehat{\vectorfont{f}}^{(r)} \right\|_2^2\right) + V_{\rbindexem{LDP,r}}}{1-\theta_{r}}} \\
					&= \frac{\theta_{r}\left(1-\simpleexpectation{\left\| \widehat{\vectorfont{f}}^{(r)} \right\|_2^2}\right) + V_{\rbindexem{LDP,r}}}{1-\theta_{r}}\\
					&= \frac{\theta_{r}\left(1-\left\| \vectorfont{f}^{(r)} \right\|_2^2 - V_{r}\right) + V_{\rbindexem{LDP,r}}}{1-\theta_{r}}\\
					&= \frac{\theta_{r}\left(1-\left\| \vectorfont{f}^{(r)} \right\|_2^2 \right) + V_{\rbindexem{LDP,r}}-\theta_{r} V_{r}}{1-\theta_{r}}\\
					&= \frac{V_{r}-\theta_{r}V_{r}}{1-\theta_{r}}\\
					&= V_{r}.
				\end{aligned}
			}
		\end{equation}
		Hence, $\widehat{V}_{r}$ is an unbiased estimator of the vector-level total variance $V_{r}$.
\end{proof}

\subsubsection{Algorithm for \basicClipping{} and Fusion-Error Reduction}
\begin{algorithm}[!t]\small
	\caption{\basicClippingTotalName{} (\basicClipping{})}
	\label{alg:to-positive-definite}
	\DontPrintSemicolon
	\KwIn{An \specialUnbiasedMatrixBTotalName{} $\widehat{\vectorfont{M}}$; the eigenvalue lower bound $\mu>0$}
	\KwOut{The \specialCorrectedMatrixCTotalName{} $\widetilde{\vectorfont{M}}$}
	Get the eigenvalue decomposition \(\vectorfont{Q}\cdot\operatorname{diag}(\vectorfont{\nu})\cdot\vectorfont{Q}^\top\gets\widehat{\vectorfont{M}}\); \\ \label{alg:clipping:decomposition}
	\For{\(\nu_k\in\vectorfont{\nu}\)}{
		\(\widetilde{\nu}_k\gets\max{(\nu_k,\mu)}\); \label{alg:clipping:max}
	}
	Combine the corrected eigenvalues \(\widetilde{\vectorfont{\nu}}\gets(\widetilde{\nu}_1,\dots,\widetilde{\nu}_{|\vectorfont{\nu}|})\);\\
	$\widetilde{\vectorfont{M}}\gets\vectorfont{Q}\cdot\shrink{diag}(\widetilde{\vectorfont{\nu}})\cdot\vectorfont{Q}^\top$;\\ \label{alg:clipping:combine}
	\Return{\(\widetilde{\vectorfont{M}}\)};
\end{algorithm}


\begin{theorem}[\basicClipping{}-Corrected Fusion-Error Reduction]\label{thm:ba-reperturb-nearly}
	For a target privacy group $PG_{H}$, let $\widetilde{\vectorfont{M}}_H=\begin{bmatrix}
		\widetilde{V}_H& \vectorfont0^\top\\
		\vectorfont 0 & \widetilde{\vectorfont{A}}_H
	\end{bmatrix}$, where $\widetilde{V}_H=\max\left\{V_{\rbindex{risk},H}, \mu\right\}$.
	Define the minimum \basicClipping{}-corrected source fusion error as $\kappa_H=\min\limits_{\substack{\vectorfont{\beta}\succeq\vectorfont{0}\\ \vectorfont{1}^\top\vectorfont{\beta}=1}}{\vectorfont{\beta}^\top\widetilde{\vectorfont{A}}_H\vectorfont{\beta}}$.
	Then, relative to the target-only corrected objective $\widetilde{V}_H$, the minimum \advancedEnhance{} fusion objective achieves a reduction of 
	\begin{equation}\notag
		\outereqsize{
			\begin{aligned}
				\widetilde{\Delta}_H=\frac{\widetilde{V}_H^2}{\widetilde{V}_H+\kappa_H}.
			\end{aligned}
		}
	\end{equation}
\end{theorem}

\begin{proof}
	Let $\vectorfont{\lambda}_{\entity{S}_H}$ denote the vector fusion weights assigned to the source groups in $\entity{S}_H$, and
	let $\alpha_H=\vectorfont{1}^\top\vectorfont{\lambda}_{\entity{S}_H}=1-\lambda_H$ denote the total source weight.
	For $\alpha_H>0$, define the normalized source-weight vector as $\vectorfont{\beta}=\frac{\vectorfont{\lambda}_{\entity{S}_H}}{\alpha_H}$, where $\vectorfont{\beta}\succeq\vectorfont{0}$ and $\vectorfont{1}^\top\vectorfont{\beta}=1$.
	The case $\alpha_H=0$ corresponds to using the target estimate alone and is included as a boundary case.
	Define the minimum \basicClipping{}-corrected source risk as 
	\begin{equation}\notag
		\outereqsize{
			\begin{aligned}
				\kappa_H=\min_{\substack{\vectorfont{\beta}\succeq\vectorfont{0}\\ \vectorfont{1}^\top\vectorfont{\beta}=1}}{\vectorfont{\beta}^\top\widetilde{\vectorfont{A}}_H\vectorfont{\beta}}.
			\end{aligned}
		}
	\end{equation}
	Under the current-estimate specialization, the \basicClipping{}-corrected risk matrix is
	$\widetilde{\vectorfont{M}}_H=\begin{bmatrix}
		\widetilde{V}_H & \vectorfont 0^\top\\
		\vectorfont 0 & \widetilde{\vectorfont{A}}_H
	\end{bmatrix}$,
	where $\widetilde{V}_H=\max\left\{V_{\rbindexem{risk},H},\mu\right\}$.
	Then we have 
	\begin{equation}\label{eq:estimation_risk}
		\outereqsize{
			\begin{aligned}
				\widetilde{R}_H(\alpha_H,\vectorfont{\beta}) &= \widetilde{V}_H(1-\alpha_H)^2+\alpha_H^2\vectorfont{\beta}^\top\widetilde{\vectorfont{A}}_H\vectorfont{\beta}.
			\end{aligned}
		}
	\end{equation}
	Let $\vectorfont{\beta}^*=\argmin\limits_{\substack{\vectorfont{\beta}\succeq\vectorfont{0}\\ \vectorfont{1}^\top\vectorfont{\beta}=1}}\vectorfont{\beta}^\top\widetilde{\vectorfont{A}}_H\vectorfont{\beta}$. By the definition of $\kappa_H$, minimizing $\widetilde{R}_H$ is therefore reduced to
	\begin{equation}\notag
		\outereqsize{
			\begin{aligned}
				\min_{0\leq\alpha_H\leq 1} \widetilde{V}_H(1-\alpha_H)^2+\alpha_H^2\kappa_H.
			\end{aligned}
		}
	\end{equation}
	Its derivative is 
	\begin{equation}\notag
		\outereqsize{
			\begin{aligned}
				\frac{\partial\widetilde{R}_H}{\partial\alpha_H} &= -2\widetilde{V}_H\left(1-\alpha_H\right)+2\alpha_H\kappa_H.
			\end{aligned}
		}
	\end{equation}
	By setting 
	\begin{equation}\notag
		\outereqsize{
			\begin{aligned}
				-2\widetilde{V}_H\left(1-\alpha_H\right)+2\alpha_H\kappa_H=0,
			\end{aligned}
		}
	\end{equation}
	we have
	\begin{equation}\notag
		\outereqsize{
			\begin{aligned}
				\alpha^*_H &= \frac{\widetilde{V}_H}{\widetilde{V}_H+\kappa_H}, &\\
				\lambda^*_H = 1-\alpha^*_H &= \frac{\kappa_H}{\widetilde{V}_H+\kappa_H},\\
				\widetilde{R}^*_H &= \frac{\widetilde{V}_H\kappa_H}{\widetilde{V}_H+\kappa_H}.
			\end{aligned}
		}
	\end{equation}
	Therefore,
	\begin{equation}\notag
		\outereqsize{
			\begin{aligned}
				\widetilde{\Delta}_H = \widetilde{V}_H - \widetilde{R}^*_H =\frac{\widetilde{V}_H^2}{\widetilde{V}_H+\kappa_H}.
			\end{aligned}
		}
	\end{equation}
\end{proof}

When $\kappa_H$ is substantially larger than $\widetilde{V}_H$, the optimal total weight assigned to the source groups approaches zero and $\widetilde{\Delta}_H$ approaches zero. \advancedEnhance{} therefore naturally falls back to the target-group estimate when the available source groups provide little transferable utility.

\subsection{Proofs of Theorems for Time Complexity}
\subsubsection{Proof of Theorem~\ref{thm:timecost_enhanced}}\label{Proof:timecost_enhanced}
\begin{proof}
	We analyze the computational complexity of \solutionA{} and \solutionB{} at an
	arbitrary time slot $t$.
	
	Note that \basicSelection{} is executed only once during initialization.
	Specifically, \basicSelection{} depends on the users' privacy-window requirements
	$\vectorfont{w}$, privacy budgets $\vectorfont{\epsilon}$, the population
	size $n$, and the domain size $d$, all of which remain unchanged
	during stream processing. Therefore, the optimized sampling size
	$z_{\rbindexem{opt}}$, recycling period $\tau_{\rbindexem{opt}}$, and
	effective privacy-budget vector
	$\vectorfont{\epsilon}'_{\rbindexem{opt}}$ are computed once and reused
	at all subsequent time slots. Hence, the one-time cost of \basicSelection{} is not
	included in the per-time-slot complexity analyzed below.
	
	In addition, the remaining population size and recycling status
	related to $\tau_{\rbindexem{opt}}$ are maintained incrementally during
	stream processing. Therefore, checking the availability of users and
	updating the recycling population incurs $O(1)$ time per expired or
	newly generated population, rather than scanning the entire
	$\tau_{\rbindexem{opt}}$-length history.

	At each time slot, \solutionA{} and \solutionB{} consist of two sub-mechanisms,
	namely the dissimilarity mechanism $\entity{M}_{\rbindexem{dis},t}$ and the
	publication mechanism $\entity{M}_{\rbindexem{pub},t}$.
	
	For $\entity{M}_{\rbindexem{dis},t}$, a user subset of size
	$z_{\rbindexem{opt}}\le n$ is sampled from the available population.
	Each selected user applies GPRR once, and the resulting report is
	inserted into the counter corresponding to its effective privacy
	budget and reported domain value. Since each user belongs to exactly
	one effective-budget group, perturbing and aggregating all selected
	reports require $O(z_{\rbindexem{opt}})$ time cost.
	
	After aggregation, PFO.E reconstructs the frequency distribution.
	For each of the $d$ domain values, it combines the estimates from
	the $m_{\rbindexem{eff}}$ effective-budget groups. Therefore, this step
	requires $O(d\cdot m_{\rbindexem{eff}})$ time cost. 
	Computing the dissimilarity between the current private
	estimate and the last publication requires an additional $O(d)$
	time cost. 
	Consequently,
	\begin{equation}\notag
		\outereqsize{
			\begin{aligned}
				O(\entity{M}_{\rbindexem{dis},t}) = O\left(z_{\rbindexem{opt}}+d\cdot m_{\rbindexem{eff}}+d\right)
				= O\left(n+d\cdot m_{\rbindexem{eff}}\right).
			\end{aligned}
		}
	\end{equation}
	
	For $\entity{M}_{\rbindexem{pub},t}$, \solutionA{} and \solutionB{} differ in how
	the publication population is allocated: \solutionA{} distributes the
	remaining population, whereas \solutionB{} determines the publication
	population through population absorption. Nevertheless, these
	population-management operations involve at most $O(n)$ users at a
	time slot and therefore do not exceed $O(n)$ time cost.
	
	If a new publication is triggered, at most $n$ users can participate
	in $\entity{M}_{\rbindexem{pub},t}$. As in $\entity{M}_{\rbindexem{dis},t}$, perturbing and
	aggregating their reports require $O(n)$ time, while the subsequent
	PFO estimation over the $d$ domain values and
	$m_{\rbindexem{eff}}$ effective-budget groups requires
	$O(d\cdot m_{\rbindexem{eff}})$ time cost. 
	Hence, $O(\entity{M}_{\rbindexem{pub},t})=O\left(n+d\cdot m_{\rbindexem{eff}}\right)$.
	If no new publication is triggered, the previous publication is
	directly reused, whose cost is no larger than the above bound.
	
	Therefore, for either \solutionA{} or \solutionB{}, the total computational cost at
	an arbitrary time slot $t$ is $O\left(n+d\cdot m_{\rbindexem{eff}}\right)$.
\end{proof}

\subsubsection{Proof of Theorem~\ref{thm:advanced_time_cost_analysis}}\label{Proof:advanced_time_cost}

\begin{proof}
	We analyze the computational complexity of \solutionC{} and \solutionD{} at an arbitrary time slot $t$.

	Similar to \basicSelection{} in \solutionA{} and \solutionB{},
	\advancedSelection{} depends only on the privacy-group information,
	including the group population sizes, privacy-window sizes, privacy
	budgets, and the domain size, all of which remain unchanged during
	stream processing. The candidate source sets
	$\{\entity{S}_H\mid H\in[K]\}$ also depend only on the privacy requirements.
	Therefore, \advancedSelection{} and the candidate source sets can be
	computed once during initialization and reused at all subsequent time
	slots. Their one-time initialization costs are thus not included in the
	following per-time-slot complexity.

	We first analyze the dissimilarity mechanism
	$\entity{M}_{\rbindexem{dis},t}$. Since the $K$ privacy groups are
	disjoint, the total number of sampled users over all groups is at most
	$n$. Hence, sampling, perturbing, and aggregating the reports require
	$O(n)$ time cost. For each privacy group, constructing its $d$-dimensional
	frequency estimate and computing its variance estimate require
	$O(d)$ time cost. Therefore, these operations over all $K$ privacy groups
	require $O(dK)$ time. The population-weighted aggregation and the
	subsequent dissimilarity computation also require at most $O(dK)$
	time. Thus,
	\begin{equation}\notag
		\outereqsize{
			\begin{aligned}
				O\left(\entity{M}_{\rbindexem{dis},t}\right)
				=
				O\left(n+dK\right).
			\end{aligned}
		}
	\end{equation}

	We next analyze the fusion-weight computation in the publication
	mechanism $\entity{M}_{\rbindexem{pub},t}$. For each target privacy
	group $PG_H$, let $\entity{C}_H=\{H\}\cup \entity{S}_H$ and
	$q_H=|\entity{C}_H|=|\entity{S}_H|+1$.

	For \advancedCorretedWeight{}  in \solutionC{}, the empirical source-risk
	block contains at most $q_H^2$ entries. Similarly, for
	\advancedCorretedWeight{} in \solutionD{}, either a full empirical
	risk matrix or a block empirical risk matrix is constructed, both of
	which contain at most $q_H^2$ entries. Since each entry involves an
	inner product between two $d$-dimensional distribution-shift vectors,
	constructing the empirical risk matrix for $PG_H$ requires
	$O(dq_H^2)$ time cost.

	After the empirical risk matrix is constructed, \basicClipping{}
	performs eigenvalue decomposition on a matrix of size at most
	$q_H\times q_H$, which requires $O\left(q_H^3\right)$ time. The corrected
	fusion weights are then obtained by the active-set simplex-QP solver.
	Let $I_{\rbindexem{QP}}\geq 1$ denote the maximum number of
	active-set iterations. Each iteration involves solving a dense linear
	system of dimension at most $q_H$, and therefore requires at most
	$O\left(q_H^3\right)$ time cost. 
	Consequently, the fusion-weight computation for target group $PG_H$ requires
	\begin{equation}\notag
		\outereqsize{
			\begin{aligned}
				O\left(d\cdot q_H^2+I_{\rbindexem{QP}}\cdot q_H^3\right),
			\end{aligned}
		}
	\end{equation}
	where the $O\left(q_H^3\right)$ cost of \basicClipping{} is absorbed into the second term because $I_{\rbindexem{QP}}\geq 1$.

	Summing over all target privacy groups, the total cost of computing
	the fusion weights is
	\begin{equation}\notag
		\outereqsize{
			\begin{aligned}
				O\left(
				d\sum_{H=1}^{K}q_H^2
				+
				I_{\rbindexem{QP}}
				\sum_{H=1}^{K}q_H^3
				\right).
			\end{aligned}
		}
	\end{equation}

	After obtaining the fusion weights, \advancedEnhance{} combines at most $q_H$
	$d$-dimensional candidate estimates for each target privacy group.
	Therefore, the total fusion cost is
	\begin{equation}\notag
		\outereqsize{
			\begin{aligned}
				O\left(
				d\sum_{H=1}^{K}q_H
				\right),
			\end{aligned}
		}
	\end{equation}
	which is dominated by $O\left(d\sum_{H=1}^{K}q_H^2\right)$.
	The potential publication error in Eq.~\eqref{eq:final_corrected_error} or Eq.~\eqref{eq:final_corrected_time_error} can
	also be evaluated within this bound.
	The projection in Line~\ref{alg:final_new_estimates} of Alg.~\ref{alg:PLPD+} takes $O(d)$ expected time, which is dominated by $O(d\sum_H q_H^2)$.

	If a new publication is triggered, sampling, perturbing, and
	aggregating the publication users over all privacy groups require at
	most $O(n)$ time cost, while constructing the group-level frequency
	estimates and variance estimates requires $O(dK)$ time cost. 
	In POST mode,
	\advancedCorretedWeight{} is invoked once
	again after the actual publication estimates become available. This
	only introduces a constant factor and therefore does not change the
	asymptotic complexity. The population-availability and recycling
	statistics can also be maintained incrementally, as in
	\solutionA{} and \solutionB{}.

	Consequently, the per-time-slot time complexity of both
	\solutionC{} and \solutionD{} is
	\begin{equation}\notag
		\outereqsize{
			\begin{aligned}
				O\left(
				n
				+
				d\sum_{H=1}^{K}q_H^2
				+
				I_{\rbindexem{QP}}
				\sum_{H=1}^{K}q_H^3
				\right).
			\end{aligned}
		}
	\end{equation}
	
	Let
	$q_{\rbindexem{max}}=\max_{H\in[K]}q_H$.
	Since $q_H\leq q_{\rbindexem{max}}\leq K$, we have $\sum_{H=1}^{K}q_H^2 \leq Kq_{\rbindexem{max}}^2$, and $\sum_{H=1}^{K}q_H^3\leq Kq_{\rbindexem{max}}^3$.
	Therefore, the complexity is upper bounded by
	\begin{equation}\notag
		\outereqsize{
			\begin{aligned}
				O\left(
				n
				+
				dKq_{\rbindexem{max}}^2
				+
				I_{\rbindexem{QP}}
				Kq_{\rbindexem{max}}^3
				\right).
			\end{aligned}
		}
	\end{equation}
	In the worst case where $q_{\rbindexem{max}}=K$, this becomes
	\begin{equation}\notag
		\outereqsize{
			\begin{aligned}
				O\left(
				n+dK^3+I_{\rbindexem{QP}}K^4
				\right).
			\end{aligned}
		}
	\end{equation}
\end{proof}

\subsection{Proofs of Theorems for Privacy Analysis}
According to relationship between the definitions of \privacyDefinitionSimpleName{}, \pldpDefinitionSimpleName{}, we have the following important three lemmas for privacy analysis.
\begin{lemma}\label{lemma:PLDP-EPLDP}
	Given any window size $w\in\mathbb{N}^+$, a mechanism $\entity{M}$ satisfies $(w\cdot \vectorfont{1}, \vectorfont{\epsilon})$-\privacyDefinitionOnlySimpleName{} if it satisfies \pldpDefinitionSimpleName{} in any window of size $w$.
\end{lemma}
\begin{proof}
	Let $S_t$ and $\widehat{S}_t$ be any $w$-neighboring stream prefixes.
	If $S_t=\widehat{S}_t$, then for any user $u_i\in\entity{U}$, it is obvious that $\Pr\left[\entity{M}\left(S_t\right)\right]\leq e^{\vectorfont{\epsilon}(u_i)}\cdot\Pr\left[\entity{M}\left(\widehat{S}_t\right)\right]$.
	Otherwise, we can find $\tau_1,\tau_2\in[t]$ satisfying  $\tau_2-\tau_1+1\leq w$,  with $S_t[\tau_1]\neq \widehat{S}_t[\tau_1]$ and $S_t[\tau_2]\neq \widehat{S}_t[\tau_2]$, meanwhile for any $\tau\in[1,\tau_1-1]\cup[\tau_2+1,t]$ it satisfies $S_t[\tau]=\widehat{S}_t[\tau]$, and for any $\tau\in[\tau_1,\tau_2]$, it satisfies $S_t[\tau]$ and $\widehat{S}_t[\tau]$ are neighboring.

	Let $\widetilde{\vectorfont{\epsilon}}=\shrink{Unique}(\vectorfont{\epsilon})$ be the unique $\epsilon$ among all users.
	Let $m$ be the number of $\widetilde{\vectorfont{\epsilon}}$.
	For any two $w$ length sub-stream segments $S_{\tau-w+1,\tau}$ and $\widehat{S}_{\tau-w+1,\tau}$ 
	satisfying $\tau-w+1\leq \tau_1\leq \tau_2\leq \tau$,
	since $\entity{M}$ satisfies \pldpDefinitionSimpleName{} within any $w$ window, then for any personalized subset-stream segment $S_{\tau-w+1,\tau}^{(k)}\subseteq S_{\tau-w+1,\tau}$ and $\widehat{S}_{\tau-w+1,\tau}^{(k)}\subseteq S_{\tau-w+1,\tau}$, we have $\forall y\in\shrink{Range}(\entity{M})$,
	\begin{equation}\notag
		\outereqsize{
			\begin{aligned}
				\Pr{\left[\entity{M}\left(S_{\tau-w+1,\tau}^{(k)}\right)=y\right]} \leq e^{\vectorfont{\epsilon}(u_i)} \Pr{\left[\entity{M}\left(\widehat{S}_{\tau-w+1,\tau}^{(k)}\right)=y\right]}.
			\end{aligned}
		}
	\end{equation}
	Therefore, for any personalized stream Prefix $S_t^{(k)}\subseteq S_t, \widehat{S}_t^{(k)}\subseteq\widehat{S}_t$ and $\forall y\in\shrink{Range}(\entity{M})$ with $y=y_{1:\tau-w}\|y_{\tau-w+1,\tau}\|y_{\tau+1,t}$, we have
	\begin{equation}\notag
		\outereqsize{
			\begin{aligned}
				&\frac{\Pr{\left[\entity{M}\left(S_{t}^{(k)}\right)=y\right]} }{\Pr{\left[\entity{M}\left(\widehat{S}_{t}^{(k)}\right)=y\right]}} \\
				\leq &\frac{\Pr{\left[\entity{M}\left(S_{1,\tau-w}^{(k)}\right)=y_{1,\tau-w}\right]} }{\Pr{\left[\entity{M}\left(\widehat{S}_{1,\tau-w}^{(k)}\right)=y_{1,\tau-w}\right]} }\times\frac{\Pr{\left[\entity{M}\left(S_{\tau-w+1,\tau}^{(k)}\right)=y_{\tau-w+1,\tau}\right]}}{\Pr{\left[\entity{M}\left(\widehat{S}_{\tau-w+1,\tau}^{(k)}\right)=y_{\tau-w+1,\tau}\right]}}\\
				&\times\frac{\Pr{\left[\entity{M}\left(S_{\tau+1,t}^{(k)}\right)=y_{\tau+1,t}\right]}}{\Pr{\left[\entity{M}\left(\widehat{S}_{\tau+1,t}^{(k)}\right)=y_{\tau+1,t}\right]}}\\
				=& \frac{\Pr{\left[\entity{M}\left(S_{\tau-w+1,\tau}^{(k)}\right)=y_{\tau-w+1,\tau}\right]}}{\Pr{\left[\entity{M}\left(\widehat{S}_{\tau-w+1,\tau}^{(k)}\right)=y_{\tau-w+1,\tau}\right]}}\\
				\leq & e^{\vectorfont{\epsilon}(u_i)}.
			\end{aligned}
		}
	\end{equation}
	Therefore, $\entity{M}$ satisfies $(w\cdot \vectorfont{1}, \vectorfont{\epsilon})$-\privacyDefinitionOnlySimpleName{}.
\end{proof}

\begin{lemma}\label{lemma:change-window}
	Given $\widehat{w}\in\constvar{N}^+$, $\entity{M}$ satisfies \privacyDefinitionSimpleName{} if it satisfies $(\widehat{w}\cdot\vectorfont{1}, \widehat{\vectorfont{\epsilon}})$-\privacyDefinitionOnlySimpleName{} where
	\begin{equation}\notag
		\outereqsize{
			\begin{aligned}
				\widehat{\vectorfont{\epsilon}}(u_i) = 
				\begin{cases}
					\vectorfont{\epsilon}(u_i), &\textrm{\em if } \vectorfont{w}(u_i)\leq \widehat{w}, \\
					\vectorfont{\epsilon}(u_i) / \left\lceil \vectorfont{w}(u_i)/\widehat{w}\right\rceil, &\textrm{\em otherwise}. \\
				\end{cases}
			\end{aligned}
		}
	\end{equation}
\end{lemma}
\begin{proof}
	All users can be classified into two cases: (1)  the users whose window sizes are no more than $\widehat{w}$; (2) the users whose window sizes are larger than $\widehat{w}$.
	
	For any user $u_i$ in case~(1), it satisfies $\vectorfont{w}(u_i)=w_k\leq \widehat{w}$. 
	Then $\forall S_t^{(k)},\widehat{S}_t^{(k)}$ with $S_t^{(k)}\sim_{w_k} \widehat{S}_t^{(k)}$, we have $S_t^{(k)}\sim_{\widehat{w}} \widehat{S}_t^{(k)}$.
	Because $\entity{M}$ satisfies $(\widehat{w}\cdot\vectorfont{1}, \widehat{\vectorfont{\epsilon}})$-\privacyDefinitionOnlySimpleName{} , we have $\forall y\in\shrink{Range}(\entity{M})$,
	\begin{equation}\notag
		\outereqsize{
			\begin{aligned}
				\frac{\Pr\left[\entity{M}\left(S_t^{(k)}\right)=y\right]}{\Pr\left[\entity{M}\left(\widehat{S}_t^{(k)}\right)=y\right]} \leq e^{\widehat{\vectorfont{\epsilon}}(u_i)} = e^{\vectorfont{\epsilon}(u_i)}.
			\end{aligned}
		}
	\end{equation}
	
	For any user $u_i$ in case~(2), it satisfies $\vectorfont{w}(u_i)=w_k> \widehat{w}$. 
	Let $S_t^{(k)},\widehat{S}_t^{(k)}$ be any two personalized stream prefixes satisfying $S_t\sim_{w_k} \widehat{S}_t$.
	If $S_t^{(k)}=\widehat{S}_t^{(k)}$, then it is obvious that $\Pr\left[\entity{M}\left(S_t^{(k)}\right)=y\right]/$ $\Pr\left[\entity{M}\left(\widehat{S}_t^{(k)}\right)=y\right] = 1 \leq e^{\vectorfont{\epsilon}(u_i)}.$
	Otherwise, we can find $\tau_1,\tau_2$ satisfying $\tau_2-\tau_1+1\leq w_k, S_t^{(k)}[\tau_1]\neq \widehat{S}_t^{(k)}[\tau_1], S_t^{(k)}[\tau_2]\neq \widehat{S}_t^{(k)}[\tau_2]$ and $\forall \tau\in[1,\tau_1-1]\cup[\tau_2+1,t]$, $S_t^{(k)}[\tau]=\widehat{S}_t^{(k)}[\tau]$.
	Let $z=\lceil w_k / \widehat{w}\rceil$, then we can construct a list of $z-1$ stream prefixes $\check{\vectorfont{S}}=\left\langle \check{S}_t^{(k,1)}, \check{S}_t^{(k,2)},\dots,\check{S}_t^{(k,z-1)}\right\rangle$ satisfying $S_t^{(k)}\sim_{\widehat{w}} \check{S}_t^{(k,1)}$, $\check{S}_t^{(k,z-1)}\sim_{\widehat{w}} \widehat{S}_t^{(k)}$ and $\forall j\in[2,z-1]$, $\check{S}_t^{(k,j-1)}\sim_{\widehat{w}} \check{S}_t^{(k,j)}$.
	By adding $S_t^{(k)}$ to the head of $\check{\vectorfont{S}}$ and $\widehat{S}_t^{(k)}$ to tail, we can get the extended list of $z+1$ stream prefixes $\check{\vectorfont{S}}^+ = \left\langle S_t^{(k)}, \check{S}_t^{(k,1)}, \check{S}_t^{(k,2)},\dots,\check{S}_t^{(k,z-1)}, \widehat{S}_t^{(k)}\right\rangle$.
	For any neighboring pair prefixes $x,x'$ in $\check{\vectorfont{S}}^+$, because $\entity{M}$ satisfies $(\widehat{w}\cdot\vectorfont{1}, \widehat{\vectorfont{\epsilon}})$-\privacyDefinitionOnlySimpleName{}, we have 
	$\forall y\in\shrink{Range}(\entity{M}),\Pr[\entity{M}(x)=y] / \Pr[\entity{M}(x')=y]\leq e^{\epsilon_k/z}$.
	Therefore, for $\forall y\in\shrink{Range}(\entity{M})$, we have
	\begin{equation}\notag
		\outereqsize{
			\begin{aligned}
				&\frac{\Pr\left[\entity{M}\left(S_{t}^{(k)}\right)=y\right]}{\Pr\left[\entity{M}\left(\widehat{S}_{t}^{(k)}\right)=y\right]} \\
				=& \frac{\Pr\left[\entity{M}\left(S_{t}^{(k)}\right)=y\right]}{\Pr\left[\entity{M}\left(\check{S}_{t}^{(k,1)}\right)=y\right]} \times \prod_{j=2}^{z-1} \frac{\Pr\left[\entity{M}\left(\check{S}_{t}^{(k,j-1}\right)=y\right]}{\Pr\left[\entity{M}\left(\check{S}_{t}^{(k,j)}\right)=y\right]}\\
				&\times \frac{\Pr\left[\entity{M}\left(\check{S}_{t}^{(k,z-1)}\right)=y\right]}{\Pr\left[\entity{M}\left(\widehat{S}_{t}^{(k)}\right)=y\right]} \\
				\leq& e^{\epsilon_k/z} \times \left(e^{\epsilon_k/z} \right)^{z-2}\times e^{\epsilon_k/z} \\
				=& e^{\epsilon_k}\\
				=& e^{\vectorfont{\epsilon}(u_i)}.
			\end{aligned}
		}
	\end{equation}
	
	Therefore, $\entity{M}$ satisfies \privacyDefinitionSimpleName{}.
\end{proof}

\begin{lemma}\label{lemma:grouped-plus}
	Let the user set be partitioned into $K$ disjoint privacy groups $U_1,\ldots,U_K$, where all users in $PG_r=(\widetilde{w}_r,\widetilde{\epsilon}_r,U_r)$ share the same privacy requirement. Suppose that, for each $PG_r$, every user
	participates at most once within any $\tau_r$ consecutive time slots and each participating user applies an $\epsilon'_r$-LDP local
	randomizer, where
	\begin{equation}\notag
		\outereqsize{
			\begin{aligned}
				\epsilon'_r =
				\frac{\widetilde{\epsilon}_r}{\left\lceil\widetilde{w}_r/\tau_r\right\rceil}.
			\end{aligned}
		}
	\end{equation}
	If all subsequent operations only process the locally private outputs, then the overall mechanism satisfies
	$(\vectorfont{w},\vectorfont{\epsilon})$-EPLDP.
\end{lemma}
\begin{proof}
	Fix an arbitrary privacy group $PG_r=(\widetilde{w}_r,\widetilde{\epsilon}_r,U_r)$.
	Since each user in $U_r$ participates at most once within any
	$\tau_r$ consecutive time slots and each report is generated by an
	$\epsilon'_r$-LDP local randomizer, the joint private output within
	such a window satisfies
	$\epsilon'_r\cdot\vectorfont{1}$-PLDP over $U_r$.
	
	According to Lemma~\ref{lemma:PLDP-EPLDP}, the mechanism restricted to $U_r$ therefore
	satisfies
	$(\tau_r\cdot\vectorfont{1},
	\epsilon'_r\cdot\vectorfont{1})$-EPLDP.
	
	Since
	$\epsilon'_r=
	\widetilde{\epsilon}_r/
	\lceil\widetilde{w}_r/\tau_r\rceil$,
	applying Lemma~\ref{lemma:change-window} to the user subset $U_r$ with
	$\widehat{w}=\tau_r$ gives
	$(\widetilde{w}_r\cdot\vectorfont{1},
	\widetilde{\epsilon}_r\cdot\vectorfont{1})$-EPLDP.
	
	Now fix any user $u_i\in U_r$.
	The remaining privacy groups contain no raw data of $u_i$.
	Moreover, after the locally private group-level reports are obtained,
	all cross-group operations, including
	\advancedCorretedWeight{},
	\basicClipping{}, \advancedEnhance{}, aggregation, reuse, and projection,
	operate only on private outputs. 
	Hence, with respect to $u_i$, conditioned on the data and locally private outputs of all other users, the final mechanism depends on $u_i$'s private stream only through $(\widetilde{w}_r,\widetilde{\epsilon}_r)$-EPLDP output generated from $PG_r$.
	Hence, the subsequent cross-group operations preserve the same privacy bound for $u_i$.

	Since every user belongs to exactly one privacy group, the above
	argument holds for every $u_i\in U$. Thus, the overall mechanism
	satisfies
	$(\vectorfont{w},\vectorfont{\epsilon})$-EPLDP.
\end{proof}

Next, we give a lemma to the privacy analysis of \solutionA{} and \solutionB{}.

\begin{lemma}\label{lemma:once_appearance}
	For both \solutionA{} and \solutionB{}, after applying \basicSelection{}, each user participates at most once within any $\tau_{\rbindex{opt}}$-length consecutive time slots, where $\tau_{\rbindex{opt}}$ is the recycling period returned by \basicSelection{}.
\end{lemma}

\begin{proof}
	We only need to prove that the total number of reporting users within any $\tau_{\rbindexem{opt}}$ consecutive time slots is no more than $n$, i.e., $\sum_{\tau=t-\tau_{\rbindexem{opt}}+1}^{t}\left|U_{\tau}\right|\leq n$. 
	
	Since the available-user set is updated by removing selected users after each report and expired users are recycled only after $\tau_{\rbindexem{opt}}$ time slots, it is sufficient to show that the total number of assigned reporting users within any $\tau_{\rbindexem{opt}}$-length window does not exceed $n$.

	For \solutionA{}, in $\entity{M}_{\rbindexem{dis},t}$, $z_{\rbindexem{opt}}$ users are allocated at each time slot.
	According to the definition of $\tau_{\rbindexem{opt}}$ in \basicSelection{} with $\rho=1/2$, $\tau_{\rbindexem{opt}} z_{\rbindexem{opt}}\le \frac n2$.
	Therefore, within any $\tau_{\rbindexem{opt}}$ consecutive time slots,
	the total number of users participating in $\entity{M}_{\rbindexem{dis},t}$ is at most $n/2$.
	In $\entity{M}_{\rbindexem{pub},t}$, \solutionA{}  either publishes with additional users $U_{\rbindexem{pub},t}$ or approximates the last release without any user assignment.
	In the former case,  there are at most $\left|U_{\rbindexem{pub},t}\right|\leq\left(n/2-\sum_{\tau=t-\tau_{\rbindexsmallem{opt}+1}}^{t-1}\left|U_{\rbindexem{pub},\tau}\right|\right)/2$ users.
	Thus, we have $0\leq\sum_{\tau=t-\tau_{\rbindexsmallem{opt}}+1}^{t}\left|U_{\rbindexem{pub},\tau}\right|\leq n/2$.
	In the latter case, $\left|U_{\rbindexem{pub},t}\right|=0$.
	Therefore, for any time slot $t$, the total number of reporting users within any $\tau_{\rbindexem{opt}}$ consecutive time slots for \solutionA{} satisfies
	\begin{equation}\notag
		\outereqsize{
			\begin{aligned}
				\sum_{\tau=t-\tau_{\rbindexsmallem{opt}}+1}^{t}\left|U_{\tau}\right|=\sum_{\tau=t-\tau_{\rbindexsmallem{opt}}+1}^{t}\left|U_{\rbindexem{dis},\tau}\right|+\sum_{\tau=t-\tau_{\rbindexsmallem{opt}}+1}^{t}\left|U_{\rbindexem{pub},\tau}\right|\leq \frac{n}{2}+\frac{n}{2} = n.
			\end{aligned}
		}
	\end{equation}

	For \solutionB{}, similarly, in $\entity{M}_{\rbindexem{dis},t}$, there are $\sum_{\tau=t-\tau_{\rbindexsmallem{opt}}+1}^{t}\left|U_{\rbindexem{dis},\tau}\right|\leq n/2$ users for any $t$ and $\tau\in[t]$.
	In $\entity{M}_{\rbindexem{pub},t}$, assume there are $c$ new publications in any $\tau_{\rbindexem{opt}}$ consecutive time slots.
	We denote these publication time slots as $\left(\tau_1, \tau_2,\dots,\tau_c\right)$.
	For any publication time slot $\tau_j \left(j\in[c]\right)$, we denote the number of its absorbed population share as $\eta_{j}$, thus after absorption, time slot $\tau_j$ will occupy $\eta_{j}+1$ population shares.
	If $c=0$, no publication users are assigned, and hence the total
	publication population is $0$.
	If $c=1$, the single publication consumes at most
	$\tau_{\rbindexem{opt}}$ population shares. Therefore,
	\begin{equation}\notag
		\outereqsize{
			\begin{aligned}
				\sum_{\tau=t-\tau_{\rbindexsmallem{opt}}+1}^{t}
				\left|U_{\rbindexem{pub},\tau}\right|
				\leq
				\tau_{\rbindexem{opt}}z_{\rbindexem{opt}}
				\leq
				\frac{n}{2}.
			\end{aligned}
		}
	\end{equation}
	It remains to consider the case $c\geq2$.
	Figure~\ref{fig:plbaPrivacyProof} illustrates an example for $c=3$ and $\tau_{\rbindexem{opt}}=9$, in which, for instance, time slot $3$ absorbs population shares from time slots $1$ and $2$,  while nullifying the population at time slots $4$ and $5$.
	\begin{figure}[h]
		\centering
		\includegraphics[width=0.4\textwidth]{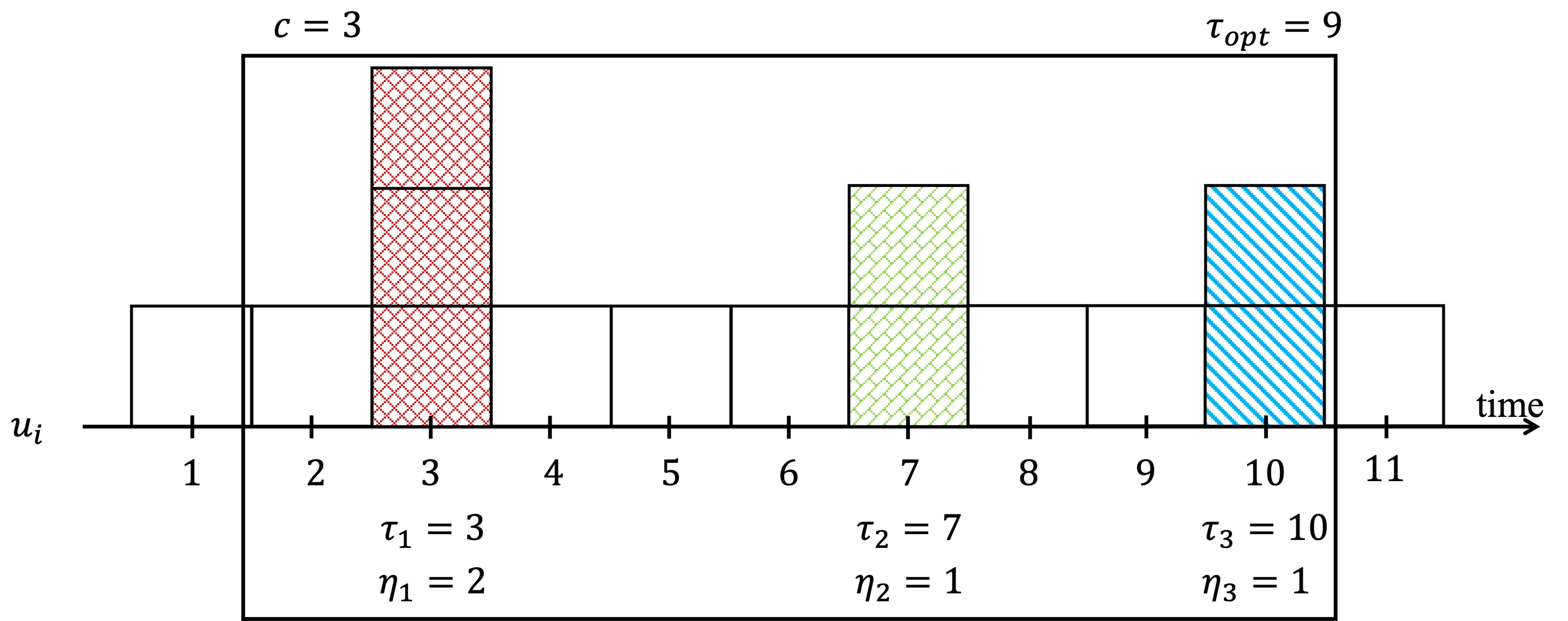}
		\caption{An example for new publications' absorption in \solutionB{}.}\label{fig:plbaPrivacyProof}
	\end{figure}
	For $c\geq2$, the absorbed and nullified time-slot lengths are symmetric with respect to each new publication. Therefore,
	\begin{equation}\notag
		\outereqsize{
			\begin{aligned}
				\tau_{\rbindexem{opt}} &\geq \sum_{j=1}^{c}\left(1+2\eta_j\right) - \eta_1 - \eta_c.
			\end{aligned}
		}
	\end{equation}
	For $c\geq 2$ and any time slot  $t$, we have
	\begin{equation}\notag
		\outereqsize{
			\begin{aligned}
				\sum_{\tau=t-\tau_{\rbindexsmallem{opt}}+1}^{t} \left|U_{\rbindexem{pub},\tau}\right| &\leq z_{\rbindexem{opt}} \cdot\sum_{j=1}^{c}\left(1+\eta_j\right)\\
				&\leq \frac{n}{2\tau_{\rbindexsmallem{opt}}} \cdot\sum_{j=1}^{c}\left(1+\eta_j\right)\\
				&\leq \frac{n\cdot \sum_{j=1}^{c}\left(1+\eta_j\right)}{2\sum_{j=1}^{c}\left(1+2\eta_j\right) - 2\eta_1 - 2\eta_c}\\
				&=  \frac{n\cdot \sum_{j=1}^{c}\left(1+\eta_j\right)}{2\sum_{j=1}^{c}\left(1+\eta_j\right) + 2\sum_{j=2}^{c-1}\eta_j }\\
				&\leq n / 2.
			\end{aligned}
		}
	\end{equation}
	
	Combining the cases $c=0$, $c=1$, and $c\geq2$, the total number of publication users within any $\tau_{\rbindexem{opt}}$-length window is at most $n/2$.
	
	Therefore, for any time slot $t$, the total number of reporting users within any $\tau_{\rbindexem{opt}}$-length window for \solutionB{} satisfies
	\begin{equation}\notag
		\outereqsize{
			\begin{aligned}
				\sum_{\tau=t-\tau_{\rbindexsmallem{opt}}+1}^{t}\left|U_{\tau}\right|=\sum_{\tau=t-\tau_{\rbindexsmallem{opt}}+1}^{t}\left|U_{\rbindexem{dis},\tau}\right|+\sum_{\tau=t-\tau_{\rbindexsmallem{opt}}+1}^{t}\left|U_{\rbindexem{pub},\tau}\right|\leq n.
			\end{aligned}
		}
	\end{equation}
\end{proof}


\subsubsection{Proof of Theorem~\ref{thm:PLP+PrivacyAnalysis}}\label{Proof:PLP+PrivacyAnalysis}
\begin{proof}
	According to Lemma~\ref{lemma:once_appearance}, for both \solutionA{} and \solutionB{}, each user
	participates at most once within any $\tau_{\rbindexem{opt}}$ consecutive
	time slots. Moreover, according to \basicSelection{}, the effective privacy budget
	of user $u_i$ is
	\[
	\epsilon'_{\rbindexem{opt},i}
	=
	\frac{\epsilon_i}
	{\left\lceil w_i/\tau_{\rbindexem{opt}}\right\rceil}.
	\]
	Since each report is perturbed by GPRR under its corresponding
	effective privacy budget, \solutionA{} and \solutionB{} satisfy
	$\vectorfont{\epsilon}'_{\rbindexem{opt}}$-PLDP within any
	$\tau_{\rbindexem{opt}}$-length window.
	
	According to Lemma~\ref{lemma:PLDP-EPLDP}, both mechanisms therefore satisfy  $(\tau_{\rbindexem{opt}}\cdot \vectorfont{1}, \vectorfont{\epsilon}_{\rbindexem{opt}}')$-\privacyDefinitionOnlySimpleName{}.

	For each user $u_i$, the effective privacy budget satisfies
	\begin{equation}\notag
		\outereqsize{
			\begin{aligned}
				\epsilon'_{\rbindexem{opt},i}
				=
				\begin{cases}
					\epsilon_i,
					& \textrm{if } w_i\le \tau_{\rbindexem{opt}},\\[1mm]
					\dfrac{\epsilon_i}
					{\left\lceil w_i/\tau_{\rbindexem{opt}}\right\rceil},
					& \textrm{otherwise}.
				\end{cases}
			\end{aligned}
		}
	\end{equation}
	Therefore, by Lemma~\ref{lemma:change-window} with
	$\widehat{w}=\tau_{\rbindexem{opt}}$, both \solutionA{} and \solutionB{} satisfy
	 \privacyDefinitionSimpleName{}.
\end{proof}

\subsubsection{Proof of Theorem~\ref{thm:plus:privacy}}\label{Proof:plus:privacy}
\begin{proof}
	We first show that, for each privacy group $PG_r$, every user
	participates at most once within any $\tau_r^*$ consecutive time
	slots in both \solutionC{} and \solutionD{}.

	For privacy group $PG_r$, let $n_r=|U_r|$. According to \advancedSelection{} with $\rho=1/2$, we have $n_{\rbindexem{sml},r}=\lfloor n_r/2\rfloor$. 
	Let $z_{\rbindexem{opt},\rbindexem{dis},r}^{*}$ and $\tau_r^*$ denote the optimal sampling size and recycling period returned by \advancedSelection{}, respectively. 
	By the definition of $\tau_r^*$, we have 
	\begin{equation}\notag
		\outereqsize{
			\begin{aligned}
				\tau_r^*
				z_{\rbindexem{opt},\rbindexem{dis},r}^{*}
				\leq
				n_{\rbindexem{sml},r}
				\leq
				\left\lfloor\frac{n_r}{2}\right\rfloor.
			\end{aligned}
		}
	\end{equation}
	Therefore, within any $\tau_r^*$ consecutive time slots, the total
	number of users assigned to the dissimilarity mechanism satisfies
	\begin{equation}\notag
		\outereqsize{
			\begin{aligned}
				\sum_{\tau=t-\tau_r^*+1}^{t}
				\left|U_{\rbindexem{dis},r,\tau}\right|
				\leq
				\left\lfloor\frac{n_r}{2}\right\rfloor.
			\end{aligned}
		}
	\end{equation}

	For \solutionC{}, the publication mechanism maintains a
	group-specific remaining publication population. At time slot $t$,
	the potential publication population satisfies
	$n_{\rbindexem{pp},r}\leq n_{\rbindexem{pub},r}$, where
	\begin{equation}\notag
		\outereqsize{
			\begin{aligned}
				n_{\rbindexem{pub},r}
				=
				\left\lfloor\frac{n_r}{2}\right\rfloor
				-
				\sum_{\tau=t-\tau_r^*+1}^{t-1}
				\left|U_{\rbindexem{pub},r,\tau}\right|.
			\end{aligned}
		}
	\end{equation}
	Hence, whether or not a new publication is generated, we have
	\begin{equation}\notag
		\outereqsize{
			\begin{aligned}
				\sum_{\tau=t-\tau_r^*+1}^{t}
				\left|U_{\rbindexem{pub},r,\tau}\right|
				\leq
				\left\lfloor\frac{n_r}{2}\right\rfloor.
			\end{aligned}
		}
	\end{equation}
	Combining the dissimilarity and publication populations gives
	\begin{equation}\notag
		\outereqsize{
			\begin{aligned}
				\sum_{\tau=t-\tau_r^*+1}^{t}
				\left(
				\left|U_{\rbindexem{dis},r,\tau}\right|
				+
				\left|U_{\rbindexem{pub},r,\tau}\right|
				\right)
				\leq
				2\left\lfloor\frac{n_r}{2}\right\rfloor
				\leq n_r.
			\end{aligned}
		}
	\end{equation}

	For \solutionD{}, the dissimilarity mechanism is identical to that
	of \solutionC{}. We therefore only need to consider the publication
	mechanism. The population-absorption procedure is applied
	independently within each privacy group and has the same structure
	as that of \solutionB{}, with the unified parameters replaced by the
	group-specific parameters
	$n_r$, $z_{\rbindexem{opt},\rbindexem{dis},r}^{*}$, and
	$\tau_r^*$.

	Fix any privacy group $PG_r$. 
	Assume that there are $c_r$ new publications of $PG_r$ within a $\tau_r^*$-length window, occurring at time slots $\tau_{r,1},\tau_{r,2},\dots,\tau_{r,c_r}$. 
	For the $j$-th publication, let $\eta_{r,j}$ denote the number of absorbed population shares.
	Thus, the $j$-th publication consumes $(1+\eta_{r,j})z_{\rbindexem{opt},\rbindexem{dis},r}^{*}$ users and nullifies $\eta_{r,j}$ subsequent population shares.
	
	If $c_r=1$, the publication consumes at most
	$\tau_r^*z_{\rbindexem{opt},\rbindexem{dis},r}^{*}\leq \lfloor n_r/2\rfloor$ users.
	For $c_r\geq2$, the symmetry between the absorbed and nullified
	population shares gives
	\begin{equation}\notag
		\outereqsize{
			\begin{aligned}
				\tau_r^*
				\geq
				\sum_{j=1}^{c_r}(1+2\eta_{r,j})
				-\eta_{r,1}-\eta_{r,c_r}.
			\end{aligned}
		}
	\end{equation}

	Thus,
	\begin{equation}\notag
		\outereqsize{
			\begin{aligned}
				\sum_{\tau=t-\tau_r^*+1}^{t}\left|U_{\rbindexem{pub},r,\tau}\right|
				&\leq z_{\rbindexem{opt},\rbindexem{dis},r}^{*}\sum_{j=1}^{c_r}(1+\eta_{r,j})\\
				&\leq \frac{n_{\rbindexem{sml},r}}{\tau_r^*}\sum_{j=1}^{c_r}(1+\eta_{r,j})\\
				&\leq n_{\rbindexem{sml},r}\frac{\sum_{j=1}^{c_r}(1+\eta_{r,j})}{\sum_{j=1}^{c_r}(1+2\eta_{r,j})-\eta_{r,1}-\eta_{r,c_r}}\\
				&= n_{\rbindexem{sml},r}\frac{\sum_{j=1}^{c_r}(1+\eta_{r,j})}{\sum_{j=1}^{c_r}(1+\eta_{r,j})+\sum_{j=2}^{c_r-1}\eta_{r,j}}\\
				&\leq n_{\rbindexem{sml},r}\\
				&\leq \left\lfloor\frac{n_r}{2}\right\rfloor.
			\end{aligned}
		}
	\end{equation}
	Therefore, the total number of dissimilarity and publication users of $PG_r$ within any $\tau_r^*$ consecutive time slots is no more than $n_r$ for both \solutionC{} and \solutionD{}.

	In both mechanisms, once a user is selected for reporting, the user
	is removed from the available-user set $U_{A,r}$ and is recycled
	only after the corresponding $\tau_r^*$-length recycling period.
	Therefore, every user in $U_r$ participates at most once within any
	$\tau_r^*$ consecutive time slots.
	

	According to \advancedSelection{}, for each privacy group
	$PG_r=(\widetilde{w}_r,\widetilde{\epsilon}_r,U_r)$, the effective
	privacy budget is
	\begin{equation}\notag
		\outereqsize{
			\begin{aligned}
				\epsilon'_r
				=
				\frac{\widetilde{\epsilon}_r}
				{\left\lceil
					\widetilde{w}_r/\tau_r^*
					\right\rceil}.
			\end{aligned}
		}
	\end{equation}
	Each selected user in $U_r$ applies GPRR with the effective privacy
	budget $\epsilon'_r$, and thus each local report satisfies
	$\epsilon'_r$-LDP.
	
	Moreover, after the locally private reports are collected, all the
	remaining operations in \solutionC{} and \solutionD{}, including
	the dissimilarity calculation, \advancedCorretedWeight{}, \basicClipping{}, \advancedEnhance{},
	population-weighted aggregation, simplex projection, and the reuse
	of historical publications, operate only on locally private outputs
	and therefore introduce no additional privacy loss.
	
	Therefore, all the conditions in Lemma~\ref{lemma:grouped-plus} are satisfied with
	$\tau_r=\tau_r^*$ for every $PG_r$. According to Lemma~\ref{lemma:grouped-plus},
	both \solutionC{} and \solutionD{} satisfy
	$(\vectorfont{w},\vectorfont{\epsilon})$-EPLDP.
\end{proof}

\subsection{Proofs of  Theorems for Utility Analysis}
\begin{lemma}\label{lemma:OPSError}
	Given  a privacy budget-count set $P=\{(\epsilon_k,n_k)\}_{k=1}^{m}$ satisfying $\sum_{k=1}^{m}n_k=n$, where $\epsilon_k$ is the $k$-th distinct privacy budget and $n_k$ is the number of users with privacy budget $\epsilon_k$.
	For a single privacy-budget group with population size $N$,
	privacy budget $\epsilon$, and sampling size $z\le N$, the error
	upper bound of the FO process is
	\begin{equation}\label{eq:FOErrorUpperBound}
		\outereqsize{
			\begin{aligned}
				\widetilde{\shrink{err}}_{\rbindexsmall{FO}}(N,z,\epsilon) &=  \frac{N-z}{z(N-1)} + \frac{d-1}{z}\cdot\frac{2e^{\epsilon}+d-2}{\left(e^{\epsilon}-1\right)^2}.
			\end{aligned}
		}
	\end{equation}
	
	Let $\epsilon_{\rbindex{min}}=\min_{k\in[m]}\epsilon_k$ denote the minimum privacy budget among all privacy-budget groups.
	The error upper bound of \shrinkem{PFO} process is:
	\begin{equation}\label{eq:PFOErrorUpperBound}
		\outereqsize{
			\begin{aligned}
				\widetilde{\shrink{err}}_{\rbindexsmall{PFO}}(P,z) &= \frac{n-z}{z(n-1)} + \frac{d-1}{z}\cdot\frac{2e^{\epsilon_{\rbindex{min}}}+d-2}{\left(e^{\epsilon_{\rbindex{min}}}-1\right)^2}.
			\end{aligned}
		}
	\end{equation}
\end{lemma}
\begin{proof}
	For the \shrink{FO} case, all $z$ sampled users are drawn from a single privacy-budget group with population size $N$.
	Hence, the GPRR perturbation variance reduces to $V_{\rbindexem{LDP}}(\epsilon, z, d)=\frac{d\lambda+\mu}{z}$.
	According to the sampling-variance upper bound in Eq.~\eqref{eq:samplevarstar}
	and the GPRR perturbation variance, we have
	\begin{equation}\notag
		\outereqsize{
			\begin{aligned}
				\shrinkem{err}_{\rbindexem{FO}}(N,z,\epsilon) &\le \frac{N-z}{z(N-1)} + \frac{1}{z}(d\lambda+\mu)\\
				&= \frac{N-z}{z(N-1)} + \frac{d-1}{z} \left(\frac{2}{e^\epsilon-1} + \frac{d}{(e^\epsilon-1)^2}\right)\\
				&= \frac{N-z}{z(N-1)} + \frac{d-1}{z}\cdot \frac{2e^\epsilon+d-2}{(e^\epsilon-1)^2}.
			\end{aligned}
		}
	\end{equation}

	For the PFO case, according to Eq.~\eqref{eq:finalvar-2}, we have
	\begin{equation}\notag
		\outereqsize{
			\begin{aligned}
				\shrinkem{err}_{\rbindexsmallem{PFO}}(P,z) &= \frac{n-z}{z(n-1)} + \frac{1/z}{\sum_{k=1}^{m}1/(da_k+b_k)}.
			\end{aligned}
		}
	\end{equation}
	For each privacy-budget group $k$, according to the definitions of
	$a_k$ and $b_k$, we have
	\begin{equation}\notag
		\outereqsize{
			\begin{aligned}
				da_k+b_k = \frac{d-1}{g_k} \left(\frac{2}{e^{\epsilon_k}-1}	+ \frac{d}{(e^{\epsilon_k}-1)^2} \right)
				= \frac{d-1}{g_k} \cdot \frac{2e^{\epsilon_k}+d-2}{(e^{\epsilon_k}-1)^2}.
			\end{aligned}
		}
	\end{equation}
	Let $h(\epsilon)=\frac{2e^\epsilon+d-2}{(e^\epsilon-1)^2}$.
	Since $h(\epsilon)$ is monotonically decreasing for $\epsilon>0$, for every $\epsilon_k\in\widetilde{\vectorfont{\epsilon}}$, we have $h(\epsilon_k)\le	h(\epsilon_{\rbindexem{min}})$.
	Therefore,
	\begin{equation}\notag
		\outereqsize{
			\begin{aligned}
				da_k+b_k \le \frac{d-1}{g_k}h(\epsilon_{\rbindexem{min}}),
			\end{aligned}
		}
	\end{equation}
	and hence
	\begin{equation}\notag
		\outereqsize{
			\begin{aligned}
				\frac{1}{da_k+b_k} \ge \frac{g_k}{(d-1)h(\epsilon_{\rbindexem{min}})}.
			\end{aligned}
		}
	\end{equation}
	Since $\sum_{k=1}^{m}g_k=1$, we obtain
	\begin{equation}\notag
		\outereqsize{
			\begin{aligned}
				\sum_{k=1}^{m} \frac{1}{da_k+b_k} \ge \frac{1}{(d-1)h(\epsilon_{\rbindexem{min}})}.
			\end{aligned}
		}
	\end{equation}
	Thus,
	\begin{equation}\notag
		\outereqsize{
			\begin{aligned}
				\shrinkem{err}_{\rbindexsmallem{PFO}}(P,z) &\le \frac{n-z}{z(n-1)} + \frac{d-1}{z}h(\epsilon_{\rbindexem{min}}) \\
				&= \frac{n-z}{z(n-1)} + \frac{d-1}{z} \cdot \frac{2e^{\epsilon_{\rbindexem{min}}}+d-2}{(e^{\epsilon_{\rbindexem{min}}}-1)^2}.
			\end{aligned}
		}
	\end{equation}
\end{proof}

\begin{lemma}\label{lemma:basic-dis-error}
	Let $\widehat{\vectorfont{f}}_{\rbindex{dis},t}$ be the PFO estimate used
	by the private dissimilarity mechanism at time slot $t$, and let $V_{\rbindex{dis},t}=\mathbb{E}\!\left[\left\|\widehat{\vectorfont{f}}_{\rbindex{dis},t}
	-\vectorfont{f}_t\right\|_2^2\right]$.
	Let $P_{\rbindex{opt}}=\{(\epsilon_k,n_k)\}_{k=1}^{m}$ denote the effective privacy-budget count set after \basicSelection{}, and let $\epsilon_{\rbindex{opt},\rbindex{min}}=\min(\widetilde{\vectorfont{\epsilon}}'_{\rbindexem{opt}})$.
	Conditioned on the publication history before time slot $t$, define $R_{\rbindex{opt}}=\frac{e^{\epsilon_{\rbindex{opt},\rbindex{min}}}+d-2}{e^{\epsilon_{\rbindex{opt},\rbindex{min}}}-1}$.
	Then the variance of the dissimilarity satisfies $\var{\shrink{dis}-\shrink{dis}^{*}}\le \frac{9R_{\rbindex{opt}}^2}{d}\widetilde{\shrink{err}}_{\rbindex{PFO}}(P_{\rbindex{opt}},z_{\rbindex{opt}})$,
	where $\widetilde{\shrink{err}}_{\rbindex{PFO}}(P_{\rbindex{opt}},z_{\rbindex{opt}})=\frac{n-z_{\rbindexmuchsmall{opt}}}{z_{\rbindexmuchsmall{opt}}(n-1)} + \frac{d-1}{z_{\rbindexmuchsmall{opt}}}\cdot\frac{2e^{\epsilon_{\rbindexmuchsmall{opt},\rbindexmuchsmall{min}}}+d-2}{\left(e^{\epsilon_{\rbindexmuchsmall{opt},\rbindexmuchsmall{min}}}-1\right)^2}$
\end{lemma}

\begin{proof}
	Let $\vectorfont{e}_t=\widehat{\vectorfont{f}}_{\rbindexem{dis},t}-\vectorfont{f}_t$, $\vectorfont{\Delta}_t=\vectorfont{f}_t-\widehat{\vectorfont{f}}_l$.
	Conditioned on the previous publication history,
	$\widehat{\vectorfont{f}}_l$ is fixed. Since
	$\widehat{\vectorfont{f}}_{\rbindexem{dis},t}$ is unbiased,
	$\simpleexpectation{\vectorfont{e}_t}=\vectorfont{0}$, and $V_{\rbindexem{dis},t}=\simpleexpectation{\|\vectorfont{e}_t\|_2^2}$.
	
	According to the definitions of $\shrinkem{dis}$ and $\shrinkem{dis}^{*}$, we have
	\begin{equation}\notag
		\outereqsize{
			\begin{aligned}
				\shrinkem{dis}-\shrinkem{dis}^{*} &= \frac{1}{d}\left\|\widehat{\vectorfont{f}}_{\rbindexem{dis},t}-\widehat{\vectorfont{f}}_l\right\|_2^2 - \frac{1}{d}V_{\rbindexem{dis},t} - \frac{1}{d}\left\|\vectorfont{\Delta}_t\right\|_2^2 \\
				&= \frac{1}{d}\left\|\widehat{\vectorfont{f}}_{\rbindexem{dis},t}-\vectorfont{f}_t+\vectorfont{f}_t-\widehat{\vectorfont{f}}_l\right\|_2^2 - \frac{1}{d}V_{\rbindexem{dis},t} - \frac{1}{d}\left\|\vectorfont{\Delta}_t\right\|_2^2\\
				&= \frac{1}{d}\left\|\vectorfont{e}_t+\vectorfont{\Delta}_t\right\|_2^2 - \frac{1}{d}V_{\rbindexem{dis},t} - \frac{1}{d}\left\|\vectorfont{\Delta}_t\right\|_2^2\\
				&=\frac{2}{d}\vectorfont{\Delta}_t^{\top}\vectorfont{e}_t+\frac{1}{d}\left(\left\|\vectorfont{e}_t\right\|_2^2
				-V_{\rbindexem{dis},t}\right).
			\end{aligned}
		}
	\end{equation}
	
	For GPRR with privacy budget $\epsilon_b$, according to Eq.~\eqref{eq:estimation-2}, we have $\widehat{f}_{b,j}=\frac{h_{b,j}-q_b}{p_b-q_b}$ and $0\leq h_{b,j}\leq 1$.
	Thus, $\widehat{f}_{b,j}\in\left[-\frac{1}{e^{\epsilon_b}-1},\frac{e^{\epsilon_b}+d-2}{e^{\epsilon_b}-1}\right]$.
	Because PFO combines the group-level estimates using non-negative
	weights $\{\eta_b\}_{b=1}^m$ with $\sum_{b=1}^m\eta_b=1$, and $\epsilon_b\ge\epsilon_{\rbindexem{opt},\rbindexem{min}}$,
	we have $\widehat{\vectorfont{f}}_{\rbindexem{dis},t}[j]=\widehat{f}_j\in\left[-\frac{1}{e^{\epsilon_{\rbindexem{opt},\rbindexem{min}}}-1},\frac{e^{\epsilon_{\rbindexem{opt},\rbindexem{min}}}+d-2}{e^{\epsilon_{\rbindexem{opt},\rbindexem{min}}}-1}\right]$.
	Since $0\leq\vectorfont{f}_t[j]\leq 1$,
	we further have
	\begin{equation}\notag
		\outereqsize{
			\begin{aligned}
				|e_t[j]| &= \left|\widehat{\vectorfont{f}}_{\rbindexem{dis},t}[j] - \vectorfont{f}_t[j]\right| \\
				&\leq \max\left\{\left|-\frac{1}{e^{\epsilon_{\rbindexmuchsmallem{opt},\rbindexmuchsmallem{min}}}-1}-1\right|,\frac{e^{\epsilon_{\rbindexmuchsmallem{opt},\rbindexmuchsmallem{min}}}+d-2}{e^{\epsilon_{\rbindexmuchsmallem{opt},\rbindexmuchsmallem{min}}}-1}\right\}\\
				&= \max\left\{\frac{e^{\epsilon_{\rbindexmuchsmallem{opt},\rbindexmuchsmallem{min}}}}{e^{\epsilon_{\rbindexmuchsmallem{opt},\rbindexmuchsmallem{min}}}-1},\frac{e^{\epsilon_{\rbindexmuchsmallem{opt},\rbindexmuchsmallem{min}}}+d-2}{e^{\epsilon_{\rbindexmuchsmallem{opt},\rbindexmuchsmallem{min}}}-1}\right\}\\
				&= R_{\rbindexem{opt}}.
			\end{aligned}
		}
	\end{equation}
	Similarly, $|\Delta_t[j]|\le R_{\rbindexem{opt}}$.
	Hence $\|\vectorfont{e}_t\|_2^2\le dR_{\rbindexem{opt}}^2$, $\|\vectorfont{\Delta}_t\|_2^2\le dR_{\rbindexem{opt}}^2$.
	
	By the Cauchy--Schwarz inequality, we have
	\begin{equation}\notag
		\outereqsize{
			\begin{aligned}
				\mathbb{E}\left[(\vectorfont{\Delta}_t^\top\vectorfont{e}_t)^2\right] \le \|\vectorfont{\Delta}_t\|_2^2 \mathbb{E}\|\vectorfont{e}_t\|_2^2
				\le dR_{\rbindexem{opt}}^2V_{\rbindexem{dis},t}.
			\end{aligned}
		}
	\end{equation}

	Since $\var{X}=\simpleexpectation{X^2}-\simpleexpectation{X}^2\leq\simpleexpectation{X^2}$, we have
	\begin{equation}\notag
		\outereqsize{
			\begin{aligned}
				\var{\|\vectorfont{e}_t\|_2^2}
				&\le \simpleexpectation{\|\vectorfont{e}_t\|_2^4}\\
				&\le dR_{\rbindexem{opt}}^2 \simpleexpectation{\|\vectorfont{e}_t\|_2^2}\\
				&= dR_{\rbindexem{opt}}^2V_{\rbindexem{dis},t}.
			\end{aligned}
		}
	\end{equation}
	
	Therefore, by the Triangle Inequality and Eq.~\eqref{eq:PFOErrorUpperBound} in Lemma~\ref{lemma:OPSError}, we have
	\begin{equation}\notag
		\outereqsize{
			\begin{aligned}
				\var{\shrinkem{dis}-\shrinkem{dis}^{*}}&=\simpleexpectation{(\shrinkem{dis}-\shrinkem{dis}^{*})^2}\\
				&\le \left(\frac{2}{d} \sqrt{\simpleexpectation{(\vectorfont{\Delta}_t^\top\vectorfont{e}_t)^2}}
				+ \frac{1}{d}\sqrt{\var{\|\vectorfont{e}_t\|_2^2}}\right)^2\\
				&\le \left(3R_{\rbindexem{opt}}\sqrt{\frac{V_{\rbindexem{dis},t}}{d}}\right)^2\\
				&= \frac{9R_{\rbindexem{opt}}^2}{d} V_{\rbindexem{dis},t}\\
				&\leq \frac{9R_{\rbindexem{opt}}^2}{d} \widetilde{\shrinkem{err}}_{\rbindexem{PFO}}(P_{\rbindexem{opt}},z_{\rbindexem{opt}})
			\end{aligned}
		}
	\end{equation}
\end{proof}

For the utility analysis of the group-aware private dissimilarity,
we first introduce several auxiliary quantities.
For each privacy group
$PG_r=(\widetilde{w}_r,\widetilde{\epsilon}_r,U_r)$, let
$n_{\rbindexem{sml},r}=\lfloor n_r/2\rfloor$, and define
\begin{equation}\notag
	\outereqsize{
		\begin{aligned}
			\overline{V}_{\rbindexem{dis},r}
			&=
			\widetilde{err}_{\rbindexem{FO}}
			\left(
			n_{\rbindexem{sml},r},
			z_{\rbindexem{opt},\rbindexem{dis},r}^{*},
			\epsilon'_r
			\right),
			\\
			\theta_{\rbindexem{dis},r}
			&=
			\frac{
				n_{\rbindexem{sml},r}
				-
				z_{\rbindexem{opt},\rbindexem{dis},r}^{*}
			}{
				z_{\rbindexem{opt},\rbindexem{dis},r}^{*}
				\left(
				n_{\rbindexem{sml},r}-1
				\right)
			},
			\\
			R_r
			&=
			\frac{
				e^{\epsilon'_r}+d-2
			}{
				e^{\epsilon'_r}-1
			}.
		\end{aligned}
	}
\end{equation}

Define the population-level variance upper bound and range
coefficient as
\begin{equation}\notag
	\outereqsize{
		\begin{aligned}
			\overline{V}_{\rbindexem{dis}}
			&=
			\sum_{r=1}^{K}
			\pi_r^2
			\overline{V}_{\rbindexem{dis},r},
			\\
			R_{\rbindexem{grp}}
			&=
			\sum_{r=1}^{K}
			\pi_rR_r.
		\end{aligned}
	}
\end{equation}
For each $PG_r$, further define
\begin{equation}\notag
	\outereqsize{
		\begin{aligned}
			Q_r
			=
			\frac{
				\theta_{\rbindexem{dis},r}
			}{
				1-\theta_{\rbindexem{dis},r}
			}
			R_r
			\sqrt{
				d\left(
				1+
				\overline{V}_{\rbindexem{dis},r}
				\right)
			}.
		\end{aligned}
	}
\end{equation}
Finally, define the vector-level private-dissimilarity penalty as
\begin{equation}\notag
	\outereqsize{
		\begin{aligned}
			D_{\rbindexem{dis}}
			=
			3R_{\rbindexem{grp}}
			\sqrt{
				d\overline{V}_{\rbindexem{dis}}
			}
			+
			\sqrt{
				\sum_{r=1}^{K}
				\pi_r^4Q_r^2
			}.
		\end{aligned}
	}
\end{equation}

\begin{lemma}\label{lemma:d-exp-plus}
	For the group-aware private dissimilarity used by
	\solutionC{} and \solutionD{}, conditioned on the publication
	history before time slot $t$, we have
	\begin{equation}\notag
		\outereqsize{
			\begin{aligned}
				d\cdot
				\simpleexpectation{
					\left|
					\shrink{dis}
					-
					\shrink{dis}^{*}
					\right|
				}
				\leq
				D_{\rbindex{dis}}.
			\end{aligned}
		}
	\end{equation}
\end{lemma}

\begin{proof}
	For each privacy group $PG_r$, define
	\begin{equation}\notag
		\outereqsize{
			\begin{aligned}
				\vectorfont{e}_{r,t}
				&=
				\widehat{\vectorfont{f}}_{\rbindexem{dis},t}^{(r)}
				-
				\vectorfont{f}_t^{(r)},
				\\
				V_{\rbindexem{dis},r,t}
				&=
				\simpleexpectation{
					\left\|
					\vectorfont{e}_{r,t}
					\right\|_2^2
				}.
			\end{aligned}
		}
	\end{equation}
	Since the group-level FO estimate is unbiased,
	$\simpleexpectation{\vectorfont{e}_{r,t}}=0$.
	The population-weighted estimation error is
	\begin{equation}\notag
		\outereqsize{
			\begin{aligned}
				\vectorfont{e}_t
				&= \widehat{\vectorfont{f}}_{\rbindexem{dis},t}-\vectorfont{f}_t\\
				&= \sum_{r=1}^{K}\pi_r\vectorfont{e}_{r,t}.
			\end{aligned}
		}
	\end{equation}
	Since different privacy groups contain disjoint users and use
	independent local randomization, their estimation errors are
	independent. Hence,
	\begin{equation}\notag
		\outereqsize{
			\begin{aligned}
				V_{\rbindexem{dis},t}
				&=
				\simpleexpectation{
					\left\|
					\vectorfont{e}_t
					\right\|_2^2
				}
				\\
				&=
				\sum_{r=1}^{K}
				\pi_r^2
				V_{\rbindexem{dis},r,t}
				\\
				&\leq
				\sum_{r=1}^{K}
				\pi_r^2
				\overline{V}_{\rbindexem{dis},r}
				=
				\overline{V}_{\rbindexem{dis}},
			\end{aligned}
		}
	\end{equation}
	where the inequality follows from Lemma~\ref{lemma:OPSError}.
	
	For the GPRR estimate of $PG_r$, each coordinate satisfies
	\begin{equation}\notag
		\outereqsize{
			\begin{aligned}
				\widehat{f}_{\rbindexem{dis},t}^{(r)}[j]
				\in
				\left[
				-\frac{1}{e^{\epsilon'_r}-1},
				\frac{
					e^{\epsilon'_r}+d-2
				}{
					e^{\epsilon'_r}-1
				}
				\right].
			\end{aligned}
		}
	\end{equation}
	Therefore,
	$|e_{r,t}[j]|\leq R_r$, and thus
	\begin{equation}\notag
		\outereqsize{
			\begin{aligned}
				|e_t[j]|
				\leq
				\sum_{r=1}^{K}
				\pi_rR_r
				=
				R_{\rbindexem{grp}}.
			\end{aligned}
		}
	\end{equation}
	Let
	$\vectorfont{\Delta}_t
	=
	\vectorfont{f}_t
	-
	\widehat{\vectorfont{f}}_l$.
	Since the released estimate is projected onto the probability
	simplex, both $\vectorfont{f}_t$ and
	$\widehat{\vectorfont{f}}_l$ are probability distributions.
	Hence,
	$|\Delta_t[j]|\leq1\leq R_{\rbindexem{grp}}$.
	Let
	\begin{equation}\notag
		\outereqsize{
			\begin{aligned}
				\widehat{V}_{\rbindexem{dis},t}
				=
				\sum_{r=1}^{K}
				\pi_r^2
				\widehat{V}_{\rbindexem{dis},r,t}.
			\end{aligned}
		}
	\end{equation}
	By Theorem~\ref{thm:precise_var},
	$\simpleexpectation{
		\widehat{V}_{\rbindexem{dis},r,t}
	}
	=
	V_{\rbindexem{dis},r,t}$, and therefore
	$\simpleexpectation{
		\widehat{V}_{\rbindexem{dis},t}
	}
	=
	V_{\rbindexem{dis},t}$.
	
	According to the definitions of $\shrinkem{dis}$ and
	$\shrinkem{dis}^{*}$,
	\begin{equation}\notag
		\outereqsize{
			\begin{aligned}
				\shrinkem{dis}
				-
				\shrinkem{dis}^{*}
				&=
				\frac{2}{d}
				\vectorfont{\Delta}_t^{\top}
				\vectorfont{e}_t
				+
				\frac{1}{d}
				\left(
				\left\|
				\vectorfont{e}_t
				\right\|_2^2
				-
				V_{\rbindexem{dis},t}
				\right)
				\\
				&\quad
				-
				\frac{1}{d}
				\left(
				\widehat{V}_{\rbindexem{dis},t}
				-
				V_{\rbindexem{dis},t}
				\right).
			\end{aligned}
		}
	\end{equation}
	
	Moreover, according to the above decomposition,
	$\simpleexpectation{\vectorfont{e}_t}=0$,
	$\simpleexpectation{
		\|\vectorfont{e}_t\|_2^2
	}
	=
	V_{\rbindexem{dis},t}$, and
	$\simpleexpectation{
		\widehat{V}_{\rbindexem{dis},t}
	}
	=
	V_{\rbindexem{dis},t}$.
	Therefore,
	\begin{equation}\notag
		\outereqsize{
			\begin{aligned}
				\simpleexpectation{
					\shrinkem{dis}
					-
					\shrinkem{dis}^{*}
				}
				=0.
			\end{aligned}
		}
	\end{equation}
	For the first term, by the Cauchy--Schwarz inequality,
	\begin{equation}\notag
		\outereqsize{
			\begin{aligned}
				\simpleexpectation{
					\left(
					\vectorfont{\Delta}_t^{\top}
					\vectorfont{e}_t
					\right)^2
				}
				&\leq
				\left\|
				\vectorfont{\Delta}_t
				\right\|_2^2
				\simpleexpectation{
					\left\|
					\vectorfont{e}_t
					\right\|_2^2
				}
				\\
				&\leq
				dR_{\rbindexem{grp}}^2
				V_{\rbindexem{dis},t}.
			\end{aligned}
		}
	\end{equation}
	Moreover,
	$\|\vectorfont{e}_t\|_2^2
	\leq dR_{\rbindexem{grp}}^2$, and thus
	\begin{equation}\notag
		\outereqsize{
			\begin{aligned}
				\var{
					\left\|
					\vectorfont{e}_t
					\right\|_2^2
				}
				&\leq
				\simpleexpectation{
					\left\|
					\vectorfont{e}_t
					\right\|_2^4
				}
				\\
				&\leq
				dR_{\rbindexem{grp}}^2
				V_{\rbindexem{dis},t}.
			\end{aligned}
		}
	\end{equation}
	
	It remains to bound the uncertainty introduced by the unbiased
	variance estimator. According to Theorem~\ref{thm:precise_var},
	\begin{equation}\notag
		\outereqsize{
			\begin{aligned}
				\widehat{V}_{\rbindexem{dis},r,t}
				=
				\frac{
					\theta_{\rbindexem{dis},r}
					\left(
					1-
					\left\|
					\widehat{\vectorfont{f}}_{\rbindexem{dis},t}^{(r)}
					\right\|_2^2
					\right)
					+
					V_{\rbindexem{LDP}}
					\left(
					\epsilon'_r,
					z_{\rbindexem{opt},\rbindexem{dis},r}^{*},
					d
					\right)
				}{
					1-\theta_{\rbindexem{dis},r}
				}.
			\end{aligned}
		}
	\end{equation}
	Therefore,
	\begin{equation}\notag
		\outereqsize{
			\begin{aligned}
				\var{
					\widehat{V}_{\rbindexem{dis},r,t}
				}
				&=
				\left(
				\frac{
					\theta_{\rbindexem{dis},r}
				}{
					1-\theta_{\rbindexem{dis},r}
				}
				\right)^2
				\var{
					\left\|
					\widehat{\vectorfont{f}}_{\rbindexem{dis},t}^{(r)}
					\right\|_2^2
				}.
			\end{aligned}
		}
	\end{equation}
	Since
	$\left|
	\widehat{f}_{\rbindexem{dis},t}^{(r)}[j]
	\right|\leq R_r$,
	we have
	\begin{equation}\notag
		\outereqsize{
			\begin{aligned}
				\var{
					\left\|
					\widehat{\vectorfont{f}}_{\rbindexem{dis},t}^{(r)}
					\right\|_2^2
				}
				&\leq
				\simpleexpectation{
					\left\|
					\widehat{\vectorfont{f}}_{\rbindexem{dis},t}^{(r)}
					\right\|_2^4
				}
				\\
				&\leq
				dR_r^2
				\simpleexpectation{
					\left\|
					\widehat{\vectorfont{f}}_{\rbindexem{dis},t}^{(r)}
					\right\|_2^2
				}
				\\
				&=
				dR_r^2
				\left(
				\left\|
				\vectorfont{f}_t^{(r)}
				\right\|_2^2
				+
				V_{\rbindexem{dis},r,t}
				\right)
				\\
				&\leq
				dR_r^2
				\left(
				1+
				\overline{V}_{\rbindexem{dis},r}
				\right).
			\end{aligned}
		}
	\end{equation}
	Hence,
	\begin{equation}\label{lemma:var-hat-v-dis-t-upper-bound}
		\outereqsize{
			\begin{aligned}
				\var{
					\widehat{V}_{\rbindexem{dis},r,t}
				}
				\leq
				Q_r^2.
			\end{aligned}
		}
	\end{equation}
	Since the group-level estimators are independent,
	\begin{equation}\notag
		\outereqsize{
			\begin{aligned}
				\var{
					\widehat{V}_{\rbindexem{dis},t}
				}
				&=
				\sum_{r=1}^{K}
				\pi_r^4
				\var{
					\widehat{V}_{\rbindexem{dis},r,t}
				}
				\\
				&\leq
				\sum_{r=1}^{K}
				\pi_r^4Q_r^2.
			\end{aligned}
		}
	\end{equation}
	
	Therefore, by the triangle inequality in $L_2$,
	\begin{equation}\notag
		\outereqsize{
			\begin{aligned}
				\sqrt{
					\var{
						\shrinkem{dis}
						-
						\shrinkem{dis}^{*}
					}
				}
				&\leq
				3R_{\rbindexem{grp}}
				\sqrt{
					\frac{
						V_{\rbindexem{dis},t}
					}{d}
				}
				+
				\frac{1}{d}
				\sqrt{
					\sum_{r=1}^{K}
					\pi_r^4Q_r^2
				}
				\\
				&\leq
				3R_{\rbindexem{grp}}
				\sqrt{
					\frac{
						\overline{V}_{\rbindexem{dis}}
					}{d}
				}
				+
				\frac{1}{d}
				\sqrt{
					\sum_{r=1}^{K}
					\pi_r^4Q_r^2
				}.
			\end{aligned}
		}
	\end{equation}
Consequently,
\begin{equation}\notag
	\outereqsize{
		\begin{aligned}
			d\cdot
			\simpleexpectation{
				\left|
				\shrinkem{dis}
				-
				\shrinkem{dis}^{*}
				\right|
			}
			&\leq
			d
			\sqrt{
				\var{
					\shrinkem{dis}
					-
					\shrinkem{dis}^{*}
				}
			}
			\\
			&\leq
			D_{\rbindexem{dis}}.
		\end{aligned}
	}
\end{equation}
\end{proof}


%

\begin{lemma}\label{lem:plpd_plus_pub_variance}
	For each privacy group $PG_r$, let 	$n_{\rbindex{sml},r}=\left\lfloor n_r/2\right\rfloor$.
	Under the group-wise uniform-publication assumption of Theorem~\ref{thm:plpd-plus-utility}, the average vector-level
	publication-variance upper bound of $PG_r$ satisfies
	\begin{equation}\notag
		\outereqsize{
			\begin{aligned}
				\overline{V}_{\rbindex{pub},r}
				={}&
				\frac{2\left(2^{s_r}-1\right)}{s_r}
				\left(
				\frac{1}{n_{\rbindex{sml},r}-1}
				+
				\frac{d-1}{n_{\rbindex{sml},r}}\cdot
				\frac{2e^{\epsilon'_r}+d-2}{\left(e^{\epsilon'_r}-1\right)^2}\right)\\
				&-
				\frac{1}{n_{\rbindex{sml},r}-1}.
			\end{aligned}
		}
	\end{equation}
\end{lemma}

\begin{proof}
	At the beginning of a recycling cycle, \solutionC{} reserves
	$n_{\rbindexem{sml},r}=\left\lfloor n_r/2 \right\rfloor$ users of $PG_r$ for the publication
	mechanism. According to Alg.~\ref{alg:PLPD+}, each new publication uses half
	of the remaining publication population. Ignoring integer rounding,
	the $j$-th publication therefore uses
	\begin{equation}\notag
		\outereqsize{
			\begin{aligned}
				z_{r,j}
				=
				\frac{n_{\rbindexem{sml},r}}{2^j},
				\qquad j\in[s_r].
			\end{aligned}
		}
	\end{equation}

	According to the FO error upper bound from Eq.~\eqref{eq:FOErrorUpperBound} in Lemma~\ref{lemma:OPSError} the
	vector-level variance of the $j$-th group-level publication estimate
	is upper-bounded by
	\begin{equation}\notag
		\outereqsize{
			\begin{aligned}
				\widetilde{err}_{\rbindexem{FO}}
				\left(
				n_{\rbindexem{sml},r},
				z_{r,j},
				\epsilon'_r
				\right)
				={}&
				\frac{
					n_{\rbindexem{sml},r}-z_{r,j}
				}{
					z_{r,j}
					\left(
					n_{\rbindexem{sml},r}-1
					\right)
				}
				+
				\frac{d-1}{z_{r,j}}\cdot
				\frac{
					2e^{\epsilon'_r}+d-2
				}{
					\left(e^{\epsilon'_r}-1\right)^2
				}.
			\end{aligned}
		}
	\end{equation}

	Substituting
	$z_{r,j}=n_{\rbindexem{sml},r}/2^j$ yields
	\begin{equation}\notag
		\outereqsize{
			\begin{aligned}
				\widetilde{err}_{\rbindexem{FO}}
				\left(
				n_{\rbindexem{sml},r},
				\frac{n_{\rbindexem{sml},r}}{2^j},
				\epsilon'_r
				\right)
				={}&
				\frac{2^j-1}
				{n_{\rbindexem{sml},r}-1}+
				\frac{2^j(d-1)}
				{n_{\rbindexem{sml},r}}\cdot
				\frac{
					2e^{\epsilon'_r}+d-2
				}{
					\left(e^{\epsilon'_r}-1\right)^2
				}.
			\end{aligned}
		}
	\end{equation}
	
	Under the uniform-publication assumption, the $s_r$ publication
	stages are equally represented in the time average. Hence,
	\begin{equation}\notag
		\outereqsize{
			\begin{aligned}
				\overline{V}_{\rbindexem{pub},r}
				=
				\frac{1}{s_r}
				\sum_{j=1}^{s_r}
				\widetilde{err}_{\rbindexem{FO}}
				\left(
				n_{\rbindexem{sml},r},
				\frac{n_{\rbindexem{sml},r}}{2^j},
				\epsilon'_r
				\right).
			\end{aligned}
		}
	\end{equation}
	
	Since
	$\sum_{j=1}^{s_r}2^j=2(2^{s_r}-1)$, we obtain
	\begin{equation}\notag
		\outereqsize{
			\begin{aligned}
				\overline{V}_{\rbindexem{pub},r}
				={}&
				\frac{2\left(2^{s_r}-1\right)}{s_r}
				\left(
				\frac{1}{n_{\rbindexem{sml},r}-1}
				+
				\frac{d-1}{n_{\rbindexem{sml},r}}\cdot
				\frac{
					2e^{\epsilon'_r}+d-2
				}{
					\left(e^{\epsilon'_r}-1\right)^2
				}
				\right)
				-
				\frac{1}{n_{\rbindexem{sml},r}-1}.
			\end{aligned}
		}
	\end{equation}
\end{proof}

\begin{lemma}[Shift-Bounded Fusion Bias]\label{lem:shift}
	Let $\Xi_t \triangleq \max_{r,s\in[K]}\|\vectorfont f^{(r)}_t-\vectorfont f^{(s)}_t\|_2\in[0,\sqrt2]$.
	For any $\vectorfont\gamma,\vectorfont\pi$ in the probability simplex,
	\begin{equation}\notag
		\outereqsize{
			\begin{aligned}
				\left\|\sum_{r=1}^K(\gamma_r-\pi_r)\vectorfont f^{(r)}_t\right\|_2
				\le \tfrac12\|\vectorfont\gamma-\vectorfont\pi\|_1\,\Xi_t \le \Xi_t .
			\end{aligned}
		}
	\end{equation}
\end{lemma}
\begin{proof}
	Let $a_r=\gamma_r-\pi_r$. Since $\sum_r a_r=0$, let
	$P=\sum_{a_r>0}a_r=\tfrac12\|\vectorfont a\|_1$; the case $P=0$ is trivial.
	Then $\sum_r a_r\vectorfont f^{(r)}_t=P(\vectorfont x-\vectorfont y)$, where
	$\vectorfont x=\sum_{a_r>0}\tfrac{a_r}{P}\vectorfont f^{(r)}_t$ and
	$\vectorfont y=\sum_{a_s<0}\tfrac{|a_s|}{P}\vectorfont f^{(s)}_t$ are convex combinations.
	Hence $\vectorfont x-\vectorfont y=\sum_{r,s}\alpha_r\beta_s(\vectorfont f^{(r)}_t-\vectorfont f^{(s)}_t)$
	with $\sum_{r,s}\alpha_r\beta_s=1$, and the triangle inequality gives
	$\|\vectorfont x-\vectorfont y\|_2\le\Xi_t$. Finally $P\le 1$.
\end{proof}

\subsubsection{Proof of Theorems~\ref{thm:PLPD+UtilityAnalysis} and~\ref{thm:PLPA+UtilityAnalysis}}\label{Proof:PLP+UtilityAnalysis}
\begin{proof}
	We first assume the $\entity{M}_{\rbindexem{dis},t}$ is not private for both \solutionA{} and \solutionB{} and get the error of $\entity{M}_{\rbindexem{pub},t}$.
	After that, we analyze the error of $\entity{M}_{\rbindexem{dis},t}$ when it is private.

	Define $A_{\rbindexem{opt}}=\frac{d-1}{n}\frac{2e^{\epsilon_{\rbindexmuchsmallem{opt},\rbindexmuchsmallem{min}}}+d-2}{(e^{\epsilon_{\rbindexmuchsmallem{opt},\rbindexmuchsmallem{min}}}-1)^2}$.
	According to Eq.~\eqref{eq:PFOErrorUpperBound} in Lemma~\ref{lemma:OPSError}, for any sampling size $z$, we have
	\begin{equation}\notag
		\outereqsize{
			\begin{aligned}
				\shrinkem{err}_{\rbindexem{PFO}}(P_{\rbindexem{opt}},z)	&\le \widetilde{\shrinkem{err}}_{\rbindexem{PFO}} (P_{\rbindexem{opt}},z)\\
				&= \frac{n-z}{z(n-1)} + \frac{d-1}{z}\cdot\frac{2e^{\epsilon_{\rbindexmuchsmallem{opt},\rbindexmuchsmallem{min}}}+d-2}{\left(e^{\epsilon_{\rbindexmuchsmallem{opt},\rbindexmuchsmallem{min}}}-1\right)^2}\\
				&= \frac nz \left(\frac1{n-1}+A_{\rbindexem{opt}}\right) - \frac1{n-1}.
			\end{aligned}
		}
	\end{equation}

	We first analyze the publication mechanism $\entity{M}_{\rbindexem{pub},t}$ in \solutionA{}. 
	Ignoring integer rounding, the population sizes of the successive new publications are
	$\frac{n}{2^2},\frac{n}{2^3},\ldots,\frac{n}{2^{s_{\rbindexmuchsmallem{opt}}+1}}$.
	Under the uniform-publication assumption, each publication represents $\tau_{\rbindexem{opt}}/s_{\rbindexmuchsmallem{opt}}$ time slots.
	Therefore, the average publication error satisfies
	\begin{equation}\notag
		\outereqsize{
			\begin{aligned}
				\overline{\shrinkem{err}}_{\rbindexem{\solutionA{}},\rbindexem{pub}} &\le \frac{1}{s_{\rbindexmuchsmallem{opt}}}\sum_{j=1}^{s_{\rbindexem{opt}}}\widetilde{\shrinkem{err}}_{\rbindexem{PFO}}\left(P_{\rbindexem{opt}},\frac{n}{2^{j+1}}\right)\\
				&= \frac1{s_{\rbindexem{opt}}}\sum_{j=1}^{s_{\rbindexem{opt}}}\left[2^{j+1}\left(\frac1{n-1}+A_{\rbindexem{opt}}\right)-\frac1{n-1}\right]\\
				&= \frac{4(2^{s_{\rbindexem{opt}}}-1)}{s_{\rbindexem{opt}}}\left(\frac1{n-1}+A_{\rbindexem{opt}}\right)-\frac{1}{n-1}.
			\end{aligned}
		}
	\end{equation}

	For the publication mechanism $\entity{M}_{\rbindexem{pub},t}$ in
	\solutionB{}, we further assume that each new publication is associated
	with the same number of skipped and nullified time slots.
	Let $B=\frac{\tau_{\rbindexem{opt}}-s_{\rbindexem{opt}}}{2s_{\rbindexem{opt}}}$	denote the number of skipped (and equivalently, nullified) time
	slots associated with each new publication, and let $C=\frac{s_{\rbindexem{opt}}}{\tau_{\rbindexem{opt}}}
	=\frac{1}{2B+1}$.
	Thus, under the uniform-publication assumption, each publication
	cycle consists of $B$ skipped time slots, one new publication, and
	$B$ nullified time slots, i.e., $2B+1$ time slots in total.
	For analytical simplicity, we assume $B\in\mathbb{N}^{+}$.
	Consider one publication cycle. According to Algorithm 2, after
	$j-1$ consecutive skipped time slots, the potential publication
	population at the $j$-th absorbable time slot is $jz_{\rbindexem{opt}}$.
	Since $\entity{M}_{\rbindexem{dis},t}$ is temporarily assumed to be
	non-private, the dissimilarity used for the publication decision
	equals the true approximation error, i.e., $\shrinkem{dis}=\shrinkem{dis}^{*}$.
	According to Alg.~\ref{alg:ops-guided-plpd-plpa}, a publication is skipped only when
	$\shrinkem{dis}\leq \shrinkem{err}$. Therefore, the approximation error
	of the $j$-th skipped time slot is no larger than its potential
	reporting error, and is bounded by
	\begin{equation}\notag
		\outereqsize{
			\begin{aligned}
				\widetilde{\shrinkem{err}}_{\rbindexem{PFO}}
				\left(P_{\rbindexem{opt}},j\cdot z_{\rbindexem{opt}}\right),
				\qquad j\in[B].
			\end{aligned}
		}
	\end{equation}
	After $B$ skipped time slots, the next time slot generates a new
	publication using $(B+1)z_{\rbindexem{opt}}$ users, whose error is
	bounded by $\widetilde{\shrinkem{err}}_{\rbindexem{PFO}}\left(P_{\rbindexem{opt}},(B+1)z_{\rbindexem{opt}}\right)$.
	Moreover, the $B$ nullified time slots associated with this
	publication contribute $B\cdot\shrinkem{err}_{\rbindexem{nlf}}$.
	Hence, the average publication error of \solutionB{} satisfies
	\begin{equation}\notag
		\outereqsize{
			\begin{aligned}
				\shrinkem{err}_{\solutionB{},\rbindexem{pub}} &\le \frac{s_{\rbindexem{opt}}}{\tau_{\rbindexem{opt}}}\left[\sum_{j=1}^{B+1}
				\widetilde{\shrinkem{err}}_{\rbindexem{PFO}} \left(P_{\rbindexem{opt}},jz_{\rbindexem{opt}}\right)
				+ B\cdot\shrinkem{err}_{\rbindexem{nlf}}\right] \\
				&= C \left[\sum_{j=1}^{B+1}\widetilde{\shrinkem{err}}_{\rbindexem{PFO}}\left(P_{\rbindexem{opt}},jz_{\rbindexem{opt}}\right)
				+ B\cdot\shrinkem{err}_{\rbindexem{nlf}}\right].
			\end{aligned}
		}
	\end{equation}
	Using
	\begin{equation}\notag
		\outereqsize{
			\begin{aligned}
				\widetilde{\shrinkem{err}}_{\rbindexem{PFO}} (P_{\rbindexem{opt}},z) = \frac{n}{z}\left(\frac{1}{n-1}+A_{\rbindexem{opt}}\right)				-\frac{1}{n-1},
			\end{aligned}
		}
	\end{equation}
	we further obtain
	\begin{equation}\notag
		\outereqsize{
			\begin{aligned}
				\shrinkem{err}_{\rbindexem{PLPA},\rbindexem{pub}} &\le C\sum_{j=1}^{B+1}\left[\frac{n}{jz_{\rbindexem{opt}}}\left(\frac{1}{n-1}+A_{\rbindexem{opt}}	\right)	- \frac{1}{n-1}\right] + BC\cdot\shrinkem{err}_{\rbindexem{nlf}}\\
				&= \frac{Cn}{z_{\rbindexem{opt}}}\left(\frac{1}{n-1}+A_{\rbindexem{opt}}\right)	H_{B+1} - \frac{C(B+1)}{n-1} + 	BC\cdot\shrinkem{err}_{\rbindexem{nlf}},
			\end{aligned}
		}
	\end{equation}
	where $H_{B+1}=\sum_{j=1}^{B+1}\frac{1}{j}$ is the $(B+1)$-th harmonic number.


	When $\entity{M}_{\rbindexem{dis},t}$ is private, the publication
	decision based on $\shrinkem{dis}$ may differ from that based on the
	non-private dissimilarity $\shrinkem{dis}^{*}$. If the two decisions are
	the same, no additional error is introduced. Otherwise, there are two
	possible choosing error cases: (1)  $\shrinkem{dis}^{*}\leq \shrinkem{err}<\shrinkem{dis}$; and (2) $\shrinkem{dis}\leq \shrinkem{err}<\shrinkem{dis}^{*}$.
	We refer to these two cases as \textit{choosing errors}, and let $\vectorfont{1}_{\operatorname{ce}}=1$ when either of the two cases happens; and $\vectorfont{1}_{\operatorname{ce}}=0$ otherwise. 
	When a choosing error occurs, the additional per-domain error-bound term is $\max\{dis^*,err\}-\min\{dis^*,err\}=|err-dis^*|$.

	For the choosing error in both cases above, we have
	\begin{equation}\notag
		\outereqsize{
			\begin{aligned}
				|\shrinkem{err}-\shrinkem{dis}^{*}| \leq |\shrinkem{dis}-\shrinkem{dis}^{*}|,
			\end{aligned}
		}
	\end{equation}
	which means, in either case, the choosing error is at most
	$|\shrinkem{dis}-\shrinkem{dis}^{*}|$. 
	Let $\shrinkem{err}_{\rbindexem{dis},t}^{(\rbindexem{per})}$ denote the expected additional per-domain error-bound term caused by a choosing error.
	Since the utility bounds above are
	expressed in vector-level total error, we have
	\begin{equation}\notag
		\outereqsize{
			\begin{aligned}
				\shrinkem{err}_{\rbindexem{dis},t} &=d\cdot\shrinkem{err}_{\rbindexem{dis},t}^{(\rbindexem{per})}\\
				 &=d\cdot\simpleexpectation{|err-dis^*|\cdot\vectorfont{1}_{\operatorname{ce}}}\\
				&\leq
				d\,\mathbb{E}
				\left[
				|\shrinkem{dis}-\shrinkem{dis}^{*}|
				\right] \\
				&\leq
				d\sqrt{
					\mathbb{E}
					\left[
					(\shrinkem{dis}-\shrinkem{dis}^{*})^2
					\right]
				} \\
				&=
				d\sqrt{
					\operatorname{Var}
					[\shrinkem{dis}-\shrinkem{dis}^{*}]
				}.
			\end{aligned}
		}
	\end{equation}
	According to Lemma~\ref{lemma:basic-dis-error}, we further have
	\begin{equation}\notag
		\outereqsize{
			\begin{aligned}
				\shrinkem{err}_{\rbindexem{dis},t}
				&\leq
				d\sqrt{
					\frac{9R_{\rbindexem{opt}}^2}{d}
					\widetilde{\shrinkem{err}}_{\rbindexem{PFO}}
					(P_{\rbindexem{opt}},z_{\rbindexem{opt}})
				} \\
				&=
				3R_{\rbindexem{opt}}
				\sqrt{
					d\,
					\widetilde{\shrinkem{err}}_{\rbindexem{PFO}}
					(P_{\rbindexem{opt}},z_{\rbindexem{opt}})
				}\\
				&= 3R_{\rbindexem{opt}}\sqrt{\frac{nd}{z_{\rbindexmuchsmallem{opt}}} \left(\frac1{n-1}+A_{\rbindexem{opt}}\right) - \frac{d}{n-1}}.
			\end{aligned}
		}
	\end{equation}

	
	Therefore, the error upper bound  is 
	\begin{equation}\notag
		\outereqsize{
			\begin{aligned}
				&3R_{\rbindexem{opt}}\sqrt{\frac{nd}{z_{\rbindexmuchsmallem{opt}}} \left(\frac1{n-1}+A_{\rbindexem{opt}}\right) - \frac{d}{n-1}}\\
				&+ \frac{4(2^{s_{\rbindexem{opt}}}-1)}{s_{\rbindexem{opt}}}\left(\frac1{n-1}+A_{\rbindexem{opt}}\right)-\frac{1}{n-1}
			\end{aligned}
		}
	\end{equation}
	 in \solutionA{},
	and 
	\begin{equation}\notag
		\outereqsize{
			\begin{aligned}
				&3R_{\rbindexem{opt}}\sqrt{\frac{nd}{z_{\rbindexmuchsmallem{opt}}} \left(\frac1{n-1}+A_{\rbindexem{opt}}\right) - \frac{d}{n-1}} + \frac{Cn}{z_{\rbindexem{opt}}}\left(\frac{1}{n-1}+A_{\rbindexem{opt}}\right)	H_{B+1} \\
				&- \frac{C(B+1)}{n-1} + 	BC\cdot\shrinkem{err}_{\rbindexem{nlf}}
			\end{aligned}
		}
	\end{equation}
	in \solutionB{}.
\end{proof}

\subsubsection{Proof of Theorem~\ref{thm:plpd-plus-utility}}\label{proof:plpa-plus-utility}

\begin{proof}
	Fix a time slot $t$ satisfying the conditions of Theorem~\ref{thm:plpd-plus-utility}, and treat the
	reuse configuration $(\vectorfont{PG}^{(A)}_t,\vectorfont{PG}^{(R)}_t,\{l_r\})$ as given.
	Write $\entity A=\vectorfont{PG}^{(A)}_t$, $\entity R=\vectorfont{PG}^{(R)}_t$,
	$\gamma_r=\gamma^{\mathrm{PLPA^+}}_{r,t}$, and let
	$\entity F_t^{\mathrm{Pre}}$ be the information available before the
	active-group publication reports are generated. By Lines~\ref{alg:plpa+:repeat-2}--\ref{alg:repop_plpa:end+} of
	Alg.~\ref{alg:PLPA+}, $\{\gamma_r\}$,
	$\{\widehat{\vectorfont f}^{(r)}_{\rbindexem{pub},l_r}\}_{r\in\entity R}$,
	$\{\widehat{\vectorfont f}^{(r)}_{\rbindexem{dis},t},\widehat V_{\rbindexem{dis},r,t}\}_{r\in[K]}$,
	$dis_t$, and $\widetilde{err}^{+,PA}_t$ are all
	$\entity F_t^{\mathrm{Pre}}$-measurable.

	\textit{Notation.} Define $\vectorfont e_r=\widehat{\vectorfont f}^{(r)}_{\rbindexem{dis},t}-\vectorfont f^{(r)}_t$
	for $r\in[K]$, $\vectorfont u_r=\widehat{\vectorfont f}^{(r)}_{\rbindexem{pub},l_r}-\vectorfont f^{(r)}_t$ for
	$r\in\entity R$, and $c_r=\gamma_r-\pi_r$ if $r\in\entity A$,
	$c_r=-\pi_r$ if $r\in\entity R$. Let
	\begin{equation}\notag
		\outereqsize{
			\begin{aligned}
				\vectorfont B_t=\sum_{r\in\entity A}(\gamma_r-\pi_r)\vectorfont f^{(r)}_t
				+\sum_{r\in\entity R}\big(\gamma_r\widehat{\vectorfont f}^{(r)}_{\rbindexem{pub},l_r}-\pi_r\vectorfont f^{(r)}_t\big),
				\qquad
				\vectorfont E_t=\sum_{r=1}^K c_r\vectorfont e_r .
			\end{aligned}
		}
	\end{equation}
	Since $\sum_r\gamma_r=\sum_r\pi_r=1$,
	\begin{equation}\label{eq:Bdecomp}
		\outereqsize{
			\begin{aligned}
				\vectorfont B_t=\sum_{r=1}^K(\gamma_r-\pi_r)\vectorfont f^{(r)}_t+\sum_{r\in\entity R}\gamma_r\vectorfont u_r ,
			\end{aligned}
		}
	\end{equation}
	and, with $\vectorfont h^{(r)}_t=\widehat{\vectorfont f}^{(r)}_{\rbindexem{dis},t}$, the plug-in bias
	vector in 
	\begin{equation}\notag
		\outereqsize{
			\begin{aligned}
				\widehat{B}^2_{t} =& \max\Big\{0,\left\| B\left(\vectorfont{H}_t,\widehat{\vectorfont{F}}_{l,t}\right) \right\|_2^2-\sum\nolimits_{r\in\vectorfont{PG}_t^{(A)}}\left(\gamma_{r,t}-\pi_r\right)^2\var{\vectorfont{h}_{t}^{(r)}}\\
				&-\sum\nolimits_{r\in\vectorfont{PG}_t^{(R)}}\left(\pi_r^2\var{\vectorfont{h}_{t}^{(r)}}+\gamma_{r,t}^2\var{\widehat{\vectorfont{f}}_{l_r}^{(r)}}\right)\Big\}
			\end{aligned}
		}
	\end{equation}
	 satisfies
	\begin{equation}\notag
		\outereqsize{
			\begin{aligned}
				\vectorfont B(\vectorfont H_t,\widehat{\vectorfont F}_{l,t})=\vectorfont B_t+\vectorfont E_t .
			\end{aligned}
		}
	\end{equation}

	\textit{Step 1 (conditional publication risk).}
	Since $\vectorfont f_t=\sum_r\pi_r\vectorfont f^{(r)}_t$ and $\sum_r\gamma_r=1$, 
	the population-level estimate obtained by Eq.~\eqref{eq:linear-combine} and Line~\ref{alg:final_new_estimates} of Alg.~\ref{alg:PLPD+} satisfies
	\begin{equation}\notag
		\outereqsize{
			\begin{aligned}
				\widetilde{\vectorfont f}_{\rbindexem{pub},t}-\vectorfont f_t
				=\sum_{r\in\entity A}\gamma_r\big(\widehat{\vectorfont f}^{(r)}_{\rbindexem{pub},t}-\vectorfont f^{(r)}_t\big)+\vectorfont B_t .
			\end{aligned}
		}
	\end{equation}
	Given $\entity F_t^{\mathrm{Pre}}$, $\vectorfont B_t$ is fixed, and the active-group
	publication errors are zero-mean and mutually independent, with variances
	upper-bounded by $V^{\mathrm{Pre}}_{\rbindexem{pub},r,t}$ according to
	Eqs.~\eqref{eq:samplevar} and~\eqref{eq:samplevarstar}. Hence
	\begin{equation}\label{eq:Rpa}
		\outereqsize{
			\begin{aligned}
				\mathbb E\big[\|\widetilde{\vectorfont f}_{\rbindexem{pub},t}-\vectorfont f_t\|_2^2\,\big|\,\entity F_t^{\mathrm{Pre}}\big]
				\le\sum_{r\in\entity A}\gamma_r^2V^{\mathrm{Pre}}_{\rbindexem{pub},r,t}+\|\vectorfont B_t\|_2^2
				\triangleq\entity R_{\rbindexem{pub},t}.
			\end{aligned}
		}
	\end{equation}
	Since the Euclidean projection onto the probability simplex (Line~\ref{alg:final_new_estimates} of Alg.~\ref{alg:PLPD+}) is
	non-expansive, the released publication satisfies the same bound.

	\textit{Step 2 (bounding $\mathbb E[\entity R_{\rbindexem{pub},t}]$).}
	Since $0\le\gamma_r\le1$, the first term of $\entity R_{\rbindexem{pub},t}$ is at most
	$\sum_{r\in\entity A}V^{\mathrm{Pre}}_{\rbindexem{pub},r,t}$. For $\vectorfont B_t$, the first
	term of Eq.~\eqref{eq:Bdecomp} is bounded by $\Xi$ according to
	Lemma~\ref{lem:shift}. For the second term, since $\gamma_r\in[0,1]$ and
	$\sum_{r\in\entity R}\gamma_r\le1$, the Cauchy--Schwarz inequality gives
	\begin{equation}\notag
		\outereqsize{
			\begin{aligned}
					\Big\|\sum_{r\in\entity R}\gamma_r\vectorfont u_r\Big\|_2^2
				\le\Big(\sum_{r\in\entity R}\gamma_r\Big)\Big(\sum_{r\in\entity R}\gamma_r\|\vectorfont u_r\|_2^2\Big)
				\le\sum_{r\in\entity R}\|\vectorfont u_r\|_2^2\triangleq U_t^2 .
			\end{aligned}
		}
	\end{equation}
	Hence $\|\vectorfont B_t\|_2\le\Xi+U_t$ holds pointwise. Since the historical
	estimate of each reused group is unbiased for $\vectorfont f^{(r)}_{l_r}$,
	$\mathbb E\|\vectorfont u_r\|_2^2=V_{\rbindexem{pub},r,l_r}+\|\vectorfont f^{(r)}_{l_r}-\vectorfont f^{(r)}_t\|_2^2$,
	and thus $\mathbb E[U_t^2]=V_{\rbindexem{hist},t}+\Gamma_t^2$. By the Minkowski
	inequality in $L^2$,
	\begin{equation}\label{eq:Bsq}
		\outereqsize{
			\begin{aligned}
				\sqrt{\mathbb E\|\vectorfont B_t\|_2^2}\le\Xi+\sqrt{\mathbb E[U_t^2]}=\beta_t ,
				\qquad
				\mathbb E[\entity R_{\rbindexem{pub},t}]\le\sum_{r\in\entity A}V^{\mathrm{Pre}}_{\rbindexem{pub},r,t}+\beta_t^2 .
			\end{aligned}
		}
	\end{equation}

	\textit{Step 3 (error of the selected action).}
	Under the population lower-bound condition, the decision in Line~\ref{line:diserrjudge_plpa+_start} of Alg.~\ref{alg:PLPA+}
	depends only on the comparison between $dis_t$ and
	$\widetilde{err}^{+,PA}_t$. If a new publication is triggered, its
	conditional error is at most $\entity R_{\rbindexem{pub},t}$ by Step~1. Otherwise,
	$dis_t\le\widetilde{err}^{+,PA}_t$, and the reuse error satisfies
	\begin{equation}\notag
		\outereqsize{
			\begin{aligned}
				d\cdot dis^*_t&\le d\cdot dis_t+d|dis_t-dis^*_t|\\
				&\le\entity R_{\rbindexem{pub},t}+\big|d\cdot\widetilde{err}^{+,PA}_t-\entity R_{\rbindexem{pub},t}\big|+d|dis_t-dis^*_t| .
			\end{aligned}
		}
	\end{equation}
	Thus, in both cases,
	\begin{equation}\label{eq:sel}
		\outereqsize{
			\begin{aligned}
				&\mathbb E\big[\|\widehat{\vectorfont f}_{\rbindexem{pub},t}-\vectorfont f_t\|_2^2\,\big|\,\entity F_t^{\mathrm{Pre}}\big]\\
				\le& \entity R_{\rbindexem{pub},t}+\big|d\cdot\widetilde{err}^{+,PA}_t-\entity R_{\rbindexem{pub},t}\big|+d|dis_t-dis^*_t| .
			\end{aligned}
		}
	\end{equation}

	\textit{Step 4 (bounding the risk-estimation error).}
	As instantiated in Line~\ref{alg:repop_plpa:end+} of Alg.~\ref{alg:PLPA+}, Eq.~\eqref{eq:final_corrected_time_error} uses
	$V^{\mathrm{Pre}}_{\rbindexem{pub},r,t}$ for active groups,
	$\widehat V_{\rbindexem{dis},r,t}$ for $\mathrm{Var}[\vectorfont h^{(r)}_t]$, and
	$\widehat V_{\rbindexem{pub},r,l_r}$ for $\mathrm{Var}[\widehat{\vectorfont f}^{(r)}_{l_r}]$. Let
	\begin{equation}\notag
		\outereqsize{
			\begin{aligned}
					X_t=\|\vectorfont B_t+\vectorfont E_t\|_2^2-\sum_{r=1}^Kc_r^2\widehat V_{\rbindexem{dis},r,t},
				\qquad
				Y_t=\sum_{r\in\entity R}\gamma_r^2\widehat V_{\rbindexem{pub},r,l_r}\ge0 .
			\end{aligned}
		}
	\end{equation}
	Then $\widehat B_t^2=\max\{0,X_t-Y_t\}$, and therefore
	$d\cdot\widetilde{err}^{+,PA}_t=\sum_{r\in\entity A}\gamma_r^2V^{\mathrm{Pre}}_{\rbindexem{pub},r,t}+\max\{X_t,Y_t\}$.
	Comparing with Eq.~\eqref{eq:Rpa}, and using the elementary inequality
	$|\max\{X,Y\}-T|\le|X-T|+Y$ for $T,Y\ge0$,\footnote{If $X\ge Y$, the left
		side equals $|X-T|$. If $X<Y$, then either $Y\ge T$, giving $Y-T\le Y$, or
		$Y<T$, giving $T-Y<T-X=|X-T|$.} we obtain
	\begin{equation}\notag
		\outereqsize{
			\begin{aligned}
				\big|d\cdot\widetilde{err}^{+,PA}_t-\entity R_{\rbindexem{pub},t}\big|
				\le\big|X_t-\|\vectorfont B_t\|_2^2\big|+Y_t ,
			\end{aligned}
		}
	\end{equation}
	where
	\begin{equation}\notag
		\outereqsize{
			\begin{aligned}
				X_t-\|\vectorfont B_t\|_2^2
				=& 2\vectorfont B_t^\top\vectorfont E_t
				+\Big(\|\vectorfont E_t\|_2^2-\sum_{r}c_r^2V_{\rbindexem{dis},r,t}\Big)\\
				&-\sum_{r}c_r^2\big(\widehat V_{\rbindexem{dis},r,t}-V_{\rbindexem{dis},r,t}\big).
			\end{aligned}
		}
	\end{equation}
	Let $\widetilde\gamma_r=\gamma_r\vectorfont{1}_{r\in\entity A}$. Since $\widetilde{\vectorfont\gamma}$
	and $\vectorfont\pi$ are non-negative with $\ell_1$-norm at most one,
	$\|\vectorfont c\|_2^2\le\|\widetilde{\vectorfont\gamma}\|_2^2+\|\vectorfont\pi\|_2^2\le2$ and
	$|c_r|\le1$. By the Cauchy--Schwarz inequality,
	$\|\vectorfont E_t\|_2^2\le\|\vectorfont c\|_2^2\sum_r\|\vectorfont e_r\|_2^2$, and hence
	$\mathbb E\|\vectorfont E_t\|_2^2\le2S_{\rbindexem{dis}}$. We bound the four terms separately:
	(i) by the Cauchy--Schwarz inequality and Eq.~\eqref{eq:Bsq},
	$2\mathbb E|\vectorfont B_t^\top\vectorfont E_t|\le2\sqrt{\mathbb E\|\vectorfont B_t\|_2^2}\sqrt{\mathbb E\|\vectorfont E_t\|_2^2}\le2\sqrt2\,\beta_t\sqrt{S_{\rbindexem{dis}}}$;
	(ii) $\mathbb E\big|\|\vectorfont E_t\|_2^2-\sum_rc_r^2V_{\rbindexem{dis},r,t}\big|\le2S_{\rbindexem{dis}}+S_{\rbindexem{dis}}=3S_{\rbindexem{dis}}$;
	(iii) by Theorem~\ref{thm:precise_var} and Eq.~\eqref{lemma:var-hat-v-dis-t-upper-bound},
	$\mathbb E\big|\sum_rc_r^2(\widehat V_{\rbindexem{dis},r,t}-V_{\rbindexem{dis},r,t})\big|
	\le\sum_r\sqrt{\mathrm{Var}[\widehat V_{\rbindexem{dis},r,t}]}\le\sum_rQ_r$;
	(iv) by Theorem~\ref{thm:precise_var}, $\mathbb E[Y_t]\le\sum_{r\in\entity R}\mathbb E[\widehat V_{\rbindexem{pub},r,l_r}]=V_{\rbindexem{hist},t}$.
	Therefore,
	\begin{equation}\label{eq:riskerr}
		\outereqsize{
			\begin{aligned}
				\mathbb E\big|d\cdot\widetilde{err}^{+,PA}_t-\entity R_{\rbindexem{pub},t}\big|
				\le2\sqrt2\,\beta_t\sqrt{S_{\rbindexem{dis}}}+3S_{\rbindexem{dis}}+\sum_{r=1}^KQ_r+V_{\rbindexem{hist},t}.
			\end{aligned}
		}
	\end{equation}

	\textit{Step 5 (combination).}
	Taking total expectation of Eq.~\eqref{eq:sel}, and applying
	Eqs.~\eqref{eq:Bsq} and \eqref{eq:riskerr} together with
	Lemma~VIII.7, i.e., $d\cdot\mathbb E|dis_t-dis^*_t|\le D_{\rbindexem{dis}}$, yields
	\begin{equation}\notag
		\outereqsize{
			\begin{aligned}
				err_t^{\mathrm{PLPA^+,Pre}}\le&
				\sum_{r\in\entity A}V^{\mathrm{Pre}}_{\rbindexem{pub},r,t}
				+\beta_t^2+2\sqrt2\,\beta_t\sqrt{S_{\rbindexem{dis}}}
				+3S_{\rbindexem{dis}}\\
				&+\sum_{r=1}^KQ_r+V_{\rbindexem{hist},t}+D_{\rbindexem{dis}},
			\end{aligned}
		}
	\end{equation}
	which is the claimed result.

\textit{\solutionC{} case.} In \solutionC{}, all privacy groups are updated synchronously,
i.e., $\entity{A}=[K]$ and $\entity{R}=\emptyset$. Repeating Steps~1--5 with
Alg.~\ref{alg:PLPD+} and Eq.~\eqref{eq:final_corrected_error} in place of Alg.~\ref{alg:PLPA+} and Eq.~\eqref{eq:final_corrected_time_error}, we have
$\Gamma_t=V_{\rbindexem{hist},t}=0$ and $\beta_t=\Xi$, and thus
\begin{equation}\notag
	\outereqsize{
		\begin{aligned}
			err_t^{\operatorname{\solutionC{},Pre}} \le \sum_{r=1}^{K} V^{\operatorname{Pre}}_{\rbindexsmallem{pub},r,t}
			+ \Xi^2 + 2\sqrt{2}\,\Xi\sqrt{S_{\rbindexsmallem{dis}}} + D .
		\end{aligned}
	}
\end{equation}
Averaging over time and applying Lemma~VIII.8 to replace
$V^{\operatorname{Pre}}_{\rbindexem{pub},r,t}$ by $\overline{V}_{\rbindexem{pub},r}$ yields the \solutionC{} bound
in Theorem~\ref{thm:plpd-plus-utility}.
\end{proof}

\subsection{Additional Experiments}
\subsubsection{Running Time}\label{appendix:experiment:timecost-extra}
\begin{figure}[h]\centering
	\subfigure{
		\scalebox{0.33}[0.33]{\includegraphics{figures/legend_main.pdf}}}\hfill\\
	\addtocounter{subfigure}{-1}
	\subfigure[][{\small \checkInDatasetName{}}]{
		\scalebox{0.295}[0.295]{\includegraphics{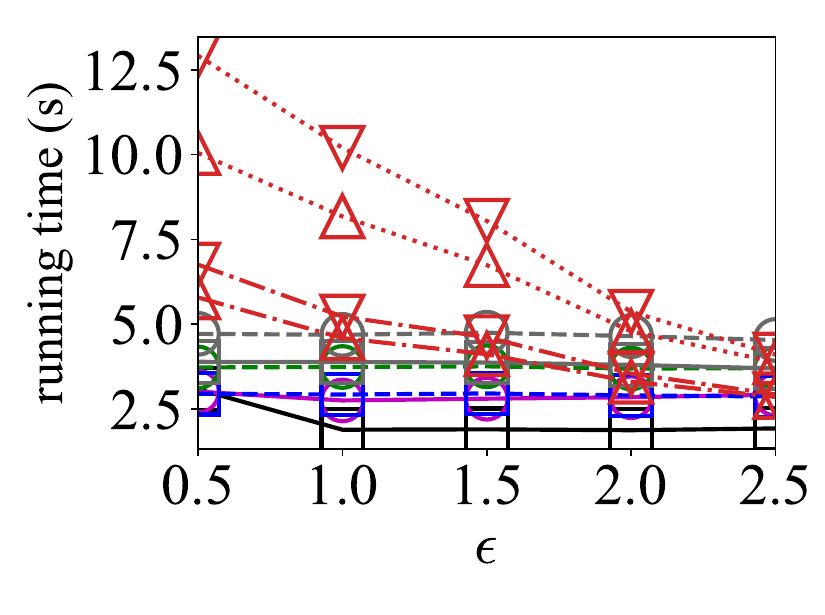}}
		\label{subfig:check_in_budget_change_running_time}}\hfill	
	\subfigure[][{\small \sinDatasetName{}}]{
		\scalebox{0.295}[0.295]{\includegraphics{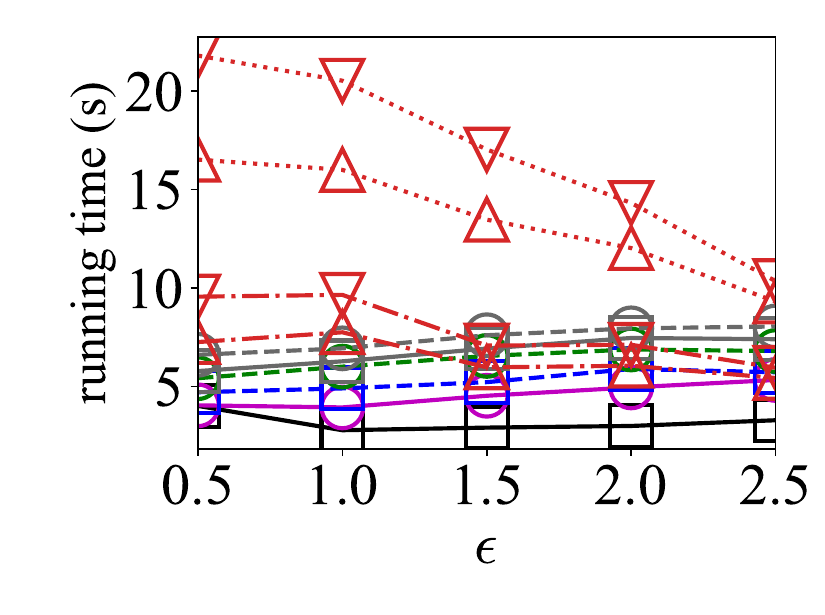}}
		\label{subfig:sin_budget_change_running_time}}\hfill 
	\subfigure[][{\small \logDatasetName{}}]{
		\scalebox{0.295}[0.295]{\includegraphics{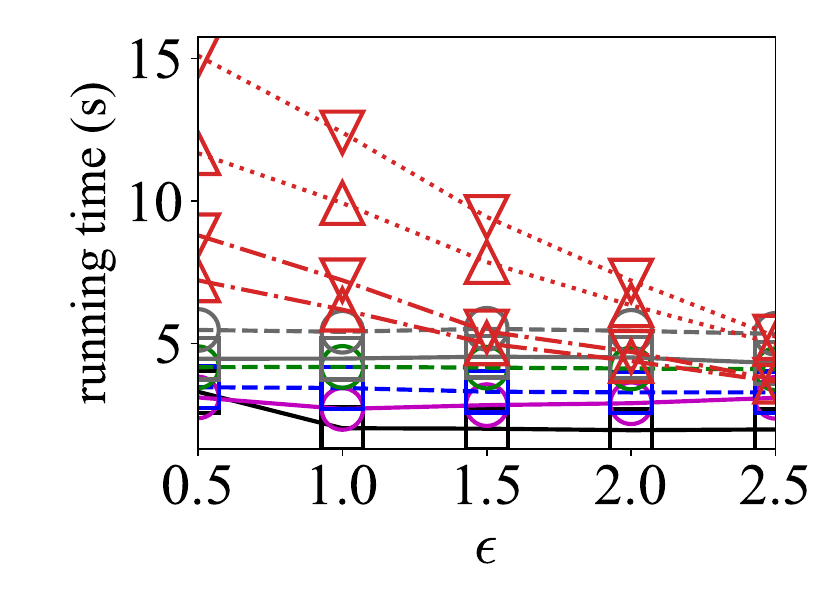}}
		\label{subfig:log_budget_change_running_time}}\hfill 
	\subfigure[][{\small \checkInDatasetName{}}]{
		\scalebox{0.295}[0.295]{\includegraphics{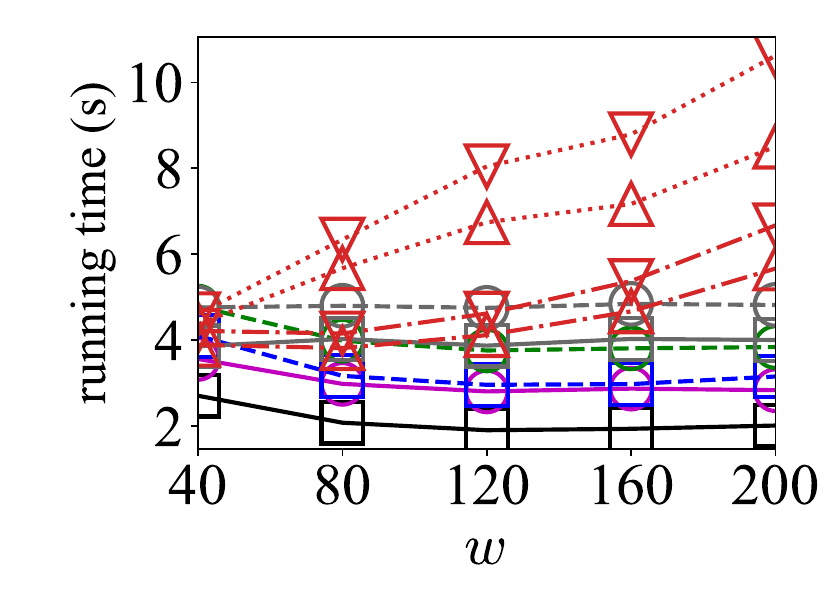}}
		\label{subfig:check_in_window_size_change_running_time}}\hfill	
	\subfigure[][{\small \sinDatasetName{}}]{
		\scalebox{0.295}[0.295]{\includegraphics{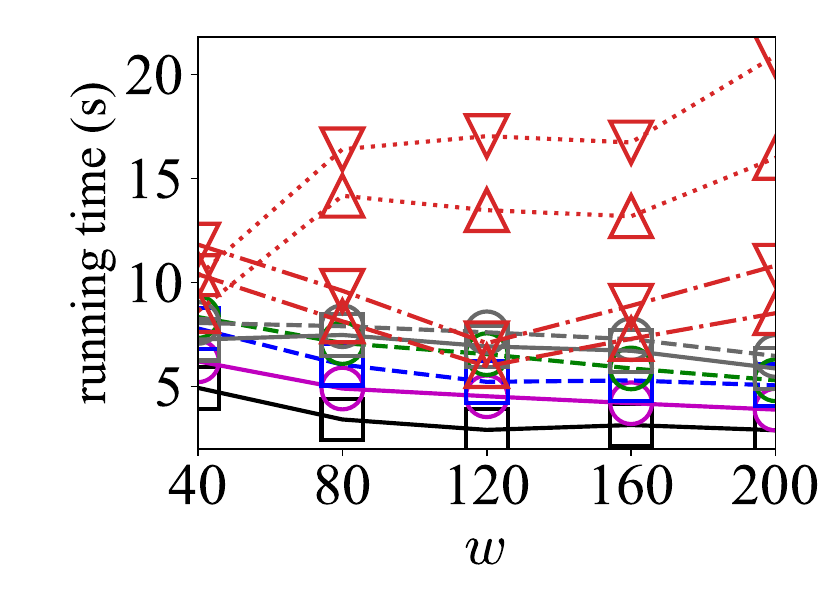}}
		\label{subfig:sin_window_size_change_running_time}}\hfill 
	\subfigure[][{\small \logDatasetName{}}]{
		\scalebox{0.295}[0.295]{\includegraphics{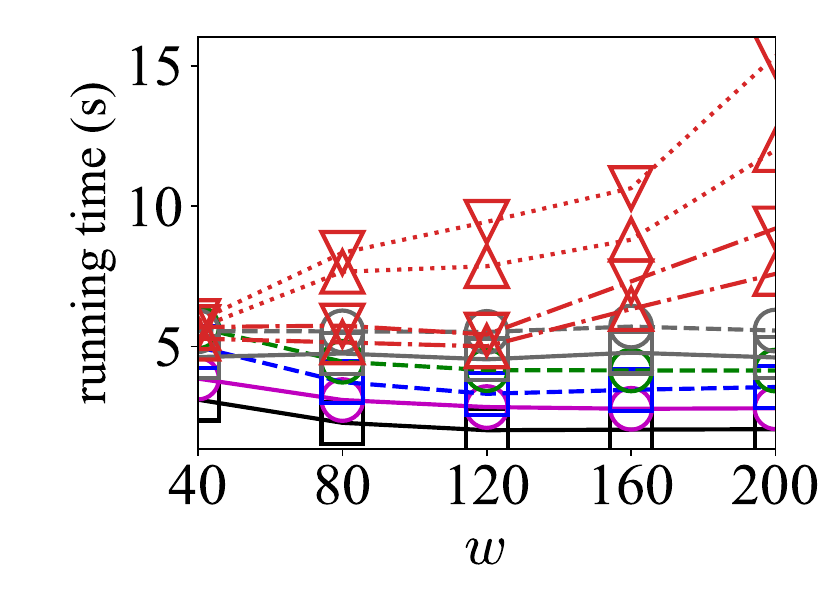}}
		\label{subfig:log_window_size_change_running_time}}\hfill 
	\caption{\small The running time with $\epsilon$ or $w$ varied.}
	\label{fig:alter_e_w_running_time-extra}
\end{figure}

Figure~\ref{fig:alter_e_w_running_time-extra} reports the running time on Foursquare, Sin, and Log, which follows the same trends as Figure~\ref{fig:alter_e_w_running_time}.

%


\subsubsection{AJSD Results}\label{appendix:extra-AJSD-result}
We further evaluate each method in terms of the \textit{Average Jensen-Shannon Divergence}~\cite{DBLP:journals/tit/Lin91} ($\shrinkem{AJSD}, \bar{D}_{JS}$), defined as:
\begin{equation}\notag
	\outereqsize{
		\begin{aligned}
			\bar{D}_{\rbindexsmallem{JS}}\left(\widehat{\vectorfont{f}}\|\vectorfont{f}\right) &= \frac{1}{T}\sum_{\tau=1}^{T}D_{\rbindexsmallem{JS}}\left(\widehat{\vectorfont{f}}\|\vectorfont{f}\right)\\
			&=\frac{1}{T}\sum_{\tau=1}^{T}\left(\frac{1}{2}D_{\rbindexsmallem{KL}}\left(\widehat{\vectorfont{f}}\|\vectorfont{v}\right)+\frac{1}{2}D_{\rbindexsmallem{KL}}\left(\vectorfont{f}\|\vectorfont{v}\right)\right) \\ &=\frac{1}{2T}\sum_{\tau=1}^{T}\sum_{j=1}^{d}\left(\widehat{\vectorfont{f}}_{j,\tau}\log{\left(\frac{\widehat{\vectorfont{f}}_{j,\tau}}{\vectorfont{v}_{j,\tau}}\right)}+\vectorfont{f}_{j,\tau}\log{\left(\frac{\vectorfont{f}_{j,\tau}}{\vectorfont{v}_{j,\tau}}\right)}\right),
		\end{aligned}
	}
\end{equation}
where $\vectorfont{v}_{j,\tau}=\frac{1}{2}\left(\widehat{\vectorfont{f}}_{j,\tau}+\vectorfont{f}_{j,\tau}\right)$.

\begin{figure*}[h]\centering
	\subfigure{
		\scalebox{0.33}[0.33]{\includegraphics{figures/legend_main.pdf}}}\hfill\\
	\addtocounter{subfigure}{-1}\vspace{-2ex}
	\subfigure[][{\small \trajectoryDatasetName{}}]{
		\scalebox{0.238}[0.238]{\includegraphics{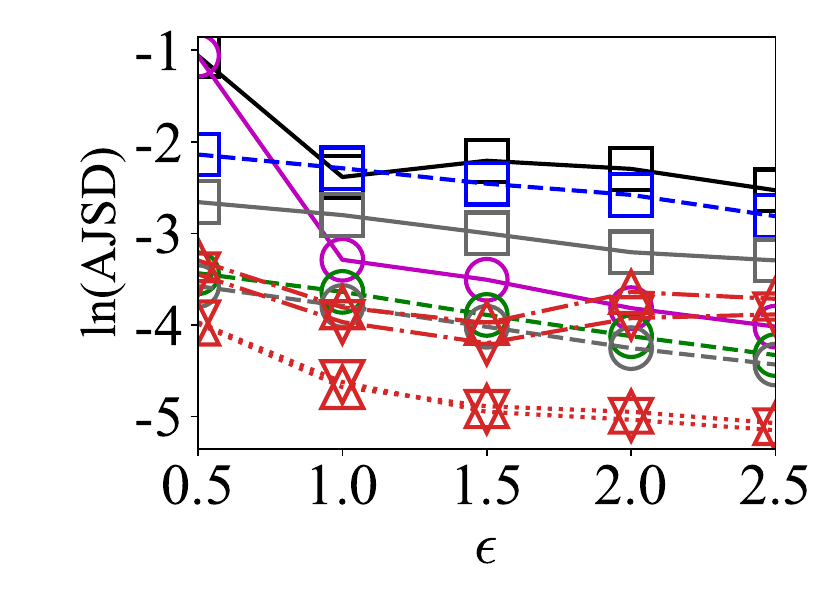}}
		\label{subfig:trajectory_budget_change_AJSD_main_ldp}}\hfill
	\subfigure[][{\small \checkInDatasetName{}}]{
		\scalebox{0.238}[0.238]{\includegraphics{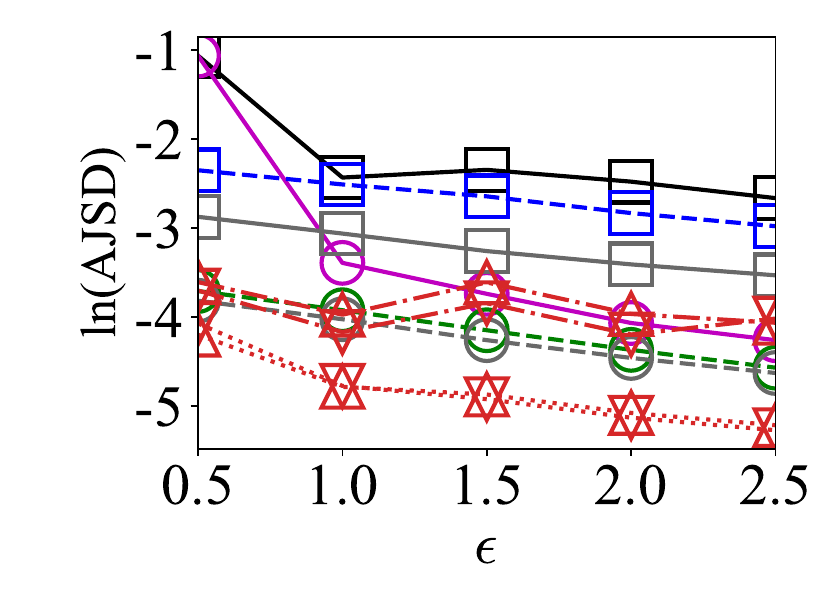}}
		\label{subfig:check_in_budget_change_AJSD_main_ldp}}\hfill	
	\subfigure[][{\small \tlnsDatasetName{}}]{
		\scalebox{0.238}[0.238]{\includegraphics{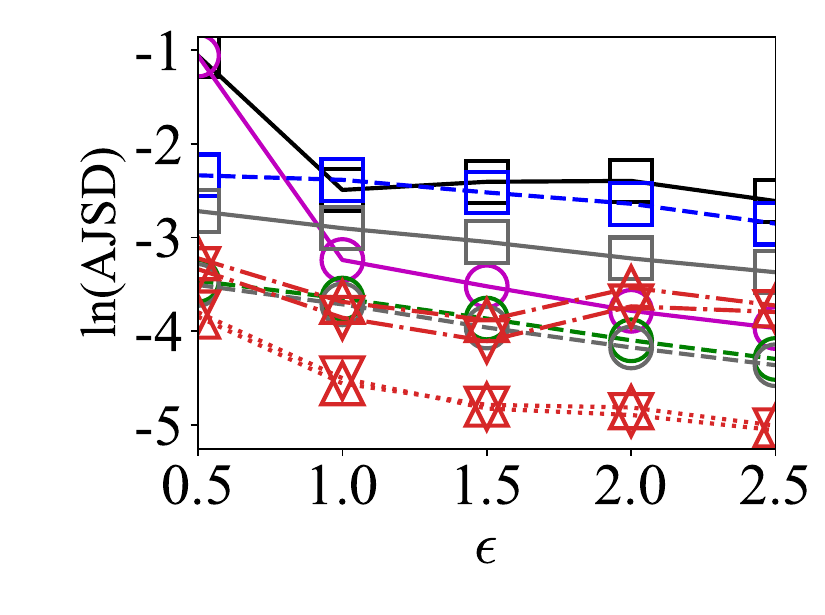}}
		\label{subfig:tlns_budget_change_AJSD_main_ldp}}\hfill 
	\subfigure[][{\small \sinDatasetName{}}]{
		\scalebox{0.238}[0.238]{\includegraphics{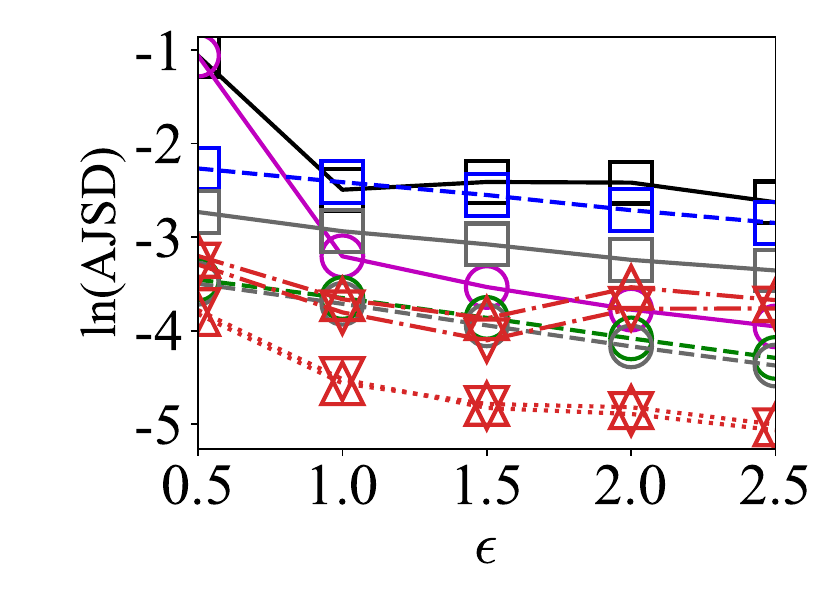}}
		\label{subfig:sin_budget_change_AJSD_main_ldp}}\hfill 
	\subfigure[][{\small \logDatasetName{}}]{
		\scalebox{0.238}[0.238]{\includegraphics{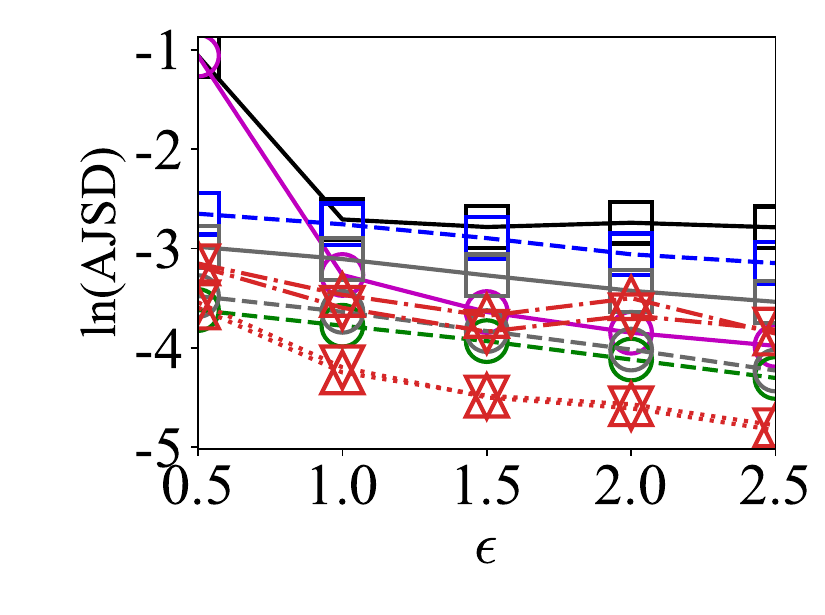}}
		\label{subfig:log_budget_change_AJSD_main_ldp}}\hfill 
	\subfigure[][{\small \trajectoryDatasetName{}}]{
		\scalebox{0.238}[0.238]{\includegraphics{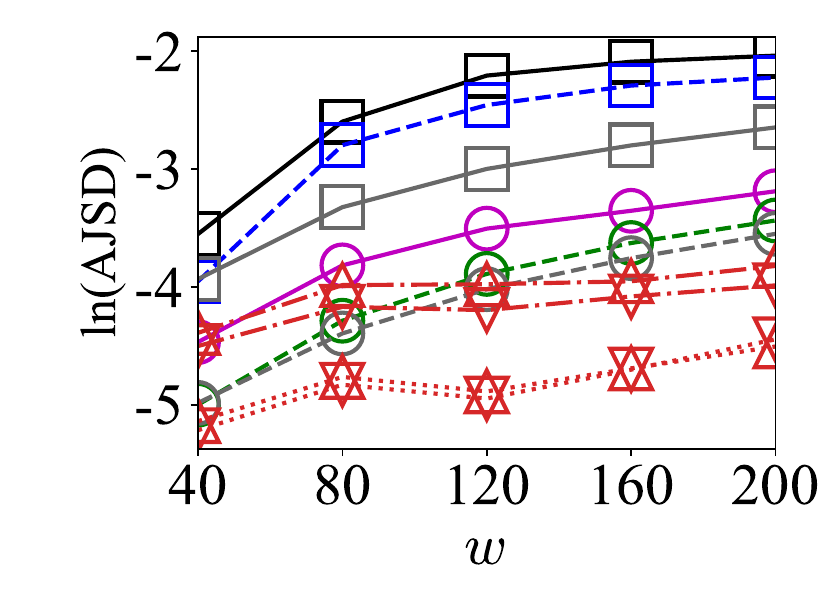}}
		\label{subfig:trajectory_window_size_change_AJSD_main_ldp}}\hfill
	\subfigure[][{\small \checkInDatasetName{}}]{
		\scalebox{0.238}[0.238]{\includegraphics{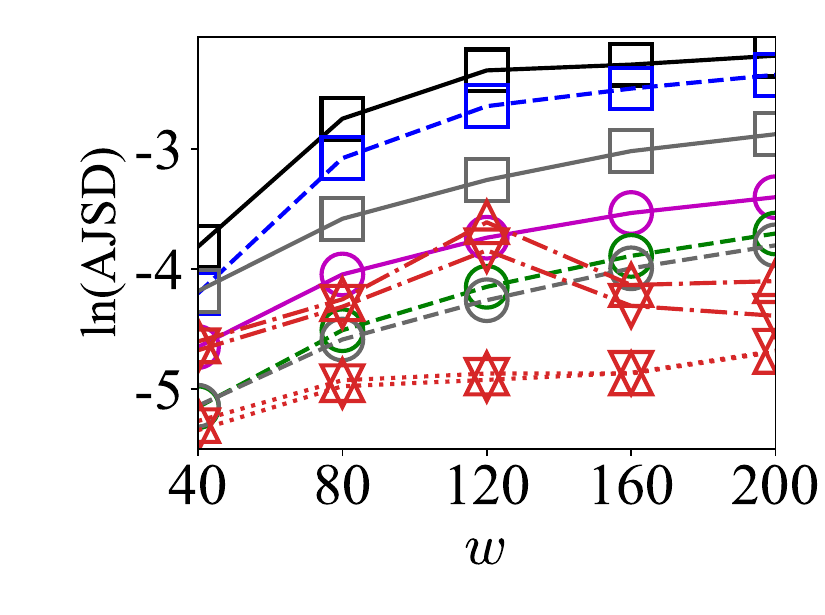}}
		\label{subfig:check_in_window_size_change_AJSD_main_ldp}}\hfill	
	\subfigure[][{\small \tlnsDatasetName{}}]{
		\scalebox{0.238}[0.238]{\includegraphics{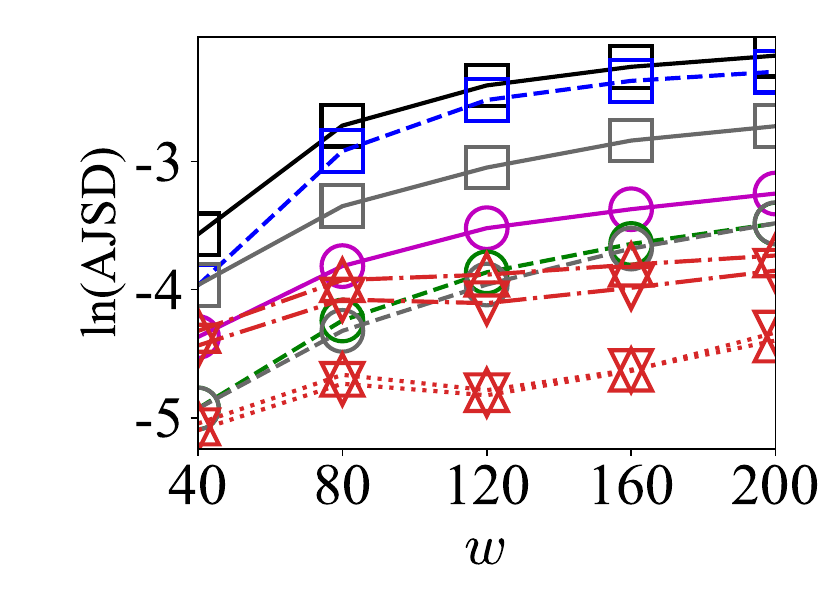}}
		\label{subfig:tlns_window_size_change_AJSD_main_ldp}}\hfill 
	\subfigure[][{\small \sinDatasetName{}}]{
		\scalebox{0.238}[0.238]{\includegraphics{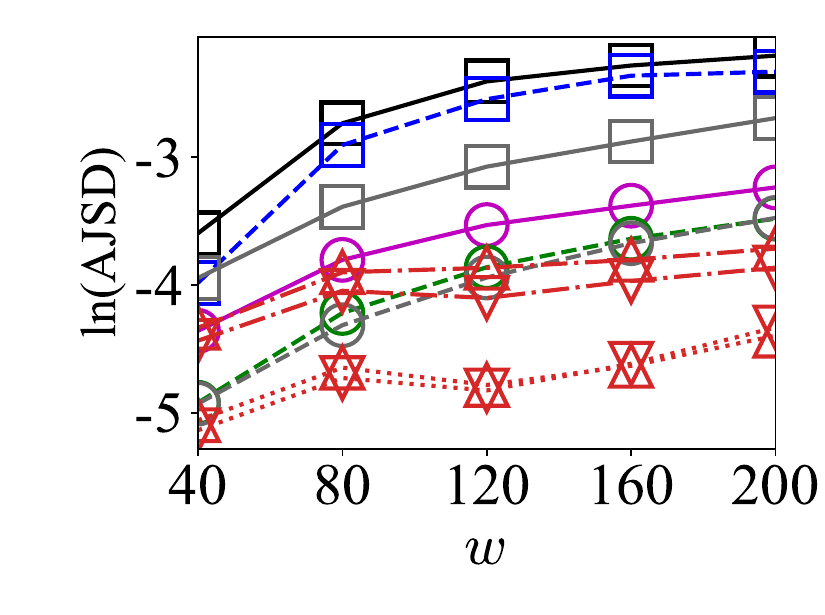}}
		\label{subfig:sin_window_size_change_AJSD_main_ldp}}\hfill 
	\subfigure[][{\small \logDatasetName{}}]{
		\scalebox{0.238}[0.238]{\includegraphics{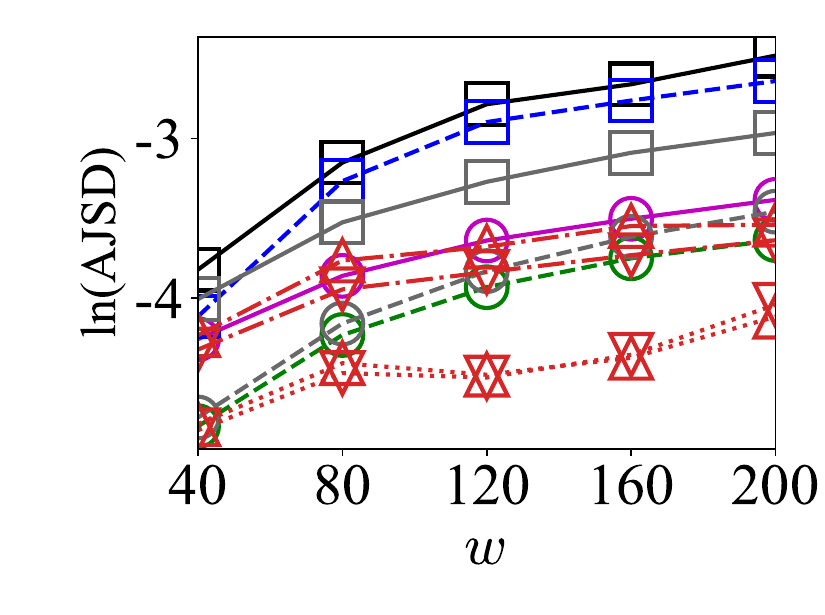}}
		\label{subfig:log_window_size_change_AJSD_main_ldp}}\hfill 
	\caption{\small $AJSD$ with $\epsilon$ or $w$ varied.}
	\label{fig:alter_e_w_AJSD_main_ldp}
\end{figure*}

The experimental results, shown in Figure~\ref{fig:alter_e_w_AJSD_main_ldp}, are consistent with that observed under the $\shrinkem{AMSE}$ metric in Figure~\ref{fig:alter_e_w_AMSE_main_ldp}.

%

\subsubsection{Ablation Study}
\begin{figure}[t]\centering
	\subfigure{
		\scalebox{0.33}[0.33]{\includegraphics{figures/legend_ablation_horizontal.pdf}}}\hfill\\
	\addtocounter{subfigure}{-1}\vspace{-2ex}
	\subfigure[][{\small \checkInDatasetName{}, $w=120$}]{
		\scalebox{0.295}[0.295]{\includegraphics{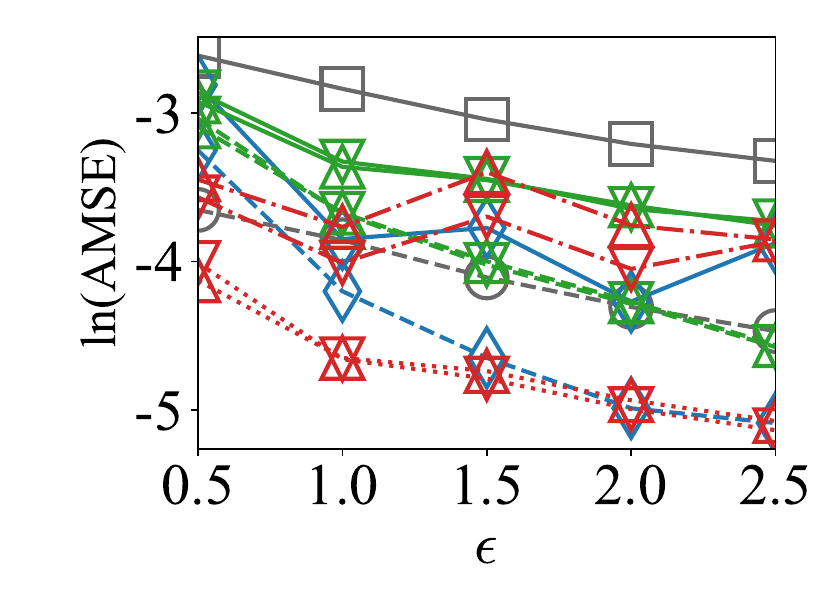}}
		\label{subfig:check_in_budget_change_AMSE_ablation_ldp}}\hfill	
	\subfigure[][{\small \sinDatasetName{}, $w=120$}]{
		\scalebox{0.295}[0.295]{\includegraphics{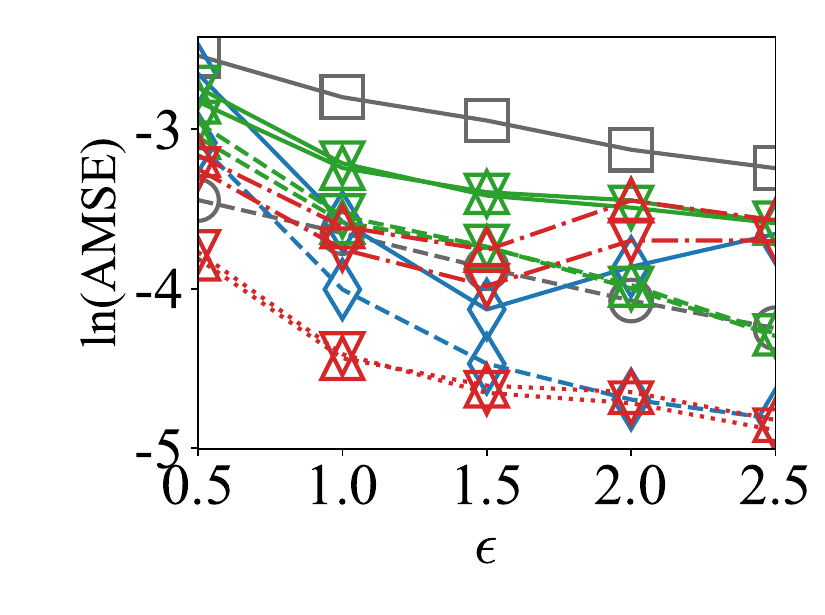}}
		\label{subfig:sin_budget_change_AMSE_ablation_ldp}}\hfill 
	\subfigure[][{\small \logDatasetName{}, $w=120$}]{
		\scalebox{0.295}[0.295]{\includegraphics{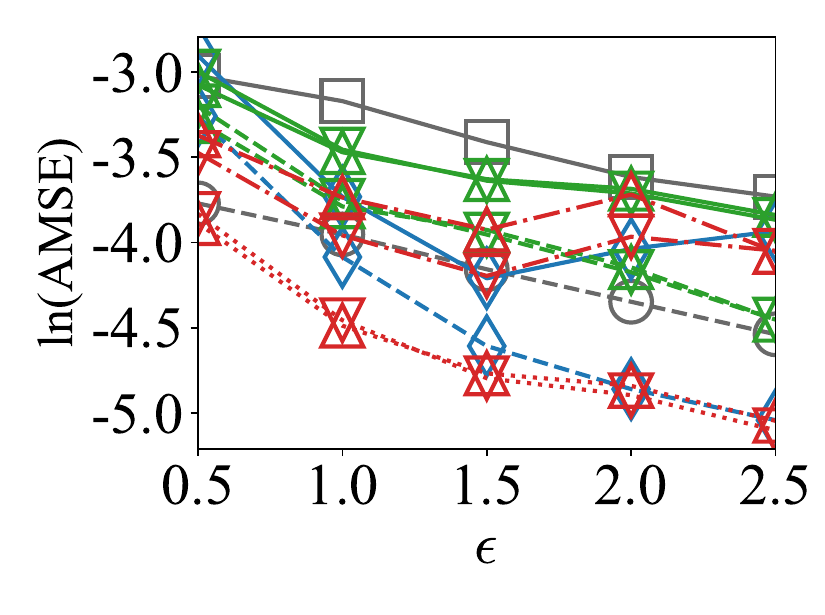}}
		\label{subfig:log_budget_change_AMSE_ablation_ldp}}\hfill 
	\subfigure[][{\small \checkInDatasetName{}, $\epsilon=1.5$}]{
		\scalebox{0.295}[0.295]{\includegraphics{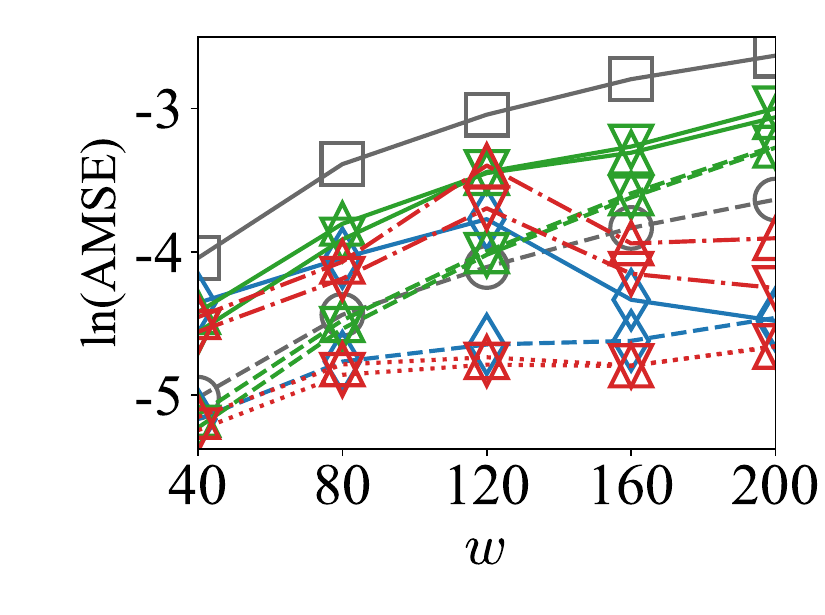}}
		\label{subfig:check_in_window_size_change_AMSE_ablation_ldp}}\hfill	
	\subfigure[][{\small \sinDatasetName{}, $\epsilon=1.5$}]{
		\scalebox{0.295}[0.295]{\includegraphics{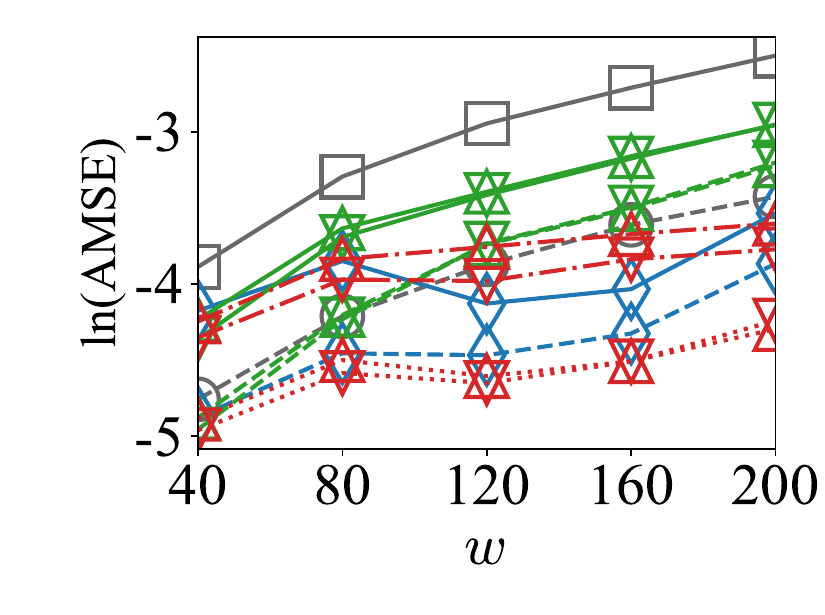}}
		\label{subfig:sin_window_size_change_AMSE_ablation_ldp}}\hfill 
	\subfigure[][{\small \logDatasetName{}, $\epsilon=1.5$}]{
		\scalebox{0.295}[0.295]{\includegraphics{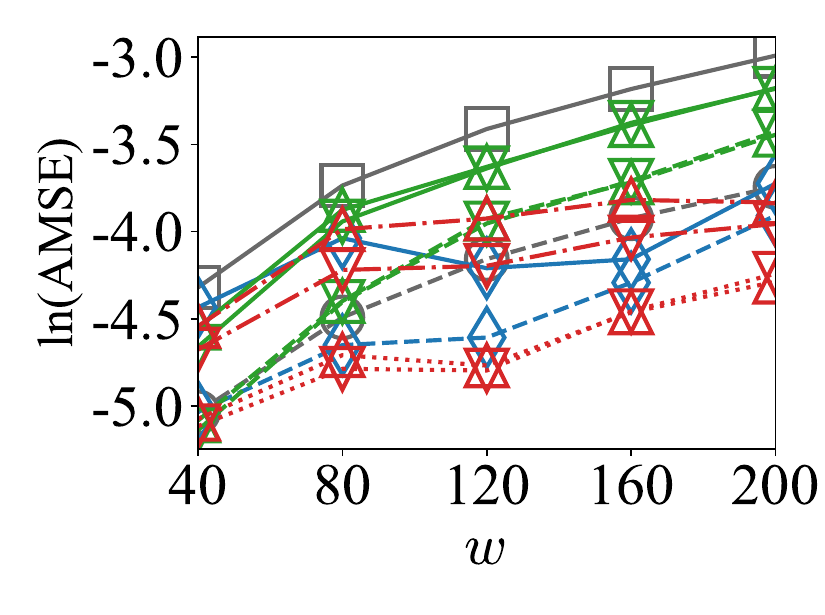}}
		\label{subfig:log_window_size_change_AMSE_ablation_ldp}}\hfill 
	\caption{\small The ablation study of $AMSE$ with different $\epsilon$ or $w$.}
	\label{fig:alter_e_w_AMSE_ablation_ldp-extra}
\end{figure}

\begin{figure*}[h]\centering
	\subfigure{
		\scalebox{0.33}[0.33]{\includegraphics{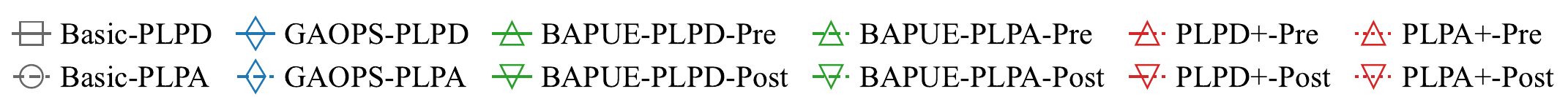}}}\hfill\\
	\addtocounter{subfigure}{-1}\vspace{-2ex}
	\subfigure[][{\small \trajectoryDatasetName{}}]{
		\scalebox{0.238}[0.238]{\includegraphics{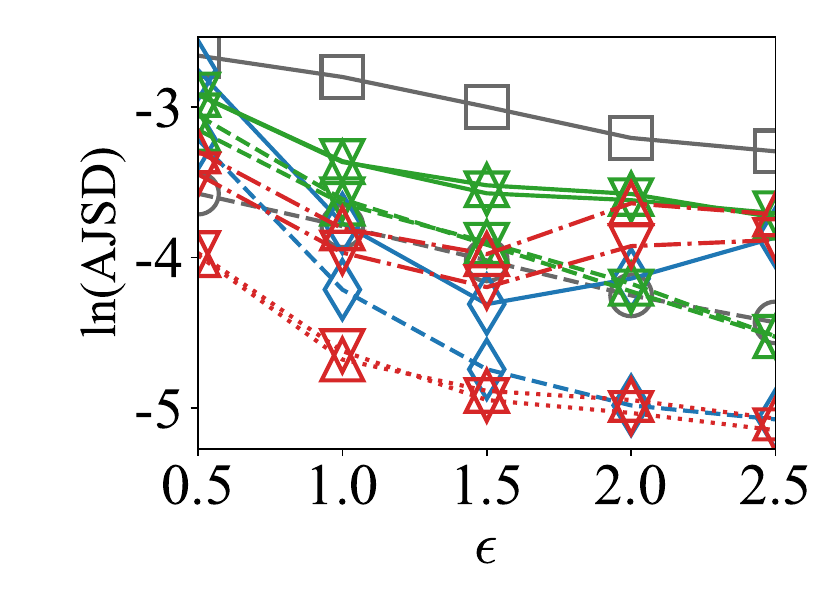}}
		\label{subfig:trajectory_budget_change_AJSD_ablation_ldp}}\hfill
	\subfigure[][{\small \checkInDatasetName{}}]{
		\scalebox{0.238}[0.238]{\includegraphics{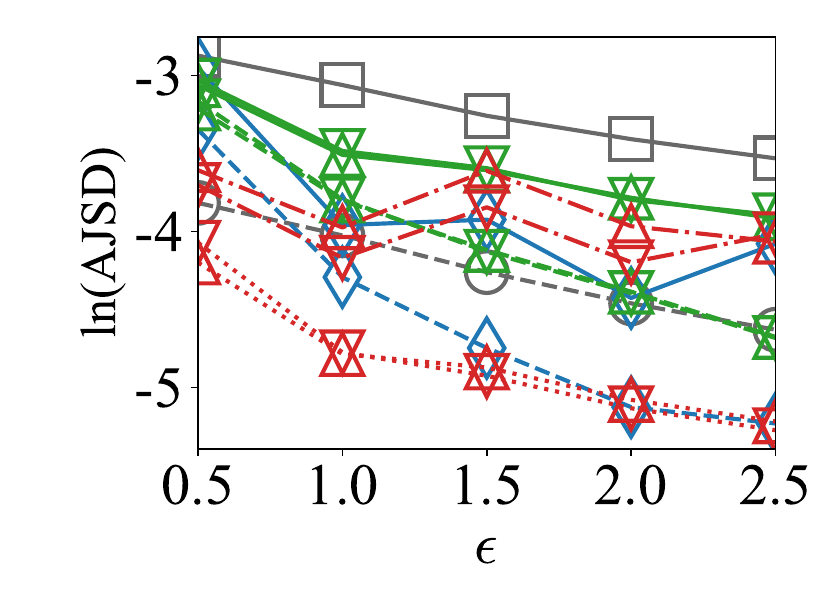}}
		\label{subfig:check_in_budget_change_AJSD_ablation_ldp}}\hfill	
	\subfigure[][{\small \tlnsDatasetName{}}]{
		\scalebox{0.238}[0.238]{\includegraphics{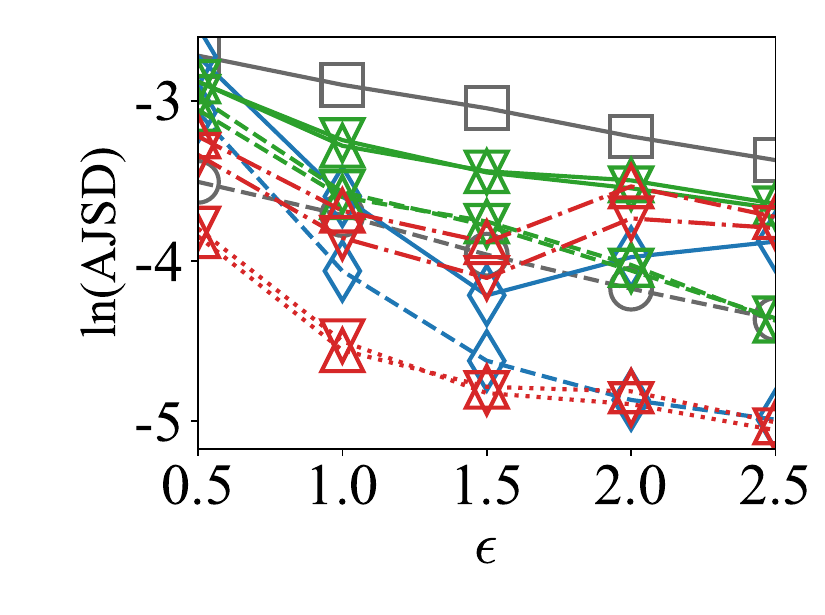}}
		\label{subfig:tlns_budget_change_AJSD_ablation_ldp}}\hfill 
	\subfigure[][{\small \sinDatasetName{}}]{
		\scalebox{0.238}[0.238]{\includegraphics{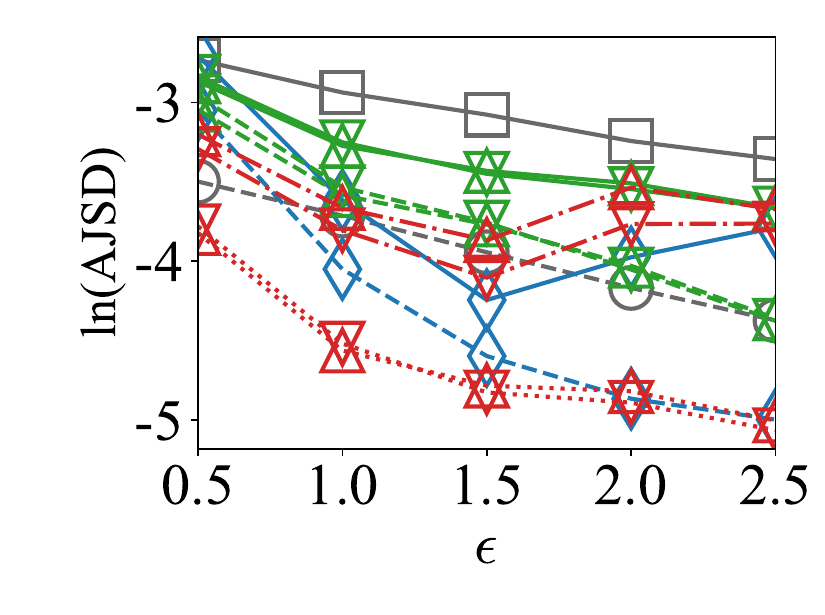}}
		\label{subfig:sin_budget_change_AJSD_ablation_ldp}}\hfill 
	\subfigure[][{\small \logDatasetName{}}]{
		\scalebox{0.238}[0.238]{\includegraphics{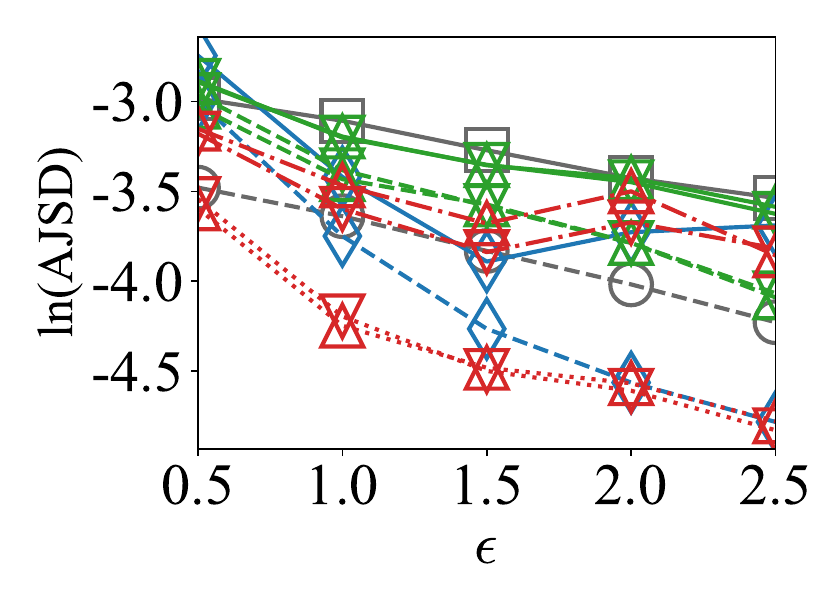}}
		\label{subfig:log_budget_change_AJSD_ablation_ldp}}\hfill 
	\subfigure[][{\small \trajectoryDatasetName{}}]{
		\scalebox{0.238}[0.238]{\includegraphics{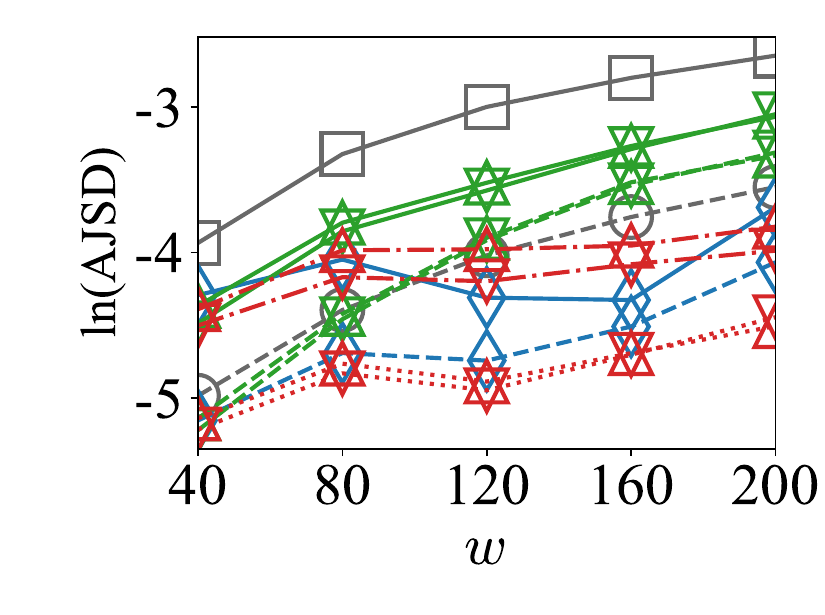}}
		\label{subfig:trajectory_window_size_change_AJSD_ablation_ldp}}\hfill
	\subfigure[][{\small \checkInDatasetName{}}]{
		\scalebox{0.238}[0.238]{\includegraphics{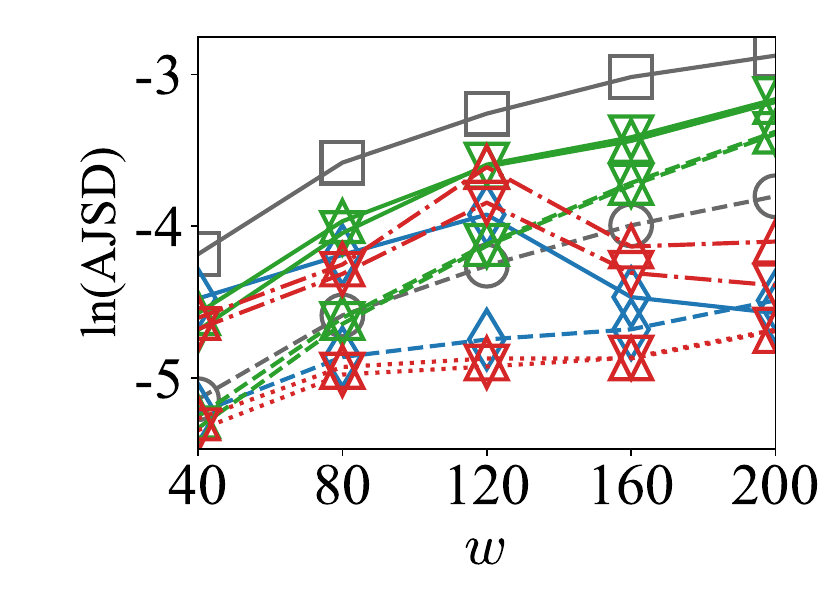}}
		\label{subfig:check_in_window_size_change_AJSD_ablation_ldp}}\hfill	
	\subfigure[][{\small \tlnsDatasetName{}}]{
		\scalebox{0.238}[0.238]{\includegraphics{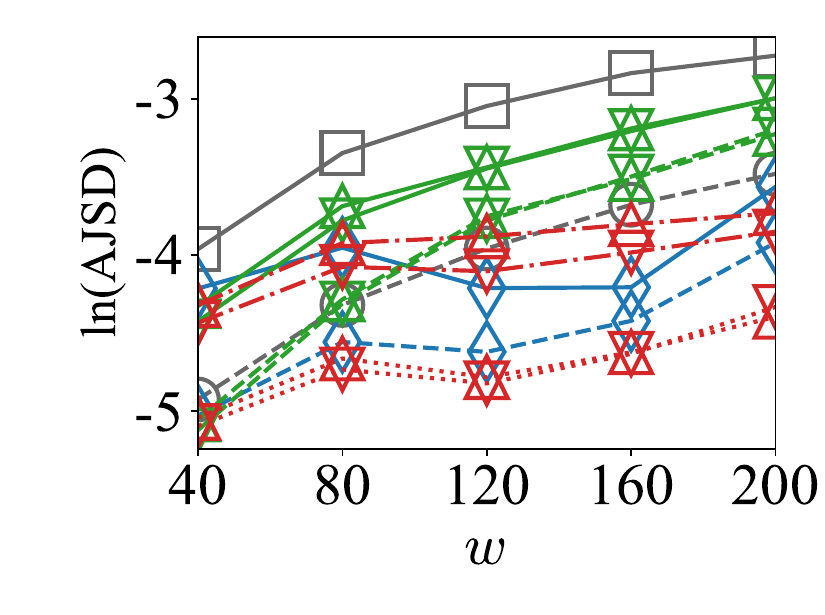}}
		\label{subfig:tlns_window_size_change_AJSD_ablation_ldp}}\hfill 
	\subfigure[][{\small \sinDatasetName{}}]{
		\scalebox{0.238}[0.238]{\includegraphics{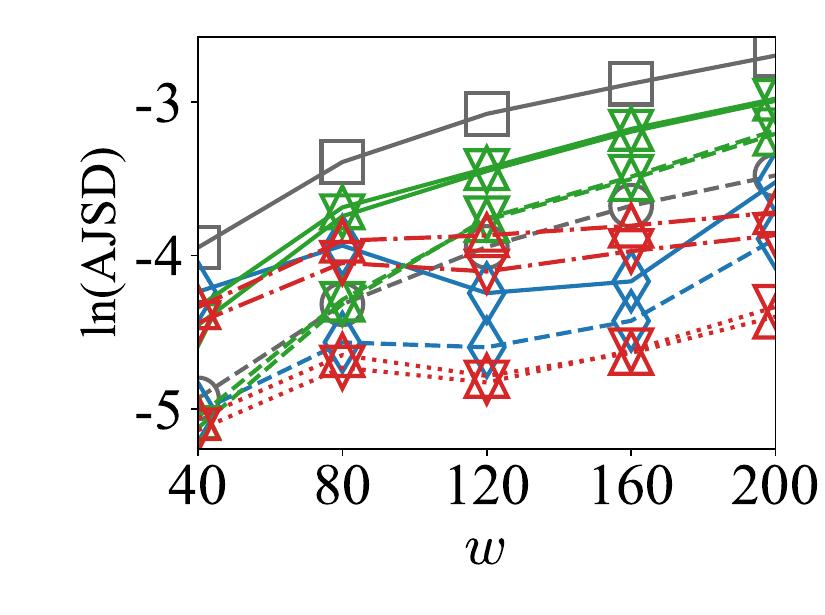}}
		\label{subfig:sin_window_size_change_AJSD_ablation_ldp}}\hfill 
	\subfigure[][{\small \logDatasetName{}}]{
		\scalebox{0.238}[0.238]{\includegraphics{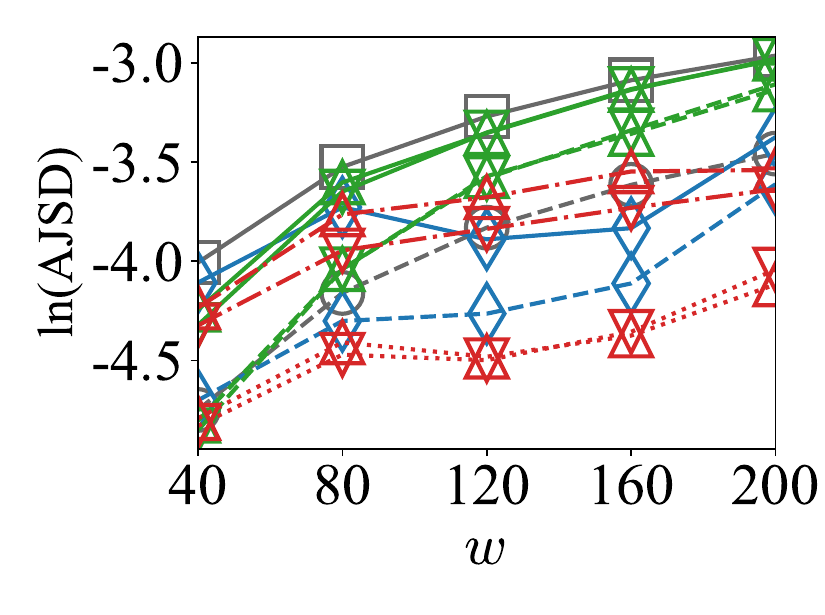}}
		\label{subfig:log_window_size_change_AJSD_ablation_ldp}}\hfill 
	\caption{\small The ablation study of $AJSD$ with $\epsilon$ or $w$ varied.}
	\label{fig:alter_e_w_AJSD_ablation_ldp}
\end{figure*}

Figure~\ref{fig:alter_e_w_AMSE_ablation_ldp-extra} reports the $\shrinkem{AMSE}$ ablation on Foursquare, Sin, and Log, and Figure~\ref{fig:alter_e_w_AJSD_ablation_ldp} reports the AJSD ablation on all datasets; the trends are consistent with Figures~\ref{fig:alter_e_w_AMSE_ablation_ldp} and~\ref{fig:alter_e_w_AMSE_ablation_ldp-extra}.

\subsection{Algorithm Details}
\subsubsection{Full \solutionA{}/\solutionB{}}
Alg.~\ref{alg:ops-guided-plpd-plpa-full} gives the full version of Alg.~\ref{alg:ops-guided-plpd-plpa}, including the PFO.P/A/E steps.

\begin{algorithm}[!t]\small  \DontPrintSemicolon
	\caption{Unified \solutionA{}/\solutionB{} Stream Estimation}
	\label{alg:ops-guided-plpd-plpa-full}
	\DontPrintSemicolon
	\KwIn{Available users $U_A$; 
		privacy requirements ($\vectorfont{w}$, $\vectorfont{\epsilon}$); 
		data publication history $(\widehat{\vectorfont{f}}_{{1}},\widehat{\vectorfont{f}}_{{2}},\ldots, \widehat{\vectorfont{f}}_{{t-1}})$; 
		mechanism mode \(\mathsf{mode}\in\{\mathrm{PLPD},\mathrm{PLPA}\}\)}
	\KwOut{The current estimated frequency $\widehat{\vectorfont{f}}_t$}
	Calculate distinct privacy budgets \(\widetilde{\boldsymbol\epsilon}\) and group statistics $\vectorfont{g}$;\\
	
	\tcp{sub-mechanism $\entity{M}_{\rbindexem{dis},t}$}
	Calculate the sampling size $\vectorfont{z}\gets\scriptmath{\left( \left\lfloor\frac{n}{2w_{{1}}}\right\rfloor,\left\lfloor\frac{n}{2w_{{2}}}\right\rfloor,\ldots,\left\lfloor\frac{n}{2w_{{n}}}\right\rfloor\right)}$; \\ 
	
	
	Calculate $\left(z_{\rbindexem{opt}},\tau_{\rbindexem{opt}},\shrinkem{err}_{\rbindexem{opt}},\vectorfont{\epsilon}_{\rbindexem{opt}}\right)\gets\shrink{OPS}(\vectorfont{z},\vectorfont{\epsilon},\vectorfont{w}, 1/2,d)$ by Alg.~\ref{alg:ops};\\ 
	
	Sample user set $U_{\rbindexem{dis,t}}$, whose data is $D_{\rbindexem{dis,t}}$ and privacy budget list is $\vectorfont{\epsilon}_{\rbindexem{opt}}'\subseteq\vectorfont{\epsilon}_{\rbindexem{opt}}$, from $U_A$ with the size of $z_{\rbindexem{opt}}$ and remove $U_{\rbindexem{dis,t}}$ from $U_A$, i.e., $U_A=U_A\backslash U_{\rbindexem{dis,t}}$; \\ 
	$\widetilde{\vectorfont{\epsilon}}'_{\rbindexem{opt}}\gets\shrink{Unique}(\vectorfont{\epsilon}'_{\rbindexem{opt}})$;\\
	Get report $\vectorfont{Y}_{\rbindexem{dis,t}}\gets\shrink{PFO.P}(D_{\rbindexem{dis,t}})$  with privacy budget list $\vectorfont{\epsilon}'_{\rbindexem{opt}}$ from users in $U_{\rbindexem{dis,t}}$; \\ 
	Get $\vectorfont{h}_{\rbindexem{dis,t,b}}\gets\shrink{PFO.A}(\vectorfont{Y}_{\rbindexem{dis,t}})$ for each group $G_{{b}}$;\\
	Estimate $\widehat{\vectorfont{f}}_{\rbindexem{dis},t}\gets\shrink{PFO.E}(\vectorfont{h}_{\rbindexem{dis,t}},\widetilde{\vectorfont{\epsilon}}'_{\rbindexem{opt}})$; \\ 
	Calculate $\shrinkem{dis}\gets \frac{1}{d}\scriptmath{\sum_{j=1}^{d}\left(\widehat{\vectorfont{f}}_{\rbindexem{dis,t}}[j]-\widehat{\vectorfont{f}}_l[j]\right)^2}-\frac{1}{d}\scriptmath{\sum_{j=1}^{d}\var{\widehat{\vectorfont{f}}_{\rbindexem{dis,t}}[j]}}$;\\ 
	\tcp{sub-mechanism $\entity{M}_{\rbindexem{pub},t}$}
	\If{$\mathsf{mode}=\shrinkem{\solutionA{}}$}{
		Calculate the optimal remaining population size $n_{\rbindexem{r,opt}}\gets n/2-\sum_{\tau=t-\tau_{\rbindexem{opt}}+1}^{t-1}|U_{{\rbindexem{pub},\ensuremath{\tau}}}|$;\\ 
		Set the optimal number of potential publication users $n_{\rbindexem{pp,opt}}\gets \left\lfloor n_{\rbindexem{r,opt}}/2\right\rfloor$;\\ 
	}
	\Else{
		Calculate the time slots to be nullified $t_{\rbindexem{N}}=\frac{\big|U_{\rbindexem{pub},l}\big|}{z_{\rbindexem{opt}}}-1$;\\ 
		\If{$t-l\leq t_{{N}}$}{
			$\widehat{\vectorfont{f}}_{{t}}\gets\widehat{\vectorfont{f}}_{{t-1}}$;\\
			\Return{$\widehat{\vectorfont{f}}_{\rbindex{pub},t}$}.
		}
		Calculate time slots that can be absorbed $t_{{A}}=t-l-t_{{N}}$;\\ 
		Set the number of potential publication users $n_{\rbindexem{pp,opt}}\gets z_{\rbindexem{opt}}\cdot\min(t_{{A}},\tau_{\rbindexem{opt}})$;\\ 
	}
	
	Sample a user set $U_{\rbindexem{pub},t}$ from $U_A$ with the size of $|U_{\rbindexem{pub},t}|=n_{\rbindexem{pp,opt}}$ and privacy budget list $\vectorfont{\epsilon}_{\rbindexem{opt}}''\subseteq\vectorfont{\epsilon}_{\rbindexem{opt}}$;\\ 
	Calculate the budget group user statistic $\vectorfont{g}_{\rbindexem{pp,opt}}$ of $\widetilde{\vectorfont{\epsilon}}''_{\rbindexem{opt}}=\operatorname{Unique}(\vectorfont{\epsilon}''_{\rbindexem{opt}})$ by $U_{\rbindexem{pub},t}$;\\ 
	Calculate group-level sampling size group $\vectorfont{z}_{\rbindexem{pp}}$, where $z_{\rbindexem{pp},r}\gets n_{\rbindexem{pp},\rbindexem{opt}}\cdot g_{\rbindexem{pp},\rbindexem{opt},r}$;\\
	Calculate the potential reporting error $\shrinkem{err}\gets \shrinkem{err}\left(\widetilde{\vectorfont{\epsilon}}''_{\rbindexem{opt}},n,\vectorfont{z}_{\rbindexem{pp}}, d\right)$ by Eq.~\eqref{eq:finalavgvar};\\ 
	\If{$\shrink{dis}>\shrink{err}$ and $n_{\rbindex{pp,opt}}\geq u_{\rbindex{min}}$}{
		Remove $U_{\rbindexem{pub},t}$ from $U_A$, i.e., $U_A\gets U_A\backslash U_{\rbindexem{pub},t}$; \\ 
		Get report $\vectorfont{Y}_{\rbindexem{pub},t}\gets\shrink{PFO.P}(D_{t})$  with privacy budget list $\vectorfont{\epsilon}''_{\rbindexem{opt}}$ from users in $U_{\rbindexem{pub},t}$; \\ 
		Get $\vectorfont{h}_{\rbindexem{pub,t,r}}\gets\shrink{PFO.A}(\vectorfont{Y}_{\rbindexem{pub},t})$ for each group $G_{{r}}$;\\
		Calculate $\widehat{\vectorfont{f}}_{\rbindexem{pub},t}\gets\shrink{PFO.E}(\vectorfont{h}_{\rbindexem{pub},t}, \widetilde{\vectorfont{\epsilon}}''_{\rbindexem{opt}})$; \\ 
	} \Else {
		Set $\widehat{\vectorfont{f}}_{\rbindexem{pub},t}\gets\widehat{\vectorfont{f}}_{\rbindexem{pub,t\em-1}}$; 
	}
	\If{$t\geq \tau_{\rbindex{opt}}$}{
		$U_A\gets U_A\cup U_{\rbindexem{dis},t-\tau_{\rbindexsmallem{opt}}+1}\cup U_{\rbindexem{pub},t-\tau_{\rbindexsmallem{opt}}+1}$; \\ 
	}
	\Return{$\widehat{\vectorfont{f}}_{\rbindex{pub},t}$}
\end{algorithm}

\subsubsection{Full \advancedCorretedWeight{}}
Alg.~\ref{alg:pa-gcw-full} details how \advancedCorretedWeight{} in Alg.~\ref{alg:pa-gcw} instantiates the auxiliary inputs of Table~\ref{tab:paacw-inputs} for active and reused groups.

\begin{algorithm}[!t]\small \DontPrintSemicolon
	\caption{\advancedCorretedWeightTotalName{} (\advancedCorretedWeight{})}
	\label{alg:pa-gcw-full}
	\KwIn{
		Mixed publication estimates
		${\widehat{\boldsymbol F}_{\rbindexem{pub},t}}$;
		current dissimilarity estimates
		${\widehat{\boldsymbol F}_{\rbindexem{dis},t}}$;
		publication-candidate variances
		${\widehat{\vectorfont{V}}_{\rbindexem{pub},t}}$;
		dissimilarity variances ${\widehat{\vectorfont{V}}_{\rbindexem{dis},t}}$;
		reused groups ${\vectorfont{PG}_{t}^{(R)}}$;
		source sets ${\vectorfont{\entity{S}}}$;
		clipping threshold ${\mu}$;
		fusion mode ${\textsf{mode}_{fsn}\in\{\operatorname{Pre},\operatorname{POST}\}}$
	}
	\KwOut{The fusion-weight matrix
		${\vectorfont{\Lambda}^{*}}$}
	
	Initialize ${\vectorfont{\Lambda}^{*}\gets\emptyset}$;\\ \label{alg:pa-gcw:construct-start}
	
	\For{${r\in[K]}$}{
		\If{${PG_r\in\vectorfont{PG}_t^{(R)}}$}{ \label{alg:pa-gcw:reused:start}
			${\tau_{r}\gets l_r}$; 
			${\widehat{\vectorfont{f}}_{\rbindexem{est},\tau_{r}}^{(r)}\gets\widehat{\vectorfont{f}}_{\rbindexem{pub},l_r}^{(r)}}$; 
			${V_{\rbindexem{est},r,\tau_{r}}\gets \widehat{V}_{\rbindexem{pub},r,l_r}}$;\\  \label{alg:PA-GCW:reused_est}\label{alg:pa-gcw:reused:end}
		} \Else{
			${\tau_{r}\gets t}$;\\ \label{alg:pa-gcw:active_Pre_start}
			\If{${\textsf{mode}_{fsn}=\operatorname{Pre}}$}{
				${\widehat{\vectorfont{f}}_{\rbindexem{est},\tau_{r}}^{(r)}\gets\widehat{\vectorfont{f}}_{\rbindexem{dis},t}^{(r)}}$;
				${V_{\rbindexem{est},r,\tau_{r}}\gets \widehat{V}_{\rbindexem{dis},r,t}}$;\\ \label{alg:PA-GCW:mode_Pre_est}\label{alg:pa-gcw:active_Pre_end}
			} \Else{ \label{alg:pa-gcw:active_post_start}
				${\widehat{\vectorfont{f}}_{\rbindexem{est},\tau_{r}}^{(r)}\gets\widehat{\vectorfont{f}}_{\rbindexem{pub},t}^{(r)}}$; 
				${V_{\rbindexem{est},r,\tau_{r}}\gets \widehat{V}_{\rbindexem{pub},r,t}}$;\\ \label{alg:PA-GCW:mode_post_est}
			} \label{alg:pa-gcw:active_post_end}
		}
	} \label{alg:pa-gcw:construct-end}
	
	\For{${H\in[K]}$}{ \label{alg:pa-gcw:enhance_start}
		Set ${\entity{C}_H\gets\{H\}\cup \entity{S}_H}$;\\
		\If{${PG_H\in\vectorfont{PG}_t^{(R)}}$}{ \label{alg:pa-gcw:enhance-reuse_start}
			\For{${r\in\entity{C}_H}$}{
				Calculate
				$\widehat{\vectorfont{\xi}}_{r\mid H,t}
				\gets
				\widehat{\vectorfont{f}}_{\rbindexem{est},\tau_{r}}^{(r)} - \widehat{\vectorfont{f}}_{\rbindexem{dis},t}^{(H)}$;\\
			}
			\For{${r,s\in\entity{C}_H}$}{
				Set $\widehat{\vectorfont{M}}_{H,t}(r,s)\gets\left\langle
				\widehat{\vectorfont{\xi}}_{r\mid H,t},
				\widehat{\vectorfont{\xi}}_{s\mid H,t}
				\right\rangle
				- \widehat{V}_{\rbindexem{dis},H,t} + \left(\widehat{V}_{\rbindexem{pub},r,\tau_{r}}-V_{\rbindexem{est},r,\tau_{r}}\right)\delta_{r,s}$;\\
			} \label{alg:pa-gcw:enhance-reuse_end}
		} \Else{  \label{alg:pa-gcw:enhance-new_start}
			\For{${r\in\entity{S}_H}$}{
				Calculate
				$\widehat{\vectorfont{\xi}}_{r\mid H,t}
				\gets
				\widehat{\vectorfont{f}}_{\rbindexem{est},\tau_{r}}^{(r)} - \widehat{\vectorfont{f}}_{\rbindexem{est},t}^{(H)}$;\\
			}
			\For{${r,s\in\entity{S}_H}$}{
				Set $\widehat{\vectorfont{A}}_{H,t}(r,s)\gets\left\langle
				\widehat{\vectorfont{\xi}}_{r\mid H,t},
				\widehat{\vectorfont{\xi}}_{s\mid H,t}
				\right\rangle
				- V_{\rbindexem{est},H,t} + \left(\widehat{V}_{\rbindexem{pub},r,\tau_{r}}-V_{\rbindexem{est},r,\tau_{r}}\right)\delta_{r,s}$;\\
			}
			$\widehat{\vectorfont{M}}_{H,t}\gets\begin{pmatrix}
				\widehat{V}_{\rbindexem{pub},H,t} & \vectorfont 0^\top\\
				\vectorfont 0 & \widehat{\vectorfont{A}}_{H,t}
			\end{pmatrix}$;\\	\label{alg:pa-gcw:enhance_end} \label{alg:pa-gcw:enhance-new_end}  
		}
		$\widetilde{\vectorfont{M}}_{H,t}
		\gets
		\operatorname{\basicClipping{}}(
		\widehat{\vectorfont{M}}_{H,t},
		\mu)$;\\ 	\label{alg:pa-gcw:pdec}
		Obtain ${\vectorfont{\lambda}_H^{*}}$ by solving the simplex-constrained quadratic program defined in Section~\ref{subsec:ba-pue} ;\\	\label{alg:pa-gcw:solve_sqp_start}
		
		Add ${\vectorfont{\lambda}_H^{*}}$ to
		${\vectorfont{\Lambda}^{*}}$;\\
	}
	
	\Return{${\vectorfont{\Lambda}^{*}}$}	\label{alg:pa-gcw:solve_sqp_end}
\end{algorithm}

\subsubsection{Ablation Algorithms for \solutionC{} and \solutionD{}}\label{appendix:ablation-algorithms}
Algorithms~\ref{alg:PLBD-AB-only-ga-ops}–\ref{alg:PLBA-BA-ba-pue} give the \advancedSelection{}-only and \advancedEnhance{}-only variants used in Section VI-E.

\begin{algorithm*}[t!]\small
	\caption{\solutionCMPI{}}
	\label{alg:PLBD-AB-only-ga-ops}
	\DontPrintSemicolon
	\KwIn{Available user set $U_A$; privacy requirement $(\vectorfont{w},\vectorfont{\epsilon})$ of all Users $U$; data domain size $d$; historical data publication $\left(\widehat{\vectorfont{f}}_{{1}},\widehat{\vectorfont{f}}_{{2}},\ldots, \widehat{\vectorfont{f}}_{{t-1}}\right)$} 
	\KwOut{$\widehat{\vectorfont{f}}_t$}
	$\widehat{\vectorfont{\entity{V}}}_{\rbindexem{dis,t}}\gets\emptyset$;
	$\widehat{\vectorfont{F}}_{\rbindexem{dis,t}}\gets\emptyset$;
	$\vectorfont{\entity{V}}_{\rbindexem{pub,t}}^{\operatorname{Pre}}\gets\emptyset$;
	$\widehat{\vectorfont{\entity{V}}}_{\rbindexem{pub,t}}\gets\emptyset$;
	$\widehat{\vectorfont{F}}_{\rbindexem{pub,t}}\gets\emptyset$;\\ \label{alg:plpd+init_start-AB-only-ga-ops}
	Get the privacy group set $\vectorfont{PG}=\{PG_r=(\widetilde{w}_r,\widetilde{\epsilon}_r,U_r)\mid r\in[K]\}$ with privacy group triple set $\vectorfont{PGT}=\{PGT_r=(\widetilde{w}_r,\widetilde{\epsilon}_r,n_r)\mid r\in[K]\}$;\\ \label{alg:pg-AB-only-ga-ops}
	\tcp{sub-mechanism $\entity{M}_{\rbindexem{dis},t}$}
	Get optimal group-aware parameters $(\vectorfont{z}^*_{\rbindexem{opt,dis}},\vectorfont{\tau}^*, \widetilde{\vectorfont{\epsilon}}')\gets\operatorname{\advancedSelection}(\vectorfont{PGT},\frac{1}{2},d)$ by Alg.~\ref{alg:adv_ops};\\ \label{alg:apply_ops-AB-only-ga-ops}\label{alg:plpd+dis_start-AB-only-ga-ops}
	\For{$PG_r\in\vectorfont{PG}$}{
		Sample user set $U_{\rbindexem{dis,r,t}}$, whose data is $D_{\rbindexem{dis,r,t}}$ and effective privacy budget is $\widetilde{\epsilon}'_{r}$, from $U_{A,r}$ with the size of $z^*_{\rbindexem{opt,dis,r}}$ and remove $U_{\rbindexem{dis,r,t}}$ from $U_{A,r}$, i.e., $U_{A,r}=U_{A,r}\backslash U_{\rbindexem{dis,r,t}}$; \\ \label{alg:sampling-AB-only-ga-ops}
		Get frequency estimation $\widehat{\vectorfont{f}}^{(r)}_{\rbindexem{dis,t}}\gets\shrink{FO}(U_{\rbindexem{dis,r,t}},\widetilde{\epsilon}'_{r})$ from users in $U_{\rbindexem{dis,r,t}}$ and calculate the unbiased vector-level total variance estimate $\widehat{V}_{\rbindexem{dis,r,t}}$ by Theorem~\ref{thm:precise_var}; \\ \label{alg:PFO.P-AB-only-ga-ops}
		Add $\widehat{\vectorfont{f}}^{(r)}_{\rbindexem{dis,t}}$ to $\widehat{\vectorfont{F}}_{\rbindexem{dis,t}}$ and $\widehat{V}_{\rbindexem{dis,r,t}}$ to  $\widehat{\vectorfont{\entity{V}}}_{\rbindexem{dis,t}}$;
	}
	Get the population-weighted global estimation $\widehat{\vectorfont{f}}_{\rbindexem{dis,t}}\gets\sum_{r\in[K]}\pi_r\widehat{\vectorfont{f}}^{(r)}_{\rbindexem{dis,t}}$;\\ \label{alg:estimate_group_combine-AB-only-ga-ops}
	Calculate $\shrinkem{dis}\gets \frac{1}{d}\sum_{j=1}^{d}\left(\widehat{\vectorfont{f}}_{\rbindexem{dis,t}}[j]-\widehat{\vectorfont{f}}_l[j]\right)^2-\frac{1}{d}\sum_{r=1}^{K}\pi_r^2 \widehat{V}_{\rbindexem{dis,r,t}}$;\\ \label{alg:dis-AB-only-ga-ops} \label{alg:plpd+dis_end-AB-only-ga-ops}
	\tcp{sub-mechanism $\entity{M}_{\rbindexem{pub},t}$}
	\For{$PG_r\in\vectorfont{PG}$}{
		Calculate the optimal remaining population size $n_{\rbindexem{pub,r}}\gets \left\lfloor n_r/2\right\rfloor-\sum_{\tau=t-\tau^*_{r}+1}^{t-1}|U_{{\rbindexem{pub},\rbindexem{r},\ensuremath{\tau}}}|$;\\ \label{alg:repop:start-AB-only-ga-ops} 
		Set the optimal number of potential publication users $n_{\rbindexem{pp,r}}\gets \left\lfloor n_{\rbindexem{pub,r}}/2\right\rfloor$;\\ \label{alg:group_half_remain-AB-only-ga-ops}
		$V_{\rbindexem{pub,r,t}}^{\operatorname{Pre}}\gets V_{\rbindexem{smpl}}\left(\left\lfloor\frac{n_r}{2}\right\rfloor,n_{\rbindexem{pp,r}}\right)+V_{\rbindexem{LDP}}\left(\epsilon'_r,n_{\rbindexem{pp,r}},d\right)$;\\
		Add $V_{\rbindexem{pub,r,t}}^{\operatorname{Pre}}$ to $\vectorfont{\entity{V}}_{\rbindexem{pub,t}}^{\operatorname{Pre}}$;
	}
	$\widetilde{\shrinkem{err}}^+\gets\frac{1}{d}\sum_{r=1}^{K}\pi_{r}^2 V_{\rbindexem{pub,r,t}}^{\operatorname{Pre}}$;\\
	
	\If{$\shrink{dis}>\widetilde{\shrink{err}}^{+}$ and $\min_{r\in[K]}n_{\rbindex{pp,r}}\geq u_{\rbindex{min}}$}{
		\For{$PG_r\in\vectorfont{PG}$}{\label{alg:begin_pub_f-AB-only-ga-ops}
			Sample a user set $U_{\rbindexem{pub,r,t}}$ from $U_{\rbindexem{A,r}}$ with the size of $|U_{\rbindexem{pub,r,t}}|=n_{\rbindexem{pp,r}}$;\\ \label{alg:sampleinadvance-AB-only-ga-ops}
			Remove $U_{\rbindexem{pub,r,t}}$ from $U_{\rbindexem{A,r}}$, i.e., $U_{\rbindexem{A,r}}\gets U_{\rbindexem{A,r}}\backslash U_{\rbindexem{pub,r,t}}$; \\ \label{alg:smpl-AB-only-ga-ops}
			Get frequency estimation $\widehat{\vectorfont{f}}^{(r)}_{\rbindexem{pub,t}}\gets\shrink{FO}(U_{\rbindexem{pub,r,t}},\widetilde{\epsilon}'_{r})$ from users in $U_{\rbindexem{pub,r,t}}$ and calculate its unbiased vector-level variance estimate $\widehat{V}_{\rbindexem{pub,r,t}}$; \\
			Add $\widehat{\vectorfont{f}}^{(r)}_{\rbindexem{pub,t}}$ to $\widehat{\vectorfont{F}}_{\rbindexem{pub,t}}$ and $\widehat{V}_{\rbindexem{pub,r,t}}$ to $\widehat{\vectorfont{\entity{V}}}_{\rbindexem{pub,t}}$;
		}
		Get the population-weighted global estimation $\widehat{\vectorfont{f}}_{\rbindexem{pub,t}}\gets\sum_{r\in[K]}\pi_r\widehat{\vectorfont{f}}^{(r)}_{\rbindexem{pub,t}}$;
	} \Else {
		Set $\widehat{\vectorfont{f}}_{\rbindexem{pub,t}}\gets\widehat{\vectorfont{f}}_{\rbindexem{pub,t-}1}$; 
	}\label{line:diserrjudge_end-AB-only-ga-ops}
	\For{$PG_r\in\vectorfont{PG}$}{
		\If{$t\geq \tau^*_{r}$}{
			$U_{A,r}\gets U_{A,r}\cup U_{\rbindexem{dis,r,t}-\tau^*_{r}+1}\cup U_{\rbindexem{pub,r,t}-\tau^*_{r}+1}$; \\ 
		}
	}
	\Return{$\widehat{\vectorfont{f}}_{\rbindex{pub,t}}$.}
\end{algorithm*}

\begin{algorithm*}[h]\small
	\caption{\solutionCMPJ{}}
	\label{alg:PLBA-AB-only-ga-ops}
	\DontPrintSemicolon
	\KwIn{Available user set $U_A$; privacy requirement $(\vectorfont{w},\vectorfont{\epsilon})$ of all Users $U$; data domain size $d$; historical data publication $\left(\widehat{\vectorfont{f}}_{{1}},\widehat{\vectorfont{f}}_{{2}},\ldots, \widehat{\vectorfont{f}}_{{t-1}}\right)$} 
	\KwOut{$\widehat{\vectorfont{f}}_t$}
	The same as Lines~\ref{alg:plpd+init_start-AB-only-ga-ops}--\ref{alg:pg-AB-only-ga-ops} in Alg.~\ref{alg:PLBD-AB-only-ga-ops};\\ 
	\tcp{sub-mechanism $\entity{M}_{\rbindexem{dis},t}$}
	The same as Lines~\ref{alg:plpd+dis_start-AB-only-ga-ops}-\ref{alg:plpd+dis_end-AB-only-ga-ops} in Alg.~\ref{alg:PLBD-AB-only-ga-ops};\\ 
	\tcp{sub-mechanism $\entity{M}_{\rbindexem{pub},t}$}
	Initialize active group set $\vectorfont{PG}_t^{(A)}\gets\emptyset$, reused group set $\vectorfont{PG}_t^{(R)}\gets\emptyset$;\\
	\For{$PG_r\in\vectorfont{PG}$}{
		Calculate the time slots to be nullified $t_{\rbindexem{N,r}}=\frac{\big|U_{\rbindexem{pub,r,l}_r}\big|}{z^*_{\rbindexem{opt,dis,r}}}-1$;\\
		\If{$t-l_r\leq t_{\rbindex{N,r}}$}{ 
			Add $\widehat{\vectorfont{f}}^{(r)}_{\rbindexem{pub,l}_r}$ to $\widehat{\vectorfont{F}}_{\rbindexem{pub,t}}$,
			$\widehat{V}_{\rbindexem{pub,r,l}_r}$ to $\vectorfont{\entity{V}}_{\rbindexem{pub,t}}^{\operatorname{Pre}}$ and $\widehat{\vectorfont{\entity{V}}}_{\rbindexem{pub,t}}$, and
			$\vectorfont{PG}_r$ to $\vectorfont{PG}_t^{(R)}$;\\ 
		} \Else{
			Calculate group-aware time slots that can be absorbed $t_{A,r}=t-l_r-t_{\rbindexem{N,r}}$;\\ 
			Set the number of potential publication users $n_{\rbindexem{pp,r}}\gets z_{\rbindexem{opt,dis,r}}^*\cdot\min(t_{\rbindexem{A,r}},\tau^*_{r})$;\\
			$V_{\rbindexem{pub,r,t}}^{\operatorname{Pre}}\gets\left(V_{\rbindexem{smpl}}\left(\left\lfloor\frac{n_r}{2}\right\rfloor,n_{\rbindexem{pp,r}}\right)+V_{\rbindexem{LDP}}\left(\epsilon'_r,n_{\rbindexem{pp,r}},d\right)\right)$;\\
			Add $V_{\rbindexem{pub,r,t}}^{\operatorname{Pre}}$ to $\vectorfont{\entity{V}}_{\rbindexem{pub,t}}^{\operatorname{Pre}}$;\\
			Add $PG_r$ to $\vectorfont{PG}_t^{(A)}$;\\ 
		}
	}
	\If{$\vectorfont{PG}_t^{(A)}=\emptyset$}{
		Set $\widehat{\vectorfont{f}}_{\rbindexem{pub,t}}\gets\widehat{\vectorfont{f}}_{\rbindexem{pub,t}-1}$; 
	} \Else{
		Calculate the staleness-aware potential reporting error 
			$\widetilde{\shrinkem{err}}^{+}$ by Eq.~\eqref{eq:final_corrected_time_error} with $\gamma_r=\pi_{r}, \forall r$;\\
		\If{$\shrink{dis}>\widetilde{\shrink{err}}^{+}$ and $\min_{PG_r\in\vectorfont{PG}_t^{(A)}}n_{\rbindex{pp,r}}\geq u_{\rbindex{min}}$}{
			\For{$PG_r\in\vectorfont{PG}_t^{(A)}$}{
				Sample a user set $U_{\rbindexem{pub,r,t}}$ from $U_{A,r}$ with the size of $|U_{\rbindexem{pub,r,t}}|=n_{\rbindexem{pp,r}}$;\\ 
				Remove $U_{\rbindexem{pub,r,t}}$ from $U_{A,r}$, i.e., $U_{A,r}\gets U_{A,r}\backslash U_{\rbindexem{pub,r,t}}$; \\ 
				Get frequency estimation $\widehat{\vectorfont{f}}^{(r)}_{\rbindexem{pub,t}}\gets\shrink{FO}(U_{\rbindexem{pub,r,t}},\widetilde{\epsilon}'_{r})$ from users in $U_{\rbindexem{pub,r,t}}$ and calculate its unbiased vector-level variance $\widehat{V}_{\rbindexem{pub,r,t}}$; \\
				Add $\widehat{\vectorfont{f}}^{(r)}_{\rbindexem{pub,t}}$ to $\widehat{\vectorfont{F}}_{\rbindexem{pub,t}}$, and $\widehat{V}_{\rbindexem{pub,r,t}}$ to $\widehat{\vectorfont{\entity{V}}}_{\rbindexem{pub,t}}$;
			}
			Get the population-weighted global estimation $\widehat{\vectorfont{f}}_{\rbindexem{pub,t}}\gets\sum_{r\in[K]}\pi_r\widehat{\vectorfont{f}}^{(r)}_{\rbindexem{pub,t}}$;
		} \Else {
			Set $\widehat{\vectorfont{f}}_{\rbindexem{pub,t}}\gets\widehat{\vectorfont{f}}_{\rbindexem{pub,t-}1}$;
		}
	}
	\For{$PG_r\in\vectorfont{PG}$}{
		\If{$t\geq \tau^*_{r}$}{
			$U_{A,r}\gets U_{A,r}\cup U_{\rbindexem{dis},r,t-\tau^*_{r}+1}\cup U_{\rbindexem{pub,r,t}-\tau^*_{r}+1}$; \\ 
		}
	}
	\Return{$\widehat{\vectorfont{f}}_{\rbindex{pub,t}}$}.
\end{algorithm*}


\begin{algorithm*}[t!]\small
	\caption{\solutionCMPG{}}
	\label{alg:PLBD-AB-ba-pue}
	\DontPrintSemicolon
	\KwIn{Available user set $U_A$; privacy requirement $(\vectorfont{w},\vectorfont{\epsilon})$ of all Users $U$; data domain size $d$; the eigenvalue lowbound $\mu$; historical data publication $\left(\widehat{\vectorfont{f}}_{{1}},\widehat{\vectorfont{f}}_{{2}},\ldots, \widehat{\vectorfont{f}}_{{t-1}}\right)$; fusion mode $\mathsf{mode_{\rbindexem{fsn}}}\in\{\operatorname{Pre},\operatorname{POST}\}$} 
	\KwOut{$\widehat{\vectorfont{f}}_t$}
	$\widehat{\vectorfont{\entity{V}}}_{\rbindexem{dis,t}}\gets\emptyset$;
	$\widehat{\vectorfont{F}}_{\rbindexem{dis,t}}\gets\emptyset$;
	$\vectorfont{\entity{V}}_{\rbindexem{pub,t}}^{\operatorname{Pre}}\gets\emptyset$;
	$\widehat{\vectorfont{\entity{V}}}_{\rbindexem{pub,t}}\gets\emptyset$;
	$\vectorfont{\Lambda}^{\operatorname{Pre}}_{\rbindexem{pub,t}}\gets\emptyset$;
	$\widehat{\vectorfont{F}}_{\rbindexem{pub,t}}\gets\emptyset$;\\ \label{alg:plpd+init_start-AB-ba-pue}
	Get the privacy group set $\vectorfont{PG}=\{PG_r=(\widetilde{w}_r,\widetilde{\epsilon}_r,U_r)\mid r\in[K]\}$ with privacy group triple set $\vectorfont{PGT}=\{PGT_r=(\widetilde{w}_r,\widetilde{\epsilon}_r,n_r)\mid r\in[K]\}$;\\ 
	Construct the candidate source sets $\left\{\entity{S}_r\mid r\in[K]\right\}$ for all groups;\\ \label{alg:plpd+init_end-AB-ba-pue}
	\tcp{sub-mechanism $\entity{M}_{\rbindexem{dis},t}$}
	$\vectorfont{z}\gets\scriptmath{\left( \left\lfloor\frac{n}{2w_{{1}}}\right\rfloor,\left\lfloor\frac{n}{2w_{{2}}}\right\rfloor,\ldots,\left\lfloor\frac{n}{2w_{{n}}}\right\rfloor\right)}$; \\ 
	Get the optimal population size, recycle window size, error and budget $z_{\rbindexem{opt}},\tau_{\rbindexem{opt}},\shrinkem{err}_{\rbindexem{opt}},\vectorfont{\epsilon}'_{\rbindexem{opt}}\gets\shrink{OPS}(\vectorfont{z},\vectorfont{\epsilon},\vectorfont{w}, 1/2,d)$ by Alg.~\ref{alg:ops};\\ \label{alg:plpd+dis_start-AB-ba-pue}
	Get optimal group privacy budget $\widetilde{\vectorfont{\epsilon}}'$ by $\vectorfont{PG}$ and $\vectorfont{\epsilon}'$;\\
	Set $\left(z_{\rbindexem{opt,dis,}1}^*\dots z_{\rbindexem{opt,dis,K}}^*\right)\gets\operatorname{LargestRemainder}\left(z_{\rbindexem{opt}},\vectorfont{\pi}\right)$;\\
	\For{$PG_r\in\vectorfont{PG}$}{
		Sample user set $U_{\rbindexem{dis,r,t}}$, whose data is $D_{\rbindexem{dis,r,t}}$, from $U_{A,r}$ with the size of $z^*_{\rbindexem{opt,dis,r}}$ and remove $U_{\rbindexem{dis,r,t}}$ from $U_{A,r}$, i.e., $U_{A,r}=U_{A,r}\backslash U_{\rbindexem{dis,r,t}}$;\\ 
		Get frequency estimation $\widehat{\vectorfont{f}}^{(r)}_{\rbindexem{dis,t}}\gets\shrink{FO}(U_{\rbindexem{dis,r,t}},\widetilde{\epsilon}'_{r})$ from users in $U_{\rbindexem{dis,r,t}}$ and calculate the unbiased vector-level total variance estimate $\widehat{V}_{\rbindexem{dis,r,t}}$ by Theorem~\ref{thm:precise_var}; \\ 
		Add $\widehat{\vectorfont{f}}^{(r)}_{\rbindexem{dis,t}}$ to $\widehat{\vectorfont{F}}_{\rbindexem{dis,t}}$ and $\widehat{V}_{\rbindexem{dis,r,t}}$ to  $\widehat{\vectorfont{\entity{V}}}_{\rbindexem{dis,t}}$;
	}
	Get the population-weighted global estimation $\widehat{\vectorfont{f}}_{\rbindexem{dis,t}}\gets\sum_{r\in[K]}\pi_r\widehat{\vectorfont{f}}^{(r)}_{\rbindexem{dis,t}}$;\\ 
	Calculate $\shrinkem{dis}\gets \frac{1}{d}\sum_{j=1}^{d}\left(\widehat{\vectorfont{f}}_{\rbindexem{dis,t}}[j]-\widehat{\vectorfont{f}}_l[j]\right)^2-\frac{1}{d}\sum_{r=1}^{K}\pi_r^2 \widehat{V}_{\rbindexem{dis,r,t}}$;\\ \label{alg:plpd+dis_end-AB-ba-pue}
	\tcp{sub-mechanism $\entity{M}_{\rbindexem{pub},t}$}
	\For{$PG_r\in\vectorfont{PG}$}{
		Calculate the optimal remaining population size $n_{\rbindexem{pub,r}}\gets \left\lfloor n_r/2\right\rfloor-\sum_{\tau=t-\tau_{\rbindexem{opt}}+1}^{t-1}|U_{{\rbindexem{pub},\rbindexem{r},\ensuremath{\tau}}}|$;\\ 
		Set the optimal number of potential publication users $n_{\rbindexem{pp,r}}\gets \left\lfloor n_{\rbindexem{pub,r}}/2\right\rfloor$;\\ 
		$V_{\rbindexem{pub,r,t}}^{\operatorname{Pre}}\gets V_{\rbindexem{smpl}}\left(\left\lfloor\frac{n_r}{2}\right\rfloor,n_{\rbindexem{pp,r}}\right)+V_{\rbindexem{LDP}}\left(\epsilon'_r,n_{\rbindexem{pp,r}},d\right)$;\\
		Add $V_{\rbindexem{pub,r,t}}^{\operatorname{Pre}}$ to $\vectorfont{\entity{V}}_{\rbindexem{pub,t}}^{\operatorname{Pre}}$;
	}
	$\vectorfont{\Lambda}^{\operatorname{Pre}}_{\rbindexem{pub,t}}\gets\operatorname{\advancedCorretedWeight{}}\left(\widehat{\vectorfont{F}}_{\rbindexem{pub,t}},\widehat{\vectorfont{F}}_{\rbindexem{dis,t}}, \vectorfont{\entity{V}}_{\rbindexem{pub,t}}^{\operatorname{Pre}},\vectorfont{\entity{V}}_{\rbindexem{dis,t}},\emptyset,\vectorfont{\entity{S}},\mu, \operatorname{Pre}\right)$ by Alg.~\ref{alg:pa-gcw-full};\\ 
	Calculate the potential publication error $\widetilde{\shrinkem{err}}^{+}$ according to Eq.~\eqref{eq:final_corrected_error}; \\ 
	\If{$\shrink{dis}>\widetilde{\shrink{err}}^{+}$ and $\min_{r\in[K]}n_{\rbindex{pp,r}}\geq u_{\rbindex{min}}$}{
		\For{$PG_r\in\vectorfont{PG}$}{
			Sample a user set $U_{\rbindexem{pub,r,t}}$ from $U_{\rbindexem{A,r}}$ with the size of $|U_{\rbindexem{pub,r,t}}|=n_{\rbindexem{pp,r}}$;\\ 
			Remove $U_{\rbindexem{pub,r,t}}$ from $U_{\rbindexem{A,r}}$, i.e., $U_{\rbindexem{A,r}}\gets U_{\rbindexem{A,r}}\backslash U_{\rbindexem{pub,r,t}}$; \\ 
			Get frequency estimation $\widehat{\vectorfont{f}}^{(r)}_{\rbindexem{pub,t}}\gets\shrink{FO}(U_{\rbindexem{pub,r,t}},\widetilde{\epsilon}'_{r})$ from users in $U_{\rbindexem{pub,r,t}}$ and calculate its unbiased vector-level variance estimate $\widehat{V}_{\rbindexem{pub,r,t}}$; \\
			Add $\widehat{\vectorfont{f}}^{(r)}_{\rbindexem{pub,t}}$ to $\widehat{\vectorfont{F}}_{\rbindexem{pub,t}}$ and $\widehat{V}_{\rbindexem{pub,r,t}}$ to $\widehat{\vectorfont{\entity{V}}}_{\rbindexem{pub,t}}$;
		}
		\If{$\mathsf{mode}_{\rbindex{fsn}}=\operatorname{POST}$}{
			$\vectorfont{\Lambda}^{(\rbindexem{post})}_{\rbindexem{pub,t}}\gets\operatorname{\advancedCorretedWeight{}}\left(\widehat{\vectorfont{F}}_{\rbindexem{pub,t}}, \widehat{\vectorfont{F}}_{\rbindexem{dis,t}},\widehat{\vectorfont{\entity{V}}}_{\rbindexem{pub,t}}, \widehat{\vectorfont{\entity{V}}}_{\rbindexem{dis,t}}, \emptyset,\vectorfont{\entity{S}},\mu,\operatorname{POST}\right)$ by Alg.~\ref{alg:pa-gcw-full};\\ 
			Set $\vectorfont{\Lambda}^{(\rbindexem{fnl})}_{\rbindexem{pub,t}}\gets\vectorfont{\Lambda}^{(\rbindexem{post})}_{\rbindexem{pub,t}}$;
		} \Else{
			Set $\vectorfont{\Lambda}^{(\rbindexem{fnl})}_{\rbindexem{pub,t}}\gets\vectorfont{\Lambda}^{\operatorname{Pre}}_{\rbindexem{pub,t}}$;
		}
		Fuse $\widetilde{\vectorfont{F}}_{\rbindexem{pub},t}$ by Eq.~\eqref{eq:linear-combine} with $\vectorfont{\Lambda}^{(\rbindexem{fnl})}_t$;\\
		Get the population-weighted global estimation $\widehat{\vectorfont{f}}_{\rbindexem{pub,t}}\gets\sum_{r\in[K]}\pi_r\widetilde{\vectorfont{f}}^{(r)}_{\rbindexem{pub,t}}$;\\		
	} \Else {
		Set $\widehat{\vectorfont{f}}_{\rbindexem{pub,t}}\gets\widehat{\vectorfont{f}}_{\rbindexem{pub,t-}1}$; 
	}
	\If{$t\geq \tau_{\rbindex{opt}}$}{
		\For{$PG_r\in\vectorfont{PG}$}{
			$U_{A,r}\gets U_{A,r}\cup U_{\rbindexem{dis,r,t}-\tau_{\rbindexem{opt}}+1}\cup U_{\rbindexem{pub,r,t}-\tau_{\rbindexem{opt}}+1}$;
		}
	}
	\Return{$\widehat{\vectorfont{f}}_{\rbindex{pub,t}}$;}
\end{algorithm*}

\begin{algorithm*}[h]\small
	\caption{\solutionCMPH{}}
	\label{alg:PLBA-BA-ba-pue}
	\DontPrintSemicolon
	\KwIn{Available user set $U_A$; privacy requirement $(\vectorfont{w},\vectorfont{\epsilon})$ of all Users $U$; data domain size $d$; the eigenvalue lowbound $\mu$; historical data publication $\left(\widehat{\vectorfont{f}}_{{1}},\widehat{\vectorfont{f}}_{{2}},\ldots, \widehat{\vectorfont{f}}_{{t-1}}\right)$; fusion mode $\mathsf{mode_{\rbindexem{fsn}}}\in\{\operatorname{Pre},\operatorname{POST}\}$} 
	\KwOut{$\widehat{\vectorfont{f}}_t$}
	The same as Lines~\ref{alg:plpd+init_start-AB-ba-pue}--\ref{alg:plpd+init_end-AB-ba-pue} in Alg.~\ref{alg:PLBD-AB-ba-pue};\\ 
	\tcp{sub-mechanism $\entity{M}_{\rbindexem{dis},t}$}
	The same as Lines~\ref{alg:plpd+dis_start-AB-ba-pue}-\ref{alg:plpd+dis_end-AB-ba-pue} in Alg.~\ref{alg:PLBD-AB-ba-pue};\\ 
	\tcp{sub-mechanism $\entity{M}_{\rbindexem{pub},t}$}
	Initialize active group set $\vectorfont{PG}_t^{(A)}\gets\emptyset$, reused group set $\vectorfont{PG}_t^{(R)}\gets\emptyset$;\\
	\For{$PG_r\in\vectorfont{PG}$}{
		Calculate the time slots to be nullified $t_{\rbindexem{N,r}}=\frac{\big|U_{\rbindexem{pub,r,l}_r}\big|}{z^*_{\rbindexem{opt,dis,r}}}-1$;\\ 
		\If{$t-l_r\leq t_{\rbindex{N,r}}$}{
			Add $\widehat{\vectorfont{f}}^{(r)}_{\rbindexem{pub,l}_r}$ to $\widehat{\vectorfont{F}}_{\rbindexem{pub,t}}$,
			$\widehat{V}_{\rbindexem{pub,r,l}_r}$ to $\vectorfont{\entity{V}}_{\rbindexem{pub,t}}^{\operatorname{Pre}}$ and $\widehat{\vectorfont{\entity{V}}}_{\rbindexem{pub,t}}$, and
			$\vectorfont{PG}_r$ to $\vectorfont{PG}_t^{(R)}$;\\ 
		} \Else{
			Calculate group-aware time slots that can be absorbed $t_{A,r}=t-l_r-t_{\rbindexem{N,r}}$;\\ 
			Set the number of potential publication users $n_{\rbindexem{pp,r}}\gets z_{\rbindexem{opt,dis,r}}^*\cdot\min(t_{\rbindexem{A,r}},\tau_{\rbindexem{opt}})$;\\ 
			$V_{\rbindexem{pub,r,t}}^{\operatorname{Pre}}\gets\left(V_{\rbindexem{smpl}}\left(\left\lfloor\frac{n_r}{2}\right\rfloor,n_{\rbindexem{pp,r}}\right)+V_{\rbindexem{LDP}}\left(\epsilon'_r,n_{\rbindexem{pp,r}},d\right)\right)$;\\
			Add $V_{\rbindexem{pub,r,t}}^{\operatorname{Pre}}$ to $\vectorfont{\entity{V}}_{\rbindexem{pub,t}}^{\operatorname{Pre}}$;\\
			Add $PG_r$ to $\vectorfont{PG}_t^{(A)}$;\\ 
		}
	}
	\If{$\vectorfont{PG}_t^{(A)}=\emptyset$}{
		Set $\widehat{\vectorfont{f}}_{\rbindexem{pub,t}}\gets\widehat{\vectorfont{f}}_{\rbindexem{pub,t}-1}$;
	} \Else{
		$\vectorfont{\Lambda}^{\operatorname{Pre}}_{\rbindexem{pub,t}}\gets\operatorname{\advancedCorretedWeight{}}\left(\widehat{\vectorfont{F}}_{\rbindexem{pub,t}},\widehat{\vectorfont{F}}_{\rbindexem{dis,t}}, \vectorfont{\entity{V}}_{\rbindexem{pub,t}}^{\operatorname{Pre}},\vectorfont{\entity{V}}_{\rbindexem{dis,t}},\vectorfont{PG}_t^{(R)},\vectorfont{\entity{S}},\mu, \operatorname{Pre}\right)$ by Alg.~\ref{alg:pa-gcw-full};\\ 
		Calculate the staleness-aware potential reporting error 
		$\widetilde{\shrinkem{err}}^{+}$ by Eq.~\eqref{eq:final_corrected_time_error}; \\ 
		\If{$\shrink{dis}>\widetilde{\shrink{err}}^{+}$ and $\min_{PG_r\in\vectorfont{PG}_t^{(A)}}n_{\rbindex{pp,r}}\geq u_{\rbindex{min}}$}{
			\For{$PG_r\in\vectorfont{PG}_t^{(A)}$}{
				Sample a user set $U_{\rbindexem{pub,r,t}}$ from $U_{A,r}$ with the size of $|U_{\rbindexem{pub,r,t}}|=n_{\rbindexem{pp,r}}$;\\ 
				Remove $U_{\rbindexem{pub,r,t}}$ from $U_{A,r}$, i.e., $U_{A,r}\gets U_{A,r}\backslash U_{\rbindexem{pub,r,t}}$; \\ 
				Get frequency estimation $\widehat{\vectorfont{f}}^{(r)}_{\rbindexem{pub,t}}\gets\shrink{FO}(U_{\rbindexem{pub,r,t}},\widetilde{\epsilon}'_{r})$ from users in $U_{\rbindexem{pub,r,t}}$ and calculate its unbiased vector-level variance $\widehat{V}_{\rbindexem{pub,r,t}}$; \\
				Add $\widehat{\vectorfont{f}}^{(r)}_{\rbindexem{pub,t}}$ to $\widehat{\vectorfont{F}}_{\rbindexem{pub,t}}$, and $\widehat{V}_{\rbindexem{pub,r,t}}$ to $\widehat{\vectorfont{\entity{V}}}_{\rbindexem{pub,t}}$;
			}
			\If{$\mathsf{mode}_{\rbindex{fsn}}=\operatorname{POST}$}{
				$\vectorfont{\Lambda}^{(\rbindexem{post})}_{\rbindexem{pub,t}}\gets\operatorname{\advancedCorretedWeight{}}\left(\widehat{\vectorfont{F}}_{\rbindexem{pub,t}}, \widehat{\vectorfont{F}}_{\rbindexem{dis,t}},\widehat{\vectorfont{\entity{V}}}_{\rbindexem{pub,t}}, \widehat{\vectorfont{\entity{V}}}_{\rbindexem{dis,t}}, \vectorfont{PG}_{\rbindexem{t}}^{(R)},\vectorfont{\entity{S}},\mu,\operatorname{POST}\right)$ by Alg.~\ref{alg:pa-gcw-full};\\ 
				Set $\vectorfont{\Lambda}^{(\rbindexem{fnl})}_{\rbindexem{pub,t}}\gets\vectorfont{\Lambda}^{(\rbindexem{post})}_{\rbindexem{pub,t}}$;
			} \Else{
				Set $\vectorfont{\Lambda}^{(\rbindexem{fnl})}_{\rbindexem{pub,t}}\gets\vectorfont{\Lambda}^{\operatorname{Pre}}_{\rbindexem{pub,t}}$;
			}
			Fuse $\widetilde{\vectorfont{F}}_{\rbindexem{pub},t}$ by Eq.~\eqref{eq:linear-combine} with $\vectorfont{\Lambda}^{(\rbindexem{fnl})}_t$;\\
			Get the population-weighted global estimation $\widehat{\vectorfont{f}}_{\rbindexem{pub,t}}\gets\sum_{r\in[K]}\pi_r\widetilde{\vectorfont{f}}^{(r)}_{\rbindexem{pub,t}}$;\\	
		} \Else {
			Set $\widehat{\vectorfont{f}}_{\rbindexem{pub,t}}\gets\widehat{\vectorfont{f}}_{\rbindexem{pub,t-}1}$; 
		}
	}
	\If{$t\geq \tau_{\rbindex{opt}}$}{
		\For{$PG_r\in\vectorfont{PG}$}{
			$U_{A,r}\gets U_{A,r}\cup U_{\rbindexem{dis,r,t}-\tau_{\rbindexem{opt}}+1}\cup U_{\rbindexem{pub,r,t}-\tau_{\rbindexem{opt}}+1}$;
		}
	}
	\Return{$\widehat{\vectorfont{f}}_{\rbindex{pub,t}}$}.
\end{algorithm*}


\end{document}